\RequirePackage{lineno}
\documentclass[aps,prl,twocolumn,superscriptaddress]{revtex4}
\usepackage{graphicx}
\begin{document}

\title{Measurement of the cross section for prompt isolated diphoton
       production using the full CDF Run II data sample}

\affiliation{Institute of Physics, Academia Sinica, Taipei, Taiwan 11529, Republic of China}
\affiliation{Argonne National Laboratory, Argonne, Illinois 60439, USA}
\affiliation{University of Athens, 157 71 Athens, Greece}
\affiliation{Institut de Fisica d'Altes Energies, ICREA, Universitat Autonoma de Barcelona, E-08193, Bellaterra (Barcelona), Spain}
\affiliation{Baylor University, Waco, Texas 76798, USA}
\affiliation{Istituto Nazionale di Fisica Nucleare Bologna, $^{ee}$University of Bologna, I-40127 Bologna, Italy}
\affiliation{University of California, Davis, Davis, California 95616, USA}
\affiliation{University of California, Los Angeles, Los Angeles, California 90024, USA}
\affiliation{Instituto de Fisica de Cantabria, CSIC-University of Cantabria, 39005 Santander, Spain}
\affiliation{Carnegie Mellon University, Pittsburgh, Pennsylvania 15213, USA}
\affiliation{Enrico Fermi Institute, University of Chicago, Chicago, Illinois 60637, USA}
\affiliation{Comenius University, 842 48 Bratislava, Slovakia; Institute of Experimental Physics, 040 01 Kosice, Slovakia}
\affiliation{Joint Institute for Nuclear Research, RU-141980 Dubna, Russia}
\affiliation{Duke University, Durham, North Carolina 27708, USA}
\affiliation{Fermi National Accelerator Laboratory, Batavia, Illinois 60510, USA}
\affiliation{University of Florida, Gainesville, Florida 32611, USA}
\affiliation{Laboratori Nazionali di Frascati, Istituto Nazionale di Fisica Nucleare, I-00044 Frascati, Italy}
\affiliation{University of Geneva, CH-1211 Geneva 4, Switzerland}
\affiliation{Glasgow University, Glasgow G12 8QQ, United Kingdom}
\affiliation{Harvard University, Cambridge, Massachusetts 02138, USA}
\affiliation{Division of High Energy Physics, Department of Physics, University of Helsinki and Helsinki Institute of Physics, FIN-00014, Helsinki, Finland}
\affiliation{University of Illinois, Urbana, Illinois 61801, USA}
\affiliation{The Johns Hopkins University, Baltimore, Maryland 21218, USA}
\affiliation{Institut f\"{u}r Experimentelle Kernphysik, Karlsruhe Institute of Technology, D-76131 Karlsruhe, Germany}
\affiliation{Center for High Energy Physics: Kyungpook National University, Daegu 702-701, Korea; Seoul National University, Seoul 151-742, Korea; Sungkyunkwan University, Suwon 440-746, Korea; Korea Institute of Science and Technology Information, Daejeon 305-806, Korea; Chonnam National University, Gwangju 500-757, Korea; Chonbuk National University, Jeonju 561-756, Korea; Ewha Womans University, Seoul, 120-750, Korea}
\affiliation{Ernest Orlando Lawrence Berkeley National Laboratory, Berkeley, California 94720, USA}
\affiliation{University of Liverpool, Liverpool L69 7ZE, United Kingdom}
\affiliation{University College London, London WC1E 6BT, United Kingdom}
\affiliation{Centro de Investigaciones Energeticas Medioambientales y Tecnologicas, E-28040 Madrid, Spain}
\affiliation{Massachusetts Institute of Technology, Cambridge, Massachusetts 02139, USA}
\affiliation{Institute of Particle Physics: McGill University, Montr\'{e}al, Qu\'{e}bec H3A~2T8, Canada; Simon Fraser University, Burnaby, British Columbia V5A~1S6, Canada; University of Toronto, Toronto, Ontario M5S~1A7, Canada; and TRIUMF, Vancouver, British Columbia V6T~2A3, Canada}
\affiliation{University of Michigan, Ann Arbor, Michigan 48109, USA}
\affiliation{Michigan State University, East Lansing, Michigan 48824, USA}
\affiliation{Institution for Theoretical and Experimental Physics, ITEP, Moscow 117259, Russia}
\affiliation{University of New Mexico, Albuquerque, New Mexico 87131, USA}
\affiliation{The Ohio State University, Columbus, Ohio 43210, USA}
\affiliation{Okayama University, Okayama 700-8530, Japan}
\affiliation{Osaka City University, Osaka 588, Japan}
\affiliation{University of Oxford, Oxford OX1 3RH, United Kingdom}
\affiliation{Istituto Nazionale di Fisica Nucleare, Sezione di Padova-Trento, $^{ff}$University of Padova, I-35131 Padova, Italy}
\affiliation{University of Pennsylvania, Philadelphia, Pennsylvania 19104, USA}
\affiliation{Istituto Nazionale di Fisica Nucleare Pisa, $^{gg}$University of Pisa, $^{hh}$University of Siena and $^{ii}$Scuola Normale Superiore, I-56127 Pisa, Italy, $^{mm}$INFN Pavia and University of Pavia, I-27100 Pavia, Italy}
\affiliation{University of Pittsburgh, Pittsburgh, Pennsylvania 15260, USA}
\affiliation{Purdue University, West Lafayette, Indiana 47907, USA}
\affiliation{University of Rochester, Rochester, New York 14627, USA}
\affiliation{The Rockefeller University, New York, New York 10065, USA}
\affiliation{Istituto Nazionale di Fisica Nucleare, Sezione di Roma 1, $^{jj}$Sapienza Universit\`{a} di Roma, I-00185 Roma, Italy}
\affiliation{Texas A\&M University, College Station, Texas 77843, USA}
\affiliation{Istituto Nazionale di Fisica Nucleare Trieste/Udine; $^{nn}$University of Trieste, I-34127 Trieste, Italy; $^{kk}$University of Udine, I-33100 Udine, Italy}
\affiliation{University of Tsukuba, Tsukuba, Ibaraki 305, Japan}
\affiliation{Tufts University, Medford, Massachusetts 02155, USA}
\affiliation{University of Virginia, Charlottesville, Virginia 22906, USA}
\affiliation{Waseda University, Tokyo 169, Japan}
\affiliation{Wayne State University, Detroit, Michigan 48201, USA}
\affiliation{University of Wisconsin, Madison, Wisconsin 53706, USA}
\affiliation{Yale University, New Haven, Connecticut 06520, USA}

\author{T.~Aaltonen}
\affiliation{Division of High Energy Physics, Department of Physics, University of Helsinki and Helsinki Institute of Physics, FIN-00014, Helsinki, Finland}
\author{S.~Amerio}
\affiliation{Istituto Nazionale di Fisica Nucleare, Sezione di Padova-Trento, $^{ff}$University of Padova, I-35131 Padova, Italy}
\author{D.~Amidei}
\affiliation{University of Michigan, Ann Arbor, Michigan 48109, USA}
\author{A.~Anastassov$^x$}
\affiliation{Fermi National Accelerator Laboratory, Batavia, Illinois 60510, USA}
\author{A.~Annovi}
\affiliation{Laboratori Nazionali di Frascati, Istituto Nazionale di Fisica Nucleare, I-00044 Frascati, Italy}
\author{J.~Antos}
\affiliation{Comenius University, 842 48 Bratislava, Slovakia; Institute of Experimental Physics, 040 01 Kosice, Slovakia}
\author{G.~Apollinari}
\affiliation{Fermi National Accelerator Laboratory, Batavia, Illinois 60510, USA}
\author{J.A.~Appel}
\affiliation{Fermi National Accelerator Laboratory, Batavia, Illinois 60510, USA}
\author{T.~Arisawa}
\affiliation{Waseda University, Tokyo 169, Japan}
\author{A.~Artikov}
\affiliation{Joint Institute for Nuclear Research, RU-141980 Dubna, Russia}
% ...

\collaboration{CDF Collaboration\footnote{With visitors from
$^a$Istituto Nazionale di Fisica Nucleare, Sezione di Cagliari, 09042 Monserrato (Cagliari), Italy,
$^b$University of California Irvine, Irvine, CA 92697, USA,
$^c$University of California Santa Barbara, Santa Barbara, CA 93106, USA,
$^d$University of California Santa Cruz, Santa Cruz, CA 95064, USA,
$^e$Institute of Physics, Academy of Sciences of the Czech Republic, 182~21, Czech Republic,
$^f$CERN, CH-1211 Geneva, Switzerland,
$^g$Cornell University, Ithaca, NY 14853, USA,
$^h$University of Cyprus, Nicosia CY-1678, Cyprus,
$^i$Office of Science, U.S. Department of Energy, Washington, DC 20585, USA,
$^j$University College Dublin, Dublin 4, Ireland,
$^k$ETH, 8092 Z\"{u}rich, Switzerland,
$^l$University of Fukui, Fukui City, Fukui Prefecture, Japan 910-0017,
$^m$Universidad Iberoamericana, Lomas de Santa Fe, M\'{e}xico, C.P. 01219, Distrito Federal,
$^n$University of Iowa, Iowa City, IA 52242, USA,
$^o$Kinki University, Higashi-Osaka City, Japan 577-8502,
$^p$Kansas State University, Manhattan, KS 66506, USA,
$^q$Brookhaven National Laboratory, Upton, NY 11973, USA,
$^r$University of Manchester, Manchester M13 9PL, United Kingdom,
$^s$Queen Mary, University of London, London, E1 4NS, United Kingdom,
$^t$University of Melbourne, Victoria 3010, Australia,
$^u$Muons, Inc., Batavia, IL 60510, USA,
$^v$Nagasaki Institute of Applied Science, Nagasaki 851-0193, Japan,
$^w$National Research Nuclear University, Moscow 115409, Russia,
$^x$Northwestern University, Evanston, IL 60208, USA,
$^y$University of Notre Dame, Notre Dame, IN 46556, USA,
$^z$Universidad de Oviedo, E-33007 Oviedo, Spain,
$^{aa}$CNRS-IN2P3, Paris, F-75205 France,
$^{bb}$Texas Tech University, Lubbock, TX 79609, USA,
$^{cc}$Universidad Tecnica Federico Santa Maria, 110v Valparaiso, Chile,
$^{dd}$Yarmouk University, Irbid 211-63, Jordan,
$^{ll}$Universite catholique de Louvain, 1348 Louvain-La-Neuve, Belgium,
$^{oo}$University of Z\"{u}rich, 8006 Z\"{u}rich, Switzerland,
$^{pp}$Massachusetts General Hospital and Harvard Medical School, Boston, MA 02114 USA,
$^{qq}$Hampton University, Hampton, VA 23668, USA,
$^{rr}$Los Alamos National Laboratory, Los Alamos, NM 87544, USA
}}
\noaffiliation

\date{\today}

\begin{abstract}
This Letter reports a measurement of the cross section for producing pairs
of central prompt isolated photons in proton-antiproton collisions at a total
energy $\sqrt{s}=1.96$ TeV using data corresponding to 9.5 fb$^{-1}$ integrated
luminosity collected with the CDF II detector at the Fermilab Tevatron. The
measured differential cross section is compared to three calculations derived
from the theory of strong interactions. These include a prediction based on a
leading order matrix element calculation merged with parton shower, a
next-to-leading order, and a next-to-next-to-leading order calculation. The
first and last calculations reproduce most aspects of the data, thus showing
the importance of higher-order contributions for understanding the theory of
strong interaction and improving measurements of the Higgs boson and searches
for new phenomena in diphoton final states.
\end{abstract}

\maketitle

The production of prompt photon pairs in hadron collisions is a significant,
irreducible background in searches for a low-mass Higgs boson decaying
into a photon pair \cite{higgs}, as well as in searches for new phenomena,
such as extra spatial dimensions \cite{dim,grav} and two-body \cite{heavy}
or cascade \cite{casc} decays of new heavy particles. Precise measurements
of the production cross sections for diphotons as functions of various
kinematic variables and their theoretical understanding are important for
these searches. The better the prompt diphoton background is understood,
the smaller uncertainties are introduced in these searches. After the
recent discovery of the Higgs boson-like particle at the LHC \cite{higgsLHC},
a better understanding of the background is important for improvements in
the precision of the measurements of the production cross section and the
decay branching ratio of this particle into a photon pair. A precise
measurement of the branching ratio is of special importance, as this decay
proceeds through a fermion loop and thus it indirectly constrains the
couplings of the Higgs boson-like particle to fermions, which are more
difficult to extract from direct decays into fermion pairs. Diphoton
production is also used to test quantum chromodynamics (QCD), the theory
of strong interaction, both in the perturbative scheme (pQCD), which is a
good approximation at high energies, and in non-perturbative schemes, such
as soft-gluon resummation methods, which provide important corrections in
certain lower-energy kinematic regions \cite{resbos}. Diphotons are expected
to be dominantly produced by quark-antiquark annihilation
$q\overline{q}\rightarrow\gamma\gamma$ and, in kinematic regions where
gluons dominate the parton distribution functions (PDF), by gluon-gluon
fusion $gg\rightarrow\gamma\gamma$ through a quark loop amplitude. Prompt
photons may also result from quark fragmentation in hard scattering,
although a strict photon isolation requirement significantly reduces the
fragmentation contributions.

Diphoton measurements have been made previously at fixed-target \cite{fixtgt}
and collider experiments \cite{ua1,ua2,cdfdip}. Recent measurements have been
made both at the Tevatron \cite{d0,cdfdiptwo} and at the LHC \cite{atlas},
which offer a consistent picture on the accuracy and limitations of the
theoretical calculations in reproducing the data. The ATLAS measurement
\cite{atlas} found diphoton production features in proton-proton collisions
at $\sqrt{s}=7$ TeV analogous to those observed in proton-antiproton
collisions at $\sqrt{s}=1.96$ TeV \cite{d0,cdfdiptwo}. The most recent
CDF measurement \cite{cdfdiptwo}, using approximately half the full CDF
data sample, compared the data with pQCD calculations at leading order (LO)
and next-to-leading order (NLO) in the expansion parameter $\alpha_{s}$, the
strong interaction coupling. Large discrepancies were found between the data
and a LO matrix-element calculation supplemented with a parton shower (PS)
model. The inclusion of photons radiated from initial- and final-state quarks
allowed by the shower model substantially improved the agreement of the PS
calculation with the data. The calculation that includes radiated photons was
recently used to predict the non-resonant background in the search for a
low-mass Higgs boson decaying into a photon pair using the full CDF data set
\cite{cdfhgg}.

This work presents the final diphoton measurements from CDF using the full
data set collected in 2001--2011 corresponding to a total integrated luminosity
of 9.5 fb$^{-1}$. The results are compared with all the available
state-of-the-art calculations under a variety of kinematic conditions
\cite{compar}, including an improved set of calculations not discussed in the
previous work \cite{cdfdiptwo}.

The reported measurement is using data collected with the Collider Detector
at Fermilab (CDF) \cite{cdf}, at the Tevatron $p\overline{p}$ collider. The
CDF detector includes a central spectrometer inside a 1.4 T axial magnetic
field, surrounded by electromagnetic and hadronic calorimeters and muon
detection chambers. The inner spectrometer measures charged particle
trajectories (tracks) with a momentum component transverse to the beam
($p_{T}$) with a precision of $\sigma_{p_{T}}/p_{T}^{2}=0.07\%({\rm GeV}/c)^{-1}$.
The pointing-tower-geometry central calorimeters cover the region
$\vert\eta\vert<1.1$, with an electromagnetic (hadronic) energy resolution
of $\sigma(E_{T})/E_{T}=13.5\%/\sqrt{E_{T}({\rm GeV})}\oplus 1.5\%$
($\sigma(E_{T})/E_{T}=50\%/\sqrt{E_{T}({\rm GeV})}\oplus 3\%$) and a tower
segmentation of $\Delta\eta\times\Delta\phi\simeq 0.1\times 15^{\circ}$, where
$E_{T}=E\sin\theta$ is the transverse energy, $\eta=-\ln[\tan(\theta/2)]$
is the pseudo-rapidity, $\theta$ is the polar angle and $\phi$ the azimuth
of the tower's axis in the coordinate system of the laboratory, with polar
axis along the proton beam direction and origin at the center of the detector.
Photons are reconstructed in clusters of up to three towers \cite{phosel}
in the central calorimeter only. The pseudorapidity of each photon in
the event is restricted to the region $\vert\eta\vert<1$, which is the most
sensitive region for diphoton measurements at the Tevatron and the LHC. A
finely-segmented detector located at a depth corresponding to the maximum
development of a typical electromagnetic shower measures the energy deposit
profile, which is required to be consistent with originating from a single
photon. The photon transverse energy is required to exceed 17 GeV for the
first photon in the event and 15 GeV for the second photon. The transverse
energy measured by the calorimeter in an isolation cone with radius in
$\eta-\phi$ space of 0.4 around each photon \cite{caliso} is required not
to exceed 2 GeV.

This measurement employs the same techniques as the previous work
\cite{cdfdiptwo}. Inclusive diphoton events are selected online by requiring
two isolated electromagnetic clusters with $E_{T}>12$ GeV each or two
electromagnetic clusters with $E_{T}>18$ GeV and no isolation requirement.
In the offline analysis additional requirements are imposed to identify a
sample rich in prompt photons. The background from events where one or both
reconstructed photons are misidentified jets is subtracted with a $4\times 4$
matrix technique using the track isolation as the discriminant between signal
and background \cite{trkiso}, defined in the same cone with the calorimetric
isolation. The matrix is constructed for each event from the $E_{T}$-dependent
efficiencies of signal and background photons passing the track isolation
criterion. This technique takes into account the full correlations between
the two photons in the event. An optimal track-isolation threshold of
1$~$GeV/$c$ is determined by maximizing the discrimination between signal
and background Monte Carlo (MC) simulation samples. The efficiencies used
in this method are determined from $\gamma$$+$jet and dijet samples generated
with {\sc pythia} \cite{pythia}, subjected to the full detector and online
event selection simulation \cite{cdfsim}, and reconstructed as the experimental
data. The probabilities of an event to be pure signal, pure background, and
a mixed photon pair are obtained for each event by multiplying the inverse of
the $4\times 4$ matrix constructed from the efficiencies with the
four-dimensional column vector of the observation values (0 or 1) for all
four combinations of the first and second photon having track isolation
larger or smaller than 1$~$GeV/$c$. The signal fraction is determined by
summing the probability of pure signal over all events and averages to
$\sim$40\% with an absolute systematic uncertainty in the range of 15--20\%.

The differential cross section for diphoton production is obtained from
the histogram of the estimated signal yield as a function of each relevant
kinematic variable. The average cross section in a bin is determined by
dividing the yield by the product of the trigger efficiency, the selection
efficiency and acceptance, the integrated luminosity, and the bin size.
The diphoton trigger efficiency is derived from data \cite{higgs}. It is
consistent with 100\% over all of the kinematic range with a flat uncertainty
of 3\%. The diphoton selection efficiency accounts for the effects from the
{\it underlying event} from collision remnants \cite{cdfdiptwo} and from
additional (pile-up) collisions overlapping with the collision that produced
the photons. The systematic uncertainty in the selection efficiency related
to the pile-up effect grows linearly from 1.8\% for $E_{T}\le 40$ GeV to 3\%
for $E_{T}=80$ GeV and remains constant above this point. A flat 3\%
uncertainty per photon accounts for possible inaccuracies in the {\sc pythia}
model for the underlying event. This is summed linearly to 6\% for two
photons, since the underlying event is not related with prompt photon
production and affects only the isolation symmetrically for the two photons,
on the average. A 6\% constant uncertainty comes from the integrated luminosity
\cite{lumi}. A 2\% difference in the photon identification efficiency between
data and MC is estimated from the $Z^{0}\rightarrow e^{+}e^{-}$ sample
\cite{higgs} and added as a systematic uncertainty to the measurement. The
electromagnetic energy scale is determined from the mass of the
$Z^{0}\rightarrow e^{+}e^{-}$ signal. The associated systematic uncertainty
is estimated to grow linearly from 0 at $E_{T}\le 40$ GeV up to 1.5\% at
$E_{T}=80$ GeV and remain constant above this point. All systematic
uncertainties are added in quadrature.

In the previous measurement \cite{cdfdiptwo}, the experimental results were
compared with three theoretical calculations: (i) the fixed NLO predictions
of the {\sc diphox} program \cite{diphox}, including non-perturbative parton
fragmentation into photons at NLO \cite{frag}, (ii) the predictions of the
{\sc resbos} program \cite{resbos} where the cross section is accurate to NLO,
but also has an analytical initial-state soft-gluon resummation, and (iii)
the predictions of the {\sc pythia} PS program \cite{pythia} including photons
radiated from initial- and final-state quarks \cite{cdfdiptwo}. Within their
known limitations, all three calculations reproduced the main features of the
data, but none of them described all aspects of the data. In this Letter, the
measurement is compared with three different calculations: (a) the fixed NLO
predictions of the {\sc mcfm} program \cite{mcfm}, including non-perturbative
parton fragmentation into photons at LO \cite{mcfmfrag}, (b) the fixed
next-to-next-to-leading order (NNLO) predictions of a recent calculation
\cite{nnlo}, and (c) the predictions of the {\sc sherpa} program \cite{sherpa},
based on a matrix element calculation merged with parton shower (ME+PS).
This calculation features a realistic representation of the physics events
including initial- and final-state radiation. The prediction of {\sc mcfm}
is an alternative calculation to {\sc diphox}, but it has not been tested
against any previous measurement. The NNLO and {\sc sherpa} predictions are
recent calculations that are expected to reproduce the data features better
than the previous calculations.

While the NLO and NNLO matrix elements for diphoton production include all
real and virtual processes at fixed order in $\alpha_{s}$, the {\sc sherpa}
matrix element includes only real processes at NNLO. However, by merging
the matrix element contribution (the hard scattering process) with those
from the parton shower (cascade radiation subprocesses from the initial-
and final-state quarks and gluons), this calculation accounts for real
processes effectively at all orders in $\alpha_{s}$. It also accounts for
some virtual effects via corrections applied in the parton shower
subprocesses. The {\sc sherpa} calculation is an extension of the {\sc pythia}
calculation including photons radiated from initial- and final-state quarks
which was introduced in the previous measurement \cite{cdfdiptwo}. In the
default {\sc sherpa} calculation the scale is adjusted to the event kinematics
automatically by the program itself \cite{sherpa}. An uncertainty of this
calculation is estimated by the difference from an alternative calculation
which uses a fixed scale. All calculations are subject to the experimental
kinematic and isolation requirements \cite{compar}. Theoretical uncertainties
are best estimated for the fixed-order NLO and NNLO calculations, where the
scale uncertainties are well-defined. The estimation is done by increasing
and decreasing the scale of each calculation by a factor of two relative to
the default scale and, for the NLO PDF uncertainties, by using the 20 CTEQ6M
eigenvectors \cite{cteq6m}. The PDF uncertainties are relatively small for
the high proton momentum fractions of the quarks and gluons involved in
prompt diphoton production calculations.

\begin{figure*}
\includegraphics[width=7.5cm]{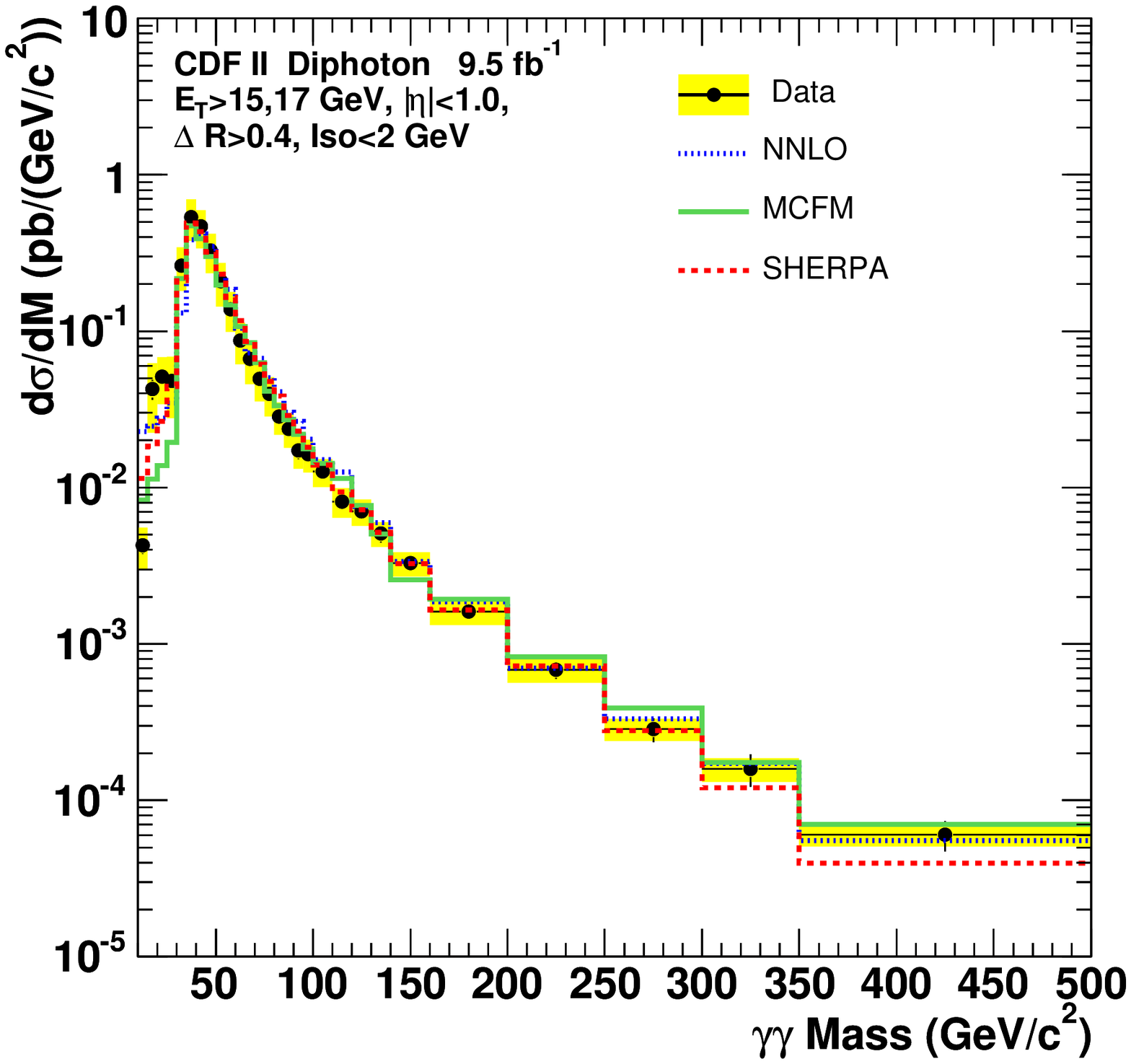}\hspace*{0cm}
\includegraphics[width=7cm]{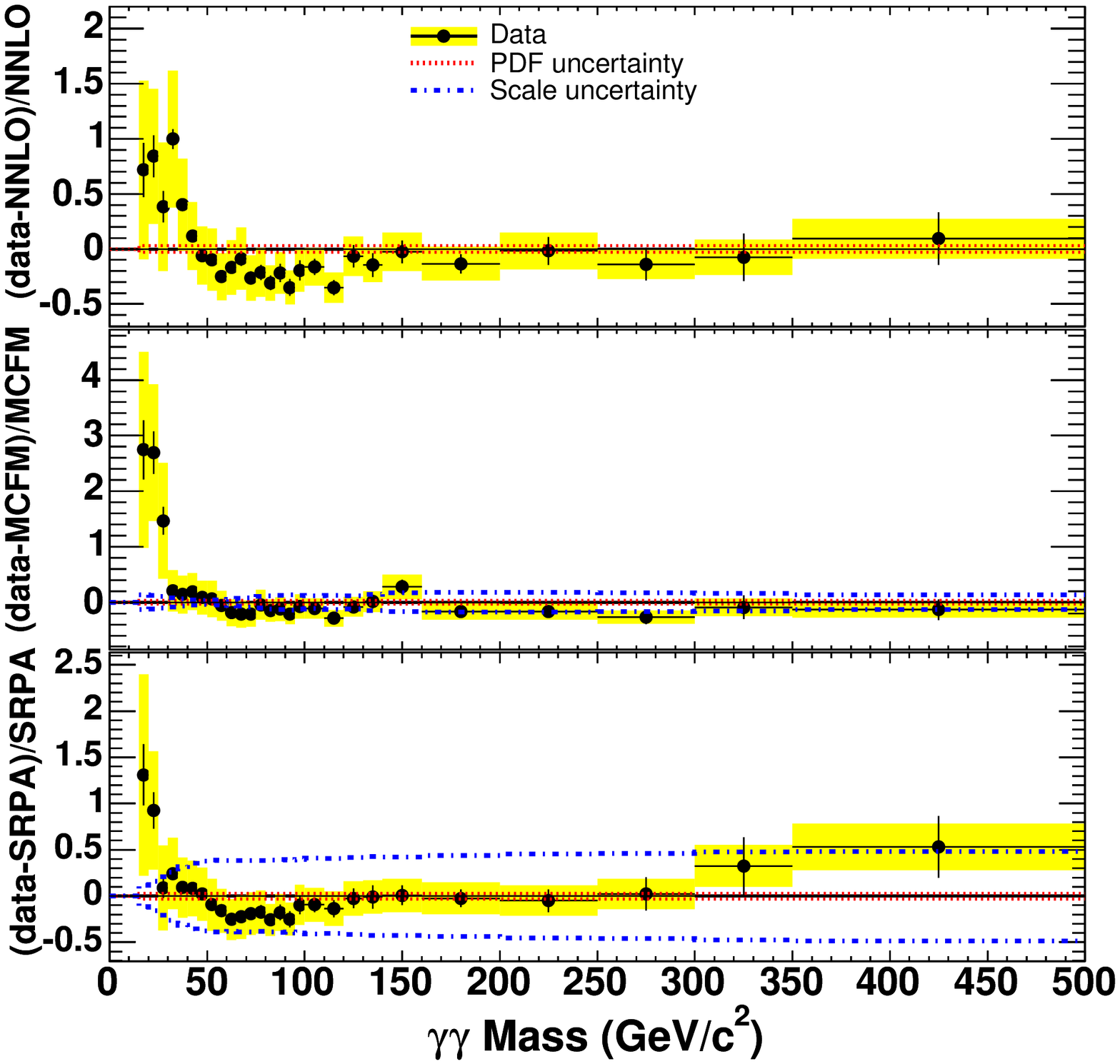}
\vspace*{0cm}
\includegraphics[width=7.5cm]{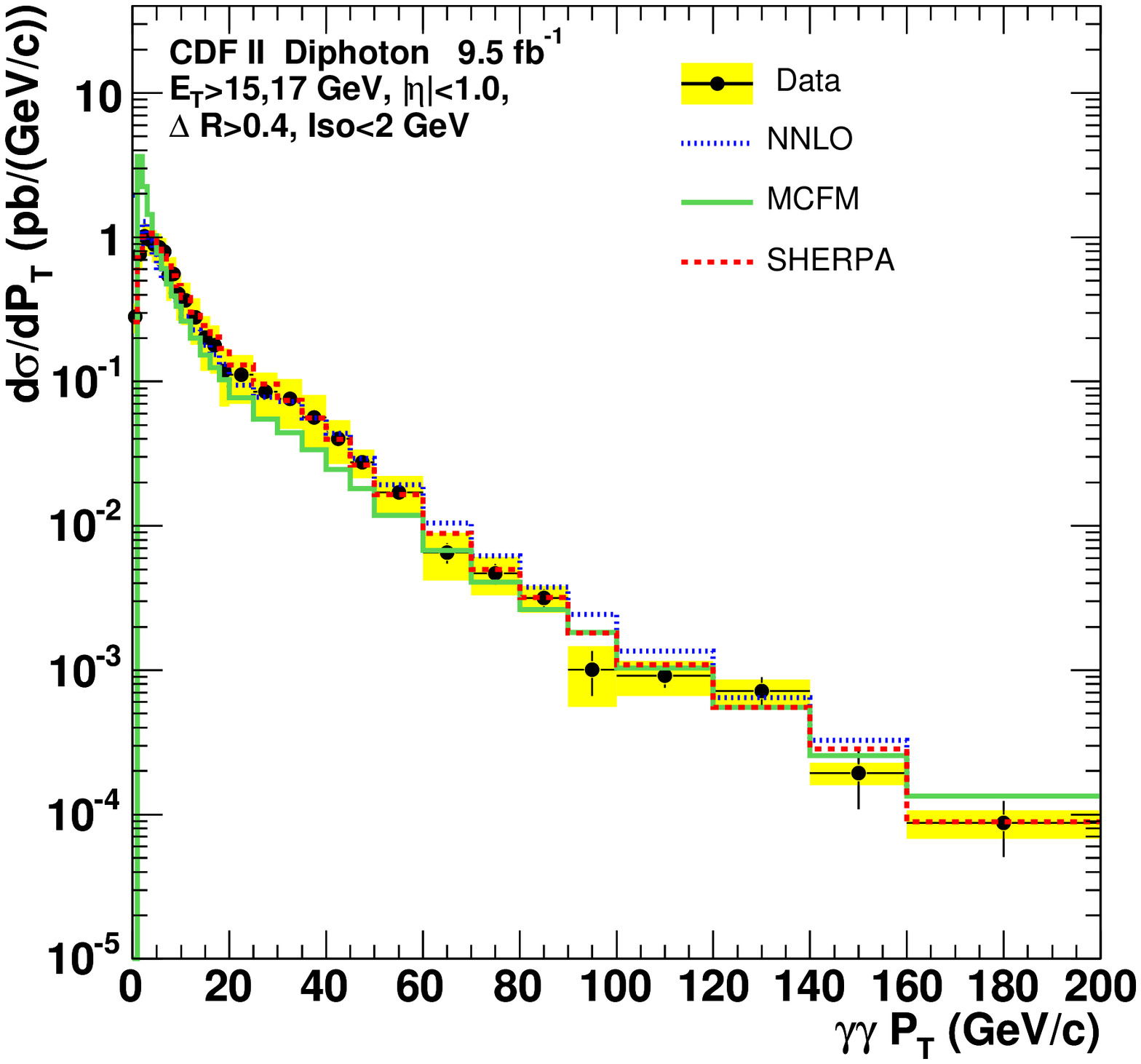}\hspace*{0cm}
\includegraphics[width=7cm]{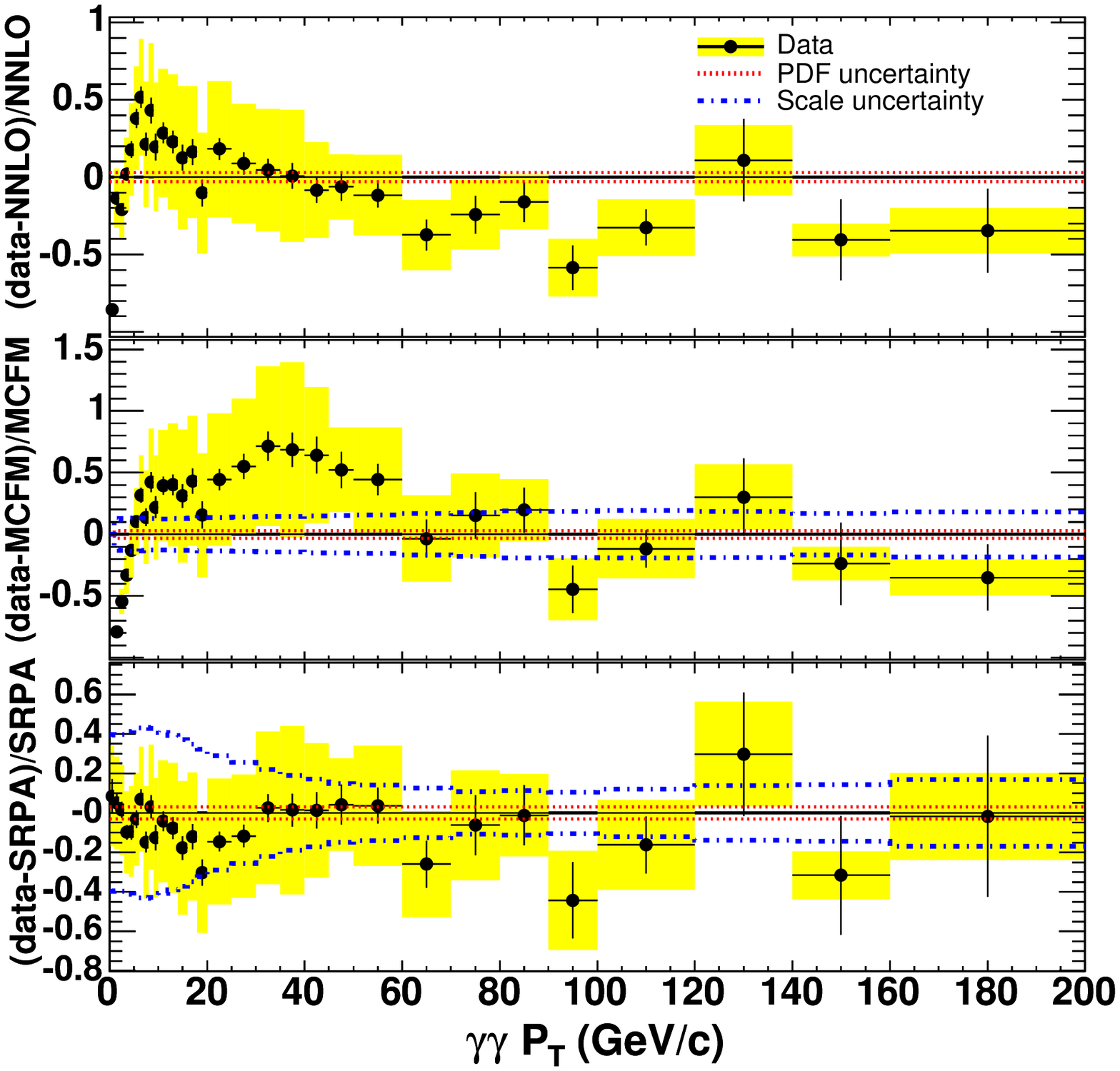}
\vspace*{0cm}
\includegraphics[width=7.5cm]{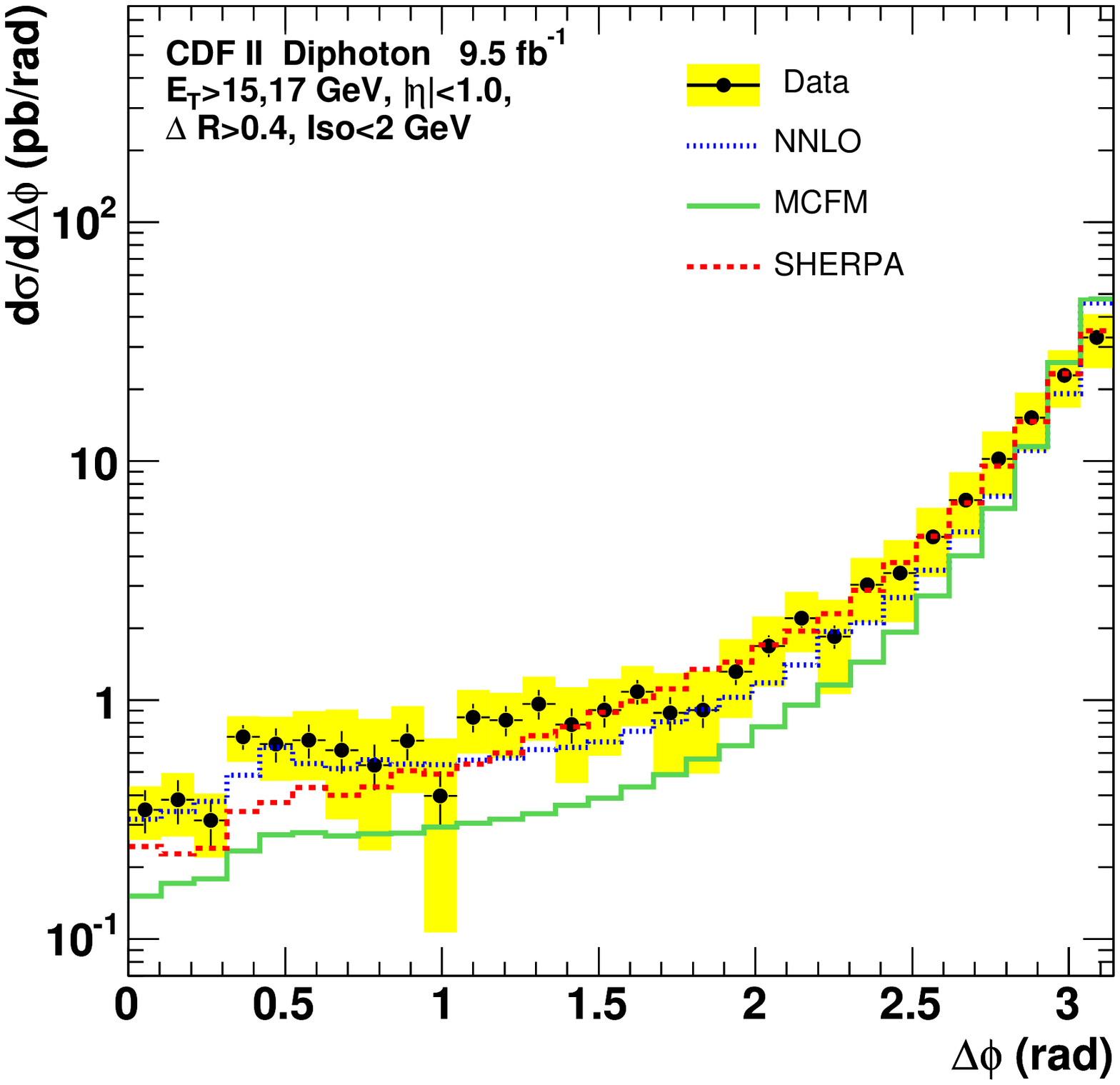}\hspace*{0cm}
\includegraphics[width=7cm]{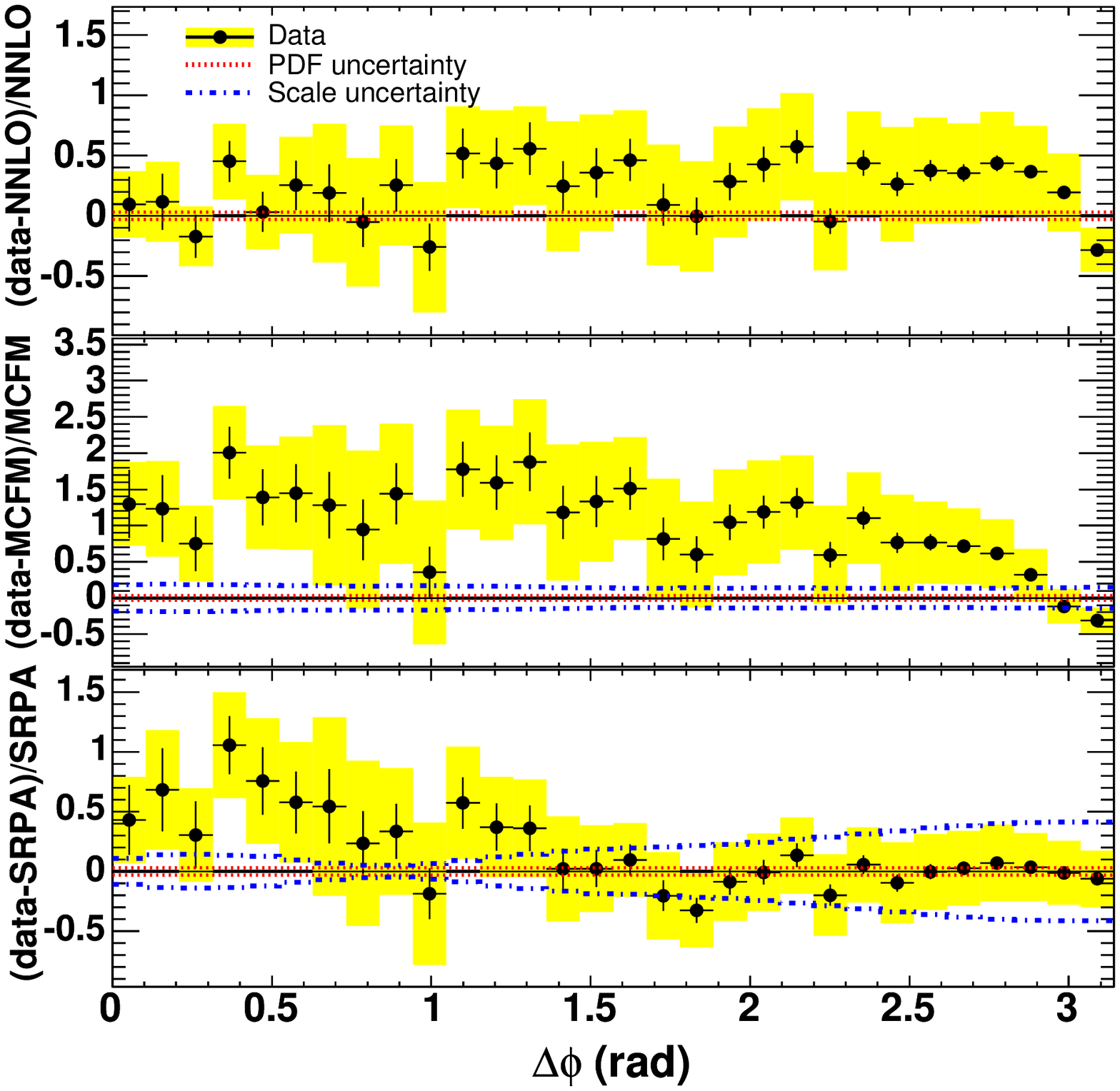}
\vspace*{-0.1cm}
\caption{Measured differential cross sections as functions of the diphoton
mass (top) and transverse momentum (middle), and of the azimuthal difference
between the photon directions (bottom), compared with three theoretical
predictions discussed in the text. The left panels show the absolute
comparisons. The lines show the predictions from {\sc sherpa} (dashed),
{\sc mcfm} (solid), and NNLO (dotted). The right panels show the fractional
deviations of the data from the theoretical predictions. The lines show the
scale uncertainty (dot-dashed) and the PDF uncertainty (dotted) of the
predictions. The vertical axis scales differ between fractional-deviation
plots. The shaded area around the data points indicates the total systematic
uncertainty of the measurement.\label{fig:results}}
\end{figure*}

The measured cross section for diphoton production integrated over the
acceptance is $12.3\pm 0.2_{\rm stat}\pm 3.5_{\rm syst}$$~$pb. The predictions
for the integrated cross section are $12.4\pm 4.4$$~$pb from {\sc sherpa},
$11.5\pm 0.3$$~$pb from {\sc mcfm}, and $11.8^{+1.7}_{-0.6}$$~$pb from the NNLO
calculation. The {\sc sherpa} scale uncertainty is the largest because it
also accounts for PS. All predictions are consistent with the measurement.
Figure \ref{fig:results} shows the comparisons between the observed and
predicted distributions in mass $M$, transverse momentum $P_{T}$ of the photon
pair, and azimuthal separation $\Delta\phi$ between the momenta of the two
photons in the event.

All predictions for the mass distribution show a reasonable agreement with the
data for all calculations above the maximum at 30 GeV/$c^{2}$, particularly
in the region around $M$$=$125 GeV/$c^{2}$ relevant to measurements of the
Higgs boson \cite{higgsLHC}. All predictions underestimate the data rate
around and below the maximum, although the NNLO prediction reproduces better
the data than the other two predictions. The {\sc sherpa} prediction tends to
underestimate the data for $M$$>$250 GeV/$c^{2}$.

In the $P_{T}$ spectrum, the {\sc mcfm} prediction underestimates the data
in the region between 30 and 60 GeV/$c$, a feature also observed in the
earlier measurements \cite{cdfdip,diphox}. The other two predictions
describe the data fairly well in this region. For $P_{T}$$<$20 GeV/$c$,
where soft gluon radiation becomes important, only the {\sc sherpa}
prediction provides a good description of the data because the parton
showering provides an effective resummation of multiple soft-gluon emission
amplitudes. The fixed-order predictions diverge in the limit of vanishing
$P_{T}$. The NNLO prediction tends to overestimate the data rate for
$P_{T}$$>$60 GeV/$c$.

Of special importance is the $\Delta\phi$ spectrum where all PS and NLO
predictions examined in the previous papers failed to describe the data over
the full range. The {\sc sherpa} model shows the best agreement at larger
$\Delta\phi$, where the diphoton system acquires substantial transverse
momentum due to multiple soft-gluon emission. However, {\sc sherpa}
progressively underestimates the data rate below 1.5 rad. The NNLO calculation
is the only prediction consistent with the data in the low $\Delta\phi$ tail,
which contains photon pairs with very low mass and relatively high $P_{T}$.
This calculation tends to underestimate the data rate above 1 rad. The
{\sc sherpa} and NNLO predictions generally are in better agreement with the
data than {\sc mcfm}. This shows that higher than NLO contributions, included
in both calculations in different ways, are needed in order to better describe
the data. More channels open at higher order, such as diphoton production
associated with the emission of two final-state partons (2$\rightarrow$4
channels), which enhance the event rate at high $P_{T}$ and low $\Delta\phi$.

The observed cross section enhancements at very low diphoton mass ($M$$<$30
GeV/$c^{2}$), moderate diphoton transverse momentum (30$<$$P_{T}$$<$60
GeV/$c$) and low $\Delta\phi$ ($<$1 rad) are correlated. The events
involved in this correlation have a topology of {\it same-side} diphotons
recoiling against at least one hard jet. For some of the contributions the
cross section is enhanced, such as when the two photons are emitted by the
same parton and are, therefore, predominantly almost collinear. Enhanced
contributions begin to appear in 2$\rightarrow$3 subprocesses. The importance
of 2$\rightarrow$3 subprocesses was shown in the previous CDF measurement
\cite{cdfdiptwo}, where the inclusion of photons radiated in hard
$\gamma$$+$jet events substantially improved the agreement of the PS
calculation with the data with respect to the simple 2$\rightarrow$2 diphoton
calculation. These subprocesses are treated in different ways at different
orders of approximation. At NLO, diphotons emitted from the same parton can
only appear in the fragmentation components \cite{diphox}. At NNLO such
contributions can result directly either from 2$\rightarrow$3 subprocesses,
where a quark loop is included in the diphoton production amplitude, or
from tree-level 2$\rightarrow$4 subprocesses \cite{nnlo}. The {\sc sherpa}
calculation also includes 2$\rightarrow$4 subprocesses \cite{sherpa}. Thus
NNLO and {\sc sherpa} describe the observed enhancement better than {\sc mcfm},
which does not include such subprocesses.

In summary, the diphoton production cross section, differential in kinematic 
variables sensitive to the parton-level processes that govern the reaction,
is measured using all data collected with the CDF II detector, corresponding
to an integrated luminosity of 9.5 fb$^{-1}$. This measurement is consistent
with the past CDF measurements \cite{cdfdip,cdfdiptwo} and supersedes them.
The measurement uses photons with $\vert\eta\vert<1$ and has sufficiently
high precision to resolve differences between state-of-the-art theoretical
predictions. The results are compared with three calculations, which apply
complementary techniques in predicting the cross section. The NNLO calculation
is generally consistent with the data, although events with very low diphoton
mass and high diphoton transverse momentum are not accurately described. The
ME+PS {\sc sherpa} calculation is also consistent with the data except in
the tails of the mass and the low $\Delta\phi$ distributions. Both NNLO and
{\sc sherpa} describe the data better than the {\sc mcfm} calculation, and
also better than the {\sc resbos}, {\sc diphox}, and {\sc pytha} calculations
\cite{compar}, in regions sensitive to diphoton production channels resulting
to nearly collinear photons. The comparisons show that parton-level processes
of order higher than NLO, which was the standard approximation in older
calculations, play an important role in diphoton production at the current
level of experimental precision. This conclusion is supported by the findings
of the recent ATLAS measurement at higher collision energy \cite{atlas}. The
inclusion of such processes in background calculations is thus important for
high precision measurements of the recently discovered Higgs boson-like
particle and searches for new phenomena in diphoton final states.

\begin{acknowledgments}
We thank the Fermilab staff and the technical staffs of the
participating institutions for their vital contributions. This work
was supported by the U.S. Department of Energy and National Science
Foundation; the Italian Istituto Nazionale di Fisica Nucleare; the
Ministry of Education, Culture, Sports, Science and Technology of
Japan; the Natural Sciences and Engineering Research Council of
Canada; the National Science Council of the Republic of China; the
Swiss National Science Foundation; the A.P. Sloan Foundation; the
Bundesministerium f\"ur Bildung und Forschung, Germany; the Korean
World Class University Program, the National Research Foundation of
Korea; the Science and Technology Facilities Council and the Royal
Society, UK; the Russian Foundation for Basic Research; the Ministerio
de Ciencia e Innovaci\'{o}n, and Programa Consolider-Ingenio 2010,
Spain; the Slovak R\&D Agency; the Academy of Finland; the Australian
Research Council (ARC); and the EU community Marie Curie Fellowship
contract 302103.
\end{acknowledgments}


\begin{thebibliography}{6}
\bibitem{caliso}
The calorimeter isolation is defined as the difference of the transverse
energy in the isolation cone minus the transverse energy in the tower
cluster of the photon. The isolation cone is defined to have a radius
$R=\sqrt{(\Delta\eta)^{2}+(\Delta\phi)^{2}}=0.4$ around the axis of the
shower profile. The isolation requirement implies that the angular separation
of the two photons is $\Delta R\ge 0.4$.
\bibitem{trkiso}
The track isolation is defined as the scalar sum of the transverse momenta
$p_{T}=\vert\vec{p}\vert\sin\theta$ of all tracks originating from the
primary vertex of the event and lying within the photon reconstruction cone.
\end{thebibliography}
\end{document}


\title{Measurement of the Cross Section for Prompt Isolated Diphoton
       Production using the Full CDF Data Sample
\newline\vskip 0.2in
{ \centering \LARGE Supplemental Material}
}

\date{\today}
\maketitle

\pagebreak
\vfill
\renewcommand{\plotname}{massSSV01}

\begin{center}
{\LARGE $M(\gamma\gamma)$}
\end{center}

\begin{figure*}[h!]
\includegraphics[width=9.0cm]{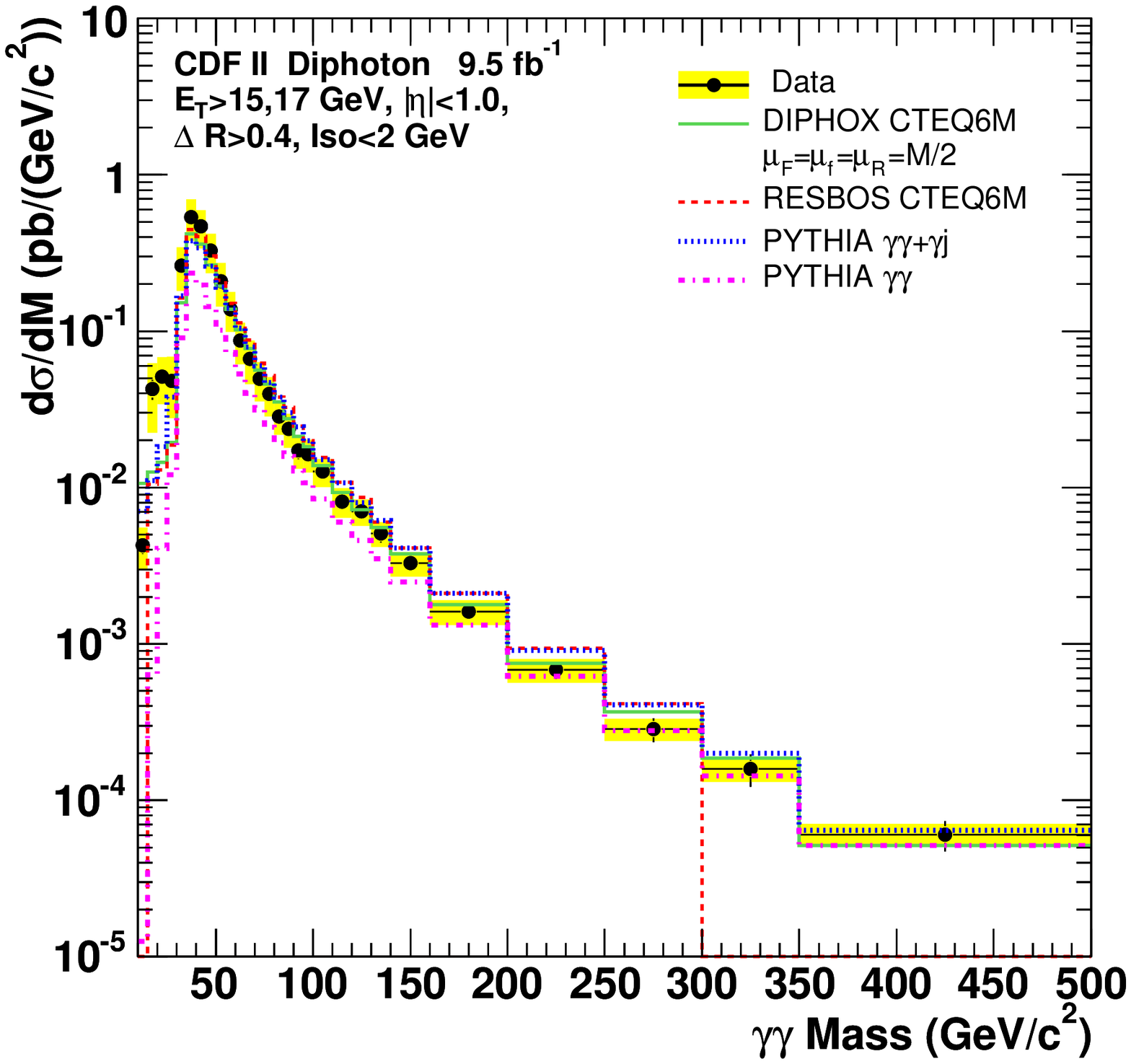}\hspace*{0cm}
\includegraphics[width=8.5cm]{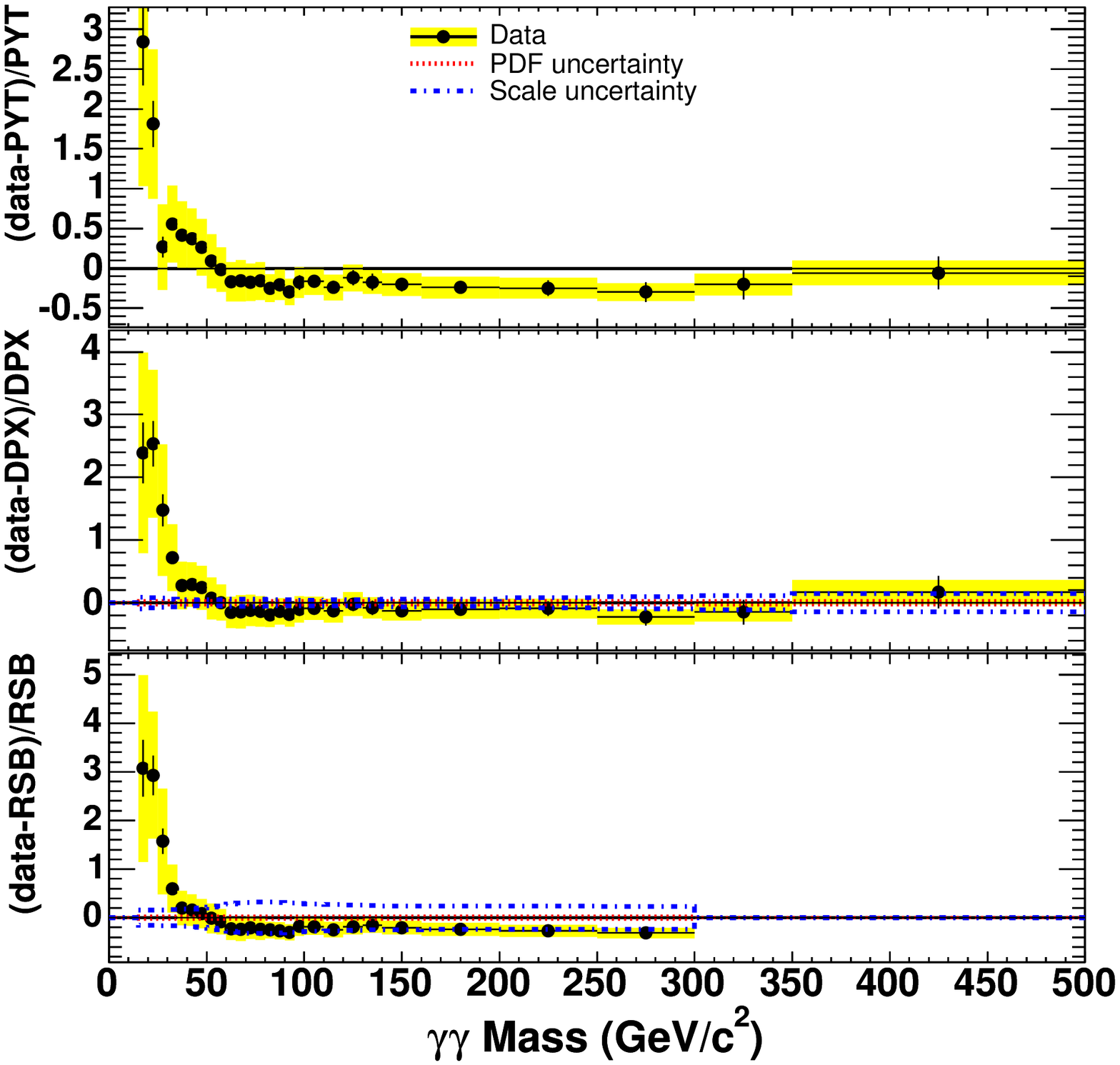}\hfill
\vspace*{0cm}
\includegraphics[width=9.0cm]{TXSec_\plotname_2_10Dec12.eps}\hspace*{0cm}
\includegraphics[width=8.5cm]{TXSec_\plotname_allratio2_23Jan13.eps}
\vspace*{-0.1cm}
\caption{The measured differential cross sections for $M(\gamma\gamma)$ 
compared with six 
theoretical predictions discussed in the text. The left windows show the
absolute comparisons and the right windows show the fractional deviations
of the data from the theoretical predictions. Note that the vertical axis
scales differ between fractional deviation plots.}
\end{figure*}

\pagebreak
\vfill

\begin{table}[!hp]
\begin{center}
\begin{tabular}{cccc}
\hline\hline
 Bin & Cross Section (pb) & Sherpa & NNLO   \\ \hline
 10.0 -  15.0 & $ 0.0043 \pm  0.00050 \pm  0.0013 $ &  0.0115 &  0.0227 \\
 15.0 -  20.0 & $ 0.043 \pm  0.0061 \pm  0.020 $ &  0.018 &  0.025 \\
 20.0 -  25.0 & $ 0.051 \pm  0.0053 \pm  0.017 $ &  0.027 &  0.028 \\
 25.0 -  30.0 & $ 0.048 \pm  0.0049 \pm  0.020 $ &  0.044 &  0.035 \\
 30.0 -  35.0 & $ 0.262 \pm  0.012 \pm  0.081 $ &  0.210 &  0.131 \\
 35.0 -  40.0 & $ 0.54 \pm  0.016 \pm  0.16 $ &  0.49 &  0.38 \\
 40.0 -  45.0 & $ 0.47 \pm  0.014 \pm  0.13 $ &  0.43 &  0.42 \\
 45.0 -  50.0 & $ 0.328 \pm  0.012 \pm  0.092 $ &  0.321 &  0.349 \\
 50.0 -  55.0 & $ 0.209 \pm  0.0091 \pm  0.065 $ &  0.231 &  0.231 \\
 55.0 -  60.0 & $ 0.138 \pm  0.0071 \pm  0.039 $ &  0.165 &  0.185 \\
 60.0 -  65.0 & $ 0.087 \pm  0.0055 \pm  0.026 $ &  0.117 &  0.104 \\
 65.0 -  70.0 & $ 0.066 \pm  0.0047 \pm  0.021 $ &  0.086 &  0.073 \\
 70.0 -  75.0 & $ 0.050 \pm  0.0039 \pm  0.014 $ &  0.061 &  0.067 \\
 75.0 -  80.0 & $ 0.040 \pm  0.0034 \pm  0.011 $ &  0.048 &  0.050 \\
 80.0 -  85.0 & $ 0.0284 \pm  0.0027 \pm  0.0066 $ &  0.0383 &  0.0410 \\
 85.0 -  90.0 & $ 0.0238 \pm  0.0025 \pm  0.0057 $ &  0.0289 &  0.0304 \\
 90.0 -  95.0 & $ 0.0173 \pm  0.0020 \pm  0.0041 $ &  0.0230 &  0.0266 \\
 95.0 - 100.0 & $ 0.0164 \pm  0.0019 \pm  0.0039 $ &  0.0181 &  0.0203 \\
100.0 - 110.0 & $ 0.0127 \pm  0.0011 \pm  0.0026 $ &  0.0140 &  0.0151 \\
110.0 - 120.0 & $ 0.0081 \pm  0.00088 \pm  0.0017 $ &  0.0094 &  0.0125 \\
120.0 - 130.0 & $ 0.0070 \pm  0.00076 \pm  0.0014 $ &  0.0072 &  0.0076 \\
130.0 - 140.0 & $ 0.00510 \pm  0.00064 \pm  0.00093 $ &  0.00515 &  0.00596 \\
140.0 - 160.0 & $ 0.00328 \pm  0.00034 \pm  0.00058 $ &  0.00325 &  0.00336 \\
160.0 - 200.0 & $ 0.00161 \pm  0.00016 \pm  0.00029 $ &  0.00165 &  0.00186 \\
200.0 - 250.0 & $ 0.00068 \pm  0.000088 \pm  0.00012 $ &  0.00072 &  0.00070 \\
250.0 - 300.0 & $ 0.000286 \pm  0.000050 \pm  0.000045 $ &  0.000279 &  0.000331 \\
300.0 - 350.0 & $ 0.000159 \pm  0.000037 \pm  0.000027 $ &  0.000120 &  0.000172 \\
350.0 - 500.0 & $ 0.000060 \pm  0.000013 \pm  0.0000100 $ &  0.000039 &  0.000055 \\
\hline\hline
\end{tabular}
\end{center}
\end{table}

\pagebreak
\vfill

\renewcommand{\plotname}{massQHSSV01}
\begin{center}
{\LARGE $M(\gamma\gamma)$ for $P_T(\gamma\gamma) > M(\gamma\gamma)$}
\end{center}

\begin{figure*}[!hp]
\includegraphics[width=9.0cm]{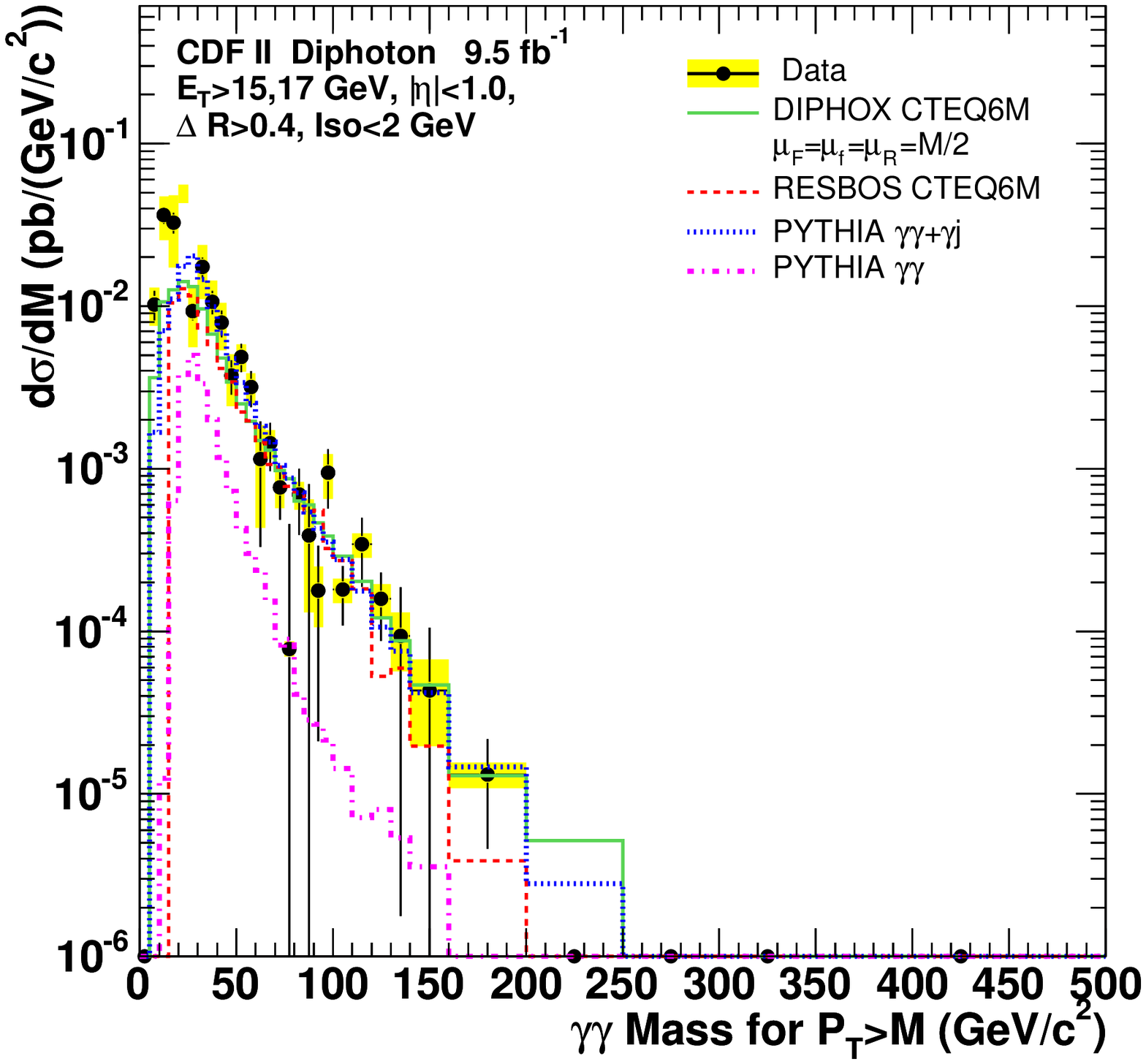}\hspace*{0cm}
\includegraphics[width=8.5cm]{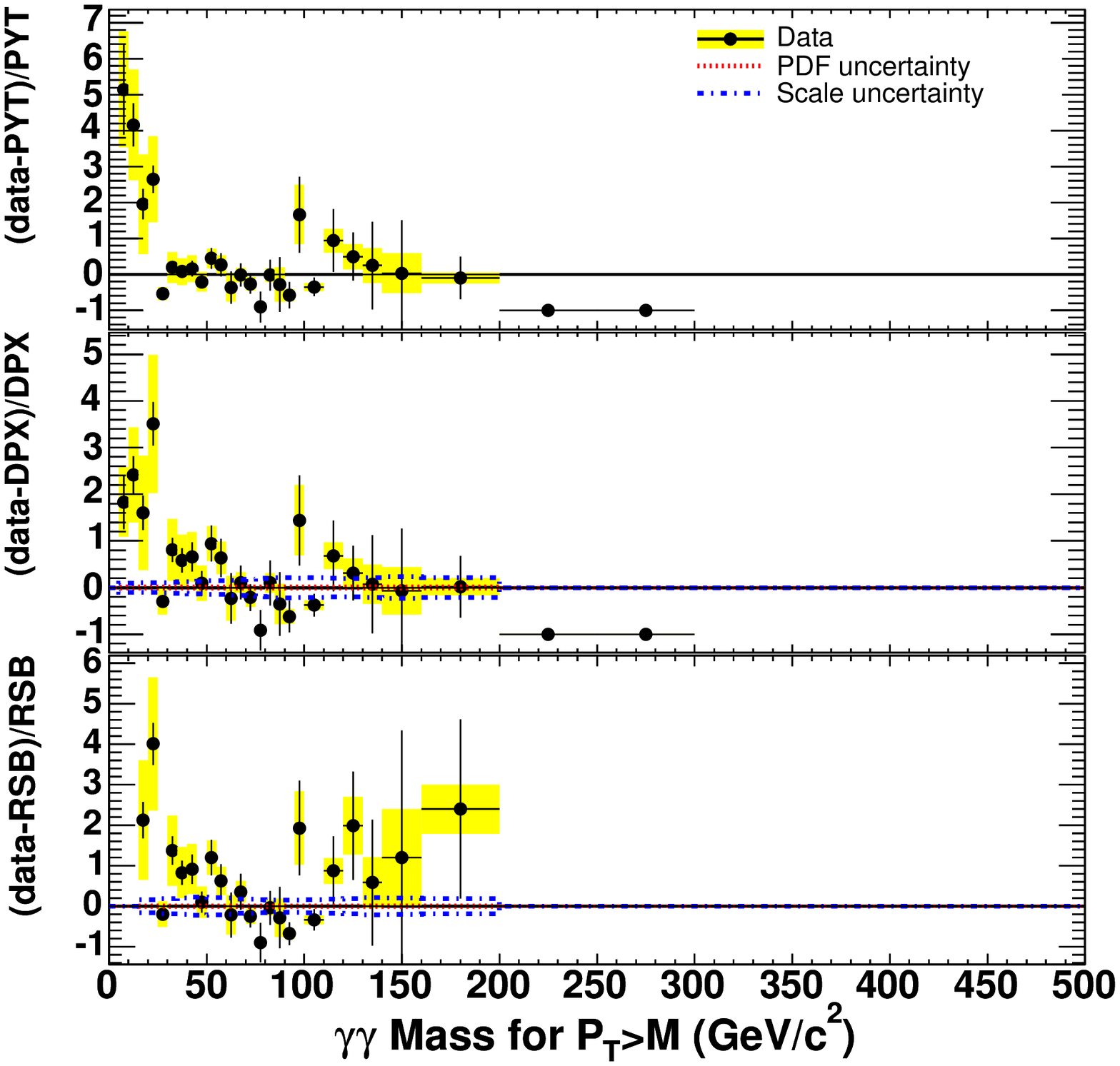}\hfill
\vspace*{0cm}
\includegraphics[width=9.0cm]{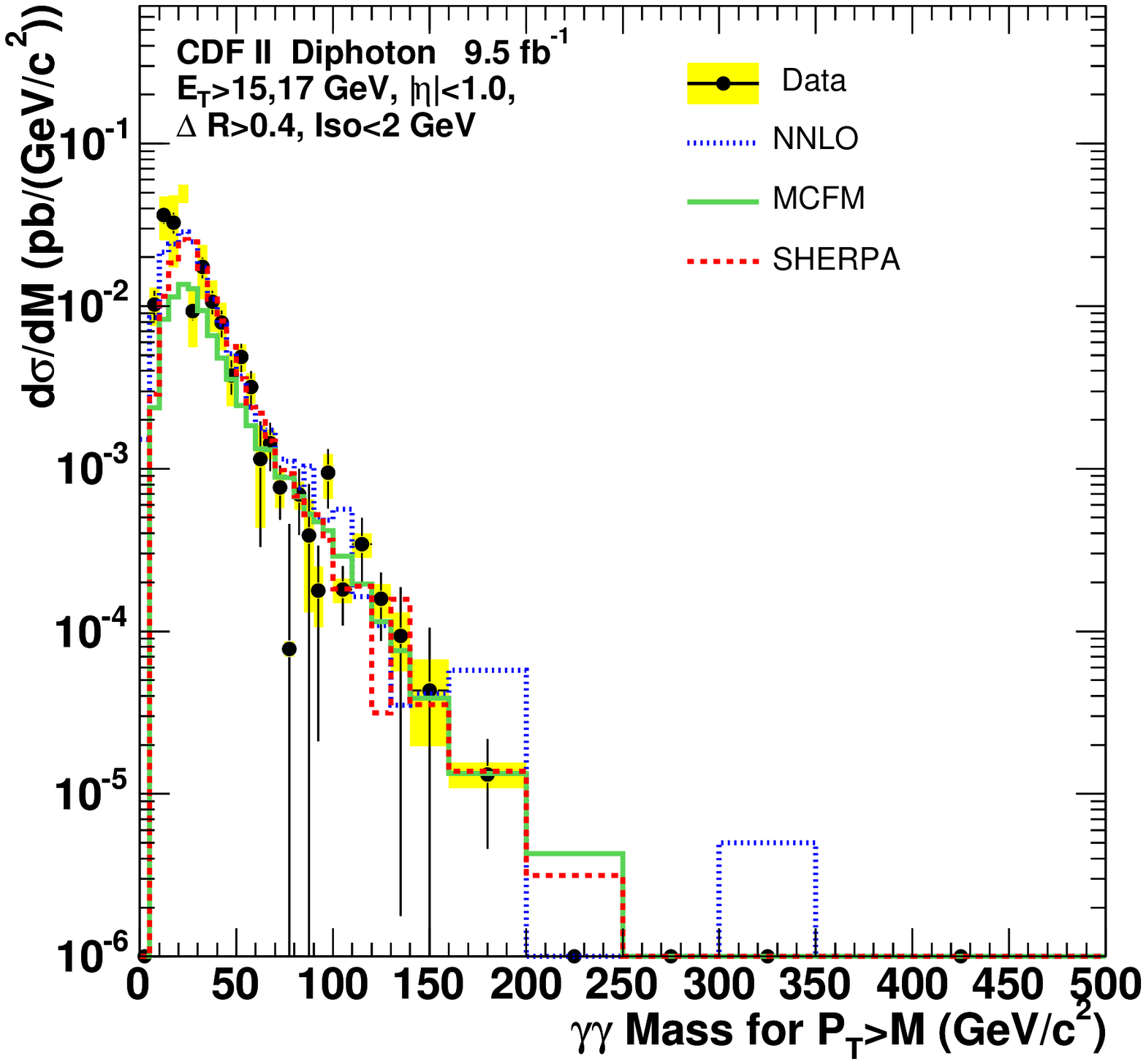}\hspace*{0cm}
\includegraphics[width=8.5cm]{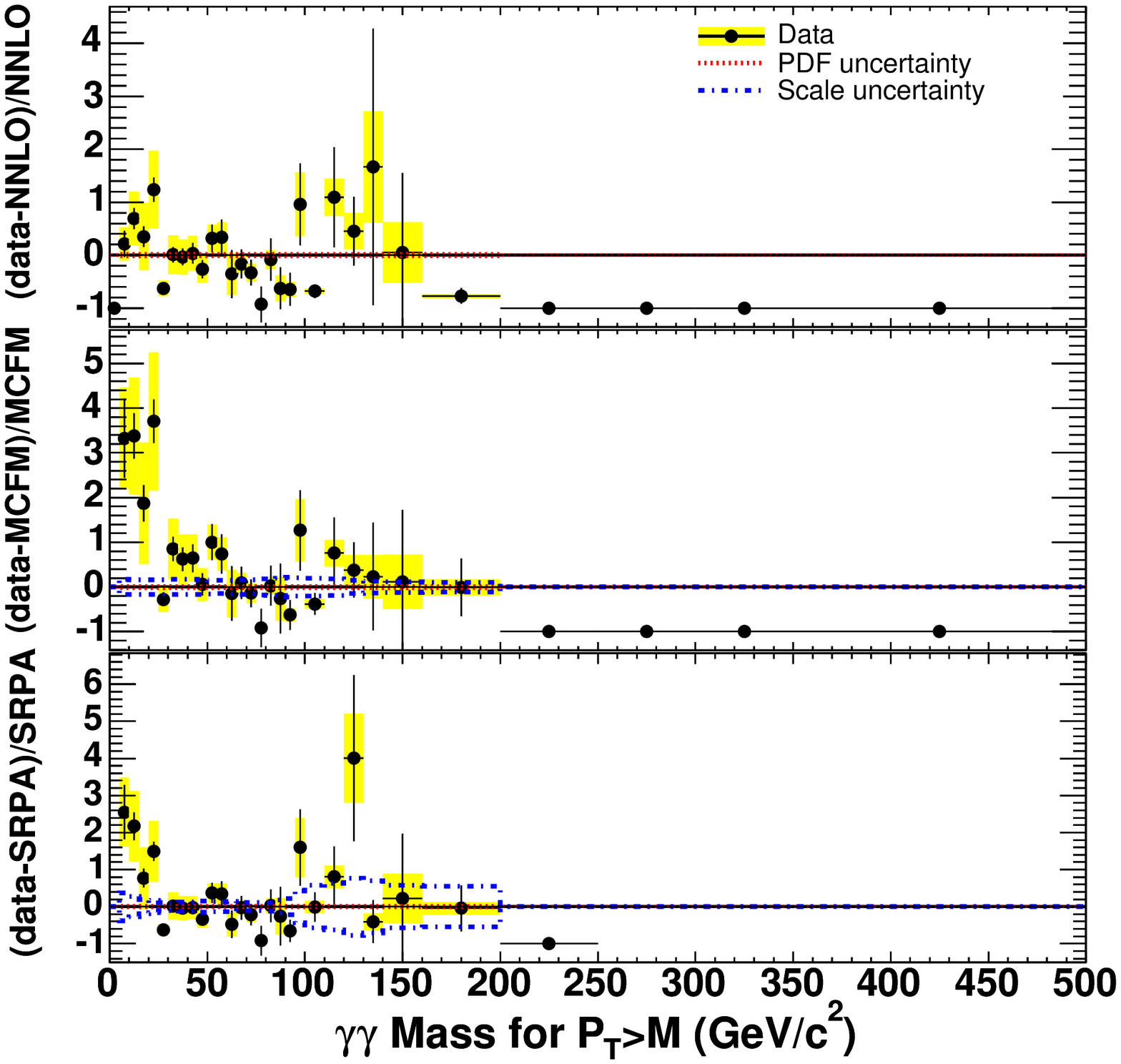}
\vspace*{-0.1cm}
\caption{The measured differential cross sections 
for $M(\gamma\gamma)$, when $P_T(\gamma\gamma) > M(\gamma\gamma)$, 
compared with six 
theoretical predictions discussed in the text. The left windows show the
absolute comparisons and the right windows show the fractional deviations
of the data from the theoretical predictions. Note that the vertical axis
scales differ between fractional deviation plots.}
\end{figure*}

\pagebreak
\vfill

\begin{table}[!hp]
\begin{center}
\begin{tabular}{cccc}
\hline\hline
 Bin & Cross Section (pb) & Sherpa & NNLO   \\ \hline
 10.0 -  15.0 & $ 0.036 \pm  0.0042 \pm  0.011 $ &  0.011 &  0.021 \\
 15.0 -  20.0 & $ 0.033 \pm  0.0047 \pm  0.015 $ &  0.018 &  0.024 \\
 20.0 -  25.0 & $ 0.064 \pm  0.0066 \pm  0.021 $ &  0.026 &  0.029 \\
 25.0 -  30.0 & $ 0.0093 \pm  0.0011 \pm  0.0037 $ &  0.0249 &  0.0249 \\
 30.0 -  35.0 & $ 0.0174 \pm  0.0025 \pm  0.0064 $ &  0.0171 &  0.0172 \\
 35.0 -  40.0 & $ 0.0107 \pm  0.0017 \pm  0.0037 $ &  0.0111 &  0.0110 \\
 40.0 -  45.0 & $ 0.0079 \pm  0.0015 \pm  0.0026 $ &  0.0081 &  0.0077 \\
 45.0 -  50.0 & $ 0.0037 \pm  0.00089 \pm  0.0013 $ &  0.0056 &  0.0051 \\
 50.0 -  55.0 & $ 0.00489 \pm  0.00097 \pm  0.00095 $ &  0.00357 &  0.00371 \\
 55.0 -  60.0 & $ 0.00319 \pm  0.00081 \pm  0.00068 $ &  0.00237 &  0.00238 \\
 60.0 -  65.0 & $ 0.00114 \pm  0.00081 \pm  0.00071 $ &  0.00219 &  0.00178 \\
 65.0 -  70.0 & $ 0.00144 \pm  0.00047 \pm  0.00029 $ &  0.00148 &  0.00173 \\
 70.0 -  75.0 & $ 0.00077 \pm  0.00028 \pm  0.00020 $ &  0.00098 &  0.00115 \\
 75.0 -  80.0 & $ 0.00008 \pm  0.00038 \pm  0.0000084 $ &  0.00093 &  0.00111 \\
 80.0 -  85.0 & $ 0.00069 \pm  0.00030 \pm  0.00013 $ &  0.00068 &  0.00076 \\
 85.0 -  90.0 & $ 0.00039 \pm  0.00041 \pm  0.00026 $ &  0.00052 &  0.00104 \\
 90.0 -  95.0 & $ 0.00018 \pm  0.00016 \pm  0.000072 $ &  0.00052 &  0.00051 \\
 95.0 - 100.0 & $ 0.00094 \pm  0.00037 \pm  0.00029 $ &  0.00036 &  0.00048 \\
100.0 - 110.0 & $ 0.000180 \pm  0.000072 \pm  0.000031 $ &  0.000182 &  0.000564 \\
110.0 - 120.0 & $ 0.00034 \pm  0.00015 \pm  0.000058 $ &  0.00019 &  0.00016 \\
120.0 - 130.0 & $ 0.000158 \pm  0.000071 \pm  0.000038 $ &  0.000032 &  0.000109 \\
130.0 - 140.0 & $ 0.000094 \pm  0.000092 \pm  0.000037 $ &  0.000158 &  0.000035 \\
140.0 - 160.0 & $ 0.000043 \pm  0.000062 \pm  0.000024 $ &  0.000036 &  0.000041 \\
160.0 - 200.0 & $ 0.0000132 \pm  0.0000086 \pm  0.0000024 $ &  0.0000138 &  0.0000578 \\
\hline\hline
\end{tabular}
\end{center}
\end{table}

\pagebreak
\vfill

\renewcommand{\plotname}{massQLSSV01}
\begin{center}
{\LARGE $M(\gamma\gamma)$ for $P_T(\gamma\gamma) < M(\gamma\gamma)$}
\end{center}

\begin{figure*}[!hp]
\includegraphics[width=9.0cm]{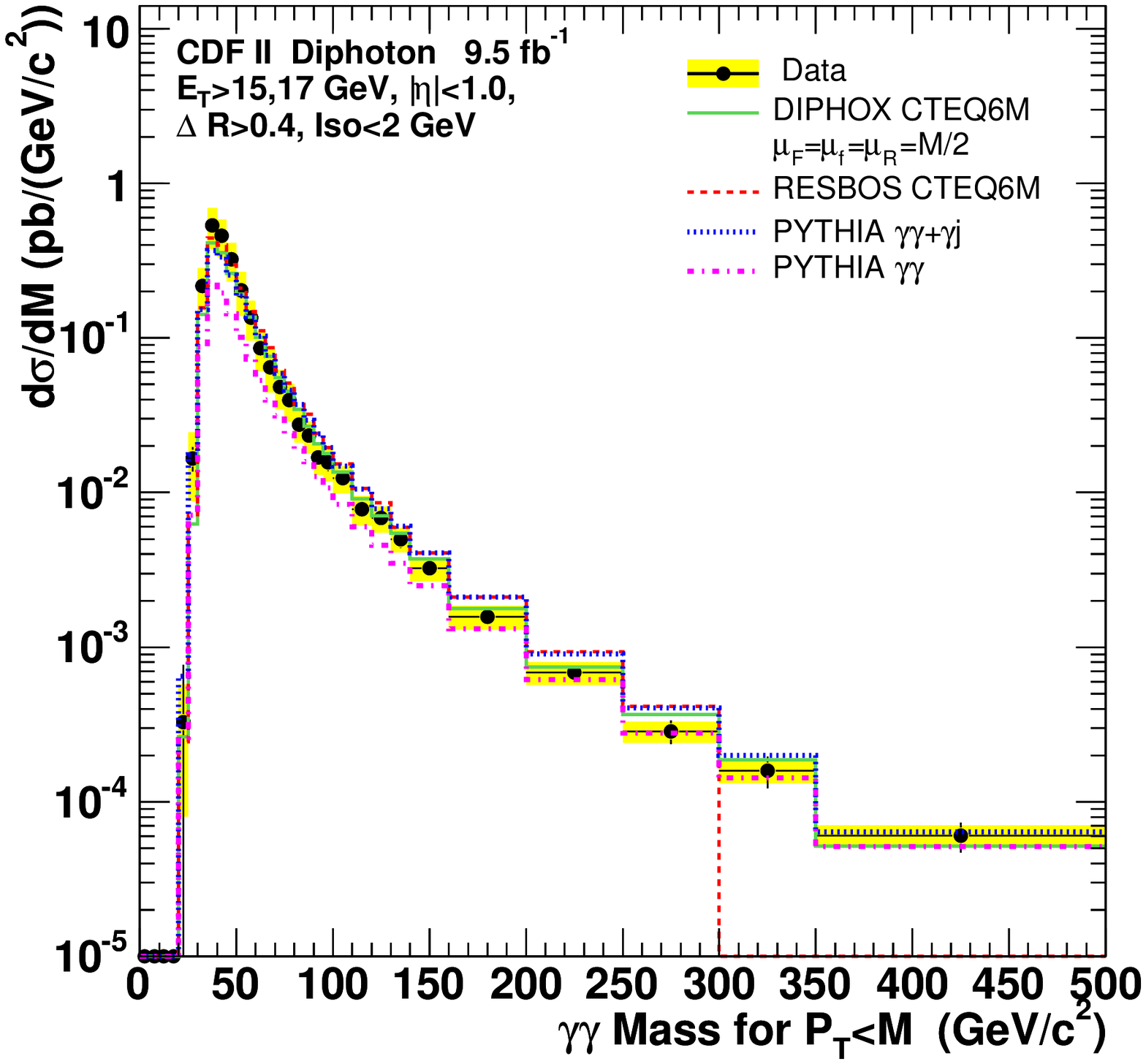}\hspace*{0cm}
\includegraphics[width=8.5cm]{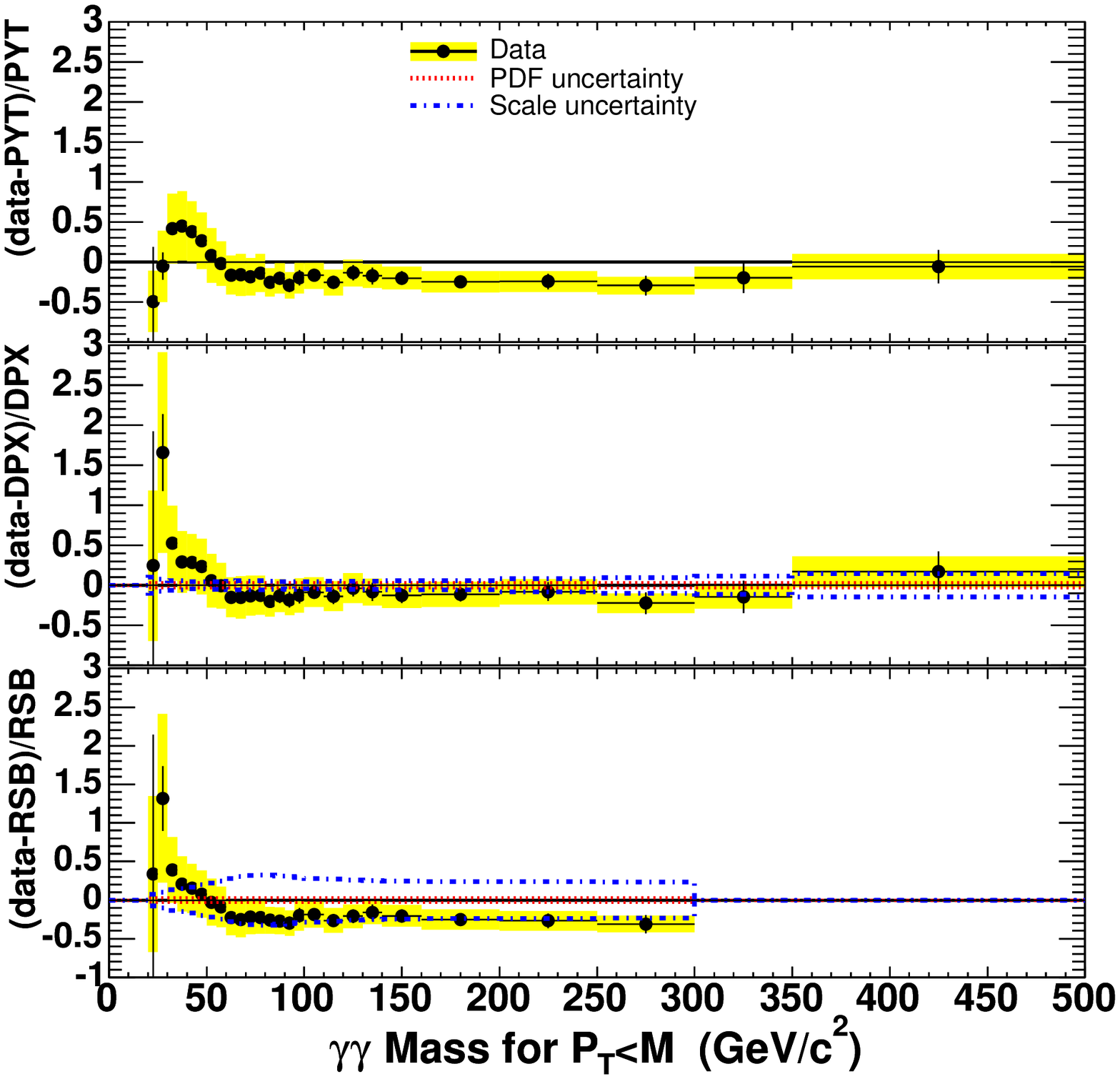}\hfill
\vspace*{0cm}
\includegraphics[width=9.0cm]{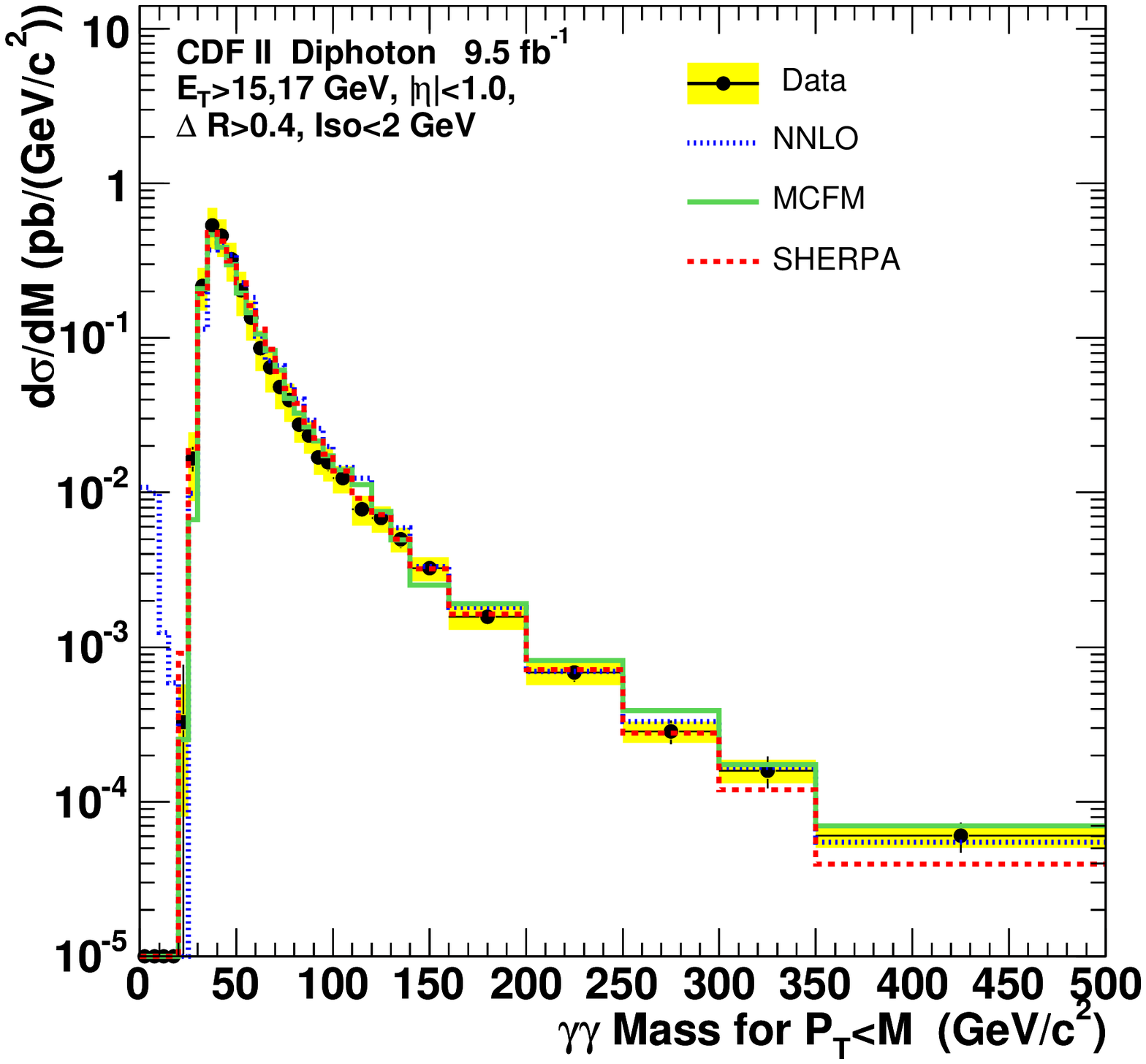}\hspace*{0cm}
\includegraphics[width=8.5cm]{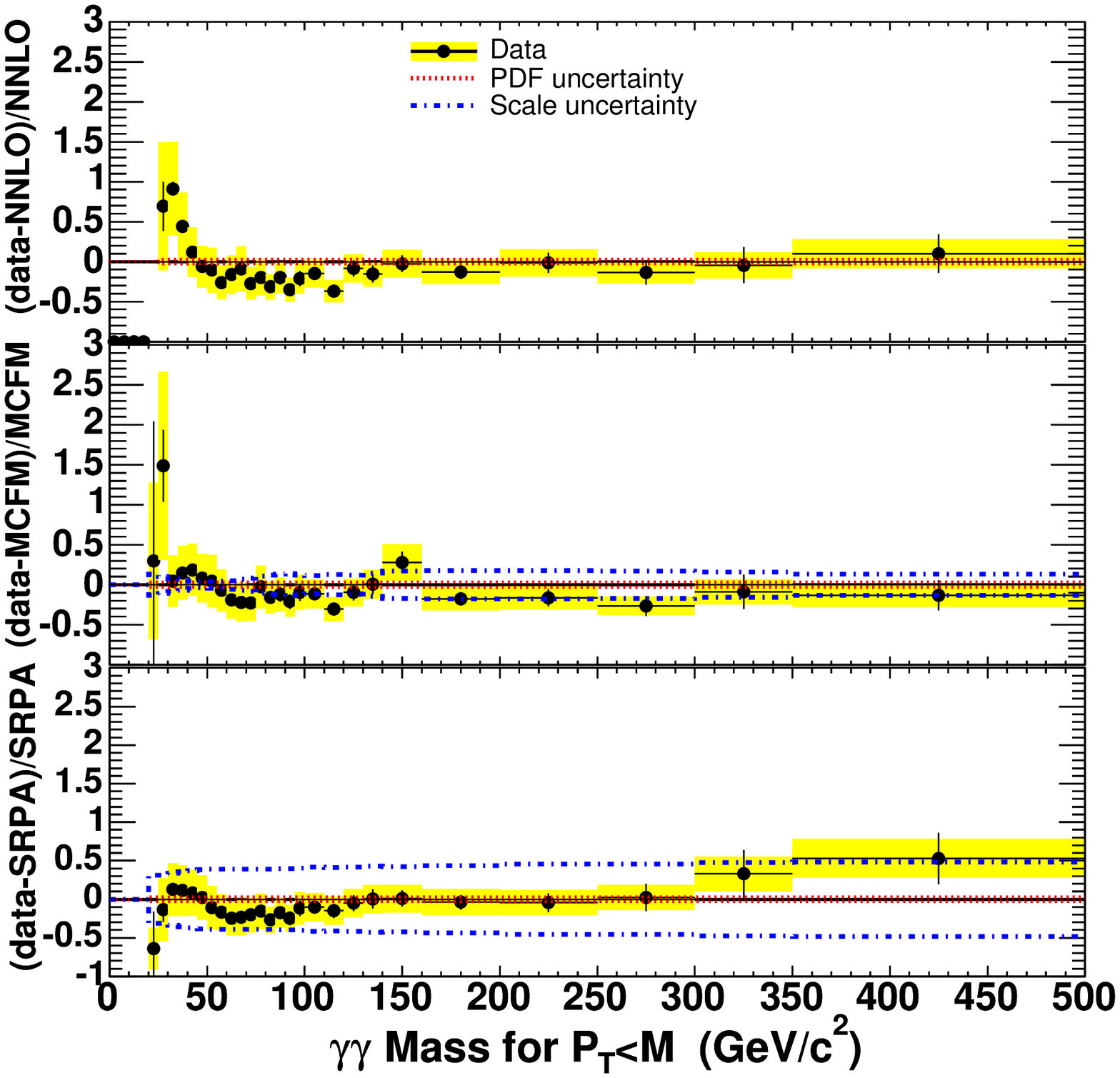}
\vspace*{-0.1cm}
\caption{The measured differential cross sections 
for $M(\gamma\gamma)$, when $P_T(\gamma\gamma) < M(\gamma\gamma)$, 
compared with six 
theoretical predictions discussed in the text. The left windows show the
absolute comparisons and the right windows show the fractional deviations
of the data from the theoretical predictions. Note that the vertical axis
scales differ between fractional deviation plots.}
\end{figure*}

\pagebreak
\vfill

\begin{table}[!hp]
\begin{center}
\begin{tabular}{cccc}
\hline\hline
 Bin & Cross Section (pb) & Sherpa & NNLO   \\ \hline
 20.0 -  25.0 & $ 0.00033 \pm  0.00044 \pm  0.00025 $ &  0.00092 & -0.00079 \\
 25.0 -  30.0 & $ 0.0167 \pm  0.0030 \pm  0.0079 $ &  0.0193 &  0.0099 \\
 30.0 -  35.0 & $ 0.217 \pm  0.010 \pm  0.067 $ &  0.193 &  0.114 \\
 35.0 -  40.0 & $ 0.53 \pm  0.016 \pm  0.16 $ &  0.48 &  0.37 \\
 40.0 -  45.0 & $ 0.46 \pm  0.014 \pm  0.13 $ &  0.42 &  0.41 \\
 45.0 -  50.0 & $ 0.322 \pm  0.011 \pm  0.090 $ &  0.315 &  0.344 \\
 50.0 -  55.0 & $ 0.203 \pm  0.0090 \pm  0.064 $ &  0.227 &  0.227 \\
 55.0 -  60.0 & $ 0.135 \pm  0.0070 \pm  0.039 $ &  0.162 &  0.182 \\
 60.0 -  65.0 & $ 0.086 \pm  0.0055 \pm  0.025 $ &  0.114 &  0.103 \\
 65.0 -  70.0 & $ 0.065 \pm  0.0047 \pm  0.020 $ &  0.084 &  0.071 \\
 70.0 -  75.0 & $ 0.048 \pm  0.0039 \pm  0.014 $ &  0.060 &  0.066 \\
 75.0 -  80.0 & $ 0.040 \pm  0.0034 \pm  0.011 $ &  0.047 &  0.049 \\
 80.0 -  85.0 & $ 0.0276 \pm  0.0026 \pm  0.0065 $ &  0.0377 &  0.0403 \\
 85.0 -  90.0 & $ 0.0234 \pm  0.0025 \pm  0.0056 $ &  0.0284 &  0.0293 \\
 90.0 -  95.0 & $ 0.0169 \pm  0.0020 \pm  0.0040 $ &  0.0225 &  0.0261 \\
 95.0 - 100.0 & $ 0.0157 \pm  0.0018 \pm  0.0038 $ &  0.0178 &  0.0199 \\
100.0 - 110.0 & $ 0.0124 \pm  0.0011 \pm  0.0025 $ &  0.0138 &  0.0145 \\
110.0 - 120.0 & $ 0.0078 \pm  0.00087 \pm  0.0017 $ &  0.0092 &  0.0124 \\
120.0 - 130.0 & $ 0.0068 \pm  0.00076 \pm  0.0013 $ &  0.0072 &  0.0074 \\
130.0 - 140.0 & $ 0.00501 \pm  0.00063 \pm  0.00091 $ &  0.00500 &  0.00593 \\
140.0 - 160.0 & $ 0.00324 \pm  0.00034 \pm  0.00057 $ &  0.00322 &  0.00332 \\
160.0 - 200.0 & $ 0.00157 \pm  0.00016 \pm  0.00028 $ &  0.00164 &  0.00180 \\
200.0 - 250.0 & $ 0.00069 \pm  0.000088 \pm  0.00012 $ &  0.00072 &  0.00070 \\
250.0 - 300.0 & $ 0.000286 \pm  0.000050 \pm  0.000045 $ &  0.000279 &  0.000331 \\
300.0 - 350.0 & $ 0.000159 \pm  0.000037 \pm  0.000027 $ &  0.000120 &  0.000167 \\
350.0 - 500.0 & $ 0.000060 \pm  0.000013 \pm  0.0000100 $ &  0.000039 &  0.000055 \\
\hline\hline
\end{tabular}
\end{center}
\end{table}

\pagebreak
\vfill
\renewcommand{\plotname}{qtSSV01}

\begin{center}
{\LARGE $P_T(\gamma\gamma)$}
\end{center}

\begin{figure*}[!hp]
\includegraphics[width=9.0cm]{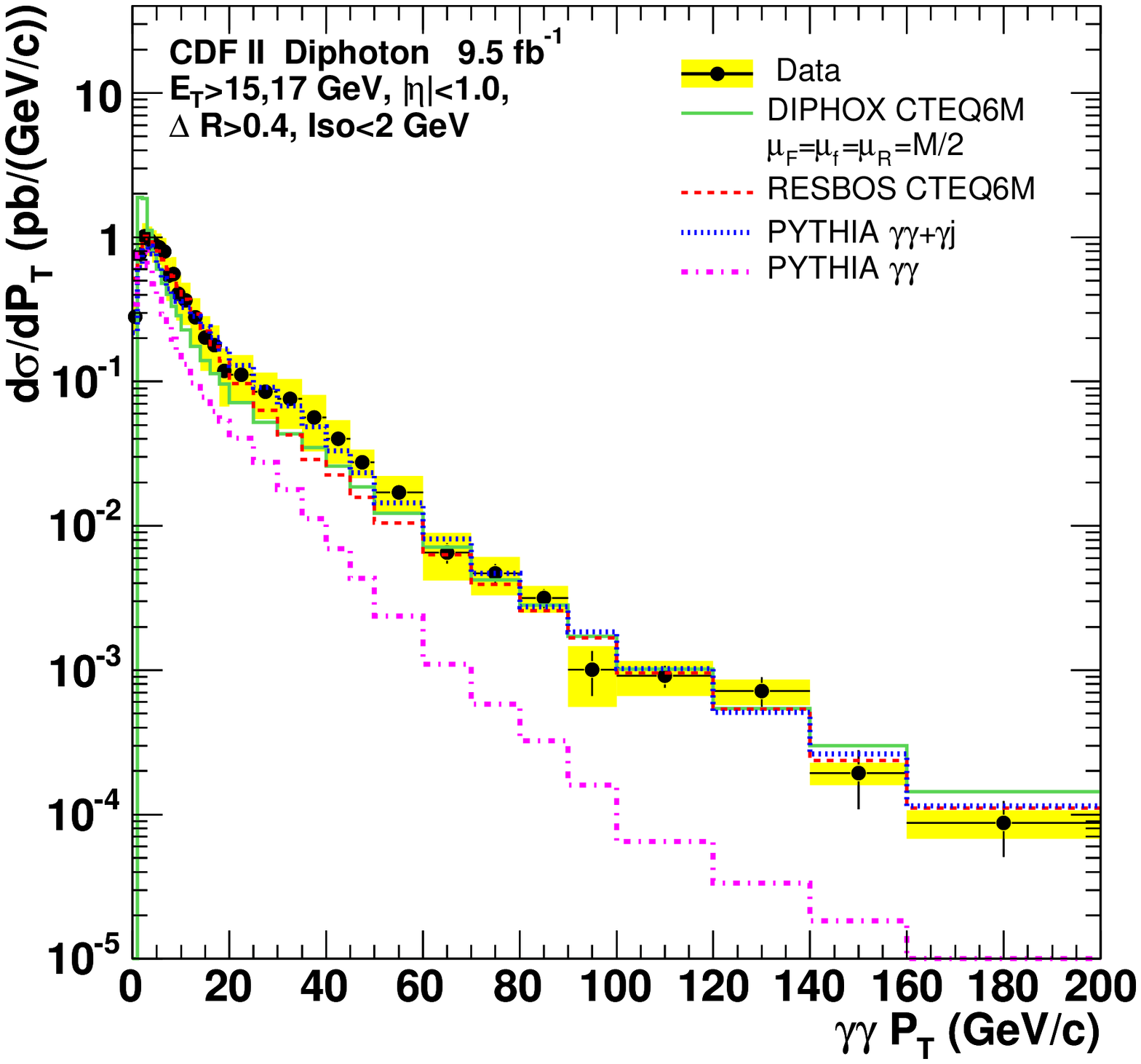}\hspace*{0cm}
\includegraphics[width=8.5cm]{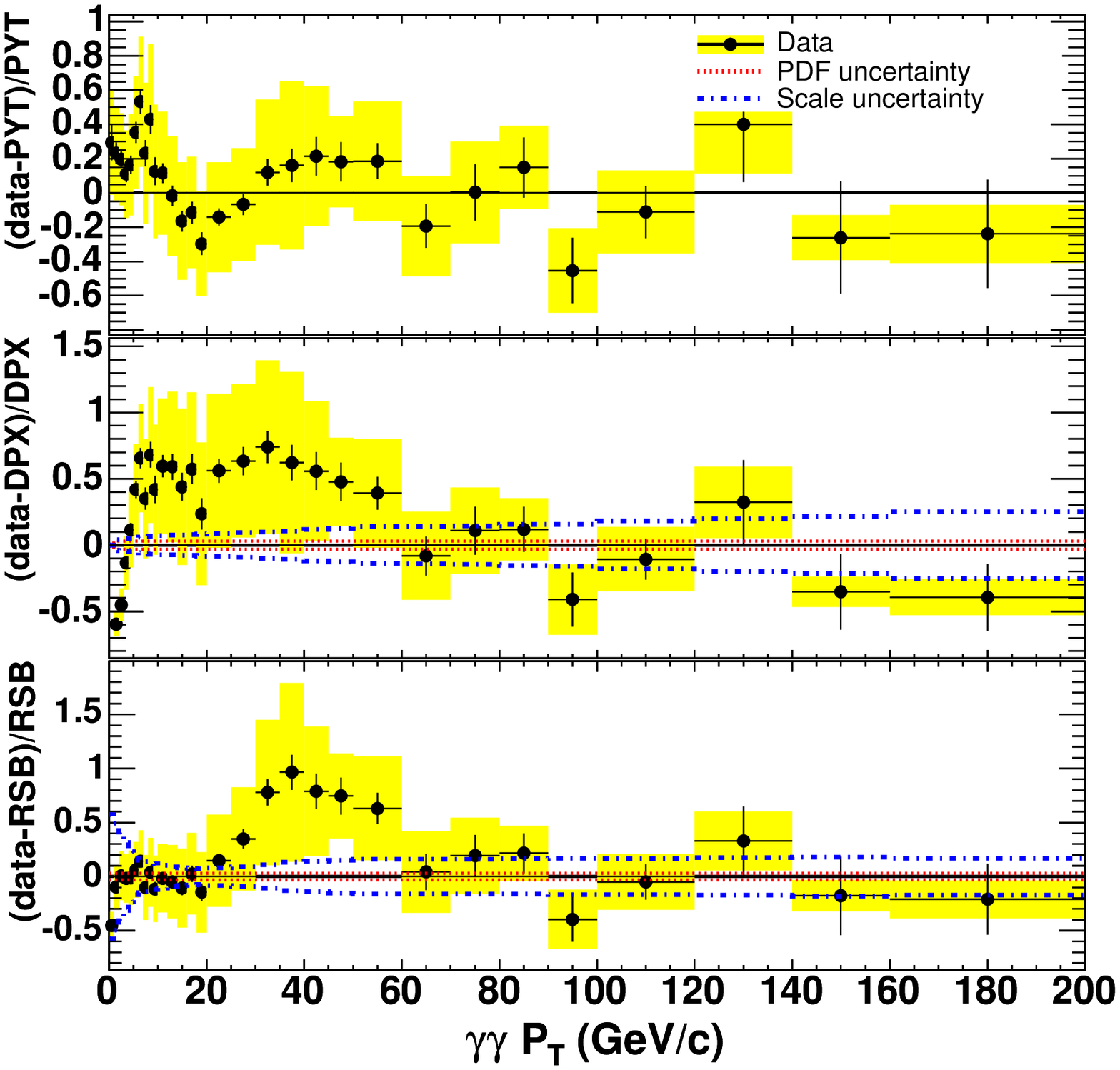}\hfill
\vspace*{0cm}
\includegraphics[width=9.0cm]{TXSec_\plotname_2_10Dec12.eps}\hspace*{0cm}
\includegraphics[width=8.5cm]{TXSec_\plotname_allratio2_23Jan13.eps}
\vspace*{-0.1cm}
\caption{The measured differential cross sections 
for $P_T(\gamma\gamma)$
compared with six 
theoretical predictions discussed in the text. The left windows show the
absolute comparisons and the right windows show the fractional deviations
of the data from the theoretical predictions. Note that the vertical axis
scales differ between fractional deviation plots.}
\end{figure*}

\pagebreak
\vfill

\begin{table}[!hp]
\begin{center}
\begin{tabular}{cccc}
\hline\hline
 Bin & Cross Section (pb) & Sherpa & NNLO   \\ \hline
  0.00 -   1.00 & $ 0.282 \pm  0.023 \pm  0.066 $ &  0.260 &  1.952 \\
  1.00 -   2.00 & $ 0.76 \pm  0.038 \pm  0.17 $ &  0.72 &  0.88 \\
  2.00 -   3.00 & $ 1.02 \pm  0.041 \pm  0.23 $ &  1.00 &  1.29 \\
  3.00 -   4.00 & $ 0.96 \pm  0.040 \pm  0.22 $ &  1.07 &  0.94 \\
  4.00 -   5.00 & $ 0.89 \pm  0.040 \pm  0.23 $ &  0.98 &  0.76 \\
  5.00 -   6.00 & $ 0.86 \pm  0.039 \pm  0.21 $ &  0.88 &  0.62 \\
  6.00 -   7.00 & $ 0.80 \pm  0.037 \pm  0.20 $ &  0.74 &  0.52 \\
  7.00 -   8.00 & $ 0.54 \pm  0.034 \pm  0.18 $ &  0.64 &  0.45 \\
  8.00 -   9.00 & $ 0.56 \pm  0.032 \pm  0.17 $ &  0.54 &  0.39 \\
  9.00 -  10.0 & $ 0.41 \pm  0.029 \pm  0.14 $ &  0.46 &  0.34 \\
 10.0 -  12.0 & $ 0.37 \pm  0.019 \pm  0.12 $ &  0.38 &  0.28 \\
 12.0 -  14.0 & $ 0.280 \pm  0.017 \pm  0.100 $ &  0.303 &  0.228 \\
 14.0 -  16.0 & $ 0.201 \pm  0.015 \pm  0.083 $ &  0.245 &  0.179 \\
 16.0 -  18.0 & $ 0.179 \pm  0.013 \pm  0.066 $ &  0.203 &  0.154 \\
 18.0 -  20.0 & $ 0.119 \pm  0.011 \pm  0.052 $ &  0.170 &  0.132 \\
 20.0 -  25.0 & $ 0.112 \pm  0.0066 \pm  0.042 $ &  0.130 &  0.094 \\
 25.0 -  30.0 & $ 0.085 \pm  0.0056 \pm  0.030 $ &  0.096 &  0.078 \\
 30.0 -  35.0 & $ 0.076 \pm  0.0053 \pm  0.029 $ &  0.074 &  0.072 \\
 35.0 -  40.0 & $ 0.057 \pm  0.0047 \pm  0.024 $ &  0.056 &  0.056 \\
 40.0 -  45.0 & $ 0.040 \pm  0.0037 \pm  0.014 $ &  0.040 &  0.044 \\
 45.0 -  50.0 & $ 0.0276 \pm  0.0027 \pm  0.0062 $ &  0.0265 &  0.0294 \\
 50.0 -  60.0 & $ 0.0171 \pm  0.0015 \pm  0.0050 $ &  0.0165 &  0.0193 \\
 60.0 -  70.0 & $ 0.0066 \pm  0.0010 \pm  0.0024 $ &  0.0089 &  0.0105 \\
 70.0 -  80.0 & $ 0.0047 \pm  0.00076 \pm  0.0014 $ &  0.0050 &  0.0062 \\
 80.0 -  90.0 & $ 0.00316 \pm  0.00048 \pm  0.00067 $ &  0.00320 &  0.00376 \\
 90.0 - 100.0 & $ 0.00101 \pm  0.00035 \pm  0.00046 $ &  0.00182 &  0.00244 \\
100.0 - 120.0 & $ 0.00091 \pm  0.00016 \pm  0.00025 $ &  0.00109 &  0.00135 \\
120.0 - 140.0 & $ 0.00072 \pm  0.00017 \pm  0.00015 $ &  0.00055 &  0.00065 \\
140.0 - 160.0 & $ 0.000194 \pm  0.000086 \pm  0.000035 $ &  0.000284 &  0.000327 \\
160.0 - 200.0 & $ 0.000087 \pm  0.000036 \pm  0.000020 $ &  0.000089 &  0.000134 \\
\hline\hline
\end{tabular}
\end{center}
\end{table}

\pagebreak
\vfill

\renewcommand{\plotname}{qtQHSSV01}
\begin{center}
{\LARGE $P_T(\gamma\gamma)$ for $P_T(\gamma\gamma) > M(\gamma\gamma)$}
\end{center}

\begin{figure*}[!hp]
\includegraphics[width=9.0cm]{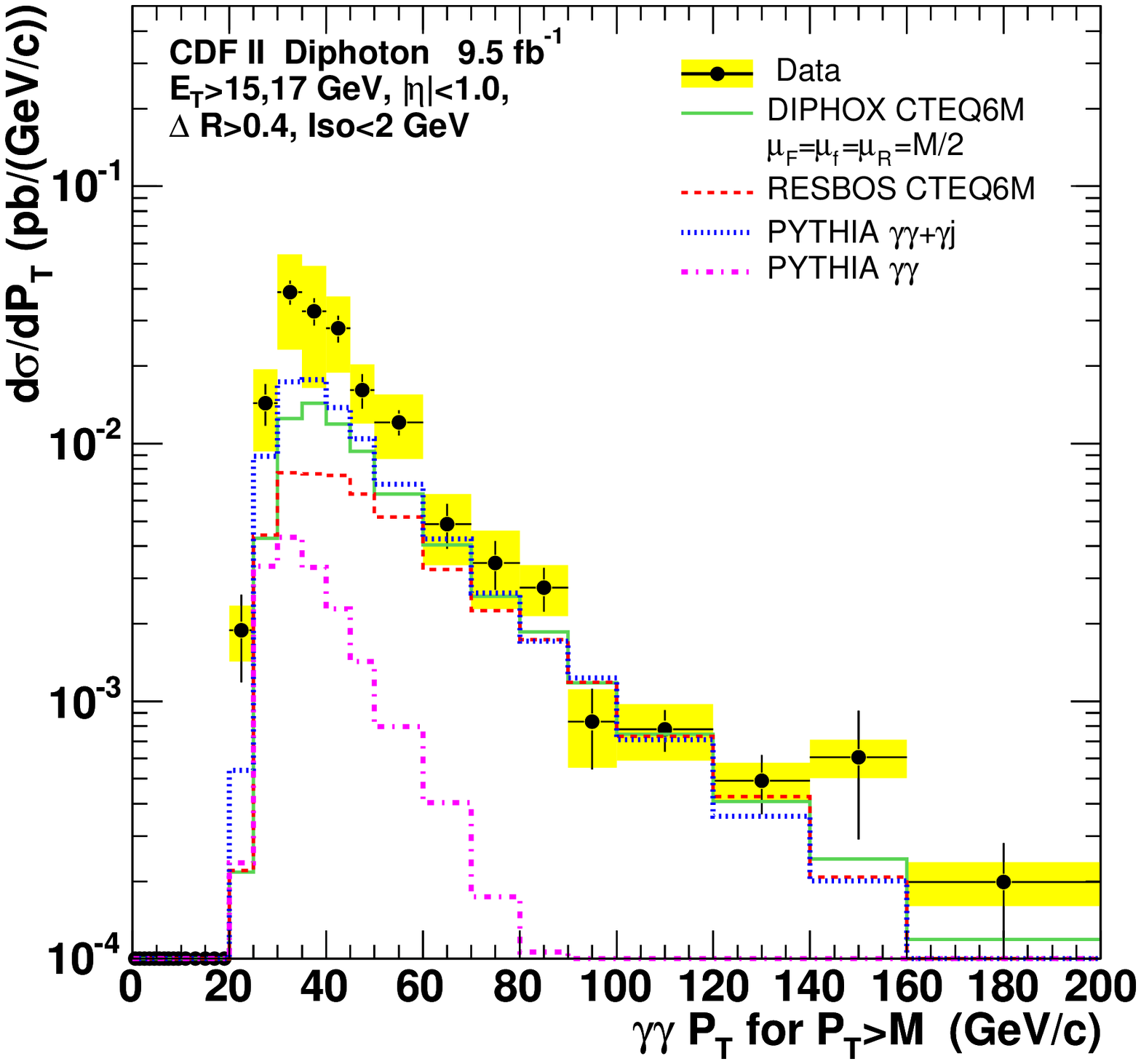}\hspace*{0cm}
\includegraphics[width=8.5cm]{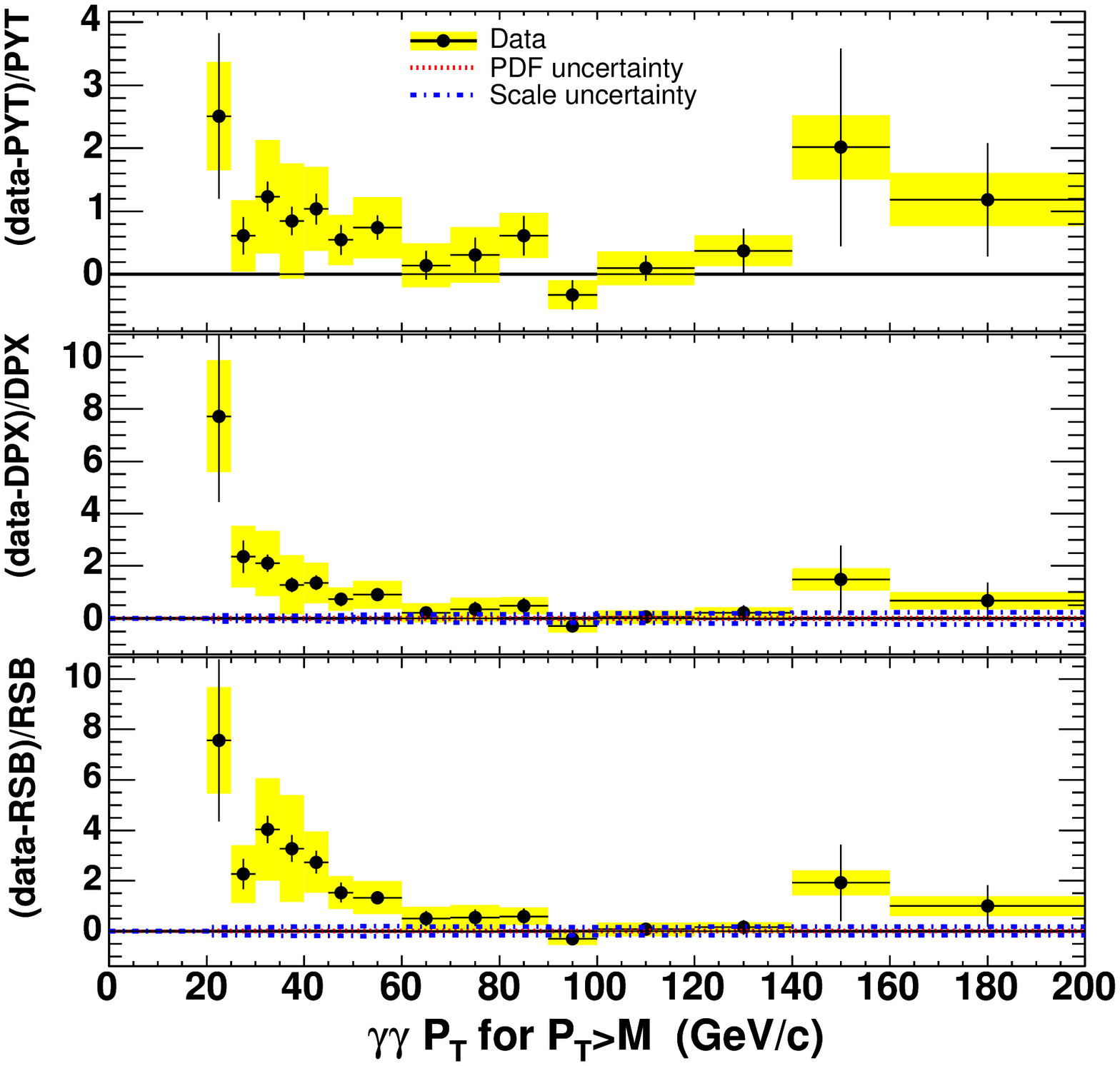}\hfill
\vspace*{0cm}
\includegraphics[width=9.0cm]{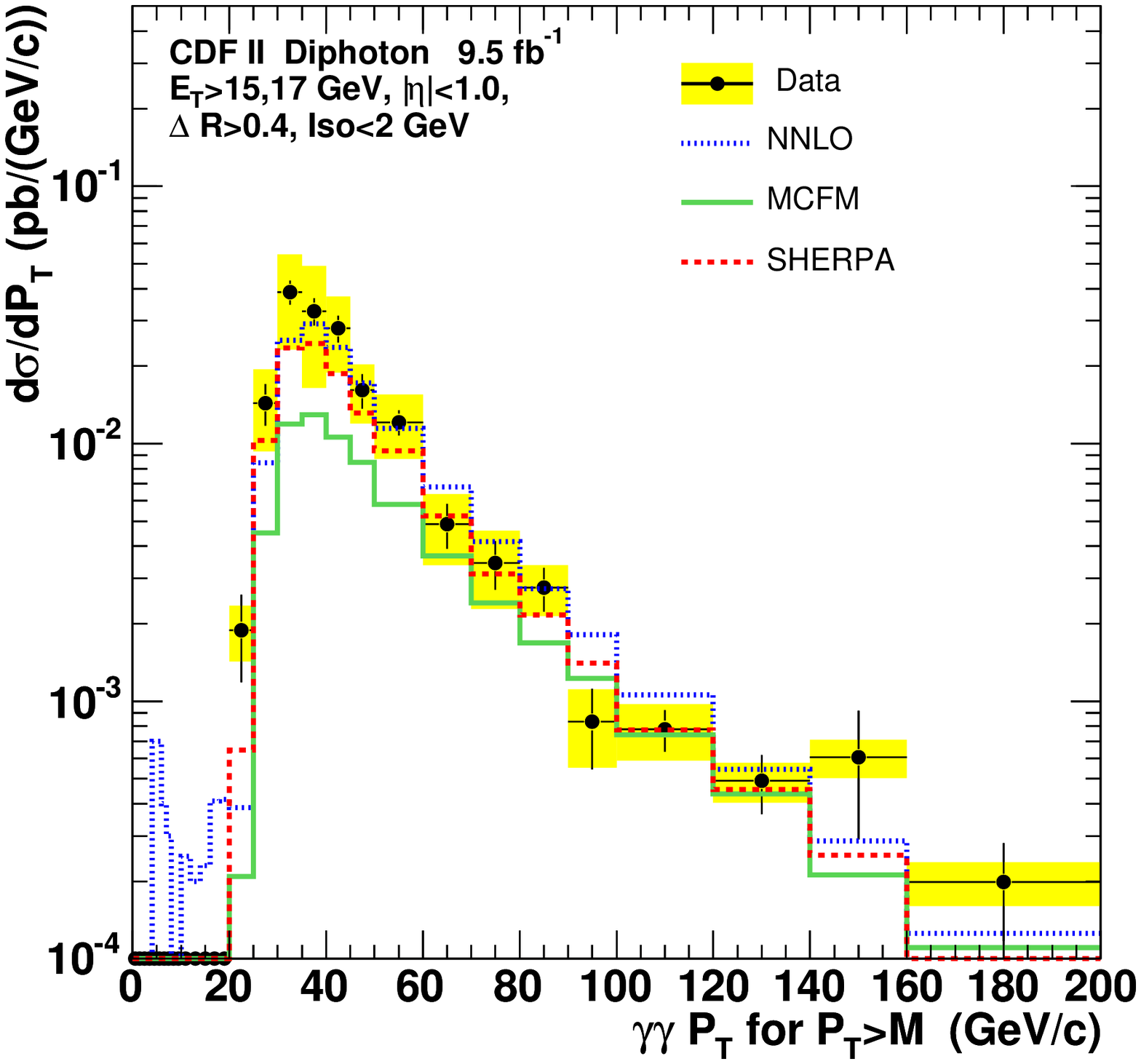}\hspace*{0cm}
\includegraphics[width=8.5cm]{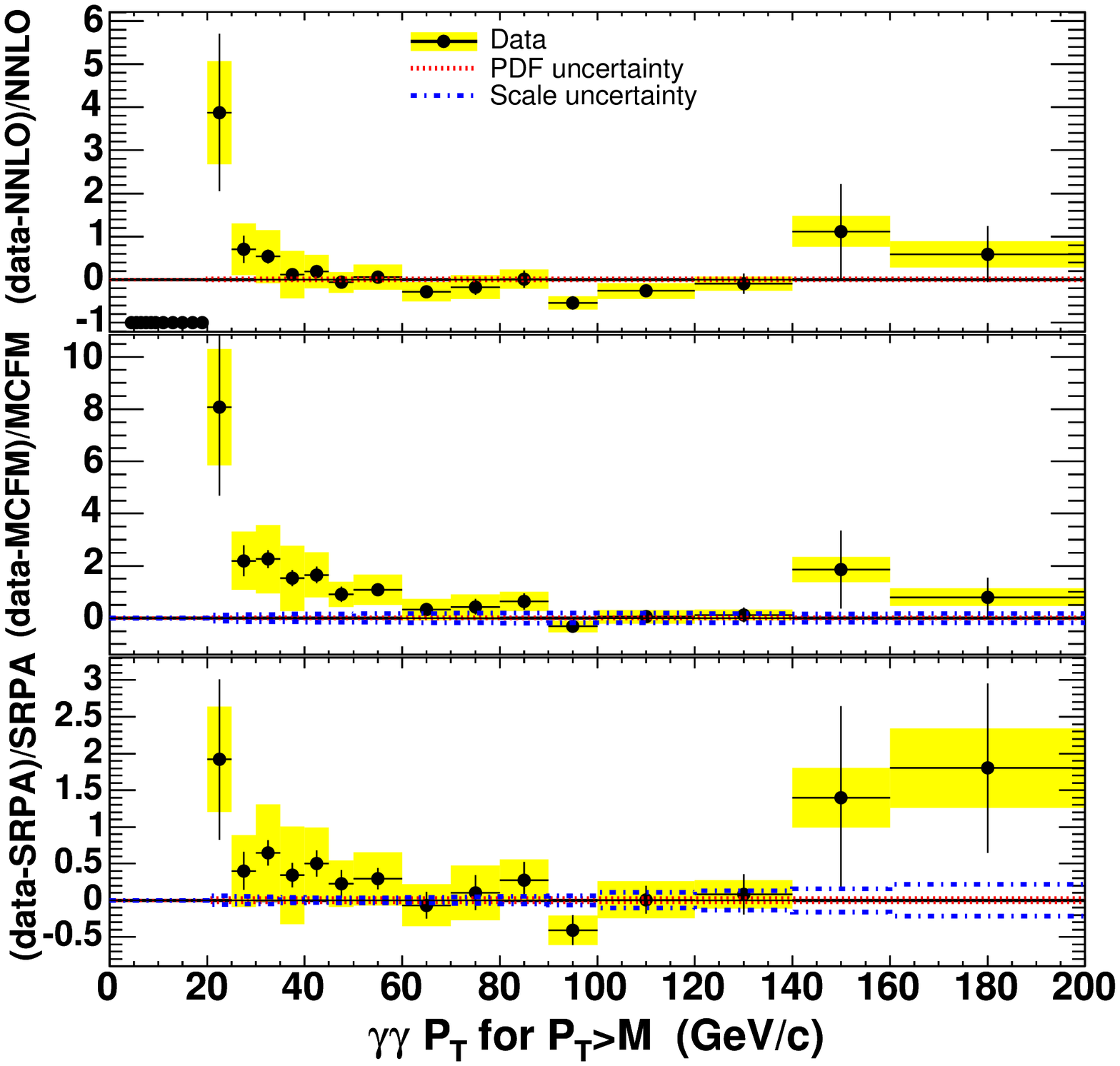}
\vspace*{-0.1cm}
\caption{The measured differential cross sections 
for $P_T(\gamma\gamma)$, when $P_T(\gamma\gamma) > M(\gamma\gamma)$, 
compared with six 
theoretical predictions discussed in the text. The left windows show the
absolute comparisons and the right windows show the fractional deviations
of the data from the theoretical predictions. Note that the vertical axis
scales differ between fractional deviation plots.}
\end{figure*}

\pagebreak
\vfill

\begin{table}[!hp]
\begin{center}
\begin{tabular}{cccc}
\hline\hline
 Bin & Cross Section (pb) & Sherpa & NNLO   \\ \hline
 20.0 -  25.0 & $ 0.00189 \pm  0.00071 \pm  0.00046 $ &  0.00065 &  0.00039 \\
 25.0 -  30.0 & $ 0.0144 \pm  0.0027 \pm  0.0050 $ &  0.0103 &  0.0084 \\
 30.0 -  35.0 & $ 0.039 \pm  0.0041 \pm  0.016 $ &  0.024 &  0.025 \\
 35.0 -  40.0 & $ 0.033 \pm  0.0040 \pm  0.016 $ &  0.024 &  0.029 \\
 40.0 -  45.0 & $ 0.0281 \pm  0.0034 \pm  0.0092 $ &  0.0187 &  0.0237 \\
 45.0 -  50.0 & $ 0.0161 \pm  0.0025 \pm  0.0041 $ &  0.0131 &  0.0172 \\
 50.0 -  60.0 & $ 0.0121 \pm  0.0013 \pm  0.0034 $ &  0.0094 &  0.0115 \\
 60.0 -  70.0 & $ 0.0049 \pm  0.00096 \pm  0.0015 $ &  0.0052 &  0.0068 \\
 70.0 -  80.0 & $ 0.0034 \pm  0.00073 \pm  0.0012 $ &  0.0031 &  0.0042 \\
 80.0 -  90.0 & $ 0.00276 \pm  0.00053 \pm  0.00061 $ &  0.00216 &  0.00273 \\
 90.0 - 100.0 & $ 0.00083 \pm  0.00029 \pm  0.00028 $ &  0.00140 &  0.00181 \\
100.0 - 120.0 & $ 0.00078 \pm  0.00014 \pm  0.00019 $ &  0.00077 &  0.00106 \\
120.0 - 140.0 & $ 0.00049 \pm  0.00013 \pm  0.000088 $ &  0.00045 &  0.00054 \\
140.0 - 160.0 & $ 0.00061 \pm  0.00031 \pm  0.00010 $ &  0.00025 &  0.00029 \\
160.0 - 200.0 & $ 0.000199 \pm  0.000082 \pm  0.000038 $ &  0.000071 &  0.000125 \\
\hline\hline
\end{tabular}
\end{center}
\end{table}

\pagebreak
\vfill

\renewcommand{\plotname}{qtQLSSV01}
\begin{center}
{\LARGE $P_T(\gamma\gamma)$ for $P_T(\gamma\gamma) < M(\gamma\gamma)$}
\end{center}

\begin{figure*}[!hp]
\includegraphics[width=9.0cm]{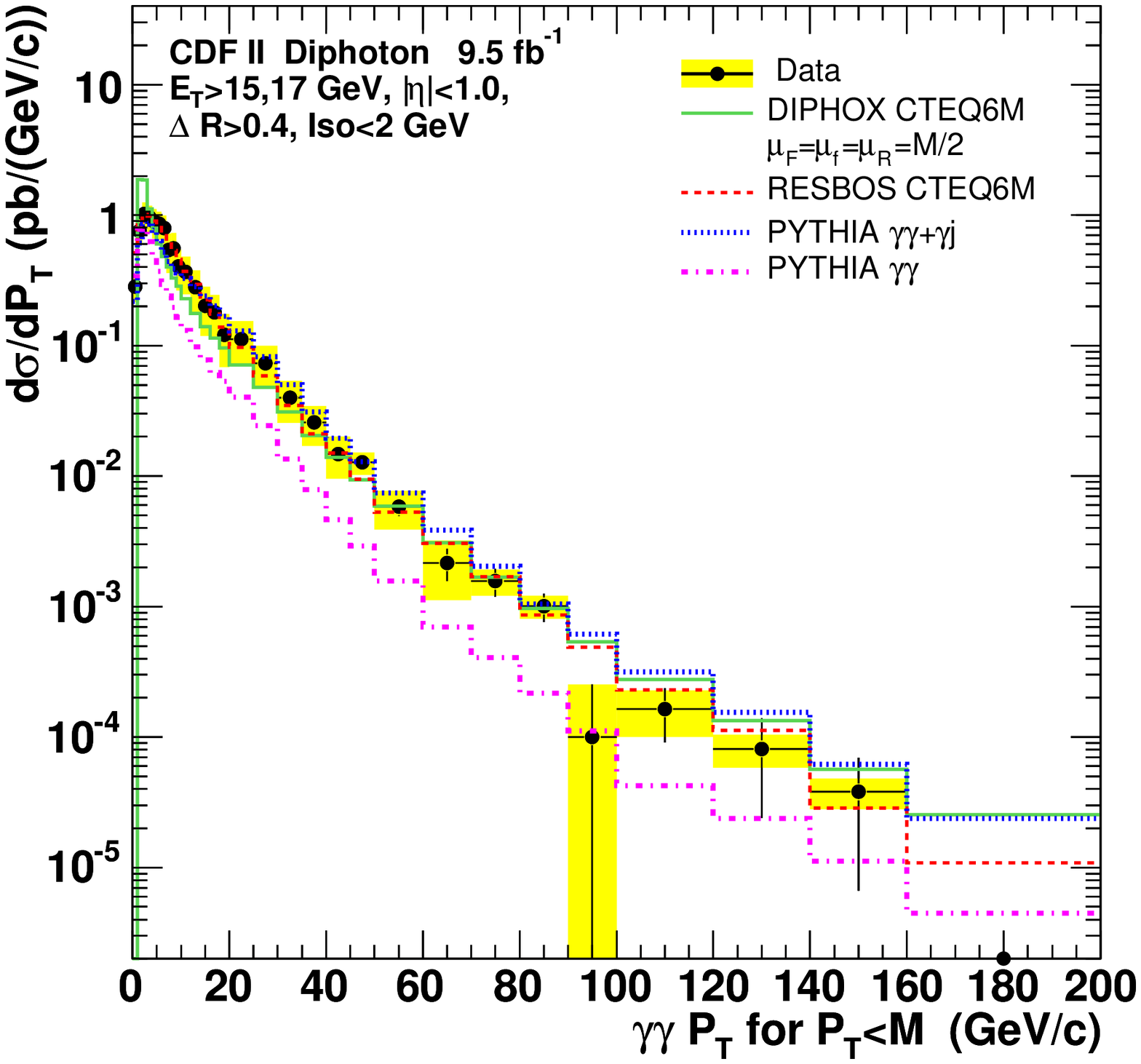}\hspace*{0cm}
\includegraphics[width=8.5cm]{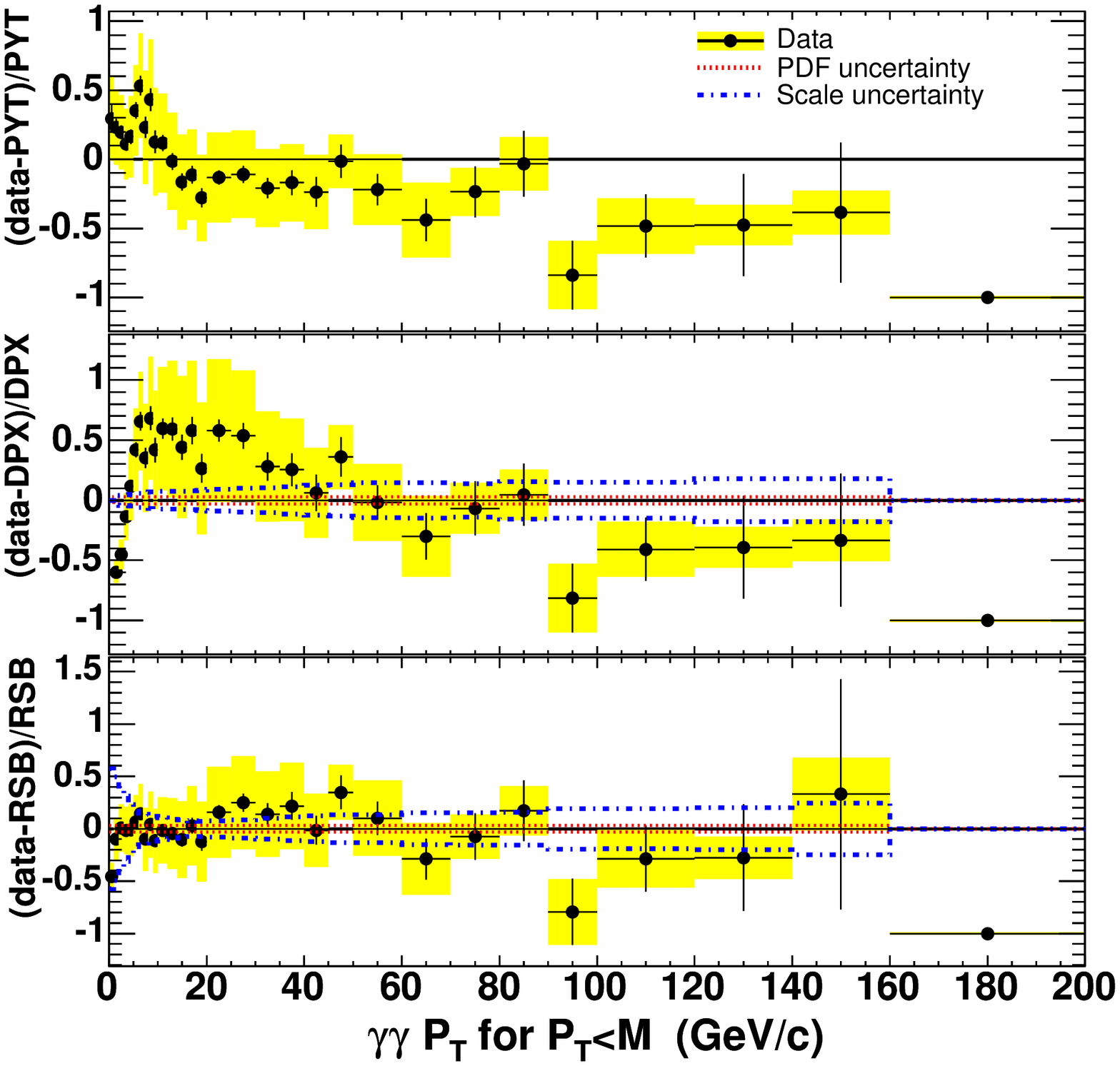}\hfill
\vspace*{0cm}
\includegraphics[width=9.0cm]{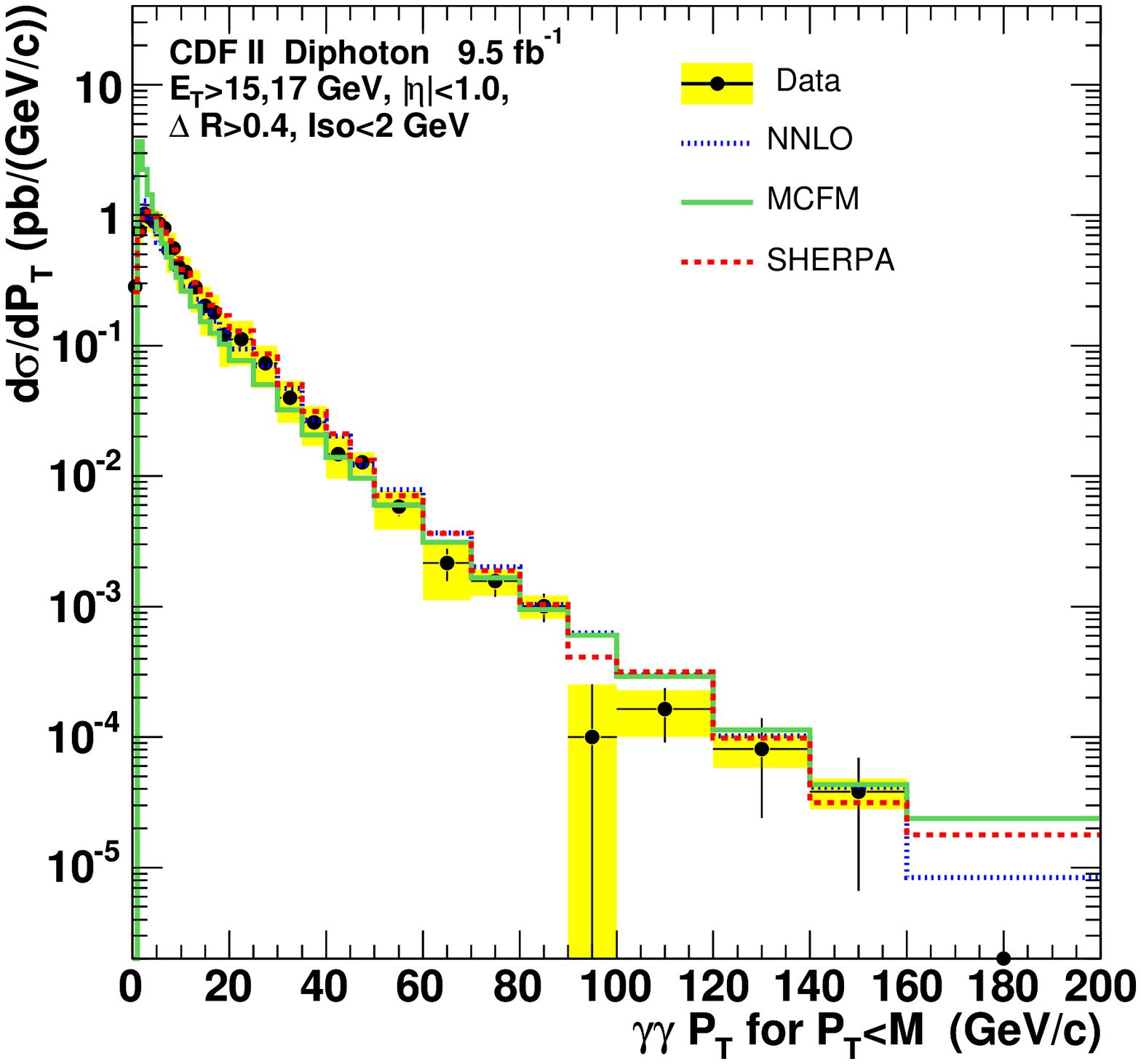}\hspace*{0cm}
\includegraphics[width=8.5cm]{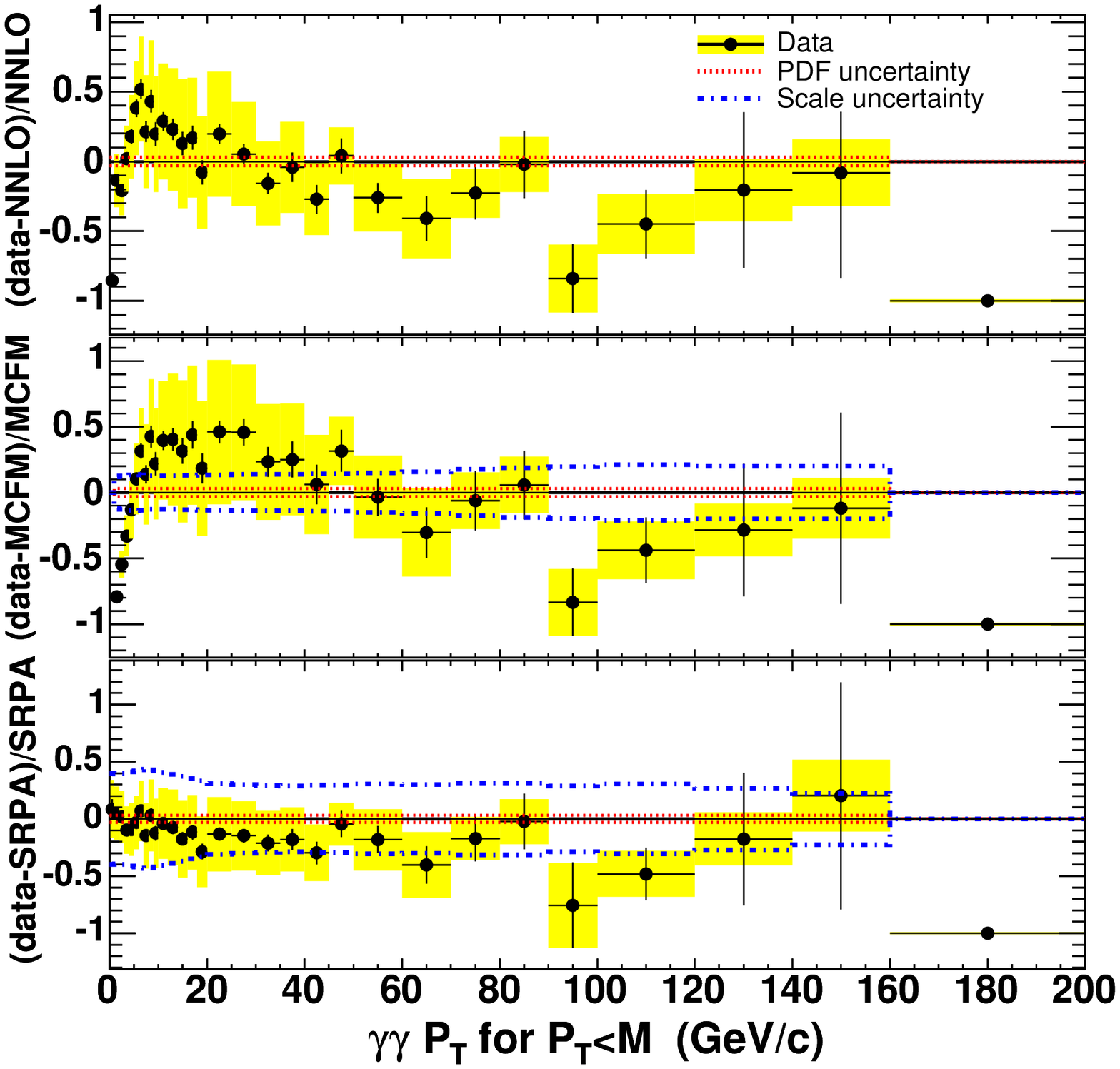}
\vspace*{-0.1cm}
\caption{The measured differential cross sections 
for $P_T(\gamma\gamma)$, when $P_T(\gamma\gamma) < M(\gamma\gamma)$, 
compared with six 
theoretical predictions discussed in the text. The left windows show the
absolute comparisons and the right windows show the fractional deviations
of the data from the theoretical predictions. Note that the vertical axis
scales differ between fractional deviation plots.}
\end{figure*}

\pagebreak
\vfill
\begin{table}[!hp]
\begin{center}
\begin{tabular}{cccc}
\hline\hline
 Bin & Cross Section (pb) & Sherpa & NNLO   \\ \hline
  0.00 -   1.00 & $ 0.282 \pm  0.023 \pm  0.066 $ &  0.260 &  1.952 \\
  1.00 -   2.00 & $ 0.76 \pm  0.038 \pm  0.17 $ &  0.72 &  0.88 \\
  2.00 -   3.00 & $ 1.02 \pm  0.041 \pm  0.23 $ &  1.00 &  1.29 \\
  3.00 -   4.00 & $ 0.96 \pm  0.040 \pm  0.22 $ &  1.07 &  0.94 \\
  4.00 -   5.00 & $ 0.89 \pm  0.040 \pm  0.23 $ &  0.98 &  0.76 \\
  5.00 -   6.00 & $ 0.86 \pm  0.039 \pm  0.21 $ &  0.88 &  0.62 \\
  6.00 -   7.00 & $ 0.80 \pm  0.037 \pm  0.20 $ &  0.74 &  0.52 \\
  7.00 -   8.00 & $ 0.54 \pm  0.034 \pm  0.18 $ &  0.64 &  0.45 \\
  8.00 -   9.00 & $ 0.56 \pm  0.032 \pm  0.17 $ &  0.54 &  0.39 \\
  9.00 -  10.0 & $ 0.41 \pm  0.029 \pm  0.14 $ &  0.46 &  0.34 \\
 10.0 -  12.0 & $ 0.37 \pm  0.019 \pm  0.12 $ &  0.38 &  0.28 \\
 12.0 -  14.0 & $ 0.280 \pm  0.017 \pm  0.100 $ &  0.303 &  0.228 \\
 14.0 -  16.0 & $ 0.201 \pm  0.015 \pm  0.083 $ &  0.245 &  0.179 \\
 16.0 -  18.0 & $ 0.180 \pm  0.013 \pm  0.066 $ &  0.203 &  0.154 \\
 18.0 -  20.0 & $ 0.121 \pm  0.012 \pm  0.053 $ &  0.170 &  0.132 \\
 20.0 -  25.0 & $ 0.112 \pm  0.0067 \pm  0.042 $ &  0.130 &  0.094 \\
 25.0 -  30.0 & $ 0.073 \pm  0.0051 \pm  0.026 $ &  0.086 &  0.070 \\
 30.0 -  35.0 & $ 0.040 \pm  0.0036 \pm  0.014 $ &  0.050 &  0.047 \\
 35.0 -  40.0 & $ 0.0258 \pm  0.0028 \pm  0.0088 $ &  0.0314 &  0.0269 \\
 40.0 -  45.0 & $ 0.0148 \pm  0.0021 \pm  0.0052 $ &  0.0211 &  0.0203 \\
 45.0 -  50.0 & $ 0.0127 \pm  0.0015 \pm  0.0025 $ &  0.0133 &  0.0122 \\
 50.0 -  60.0 & $ 0.0058 \pm  0.00085 \pm  0.0019 $ &  0.0071 &  0.0079 \\
 60.0 -  70.0 & $ 0.0022 \pm  0.00060 \pm  0.0010 $ &  0.0036 &  0.0037 \\
 70.0 -  80.0 & $ 0.00156 \pm  0.00038 \pm  0.00036 $ &  0.00189 &  0.00202 \\
 80.0 -  90.0 & $ 0.00101 \pm  0.00025 \pm  0.00020 $ &  0.00103 &  0.00103 \\
 90.0 - 100.0 & $ 0.00010 \pm  0.00015 \pm  0.00015 $ &  0.00041 &  0.00063 \\
100.0 - 120.0 & $ 0.000164 \pm  0.000073 \pm  0.000064 $ &  0.000316 &  0.000297 \\
120.0 - 140.0 & $ 0.000081 \pm  0.000057 \pm  0.000023 $ &  0.000099 &  0.000102 \\
140.0 - 160.0 & $ 0.000038 \pm  0.000031 \pm  0.0000099 $ &  0.000032 &  0.000041 \\
\hline\hline
\end{tabular}
\end{center}
\end{table}

\pagebreak
\vfill
\renewcommand{\plotname}{dphiSS01}

\begin{center}
{\LARGE $\Delta\Phi(\gamma\gamma)$}
\end{center}

\begin{figure*}[!hp]
\includegraphics[width=9.0cm]{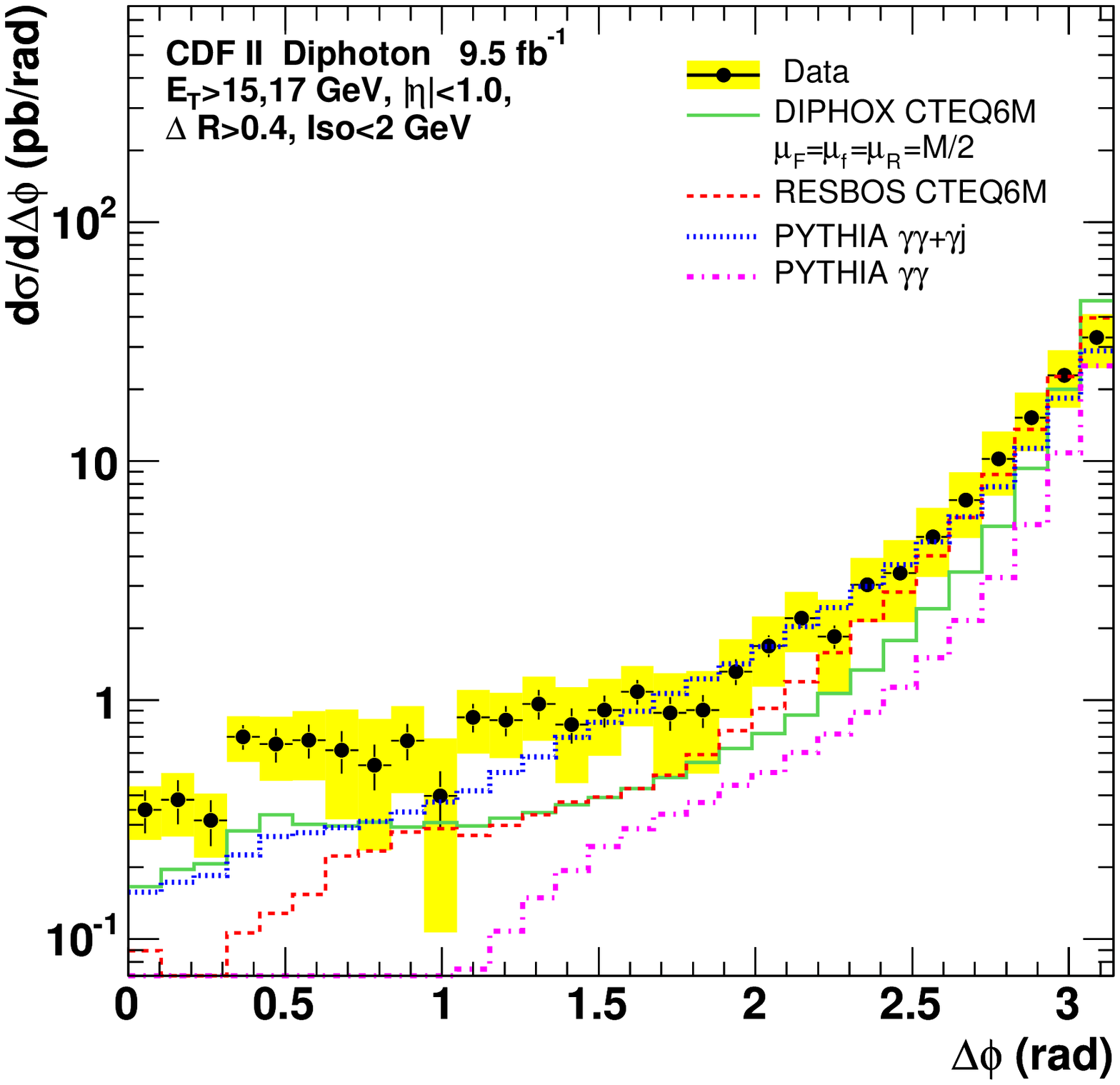}\hspace*{0cm}
\includegraphics[width=8.5cm]{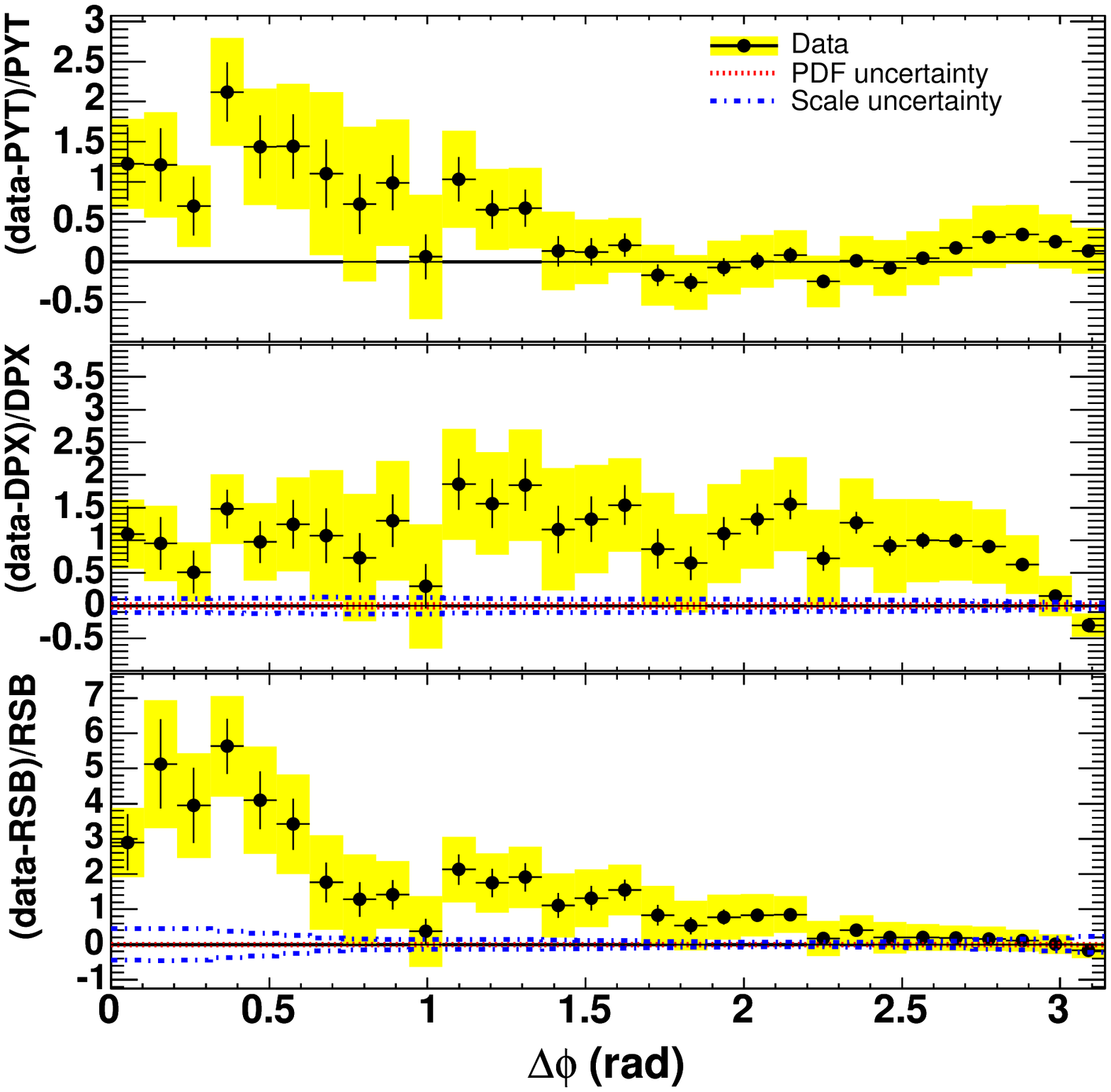}\hfill
\vspace*{0cm}
\includegraphics[width=9.0cm]{TXSec_\plotname_2_10Dec12.eps}\hspace*{0cm}
\includegraphics[width=8.5cm]{TXSec_\plotname_allratio2_23Jan13.eps}
\vspace*{-0.1cm}
\caption{The measured differential cross sections 
for $\Delta\Phi(\gamma\gamma)$ 
compared with six 
theoretical predictions discussed in the text. The left windows show the
absolute comparisons and the right windows show the fractional deviations
of the data from the theoretical predictions. Note that the vertical axis
scales differ between fractional deviation plots.}
\end{figure*}

\pagebreak
\vfill

\begin{table}[!hp]
\begin{center}
\begin{tabular}{cccc}
\hline\hline
 Bin & Cross Section (pb) & Sherpa & NNLO   \\ \hline
  0.00 -   0.105 & $ 0.348 \pm  0.071 \pm  0.088 $ &  0.243 &  0.317 \\
  0.105 -   0.209 & $ 0.38 \pm  0.079 \pm  0.11 $ &  0.23 &  0.34 \\
  0.209 -   0.314 & $ 0.313 \pm  0.067 \pm  0.094 $ &  0.240 &  0.377 \\
  0.314 -   0.419 & $ 0.70 \pm  0.083 \pm  0.15 $ &  0.34 &  0.48 \\
  0.419 -   0.524 & $ 0.65 \pm  0.11 \pm  0.20 $ &  0.37 &  0.63 \\
  0.524 -   0.628 & $ 0.68 \pm  0.11 \pm  0.22 $ &  0.43 &  0.54 \\
  0.628 -   0.733 & $ 0.62 \pm  0.12 \pm  0.30 $ &  0.40 &  0.52 \\
  0.733 -   0.838 & $ 0.53 \pm  0.12 \pm  0.30 $ &  0.43 &  0.56 \\
  0.838 -   0.942 & $ 0.68 \pm  0.12 \pm  0.27 $ &  0.51 &  0.54 \\
  0.942 -   1.05 & $ 0.40 \pm  0.10 \pm  0.29 $ &  0.49 &  0.54 \\
  1.05 -   1.15 & $ 0.85 \pm  0.12 \pm  0.25 $ &  0.54 &  0.56 \\
  1.15 -   1.26 & $ 0.82 \pm  0.12 \pm  0.25 $ &  0.60 &  0.57 \\
  1.26 -   1.36 & $ 0.96 \pm  0.13 \pm  0.29 $ &  0.71 &  0.62 \\
  1.36 -   1.47 & $ 0.79 \pm  0.13 \pm  0.34 $ &  0.77 &  0.63 \\
  1.47 -   1.57 & $ 0.91 \pm  0.14 \pm  0.32 $ &  0.89 &  0.67 \\
  1.57 -   1.68 & $ 1.08 \pm  0.13 \pm  0.31 $ &  0.99 &  0.74 \\
  1.68 -   1.78 & $ 0.88 \pm  0.14 \pm  0.41 $ &  1.11 &  0.81 \\
  1.78 -   1.88 & $ 0.91 \pm  0.14 \pm  0.41 $ &  1.35 &  0.91 \\
  1.88 -   1.99 & $ 1.32 \pm  0.16 \pm  0.47 $ &  1.44 &  1.03 \\
  1.99 -   2.09 & $ 1.69 \pm  0.17 \pm  0.55 $ &  1.70 &  1.18 \\
  2.09 -   2.20 & $ 2.20 \pm  0.19 \pm  0.62 $ &  1.94 &  1.40 \\
  2.20 -   2.30 & $ 1.84 \pm  0.21 \pm  0.78 $ &  2.30 &  1.93 \\
  2.30 -   2.41 & $ 3.03 \pm  0.22 \pm  0.91 $ &  2.87 &  2.10 \\
  2.41 -   2.51 & $ 3.4 \pm  0.27 \pm  1.3 $ &  3.8 &  2.7 \\
  2.51 -   2.62 & $ 4.8 \pm  0.30 \pm  1.5 $ &  4.8 &  3.5 \\
  2.62 -   2.72 & $ 6.9 \pm  0.37 \pm  2.1 $ &  6.7 &  5.1 \\
  2.72 -   2.83 & $10.2 \pm  0.46 \pm  3.0 $ &  9.5 &  7.1 \\
  2.83 -   2.93 & $15.1 \pm  0.54 \pm  4.2 $ & 14.6 & 11.1 \\
  2.93 -   3.04 & $22.9 \pm  0.67 \pm  6.2 $ & 23.2 & 19.2 \\
  3.04 -   3.14 & $32.9 \pm  0.75 \pm  8.3 $ & 35.1 & 45.8 \\
\hline\hline
\end{tabular}
\end{center}
\end{table}

\pagebreak
\vfill

\renewcommand{\plotname}{dphiQHSS01}
\begin{center}
{\LARGE $\Delta\Phi(\gamma\gamma)$ for $P_T(\gamma\gamma) > M(\gamma\gamma)$}
\end{center}

\begin{figure*}[!hp]
\includegraphics[width=9.0cm]{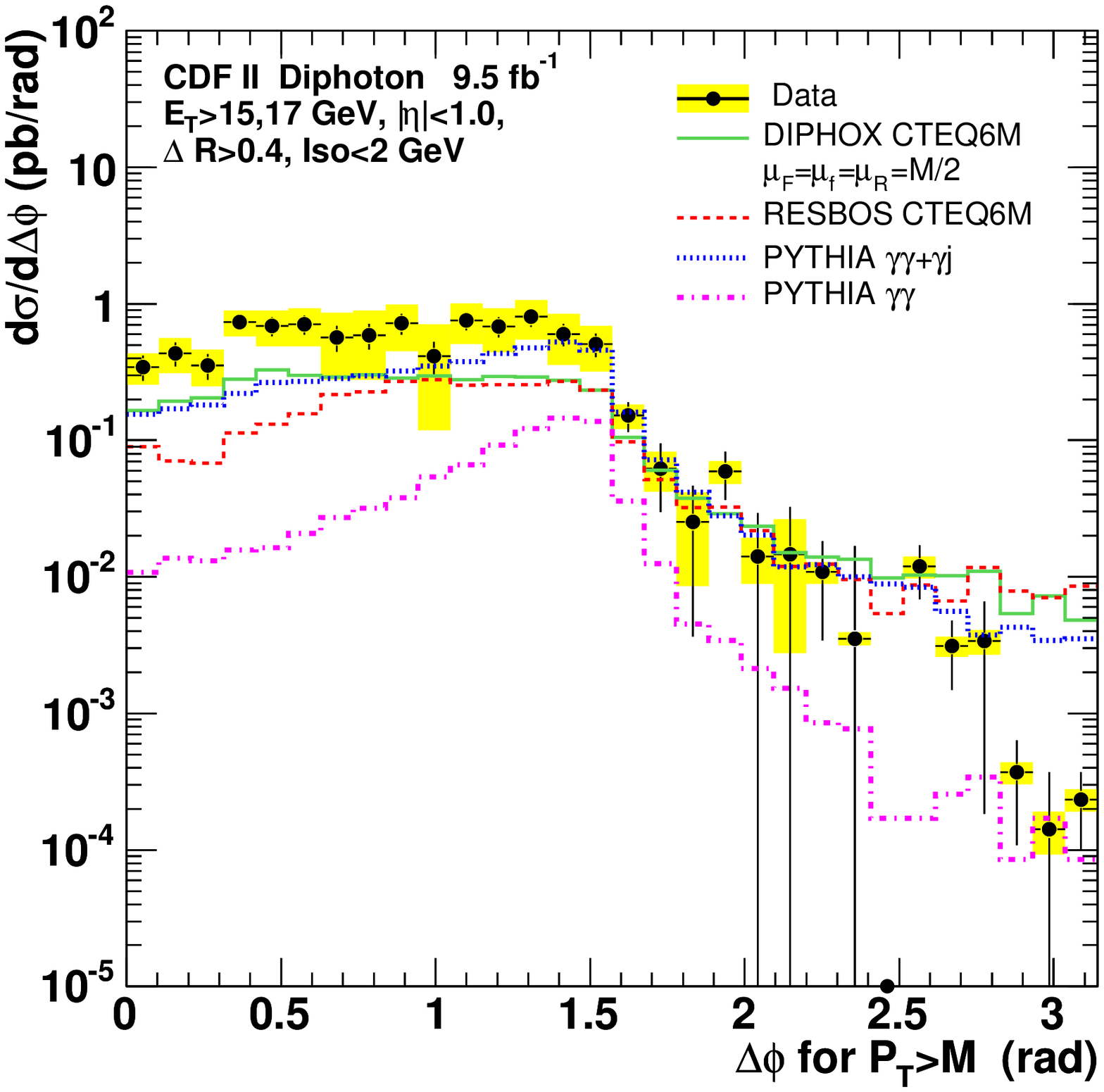}\hspace*{0cm}
\includegraphics[width=8.5cm]{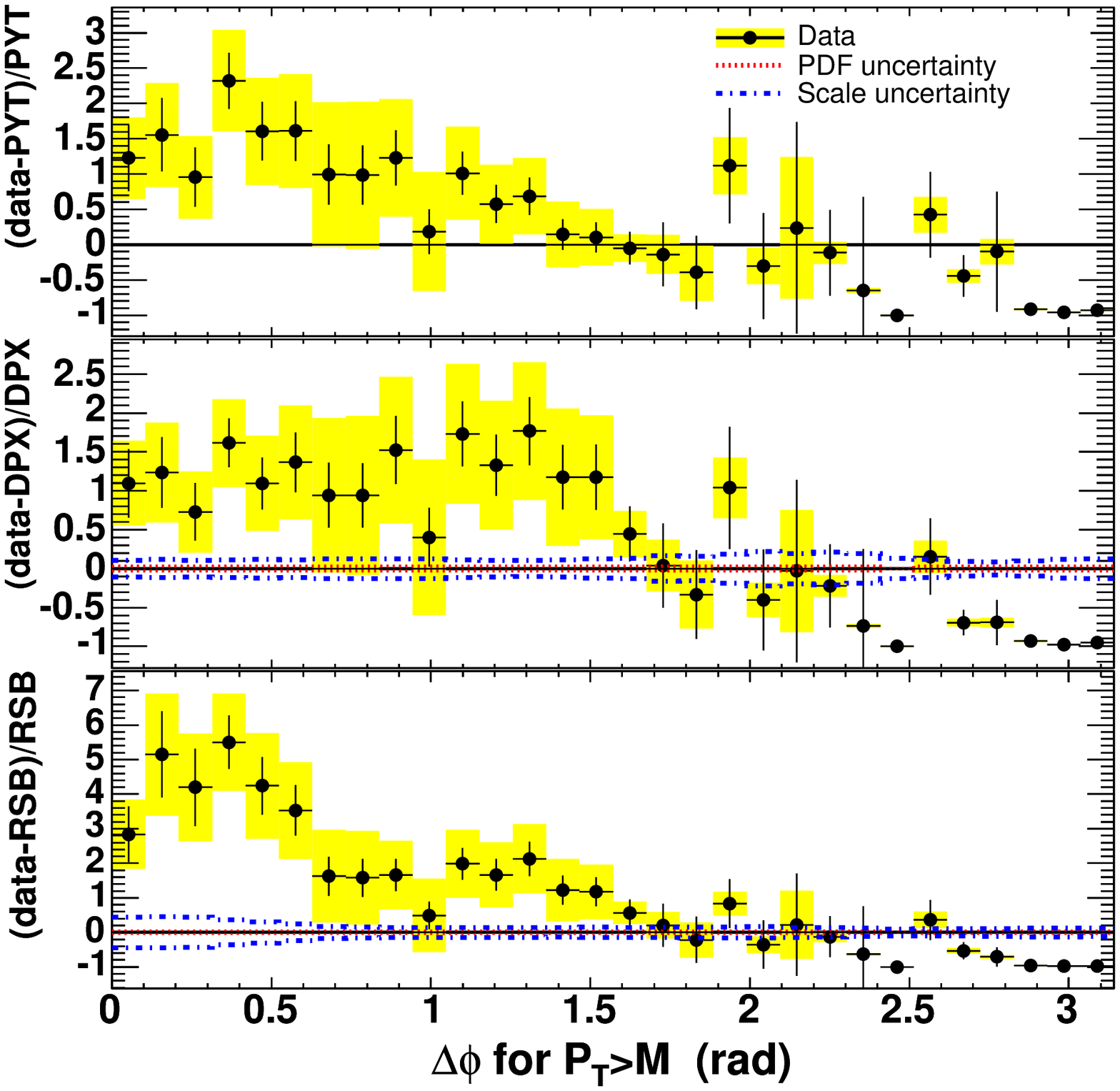}\hfill
\vspace*{0cm}
\includegraphics[width=9.0cm]{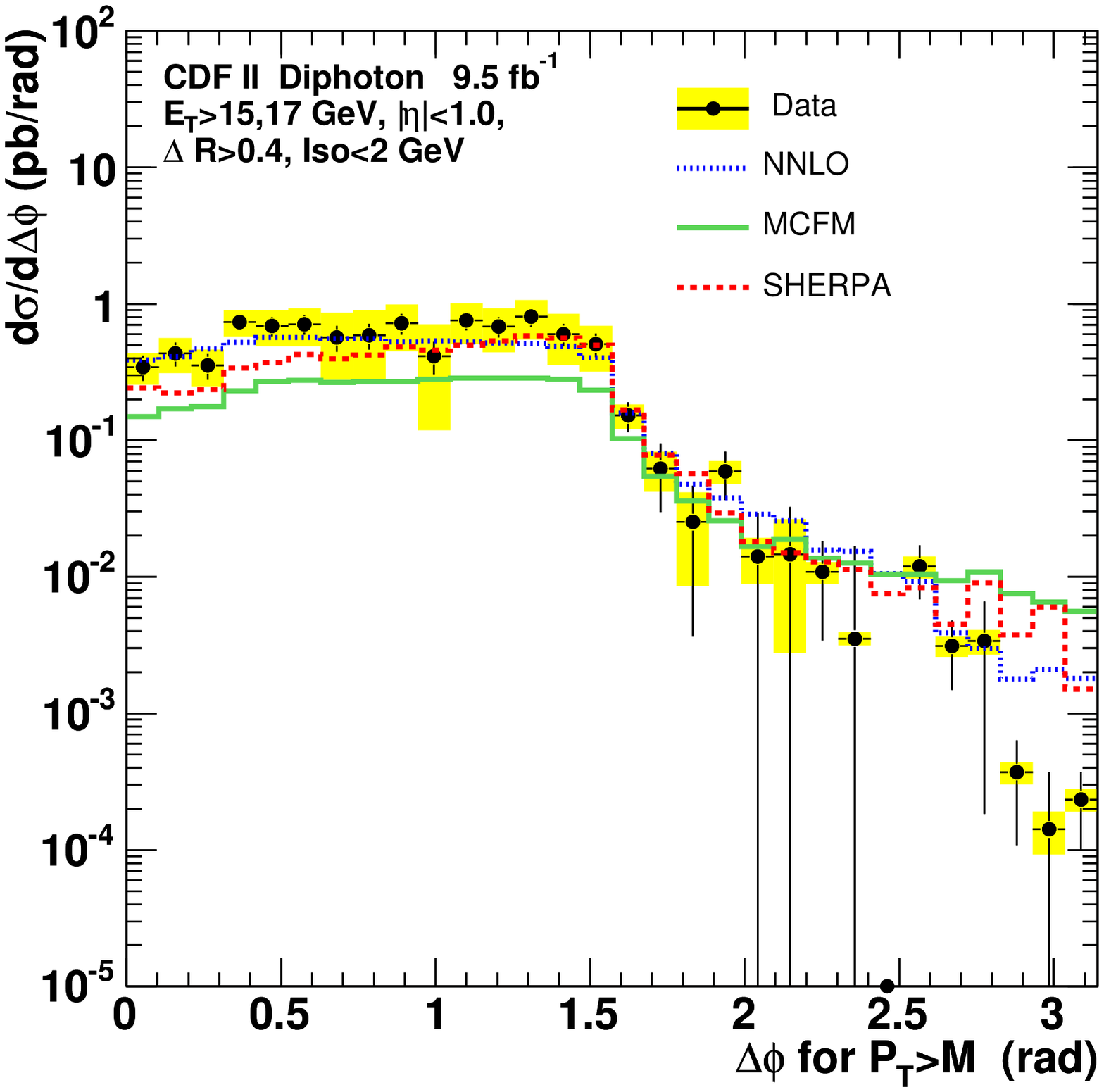}\hspace*{0cm}
\includegraphics[width=8.5cm]{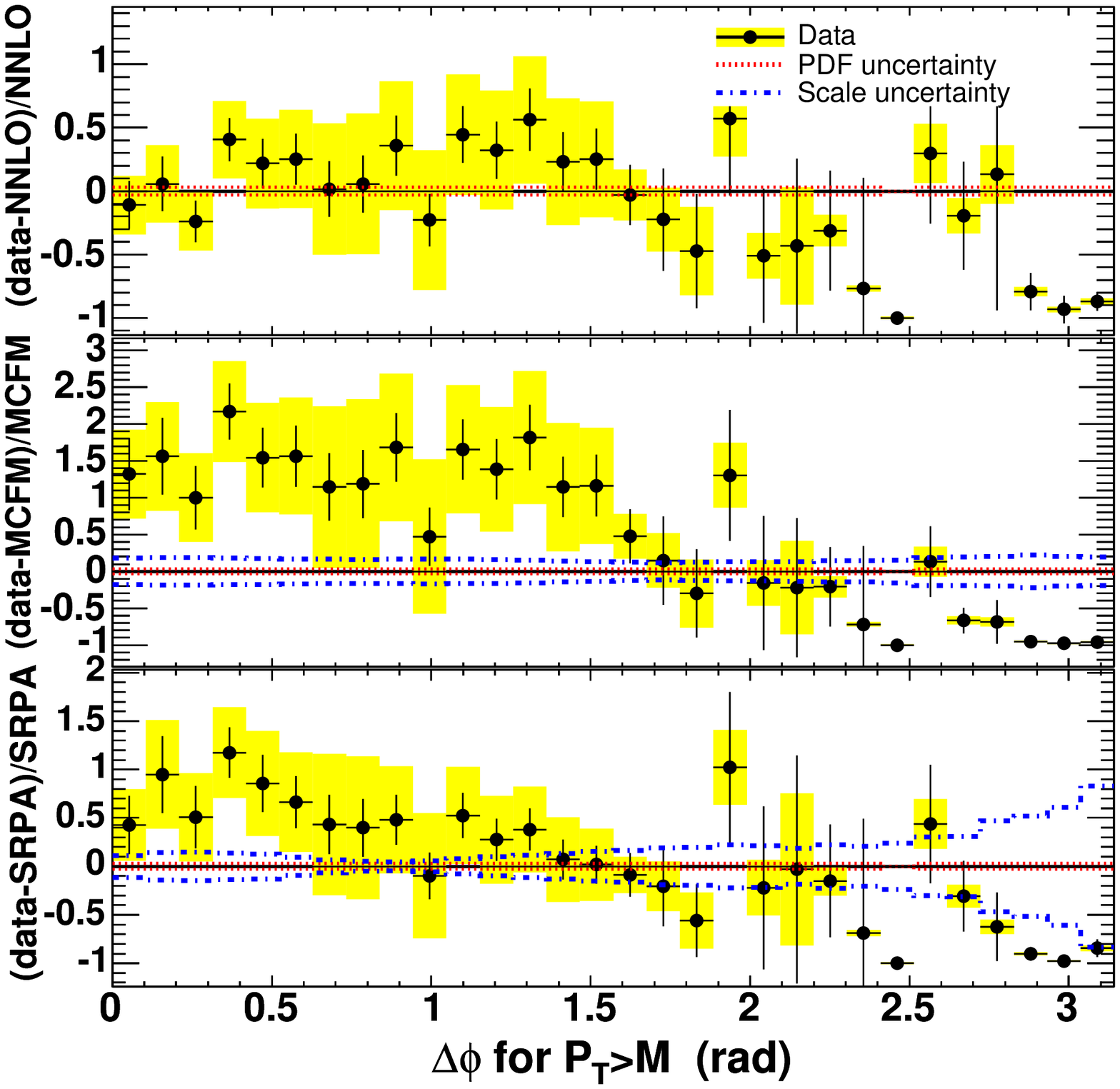}
\vspace*{-0.1cm}
\caption{The measured differential cross sections 
for $\Delta\Phi(\gamma\gamma)$, when $P_T(\gamma\gamma) > M(\gamma\gamma)$,
compared with six 
theoretical predictions discussed in the text. The left windows show the
absolute comparisons and the right windows show the fractional deviations
of the data from the theoretical predictions. Note that the vertical axis
scales differ between fractional deviation plots.}
\end{figure*}

\pagebreak
\vfill
\begin{table}[!hp]
\begin{center}
\begin{tabular}{cccc}
\hline\hline
 Bin & Cross Section (pb) & Sherpa & NNLO   \\ \hline
  0.00 -   0.105 & $ 0.345 \pm  0.073 \pm  0.090 $ &  0.242 &  0.388 \\
  0.105 -   0.209 & $ 0.43 \pm  0.089 \pm  0.13 $ &  0.22 &  0.41 \\
  0.209 -   0.314 & $ 0.35 \pm  0.076 \pm  0.11 $ &  0.24 &  0.46 \\
  0.314 -   0.419 & $ 0.74 \pm  0.088 \pm  0.16 $ &  0.34 &  0.52 \\
  0.419 -   0.524 & $ 0.69 \pm  0.11 \pm  0.20 $ &  0.37 &  0.57 \\
  0.524 -   0.628 & $ 0.71 \pm  0.11 \pm  0.22 $ &  0.43 &  0.56 \\
  0.628 -   0.733 & $ 0.57 \pm  0.12 \pm  0.29 $ &  0.40 &  0.56 \\
  0.733 -   0.838 & $ 0.59 \pm  0.12 \pm  0.31 $ &  0.42 &  0.56 \\
  0.838 -   0.942 & $ 0.72 \pm  0.12 \pm  0.27 $ &  0.49 &  0.53 \\
  0.942 -   1.05 & $ 0.41 \pm  0.11 \pm  0.30 $ &  0.46 &  0.54 \\
  1.05 -   1.15 & $ 0.76 \pm  0.12 \pm  0.25 $ &  0.50 &  0.52 \\
  1.15 -   1.26 & $ 0.68 \pm  0.12 \pm  0.24 $ &  0.53 &  0.52 \\
  1.26 -   1.36 & $ 0.80 \pm  0.13 \pm  0.26 $ &  0.58 &  0.51 \\
  1.36 -   1.47 & $ 0.60 \pm  0.11 \pm  0.24 $ &  0.56 &  0.49 \\
  1.47 -   1.57 & $ 0.51 \pm  0.098 \pm  0.18 $ &  0.50 &  0.40 \\
  1.57 -   1.68 & $ 0.153 \pm  0.038 \pm  0.031 $ &  0.167 &  0.157 \\
  1.68 -   1.78 & $ 0.062 \pm  0.033 \pm  0.020 $ &  0.078 &  0.080 \\
  1.78 -   1.88 & $ 0.025 \pm  0.022 \pm  0.017 $ &  0.057 &  0.048 \\
  1.88 -   1.99 & $ 0.059 \pm  0.023 \pm  0.011 $ &  0.029 &  0.038 \\
  1.99 -   2.09 & $ 0.014 \pm  0.015 \pm  0.0052 $ &  0.018 &  0.029 \\
  2.09 -   2.20 & $ 0.015 \pm  0.018 \pm  0.012 $ &  0.015 &  0.026 \\
  2.20 -   2.30 & $ 0.0109 \pm  0.0075 \pm  0.0020 $ &  0.0128 &  0.0158 \\
  2.30 -   2.41 & $ 0.004 \pm  0.013 \pm  0.00038 $ &  0.011 &  0.015 \\
  2.41 -   2.51 & $ 0.0000 \pm  0.0035 \pm  0.00000000001 $ &  0.0075 &  0.0106 \\
  2.51 -   2.62 & $ 0.0119 \pm  0.0051 \pm  0.0021 $ &  0.0083 &  0.0092 \\
  2.62 -   2.72 & $ 0.0031 \pm  0.0016 \pm  0.00054 $ &  0.0045 &  0.0039 \\
  2.72 -   2.83 & $ 0.0034 \pm  0.0032 \pm  0.00069 $ &  0.0090 &  0.0030 \\
  2.83 -   2.93 & $ 0.00037 \pm  0.00026 \pm  0.000067 $ &  0.00377 &  0.00179 \\
  2.93 -   3.04 & $ 0.00014 \pm  0.00023 \pm  0.000049 $ &  0.00603 &  0.00209 \\
  3.04 -   3.14 & $ 0.00023 \pm  0.00014 \pm  0.000045 $ &  0.00151 &  0.00181 \\
\hline\hline
\end{tabular}
\end{center}
\end{table}

\pagebreak
\vfill

\renewcommand{\plotname}{dphiQLSS01}
\begin{center}
{\LARGE $\Delta\Phi(\gamma\gamma)$ for $P_T(\gamma\gamma) < M(\gamma\gamma)$}
\end{center}

\begin{figure*}[!hp]
\includegraphics[width=9.0cm]{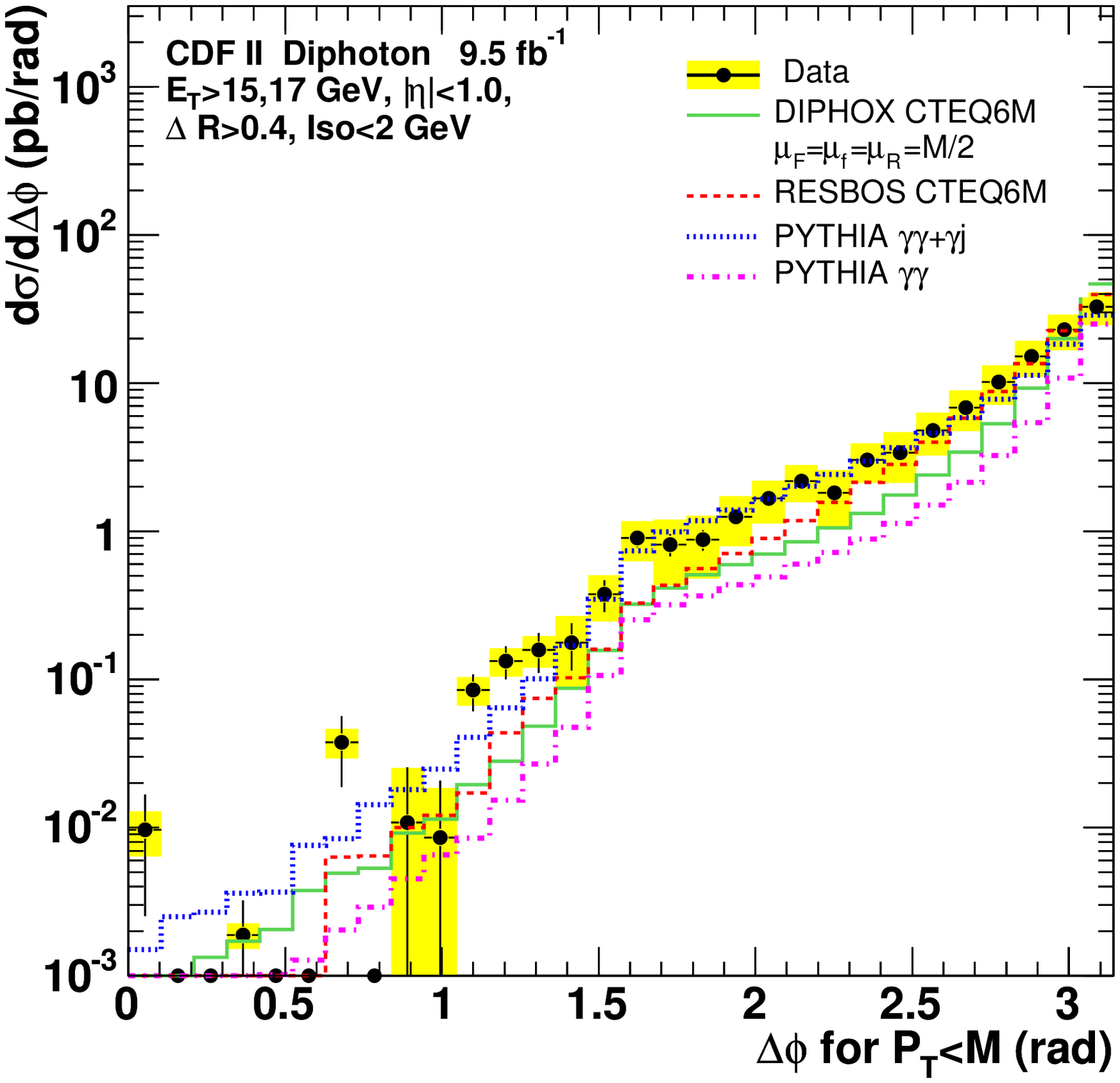}\hspace*{0cm}
\includegraphics[width=8.5cm]{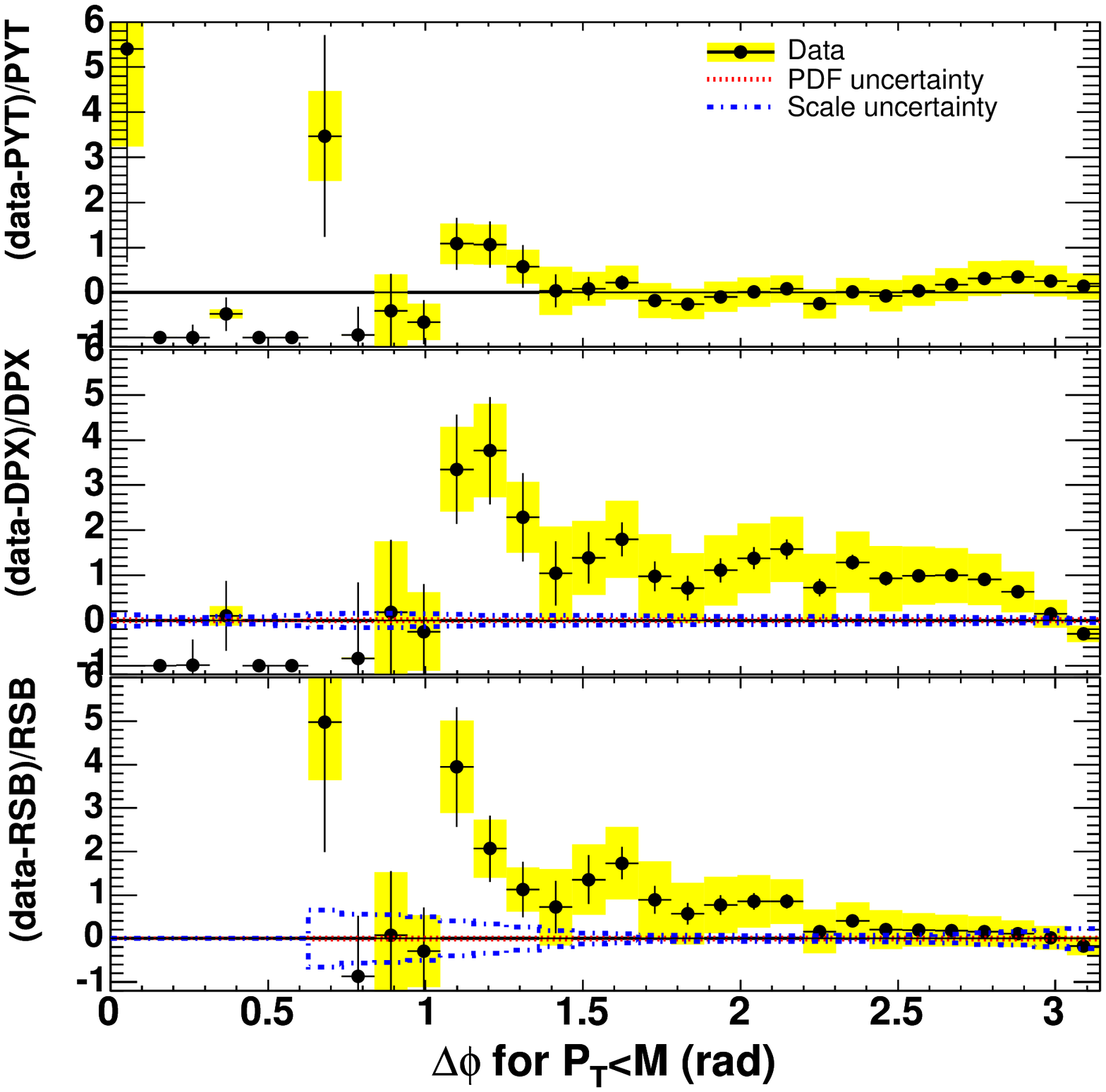}\hfill
\vspace*{0cm}
\includegraphics[width=9.0cm]{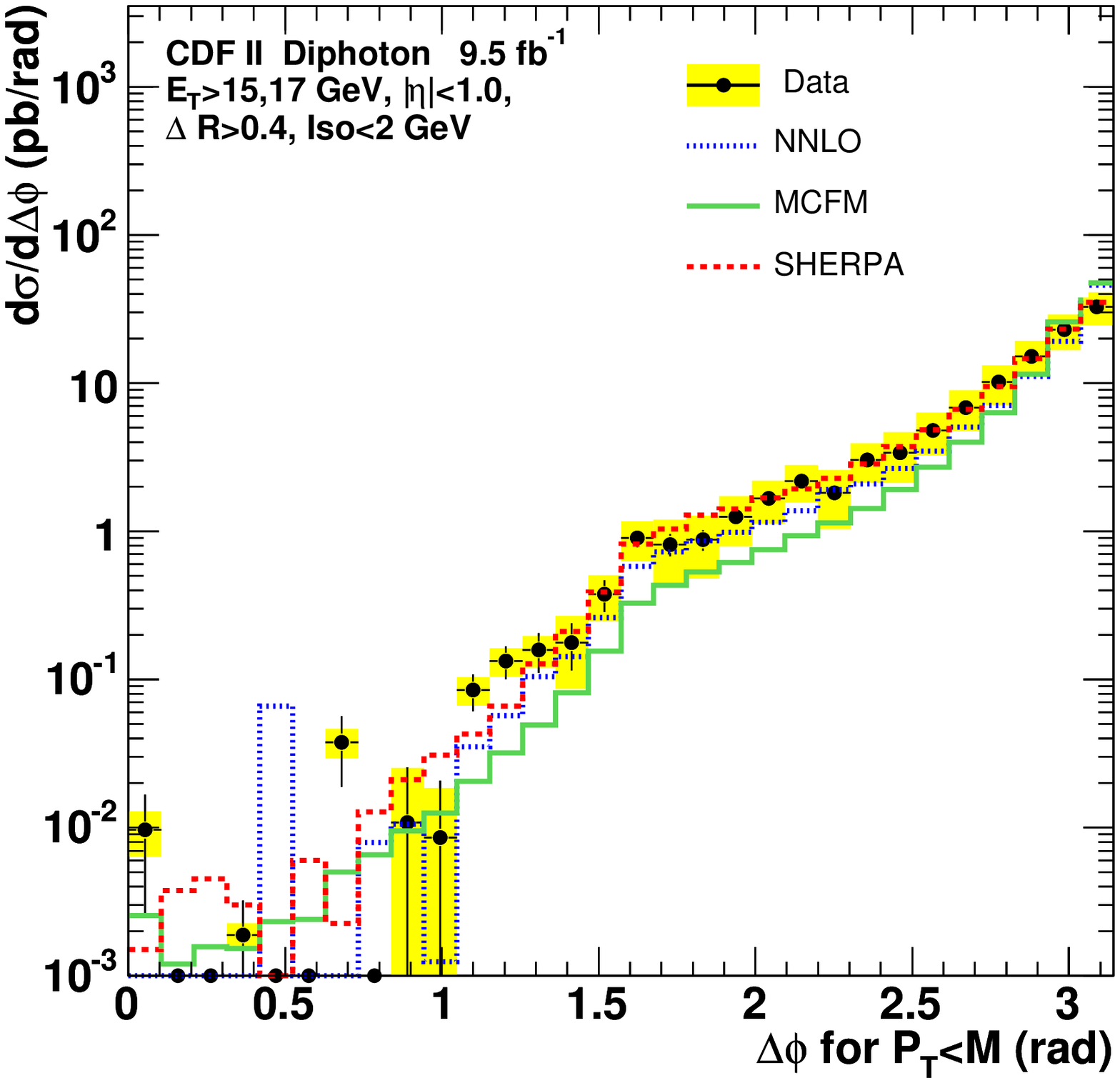}\hspace*{0cm}
\includegraphics[width=8.5cm]{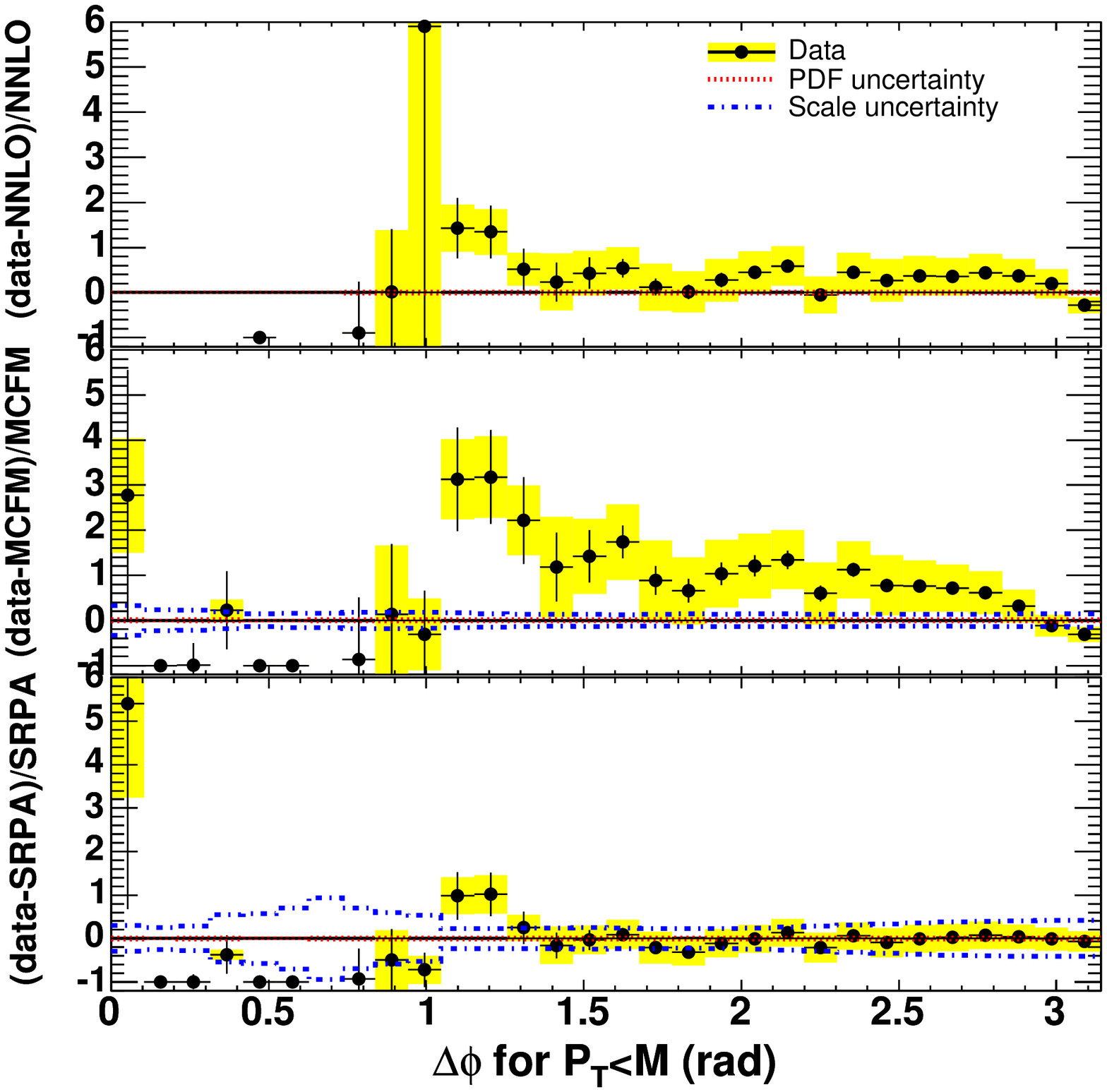}
\vspace*{-0.1cm}
\caption{The measured differential cross sections 
for $\Delta\Phi(\gamma\gamma)$, when $P_T(\gamma\gamma) < M(\gamma\gamma)$,
compared with six 
theoretical predictions discussed in the text. The left windows show the
absolute comparisons and the right windows show the fractional deviations
of the data from the theoretical predictions. Note that the vertical axis
scales differ between fractional deviation plots.}
\end{figure*}

\pagebreak
\vfill

\begin{table}[!hp]
\begin{center}
\begin{tabular}{cccc}
\hline\hline
 Bin & Cross Section (pb) & Sherpa & NNLO   \\ \hline
  0.00 -   0.105 & $ 0.0097 \pm  0.0071 \pm  0.0033 $ &  0.0015 & -0.0707 \\
  0.105 -   0.209 & $ 0.0000 \pm  0.0020 \pm  0.00000000001 $ &  0.0038 & -0.0700 \\
  0.209 -   0.314 & $ 0.00000 \pm  0.00077 \pm  0.00000046 $ &  0.00452 & -0.08783 \\
  0.314 -   0.419 & $ 0.0019 \pm  0.0013 \pm  0.00037 $ &  0.0030 & -0.0397 \\
  0.419 -   0.524 & $ 0.0000 \pm  0.0046 \pm  0.00000000001 $ &  0.0008 &  0.0664 \\
  0.524 -   0.628 & $ 0.0000 \pm  0.0065 \pm  0.00000000001 $ &  0.0060 & -0.0227 \\
  0.628 -   0.733 & $ 0.038 \pm  0.019 \pm  0.0084 $ &  0.002 & -0.041 \\
  0.733 -   0.838 & $ 0.0008 \pm  0.0090 \pm  0.000089 $ &  0.0128 &  0.0079 \\
  0.838 -   0.942 & $ 0.011 \pm  0.015 \pm  0.014 $ &  0.021 &  0.011 \\
  0.942 -   1.05 & $ 0.009 \pm  0.012 \pm  0.0100 $ &  0.031 &  0.001 \\
  1.05 -   1.15 & $ 0.085 \pm  0.024 \pm  0.018 $ &  0.043 &  0.035 \\
  1.15 -   1.26 & $ 0.134 \pm  0.033 \pm  0.029 $ &  0.066 &  0.057 \\
  1.26 -   1.36 & $ 0.159 \pm  0.047 \pm  0.038 $ &  0.127 &  0.105 \\
  1.36 -   1.47 & $ 0.177 \pm  0.062 \pm  0.091 $ &  0.211 &  0.143 \\
  1.47 -   1.57 & $ 0.38 \pm  0.090 \pm  0.13 $ &  0.39 &  0.26 \\
  1.57 -   1.68 & $ 0.90 \pm  0.12 \pm  0.27 $ &  0.82 &  0.58 \\
  1.68 -   1.78 & $ 0.82 \pm  0.14 \pm  0.38 $ &  1.03 &  0.73 \\
  1.78 -   1.88 & $ 0.88 \pm  0.14 \pm  0.40 $ &  1.29 &  0.86 \\
  1.88 -   1.99 & $ 1.26 \pm  0.16 \pm  0.46 $ &  1.41 &  0.99 \\
  1.99 -   2.09 & $ 1.67 \pm  0.17 \pm  0.54 $ &  1.68 &  1.15 \\
  2.09 -   2.20 & $ 2.19 \pm  0.19 \pm  0.61 $ &  1.93 &  1.37 \\
  2.20 -   2.30 & $ 1.82 \pm  0.21 \pm  0.78 $ &  2.29 &  1.91 \\
  2.30 -   2.41 & $ 3.02 \pm  0.22 \pm  0.90 $ &  2.86 &  2.09 \\
  2.41 -   2.51 & $ 3.4 \pm  0.27 \pm  1.3 $ &  3.7 &  2.7 \\
  2.51 -   2.62 & $ 4.8 \pm  0.30 \pm  1.5 $ &  4.8 &  3.5 \\
  2.62 -   2.72 & $ 6.8 \pm  0.37 \pm  2.1 $ &  6.7 &  5.1 \\
  2.72 -   2.83 & $10.2 \pm  0.46 \pm  3.0 $ &  9.5 &  7.1 \\
  2.83 -   2.93 & $15.1 \pm  0.54 \pm  4.2 $ & 14.6 & 11.1 \\
  2.93 -   3.04 & $22.9 \pm  0.67 \pm  6.2 $ & 23.2 & 19.1 \\
  3.04 -   3.14 & $32.8 \pm  0.75 \pm  8.3 $ & 35.1 & 45.8 \\
\hline\hline
\end{tabular}
\end{center}
\end{table}

\pagebreak
\vfill
\renewcommand{\plotname}{yggSS01}

\begin{center}
{\LARGE $Y(\gamma\gamma)$}
\end{center}

\begin{figure*}[!hp]
\includegraphics[width=9.0cm]{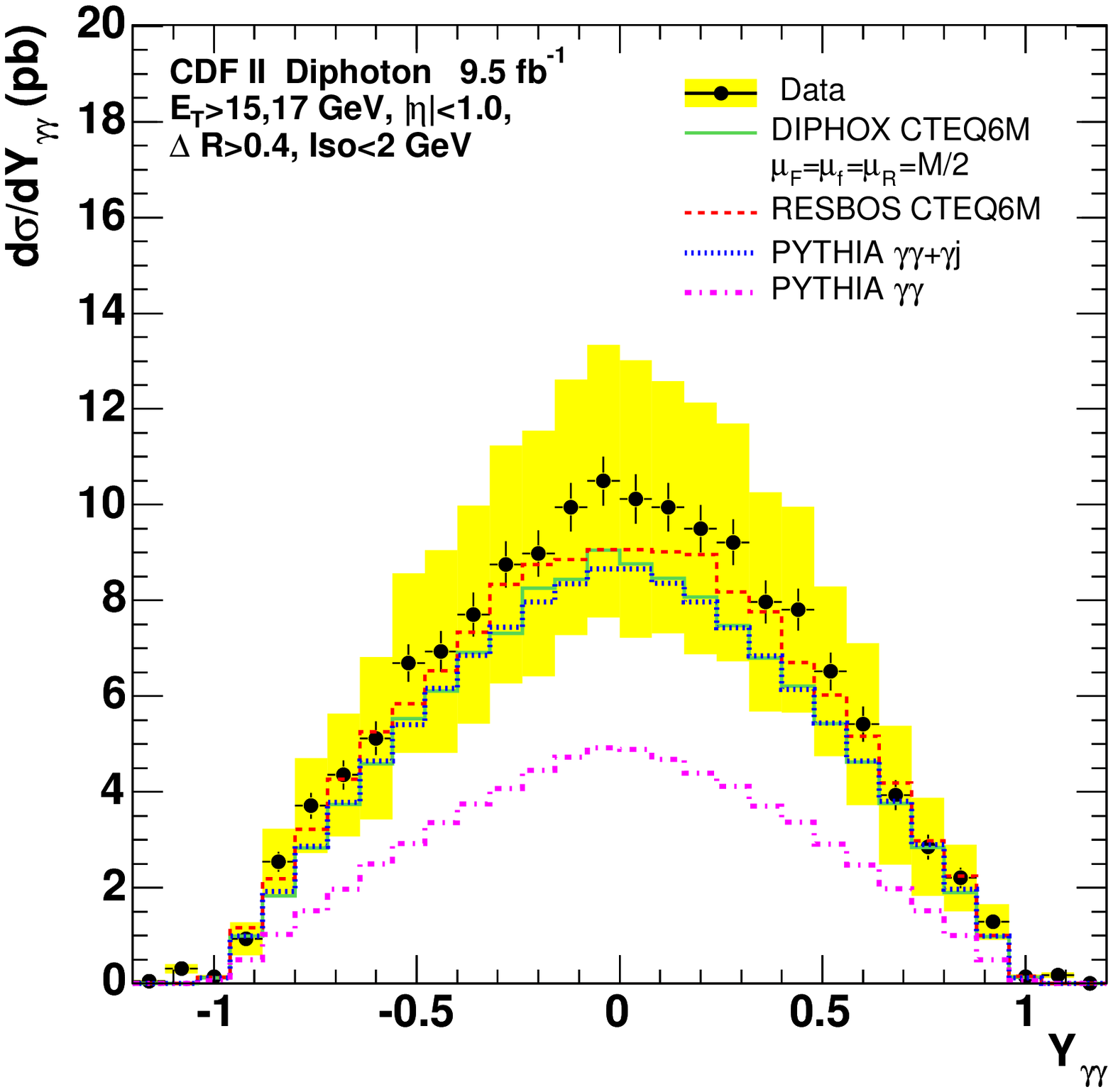}\hspace*{0cm}
\includegraphics[width=8.5cm]{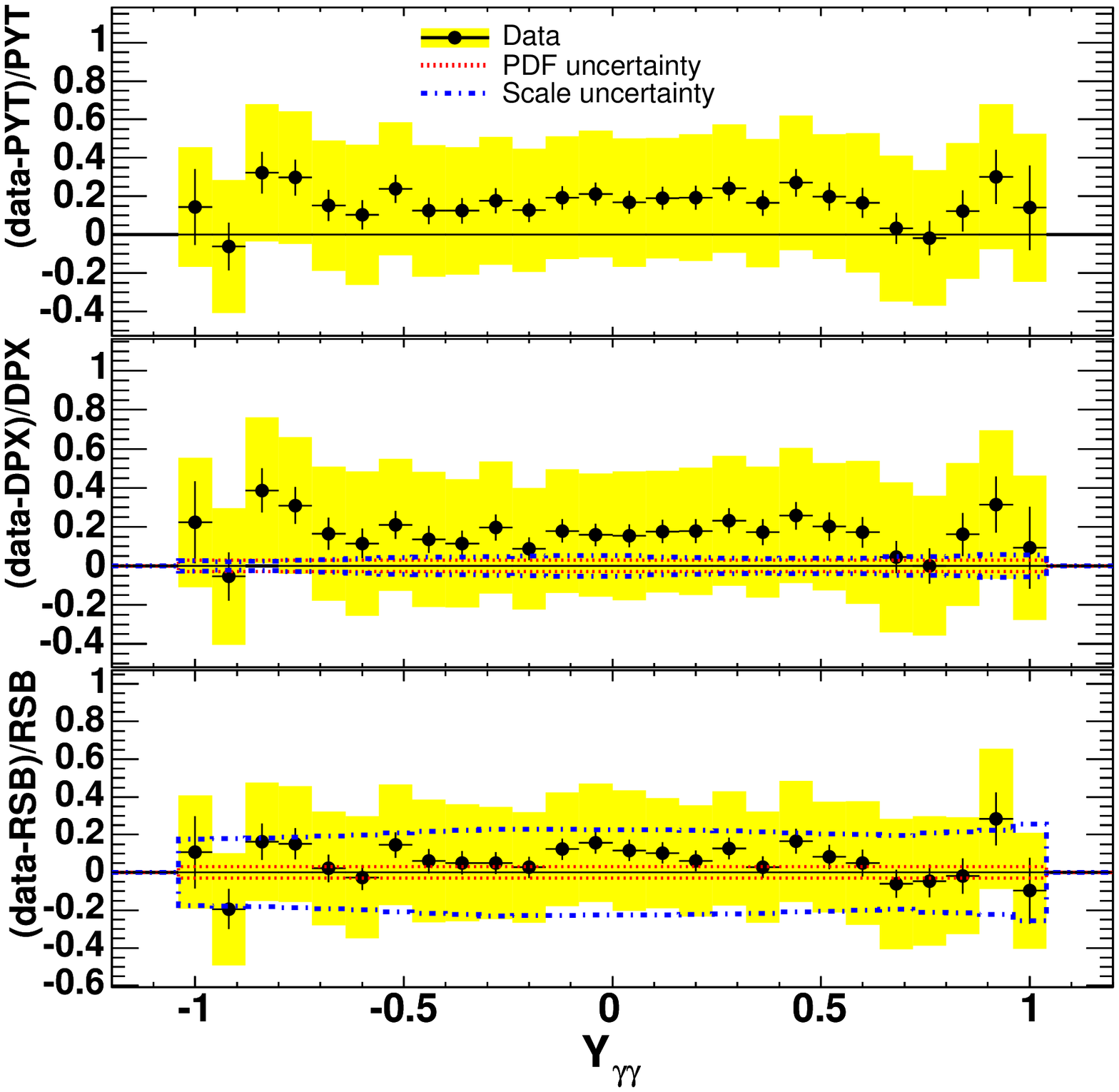}\hfill
\vspace*{0cm}
\includegraphics[width=9.0cm]{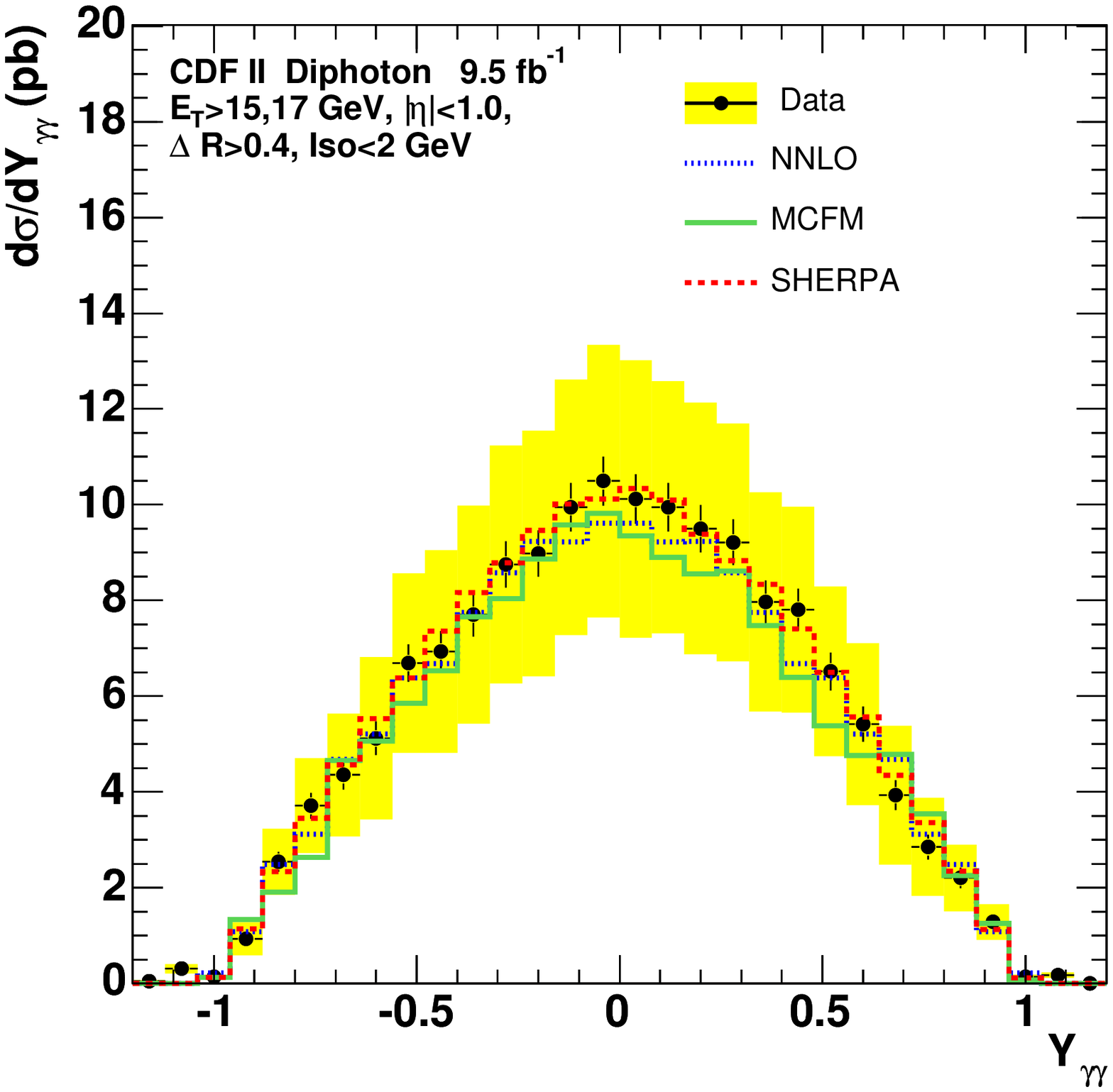}\hspace*{0cm}
\includegraphics[width=8.5cm]{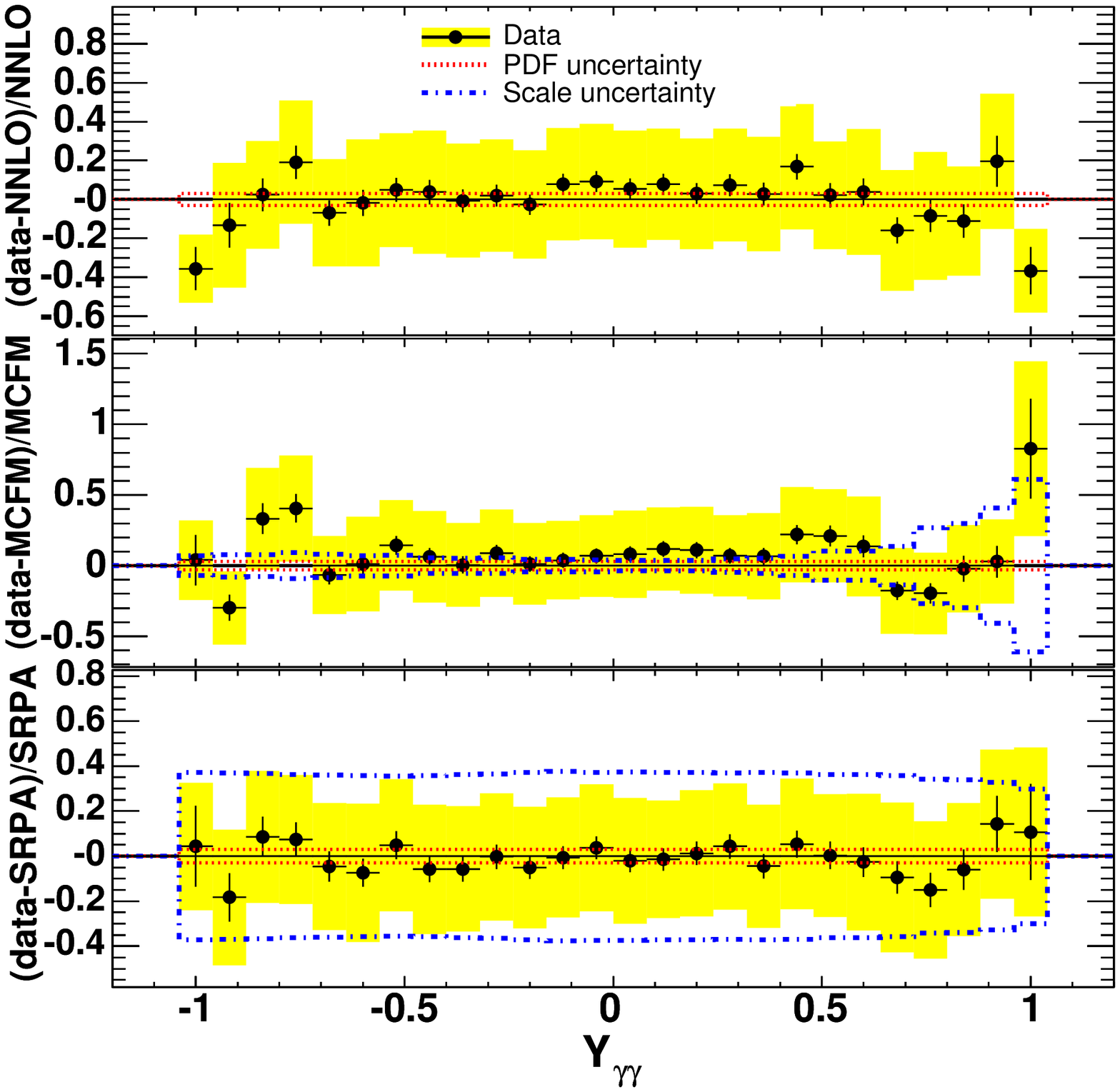}
\vspace*{-0.1cm}
\caption{The measured differential cross sections 
for $Y(\gamma\gamma)$, the rapidity of the diphoton system, 
compared with six 
theoretical predictions discussed in the text. The left windows show the
absolute comparisons and the right windows show the fractional deviations
of the data from the theoretical predictions. Note that the vertical axis
scales differ between fractional deviation plots.}
\end{figure*}

\pagebreak
\vfill

\begin{table}[!hp]
\begin{center}
\begin{tabular}{cccc}
\hline\hline
 Bin & Cross Section (pb) & Sherpa & NNLO   \\ \hline
 -1.20 -  -1.12 & $ 0.055 \pm  0.023 \pm  0.016 $ &  0.000 &  0.000 \\
 -1.12 -  -1.04 & $ 0.309 \pm  0.084 \pm  0.094 $ &  0.000 &  0.000 \\
 -1.04 -  -0.960 & $ 0.140 \pm  0.024 \pm  0.038 $ &  0.134 &  0.217 \\
 -0.960 -  -0.880 & $ 0.94 \pm  0.12 \pm  0.35 $ &  1.15 &  1.08 \\
 -0.880 -  -0.800 & $ 2.54 \pm  0.21 \pm  0.69 $ &  2.34 &  2.48 \\
 -0.800 -  -0.720 & $ 3.71 \pm  0.27 \pm  0.99 $ &  3.46 &  3.12 \\
 -0.720 -  -0.640 & $ 4.4 \pm  0.31 \pm  1.3 $ &  4.6 &  4.7 \\
 -0.640 -  -0.560 & $ 5.1 \pm  0.35 \pm  1.7 $ &  5.5 &  5.2 \\
 -0.560 -  -0.480 & $ 6.7 \pm  0.39 \pm  1.9 $ &  6.4 &  6.4 \\
 -0.480 -  -0.400 & $ 6.9 \pm  0.42 \pm  2.1 $ &  7.4 &  6.7 \\
 -0.400 -  -0.320 & $ 7.7 \pm  0.45 \pm  2.3 $ &  8.2 &  7.8 \\
 -0.320 -  -0.240 & $ 8.8 \pm  0.48 \pm  2.5 $ &  8.8 &  8.6 \\
 -0.240 -  -0.160 & $ 9.0 \pm  0.49 \pm  2.6 $ &  9.5 &  9.2 \\
 -0.160 -  -0.0800 & $10.0 \pm  0.50 \pm  2.7 $ & 10.0 &  9.2 \\
 -0.0800 -   0.00 & $10.5 \pm  0.52 \pm  2.8 $ & 10.1 &  9.6 \\
  0.00 -   0.0800 & $10.1 \pm  0.51 \pm  2.9 $ & 10.3 &  9.6 \\
  0.0800 -   0.160 & $ 9.9 \pm  0.50 \pm  2.6 $ & 10.1 &  9.2 \\
  0.160 -   0.240 & $ 9.5 \pm  0.49 \pm  2.6 $ &  9.4 &  9.2 \\
  0.240 -   0.320 & $ 9.2 \pm  0.48 \pm  2.5 $ &  8.8 &  8.6 \\
  0.320 -   0.400 & $ 8.0 \pm  0.45 \pm  2.3 $ &  8.3 &  7.8 \\
  0.400 -   0.480 & $ 7.8 \pm  0.44 \pm  2.1 $ &  7.4 &  6.7 \\
  0.480 -   0.560 & $ 6.5 \pm  0.40 \pm  1.8 $ &  6.5 &  6.4 \\
  0.560 -   0.640 & $ 5.4 \pm  0.37 \pm  1.7 $ &  5.6 &  5.2 \\
  0.640 -   0.720 & $ 3.9 \pm  0.31 \pm  1.4 $ &  4.3 &  4.7 \\
  0.720 -   0.800 & $ 2.9 \pm  0.26 \pm  1.0 $ &  3.4 &  3.1 \\
  0.800 -   0.880 & $ 2.21 \pm  0.21 \pm  0.69 $ &  2.35 &  2.48 \\
  0.880 -   0.960 & $ 1.29 \pm  0.14 \pm  0.37 $ &  1.13 &  1.08 \\
  0.960 -   1.04 & $ 0.138 \pm  0.027 \pm  0.047 $ &  0.124 &  0.217 \\
  1.04 -   1.12 & $ 0.174 \pm  0.076 \pm  0.055 $ &  0.000 &  0.000 \\
  1.12 -   1.20 & $ 0.003 \pm  0.023 \pm  0.00033 $ &  0.000 &  0.000 \\
\hline\hline
\end{tabular}
\end{center}
\end{table}

\pagebreak
\vfill

\renewcommand{\plotname}{yggQHSS01}
\begin{center}
{\LARGE $Y(\gamma\gamma)$ for $P_T(\gamma\gamma) > M(\gamma\gamma)$}
\end{center}

\begin{figure*}[!hp]
\includegraphics[width=9.0cm]{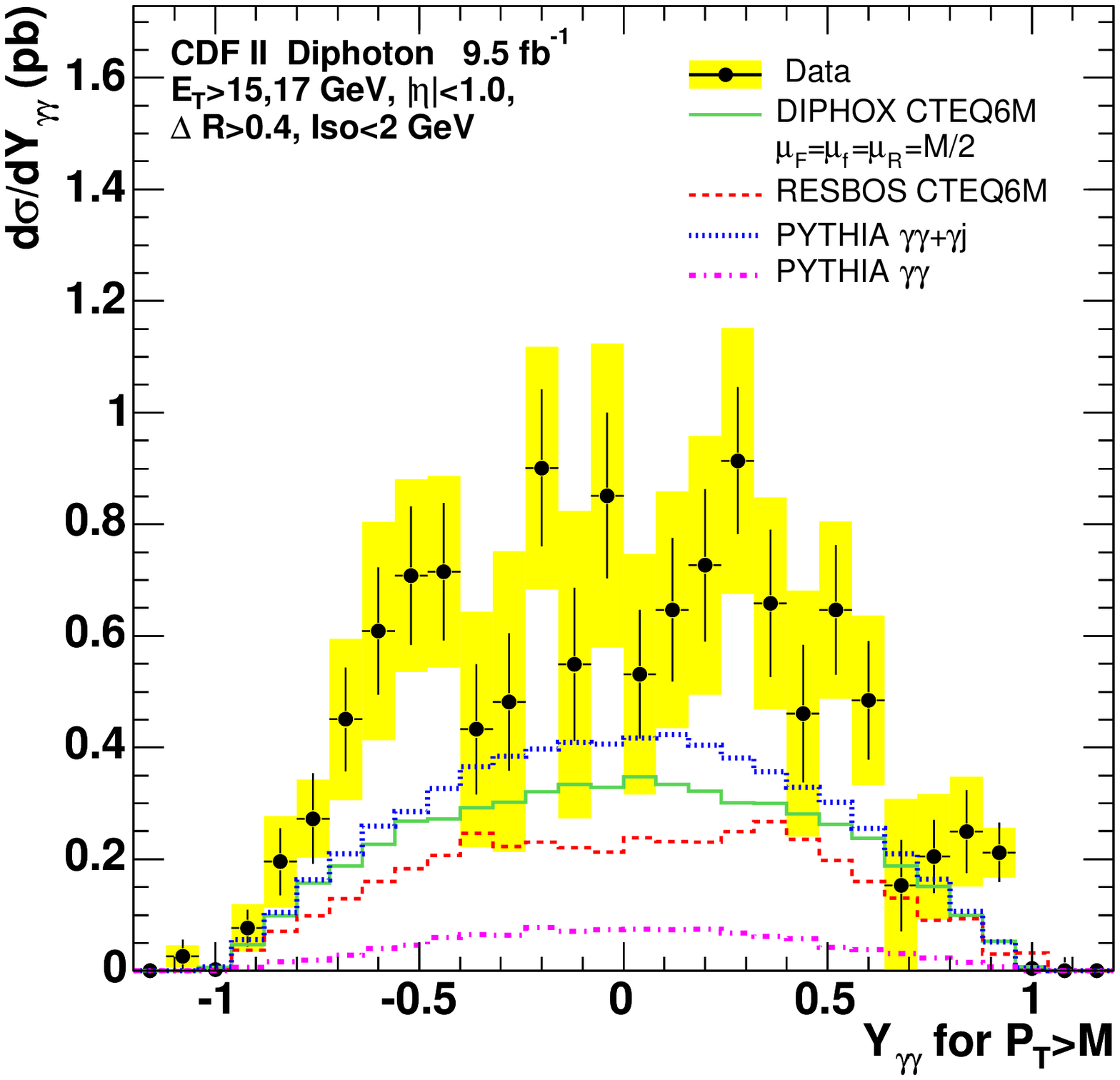}\hspace*{0cm}
\includegraphics[width=8.5cm]{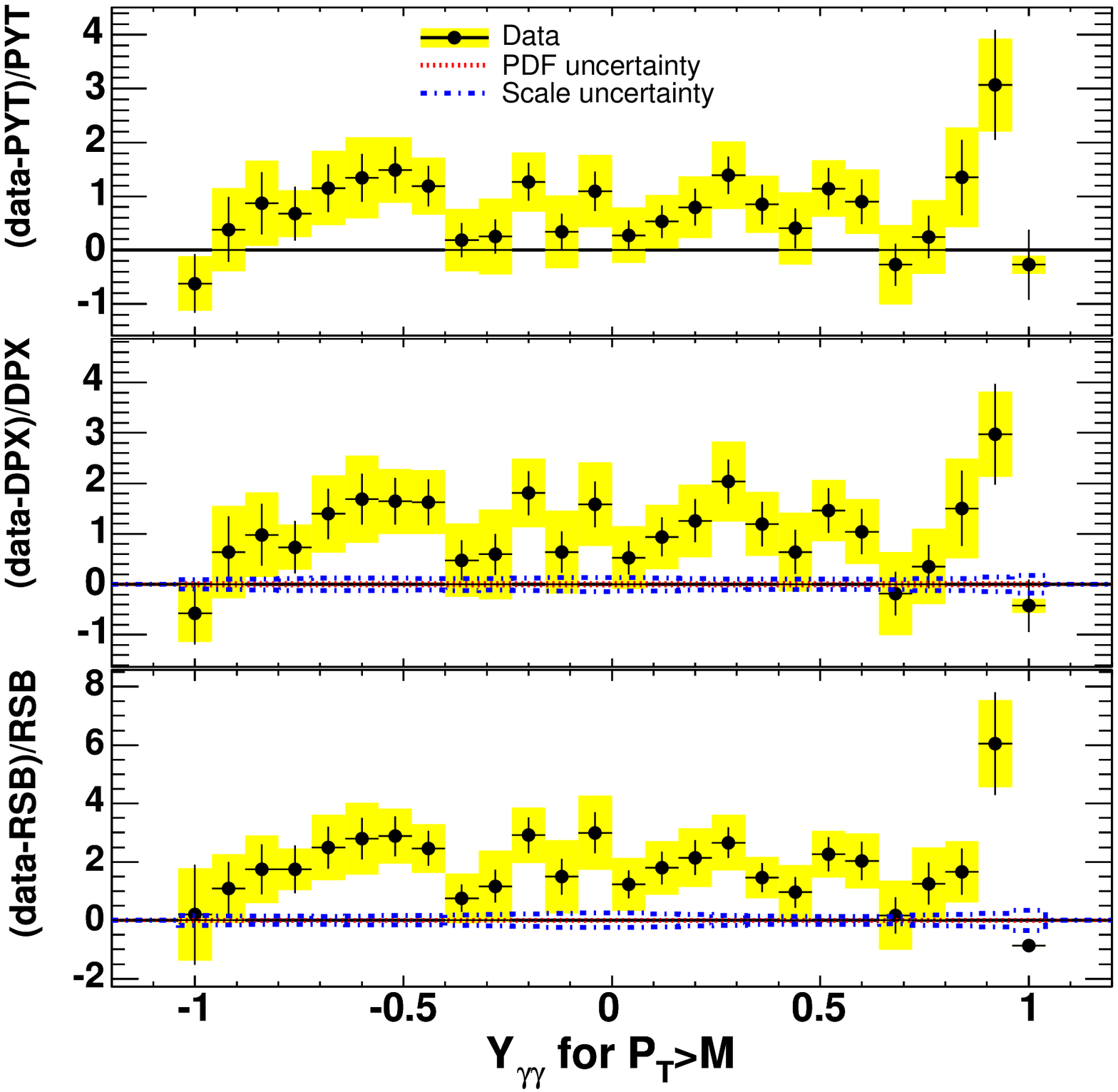}\hfill
\vspace*{0cm}
\includegraphics[width=9.0cm]{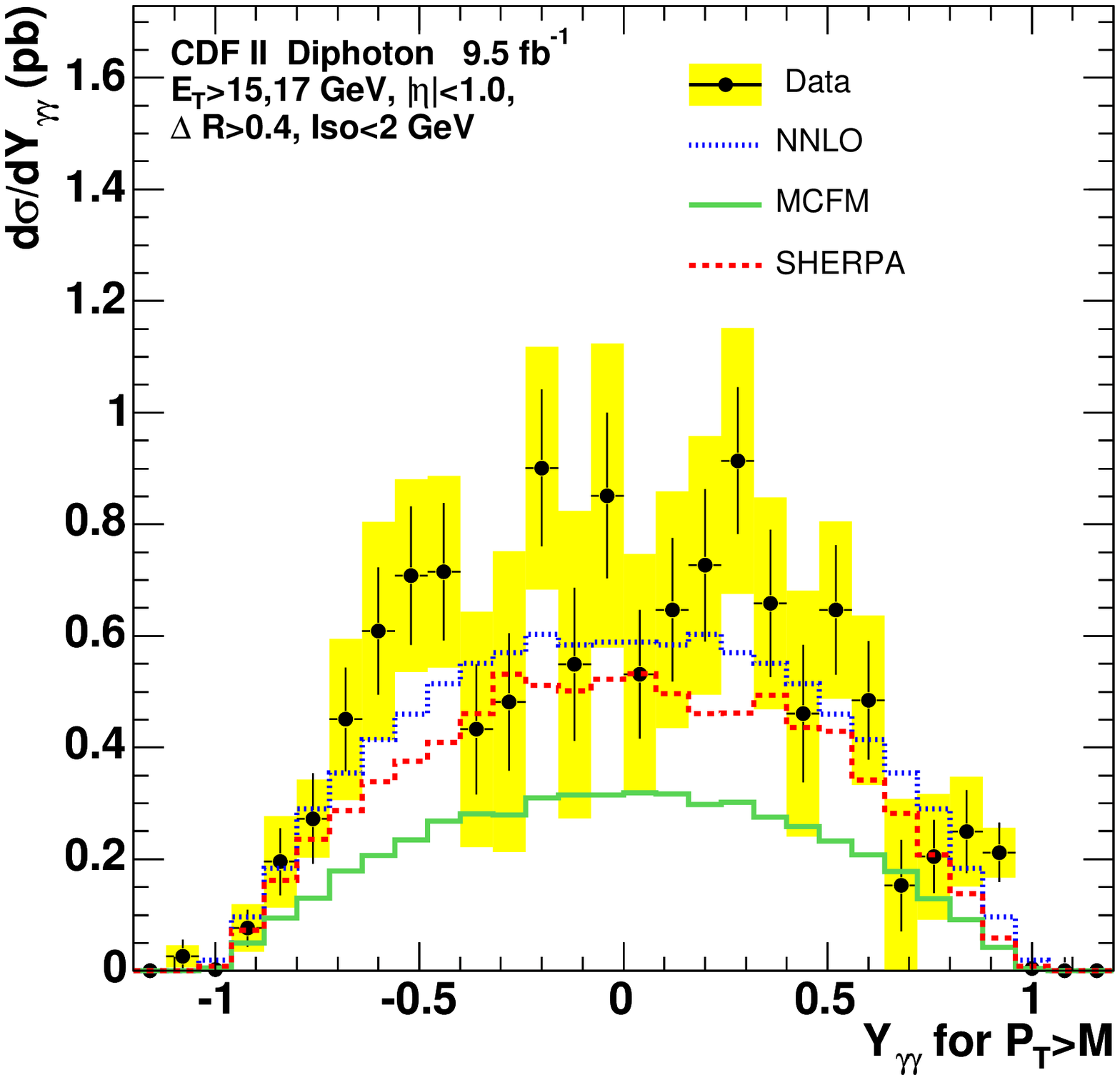}\hspace*{0cm}
\includegraphics[width=8.5cm]{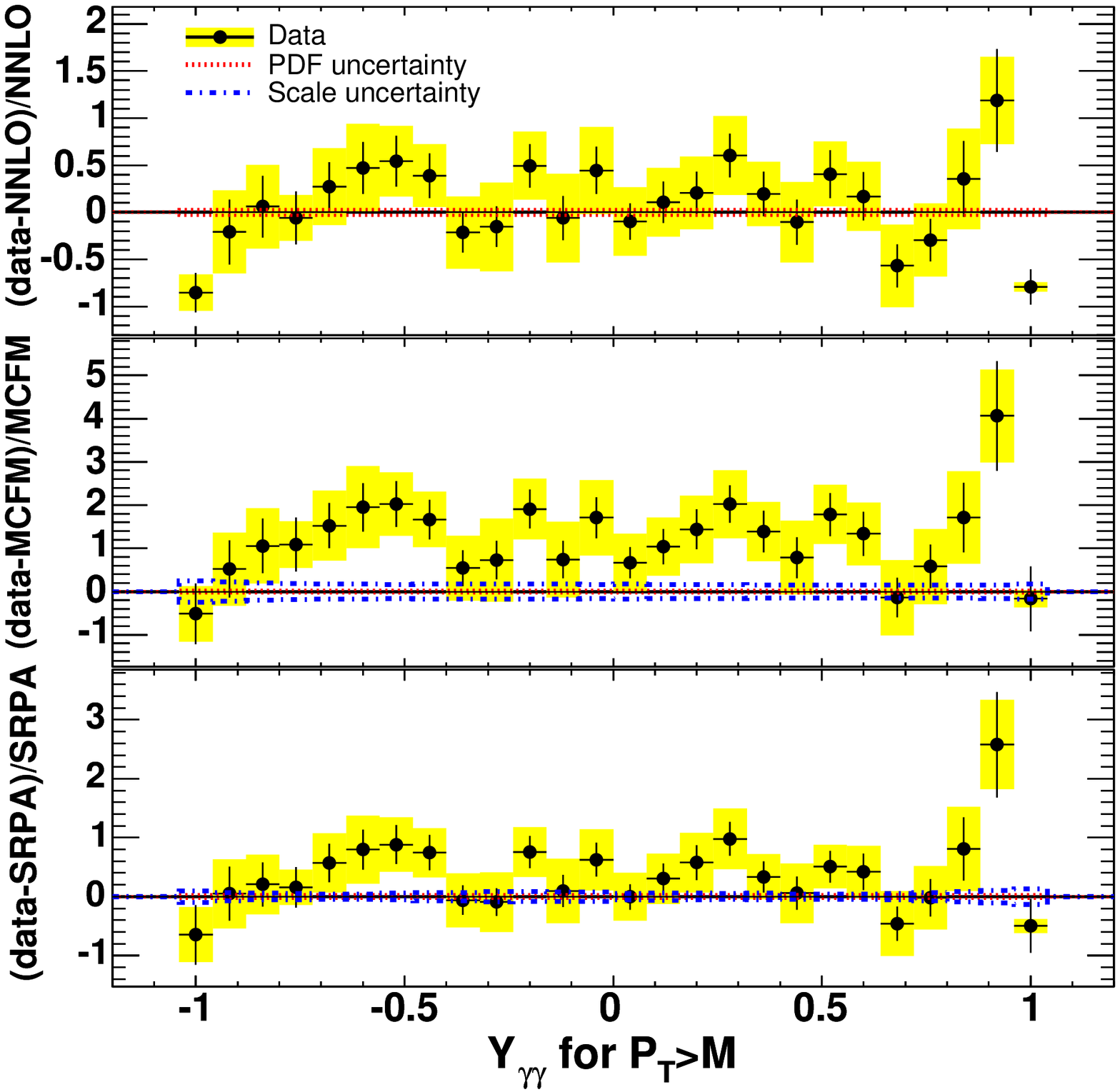}
\vspace*{-0.1cm}
\caption{The measured differential cross sections 
for $Y(\gamma\gamma)$, the rapidity of the diphoton system, 
when $P_T(\gamma\gamma) > M(\gamma\gamma)$, 
compared with six 
theoretical predictions discussed in the text. The left windows show the
absolute comparisons and the right windows show the fractional deviations
of the data from the theoretical predictions. Note that the vertical axis
scales differ between fractional deviation plots.}
\end{figure*}

\pagebreak
\vfill

\begin{table}[!hp]
\begin{center}
\begin{tabular}{cccc}
\hline\hline
 Bin & Cross Section (pb) & Sherpa & NNLO   \\ \hline
 -1.20 -  -1.12 & $ 0.0003 \pm  0.0098 \pm  0.000033 $ &  0.0000 &  0.0000 \\
 -1.12 -  -1.04 & $ 0.026 \pm  0.029 \pm  0.020 $ &  0.000 &  0.000 \\
 -1.04 -  -0.960 & $ 0.0028 \pm  0.0040 \pm  0.0037 $ &  0.0079 &  0.0192 \\
 -0.960 -  -0.880 & $ 0.077 \pm  0.033 \pm  0.043 $ &  0.073 &  0.097 \\
 -0.880 -  -0.800 & $ 0.195 \pm  0.060 \pm  0.082 $ &  0.162 &  0.184 \\
 -0.800 -  -0.720 & $ 0.273 \pm  0.081 \pm  0.070 $ &  0.236 &  0.290 \\
 -0.720 -  -0.640 & $ 0.45 \pm  0.093 \pm  0.14 $ &  0.29 &  0.35 \\
 -0.640 -  -0.560 & $ 0.61 \pm  0.11 \pm  0.20 $ &  0.34 &  0.41 \\
 -0.560 -  -0.480 & $ 0.71 \pm  0.12 \pm  0.17 $ &  0.38 &  0.46 \\
 -0.480 -  -0.400 & $ 0.71 \pm  0.12 \pm  0.17 $ &  0.41 &  0.51 \\
 -0.400 -  -0.320 & $ 0.43 \pm  0.12 \pm  0.21 $ &  0.46 &  0.55 \\
 -0.320 -  -0.240 & $ 0.48 \pm  0.12 \pm  0.27 $ &  0.53 &  0.57 \\
 -0.240 -  -0.160 & $ 0.90 \pm  0.14 \pm  0.22 $ &  0.51 &  0.60 \\
 -0.160 -  -0.0800 & $ 0.55 \pm  0.14 \pm  0.28 $ &  0.50 &  0.58 \\
 -0.0800 -   0.00 & $ 0.85 \pm  0.15 \pm  0.27 $ &  0.52 &  0.59 \\
  0.00 -   0.0800 & $ 0.53 \pm  0.11 \pm  0.22 $ &  0.53 &  0.59 \\
  0.0800 -   0.160 & $ 0.65 \pm  0.13 \pm  0.21 $ &  0.50 &  0.58 \\
  0.160 -   0.240 & $ 0.73 \pm  0.14 \pm  0.23 $ &  0.46 &  0.60 \\
  0.240 -   0.320 & $ 0.91 \pm  0.13 \pm  0.24 $ &  0.46 &  0.57 \\
  0.320 -   0.400 & $ 0.66 \pm  0.13 \pm  0.19 $ &  0.49 &  0.55 \\
  0.400 -   0.480 & $ 0.46 \pm  0.12 \pm  0.22 $ &  0.44 &  0.51 \\
  0.480 -   0.560 & $ 0.65 \pm  0.12 \pm  0.16 $ &  0.43 &  0.46 \\
  0.560 -   0.640 & $ 0.48 \pm  0.11 \pm  0.15 $ &  0.34 &  0.41 \\
  0.640 -   0.720 & $ 0.15 \pm  0.082 \pm  0.16 $ &  0.28 &  0.35 \\
  0.720 -   0.800 & $ 0.20 \pm  0.065 \pm  0.11 $ &  0.21 &  0.29 \\
  0.800 -   0.880 & $ 0.250 \pm  0.074 \pm  0.098 $ &  0.138 &  0.184 \\
  0.880 -   0.960 & $ 0.212 \pm  0.053 \pm  0.045 $ &  0.059 &  0.097 \\
  0.960 -   1.04 & $ 0.0040 \pm  0.0036 \pm  0.00094 $ &  0.0079 &  0.0192 \\
\hline\hline
\end{tabular}
\end{center}
\end{table}

\pagebreak
\vfill

\renewcommand{\plotname}{yggQLSS01}
\begin{center}
{\LARGE $Y(\gamma\gamma)$ for $P_T(\gamma\gamma) < M(\gamma\gamma)$}
\end{center}

\begin{figure*}[!hp]
\includegraphics[width=9.0cm]{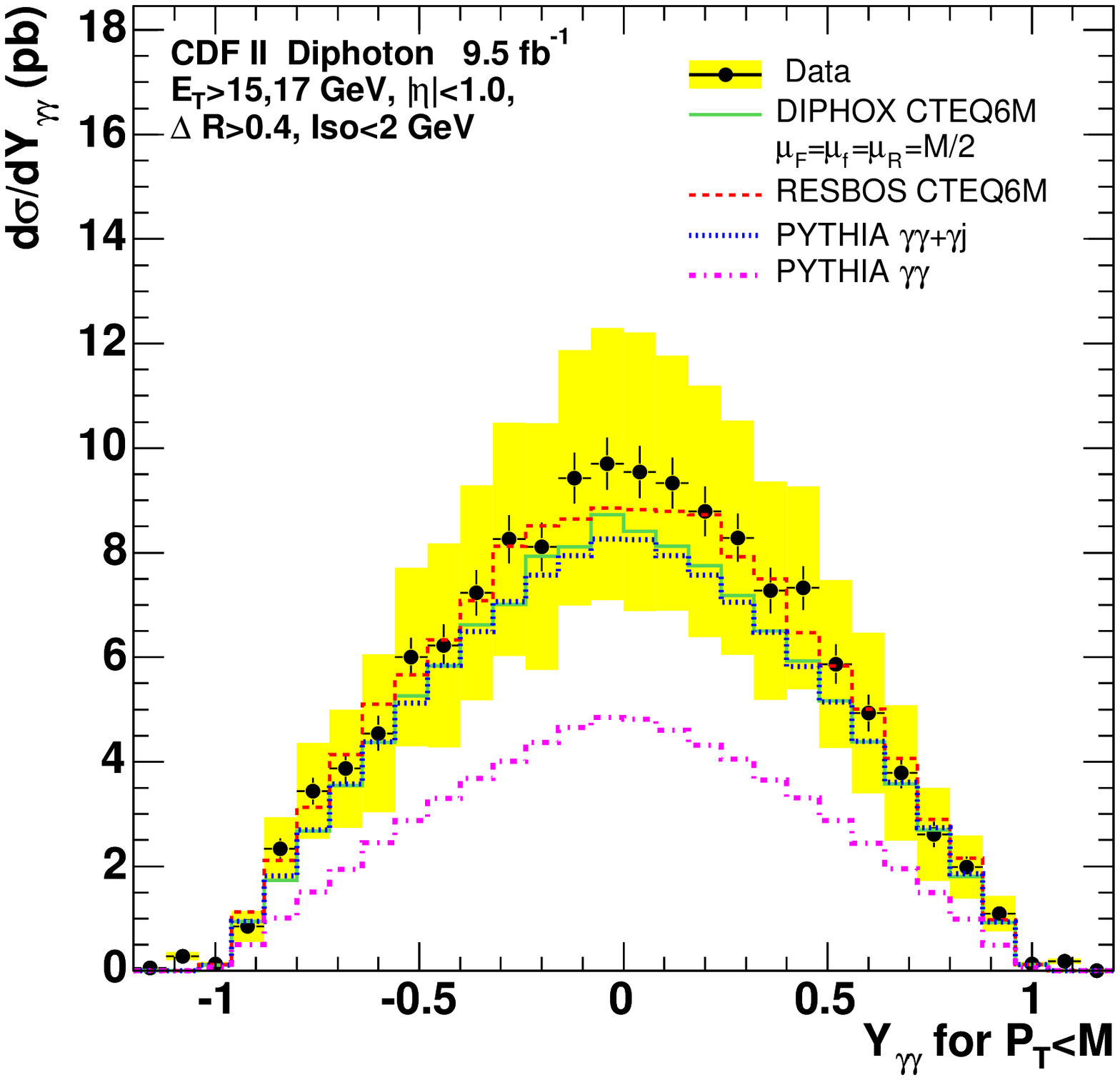}\hspace*{0cm}
\includegraphics[width=8.5cm]{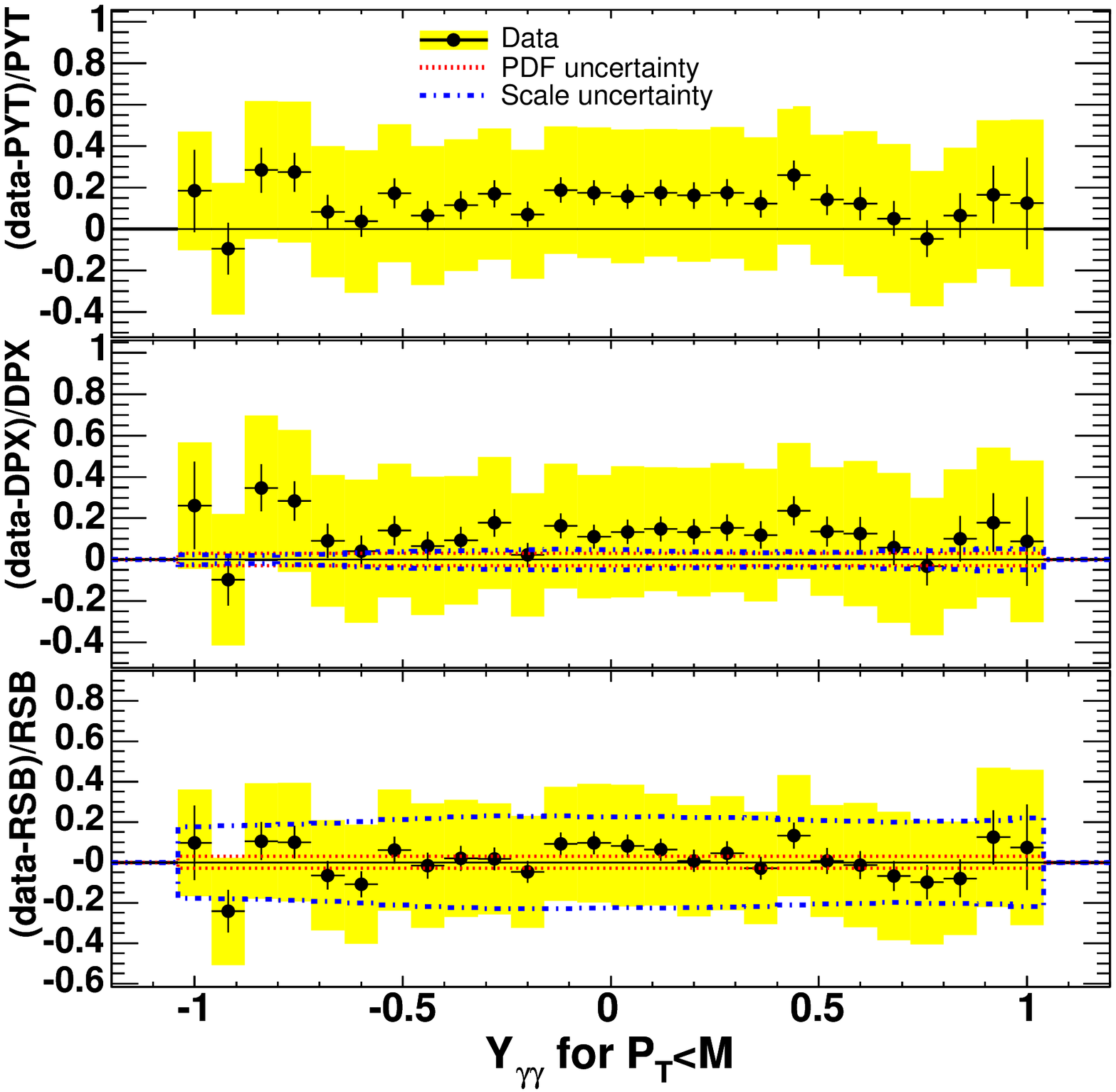}\hfill
\vspace*{0cm}
\includegraphics[width=9.0cm]{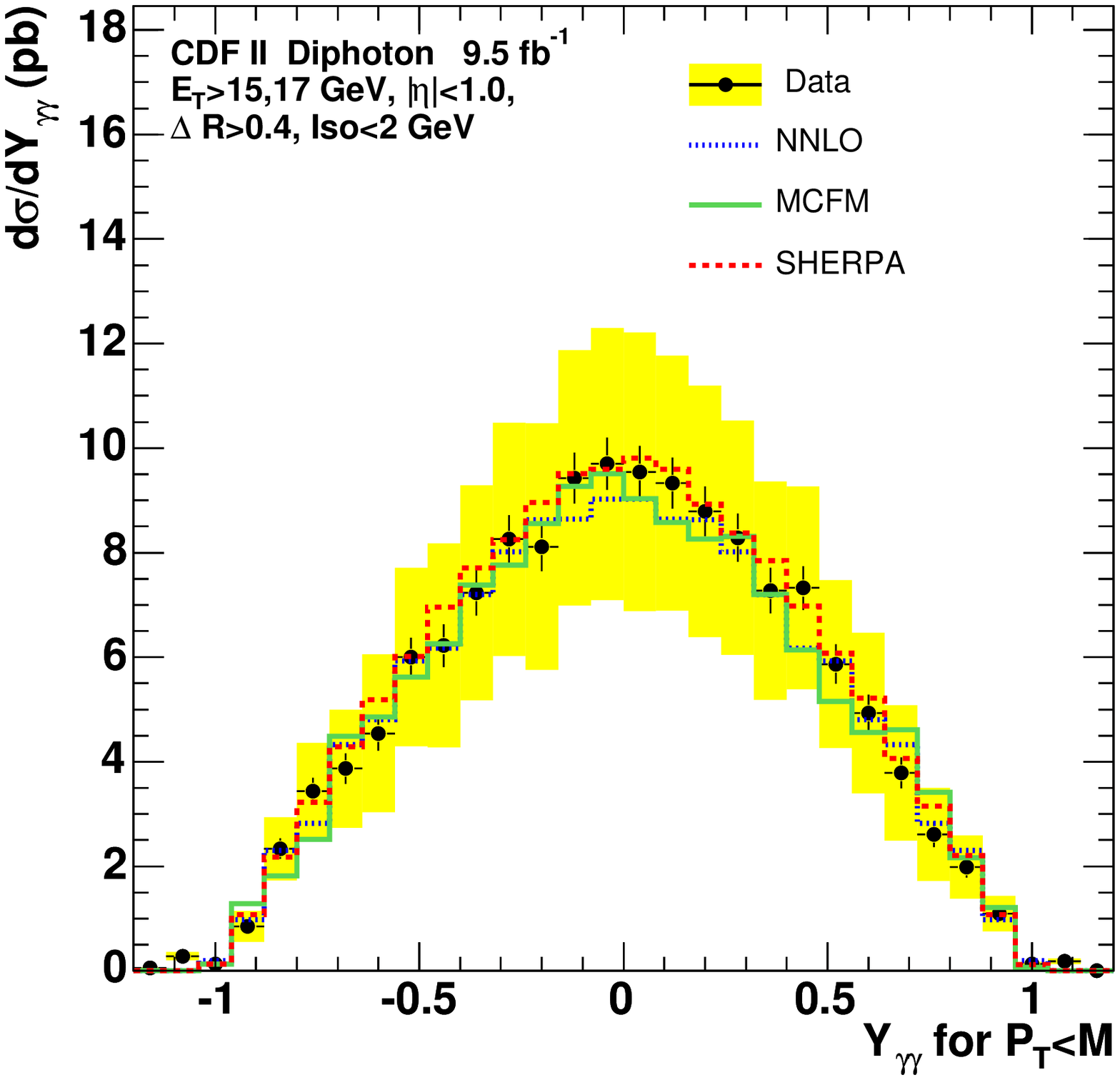}\hspace*{0cm}
\includegraphics[width=8.5cm]{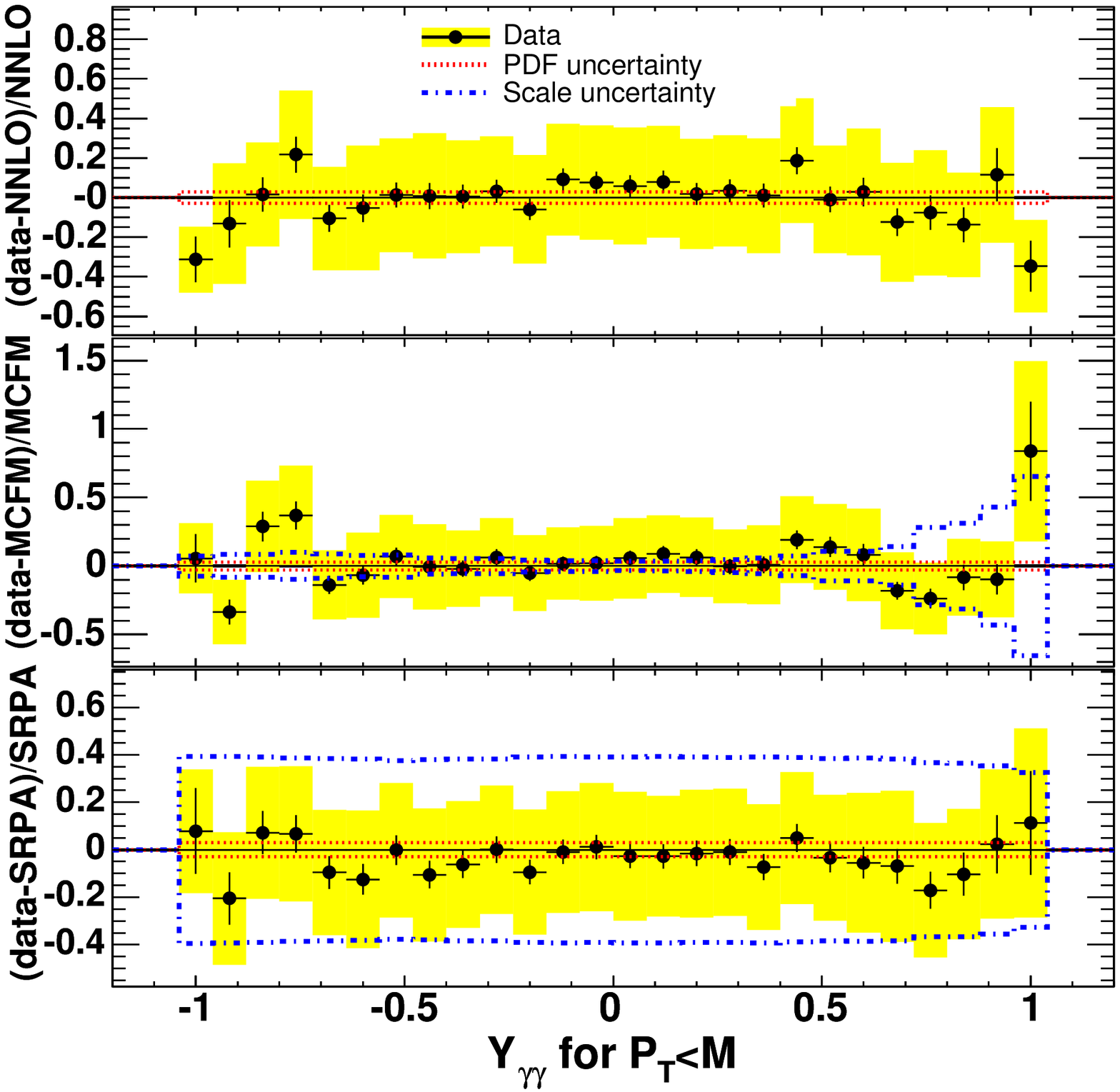}
\vspace*{-0.1cm}
\caption{The measured differential cross sections 
for $Y(\gamma\gamma)$, the rapidity of the diphoton system, 
when $P_T(\gamma\gamma) < M(\gamma\gamma)$, 
compared with six 
theoretical predictions discussed in the text. The left windows show the
absolute comparisons and the right windows show the fractional deviations
of the data from the theoretical predictions. Note that the vertical axis
scales differ between fractional deviation plots.}
\end{figure*}

\pagebreak
\vfill

\begin{table}[!hp]
\begin{center}
\begin{tabular}{cccc}
\hline\hline
 Bin & Cross Section (pb) & Sherpa & NNLO   \\ \hline
 -1.20 -  -1.12 & $ 0.054 \pm  0.021 \pm  0.018 $ &  0.000 &  0.000 \\
 -1.12 -  -1.04 & $ 0.283 \pm  0.078 \pm  0.077 $ &  0.000 &  0.000 \\
 -1.04 -  -0.960 & $ 0.136 \pm  0.023 \pm  0.033 $ &  0.126 &  0.198 \\
 -0.960 -  -0.880 & $ 0.85 \pm  0.12 \pm  0.30 $ &  1.07 &  0.98 \\
 -0.880 -  -0.800 & $ 2.34 \pm  0.20 \pm  0.61 $ &  2.18 &  2.30 \\
 -0.800 -  -0.720 & $ 3.44 \pm  0.25 \pm  0.92 $ &  3.22 &  2.83 \\
 -0.720 -  -0.640 & $ 3.9 \pm  0.29 \pm  1.1 $ &  4.3 &  4.3 \\
 -0.640 -  -0.560 & $ 4.5 \pm  0.33 \pm  1.5 $ &  5.2 &  4.8 \\
 -0.560 -  -0.480 & $ 6.0 \pm  0.37 \pm  1.7 $ &  6.0 &  5.9 \\
 -0.480 -  -0.400 & $ 6.2 \pm  0.40 \pm  2.0 $ &  7.0 &  6.2 \\
 -0.400 -  -0.320 & $ 7.2 \pm  0.44 \pm  2.1 $ &  7.7 &  7.2 \\
 -0.320 -  -0.240 & $ 8.3 \pm  0.46 \pm  2.2 $ &  8.2 &  8.0 \\
 -0.240 -  -0.160 & $ 8.1 \pm  0.47 \pm  2.4 $ &  9.0 &  8.6 \\
 -0.160 -  -0.0800 & $ 9.4 \pm  0.49 \pm  2.4 $ &  9.5 &  8.6 \\
 -0.0800 -   0.00 & $ 9.7 \pm  0.50 \pm  2.6 $ &  9.6 &  9.0 \\
  0.00 -   0.0800 & $ 9.5 \pm  0.50 \pm  2.7 $ &  9.8 &  9.0 \\
  0.0800 -   0.160 & $ 9.3 \pm  0.49 \pm  2.4 $ &  9.6 &  8.6 \\
  0.160 -   0.240 & $ 8.8 \pm  0.47 \pm  2.4 $ &  8.9 &  8.6 \\
  0.240 -   0.320 & $ 8.3 \pm  0.46 \pm  2.2 $ &  8.4 &  8.0 \\
  0.320 -   0.400 & $ 7.3 \pm  0.43 \pm  2.1 $ &  7.8 &  7.2 \\
  0.400 -   0.480 & $ 7.3 \pm  0.42 \pm  1.9 $ &  7.0 &  6.2 \\
  0.480 -   0.560 & $ 5.9 \pm  0.38 \pm  1.6 $ &  6.1 &  5.9 \\
  0.560 -   0.640 & $ 4.9 \pm  0.35 \pm  1.5 $ &  5.2 &  4.8 \\
  0.640 -   0.720 & $ 3.8 \pm  0.30 \pm  1.3 $ &  4.1 &  4.3 \\
  0.720 -   0.800 & $ 2.61 \pm  0.25 \pm  0.89 $ &  3.15 &  2.83 \\
  0.800 -   0.880 & $ 1.98 \pm  0.20 \pm  0.61 $ &  2.21 &  2.30 \\
  0.880 -   0.960 & $ 1.10 \pm  0.13 \pm  0.34 $ &  1.07 &  0.98 \\
  0.960 -   1.04 & $ 0.130 \pm  0.025 \pm  0.046 $ &  0.116 &  0.198 \\
  1.04 -   1.12 & $ 0.183 \pm  0.072 \pm  0.050 $ &  0.000 &  0.000 \\
  1.12 -   1.20 & $ 0.003 \pm  0.023 \pm  0.00033 $ &  0.000 &  0.000 \\
\hline\hline
\end{tabular}
\end{center}
\end{table}

\pagebreak
\vfill
\renewcommand{\plotname}{zSS01}

\begin{center}
{\LARGE $E_T^{\gamma 2}/E_T^{\gamma 1}$}
\end{center}

\begin{figure*}[!hp]
\includegraphics[width=9.0cm]{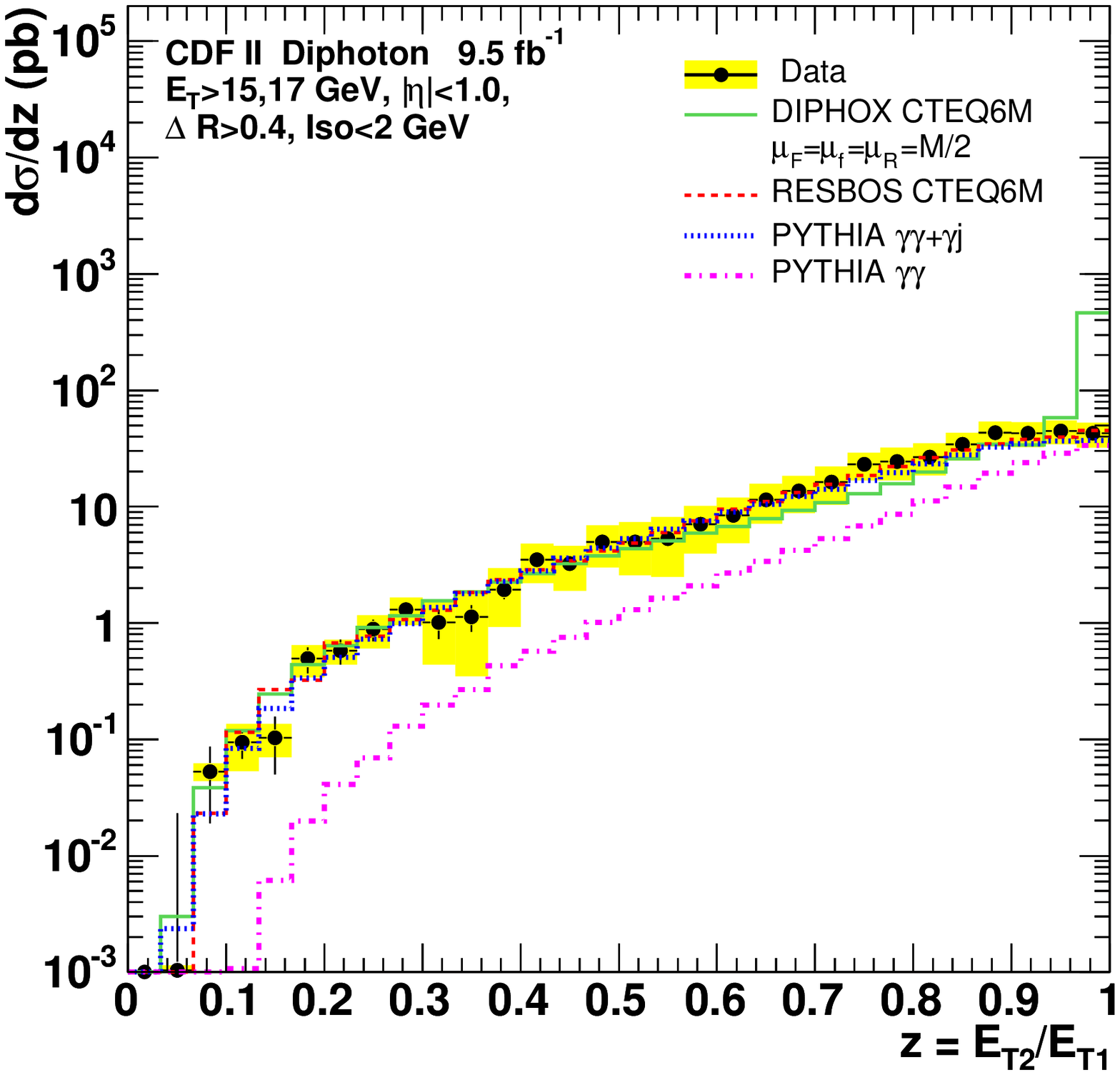}\hspace*{0cm}
\includegraphics[width=8.5cm]{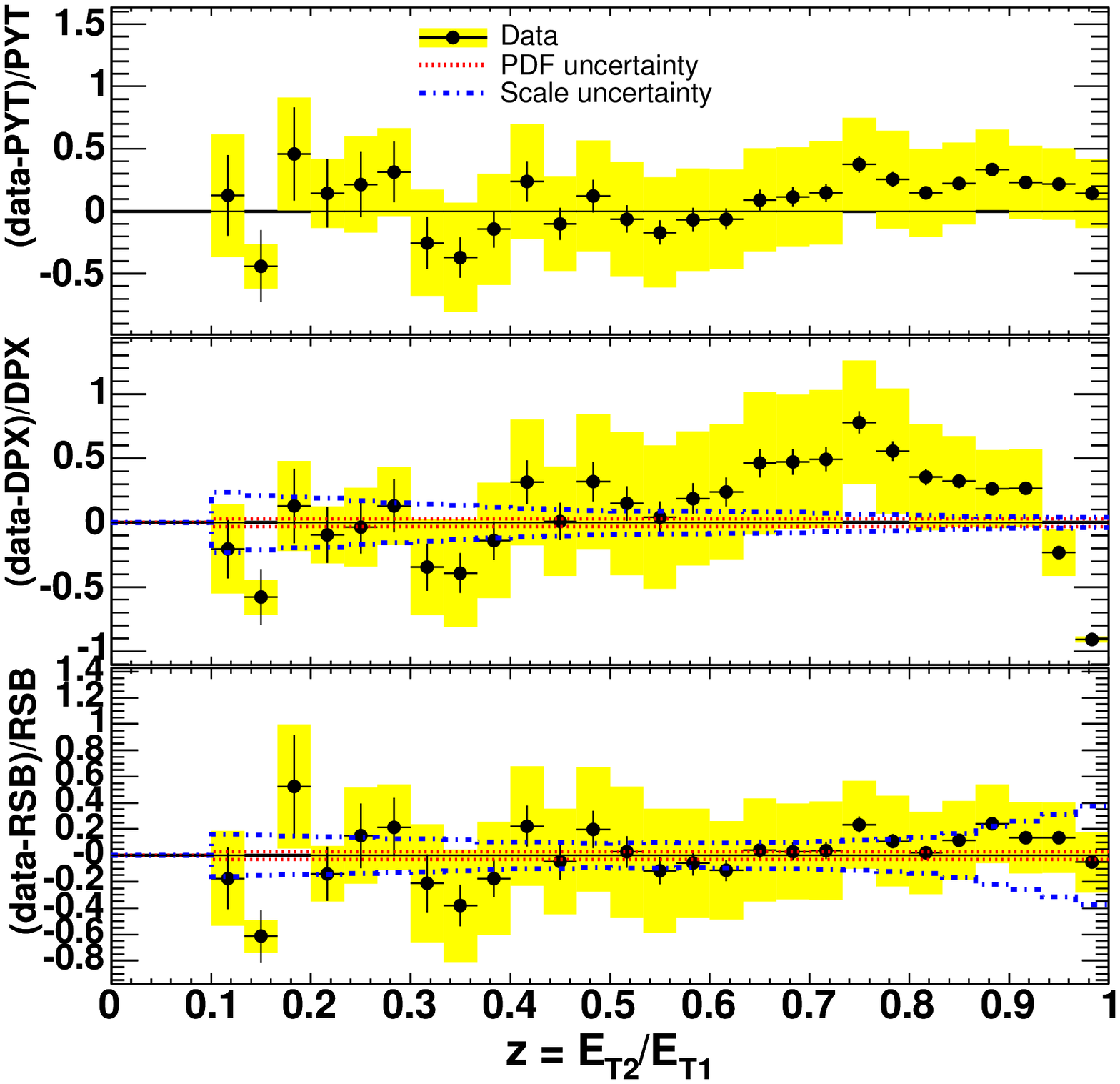}\hfill
\vspace*{0cm}
\includegraphics[width=9.0cm]{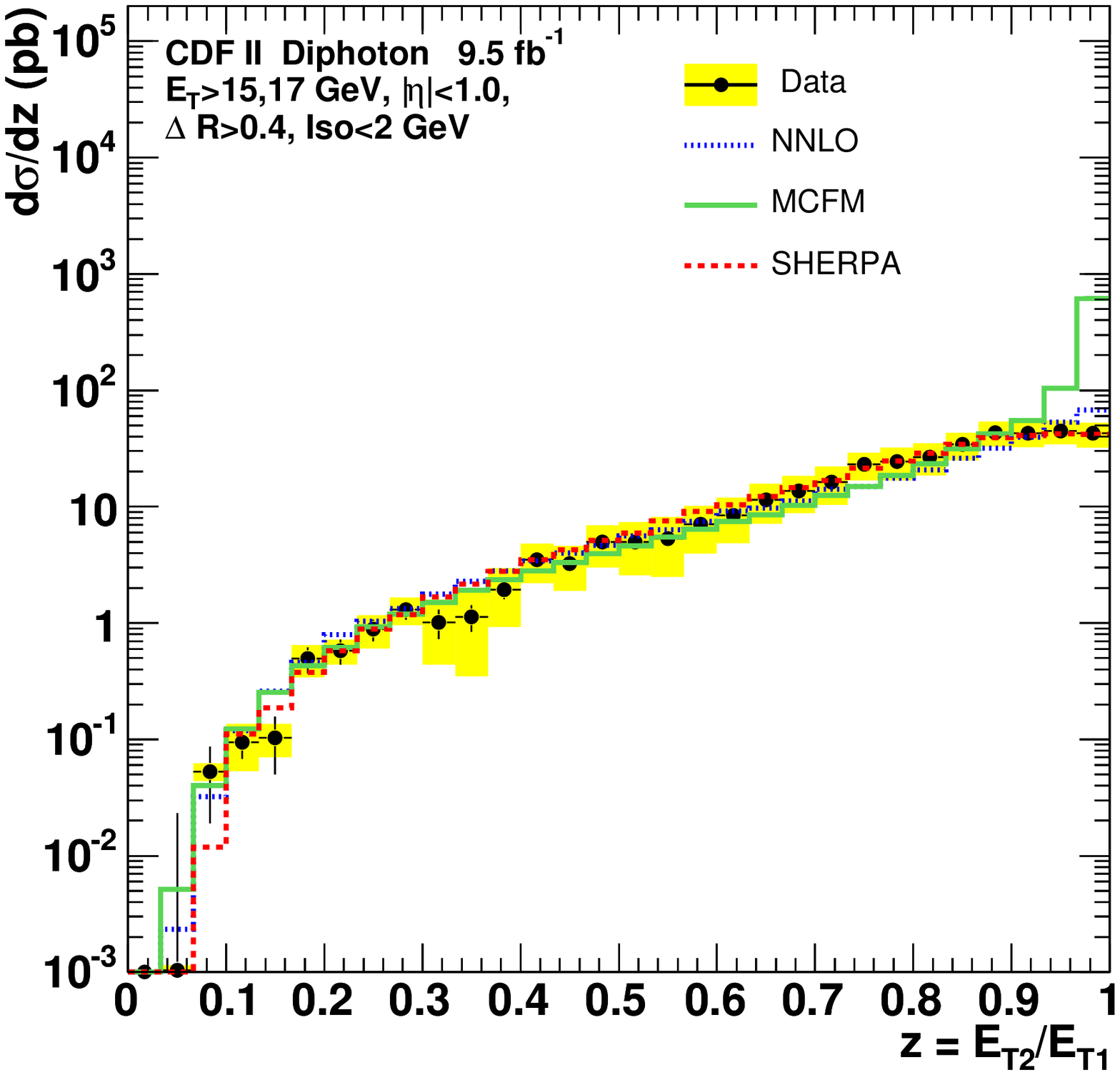}\hspace*{0cm}
\includegraphics[width=8.5cm]{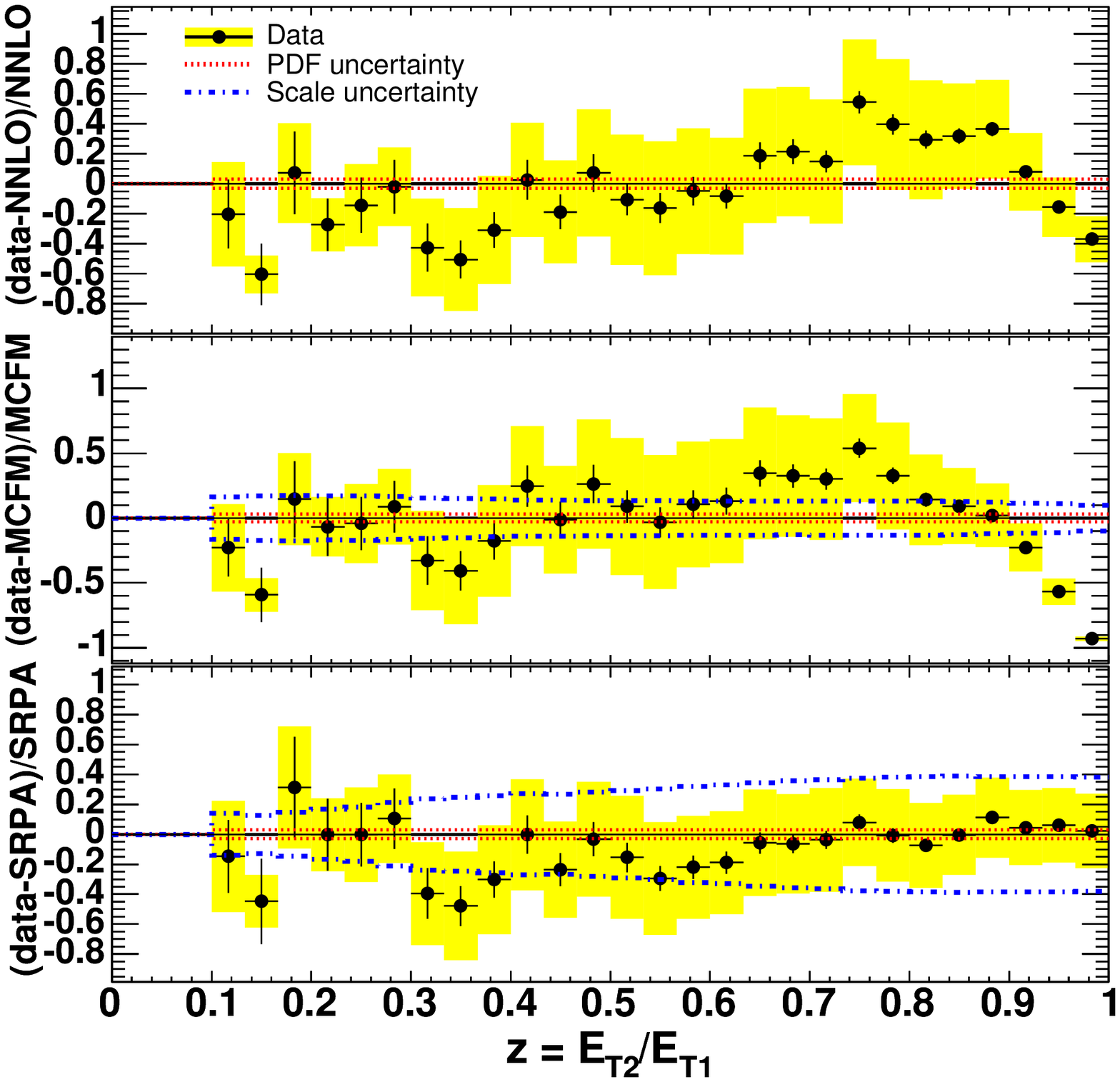}
\vspace*{-0.1cm}
\caption{The measured differential cross sections 
for $E_T^{\gamma 2}/E_T^{\gamma 1}$ 
compared with six 
theoretical predictions discussed in the text. The left windows show the
absolute comparisons and the right windows show the fractional deviations
of the data from the theoretical predictions. Note that the vertical axis
scales differ between fractional deviation plots.}
\end{figure*}

\pagebreak
\vfill
\begin{table}[!hp]
\begin{center}
\begin{tabular}{cccc}
\hline\hline
 Bin & Cross Section (pb) & Sherpa & NNLO   \\ \hline
  0.0333 -   0.0667 & $ 0.001 \pm  0.022 \pm  0.00011 $ &  0.000 &  0.002 \\
  0.0667 -   0.100 & $ 0.053 \pm  0.034 \pm  0.0093 $ &  0.012 &  0.032 \\
  0.100 -   0.133 & $ 0.095 \pm  0.027 \pm  0.041 $ &  0.111 &  0.119 \\
  0.133 -   0.167 & $ 0.103 \pm  0.053 \pm  0.033 $ &  0.187 &  0.261 \\
  0.167 -   0.200 & $ 0.49 \pm  0.13 \pm  0.15 $ &  0.38 &  0.46 \\
  0.200 -   0.233 & $ 0.58 \pm  0.14 \pm  0.14 $ &  0.58 &  0.80 \\
  0.233 -   0.267 & $ 0.89 \pm  0.19 \pm  0.28 $ &  0.89 &  1.03 \\
  0.267 -   0.300 & $ 1.30 \pm  0.24 \pm  0.35 $ &  1.18 &  1.33 \\
  0.300 -   0.333 & $ 1.01 \pm  0.28 \pm  0.57 $ &  1.68 &  1.76 \\
  0.333 -   0.367 & $ 1.13 \pm  0.29 \pm  0.78 $ &  2.17 &  2.28 \\
  0.367 -   0.400 & $ 1.9 \pm  0.34 \pm  1.0 $ &  2.8 &  2.8 \\
  0.400 -   0.433 & $ 3.5 \pm  0.45 \pm  1.3 $ &  3.5 &  3.4 \\
  0.433 -   0.467 & $ 3.3 \pm  0.47 \pm  1.4 $ &  4.3 &  4.0 \\
  0.467 -   0.500 & $ 5.0 \pm  0.58 \pm  2.0 $ &  5.1 &  4.6 \\
  0.500 -   0.533 & $ 5.0 \pm  0.58 \pm  2.4 $ &  5.9 &  5.6 \\
  0.533 -   0.567 & $ 5.3 \pm  0.63 \pm  2.8 $ &  7.6 &  6.4 \\
  0.567 -   0.600 & $ 7.1 \pm  0.70 \pm  3.1 $ &  9.1 &  7.4 \\
  0.600 -   0.633 & $ 8.4 \pm  0.76 \pm  3.6 $ & 10.4 &  9.2 \\
  0.633 -   0.667 & $11.5 \pm  0.87 \pm  4.3 $ & 12.2 &  9.7 \\
  0.667 -   0.700 & $13.6 \pm  0.93 \pm  4.8 $ & 14.5 & 11.2 \\
  0.700 -   0.733 & $16.2 \pm  1.0 \pm  5.9 $ & 16.9 & 14.1 \\
  0.733 -   0.767 & $23.0 \pm  1.1 \pm  6.2 $ & 21.3 & 14.9 \\
  0.767 -   0.800 & $24.5 \pm  1.2 \pm  7.6 $ & 24.7 & 17.6 \\
  0.800 -   0.833 & $26.8 \pm  1.3 \pm  8.2 $ & 28.9 & 20.7 \\
  0.833 -   0.867 & $34.2 \pm  1.4 \pm  9.2 $ & 34.4 & 26.0 \\
  0.867 -   0.900 & $43.3 \pm  1.5 \pm 10.5 $ & 39.0 & 31.8 \\
  0.900 -   0.933 & $42.8 \pm  1.5 \pm 10.3 $ & 41.0 & 39.7 \\
  0.933 -   0.967 & $44.8 \pm  1.6 \pm 10.5 $ & 42.2 & 53.1 \\
  0.967 -   1.00 & $42.6 \pm  1.6 \pm 10.3 $ & 41.7 & 67.6 \\
\hline\hline
\end{tabular}
\end{center}
\end{table}

\pagebreak
\vfill

\renewcommand{\plotname}{zQHSS01}
\begin{center}
{\LARGE $E_T^{\gamma 2}/E_T^{\gamma 1}$ for $P_T(\gamma\gamma) > M(\gamma\gamma)$}
\end{center}

\begin{figure*}[!hp]
\includegraphics[width=9.0cm]{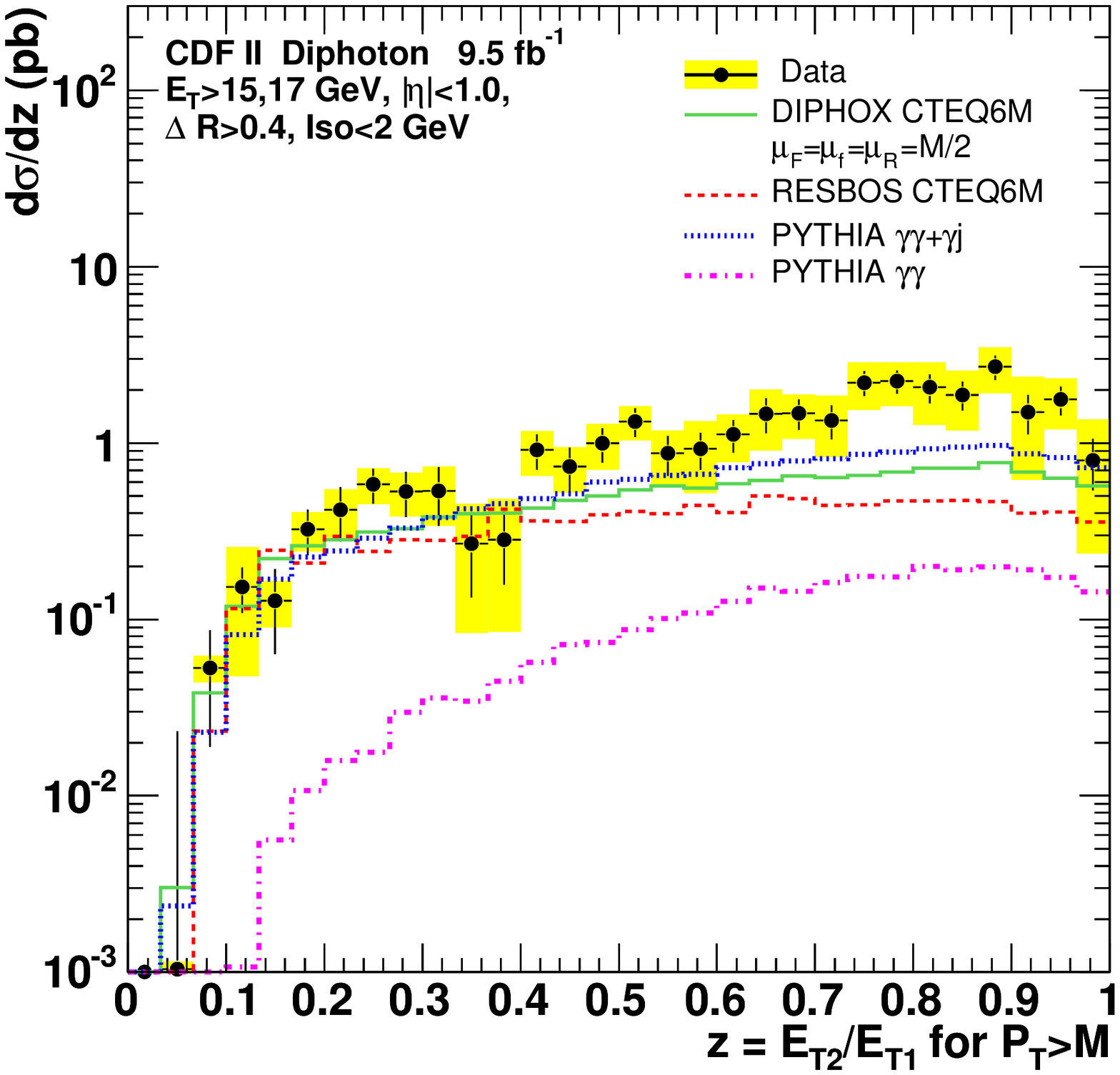}\hspace*{0cm}
\includegraphics[width=8.5cm]{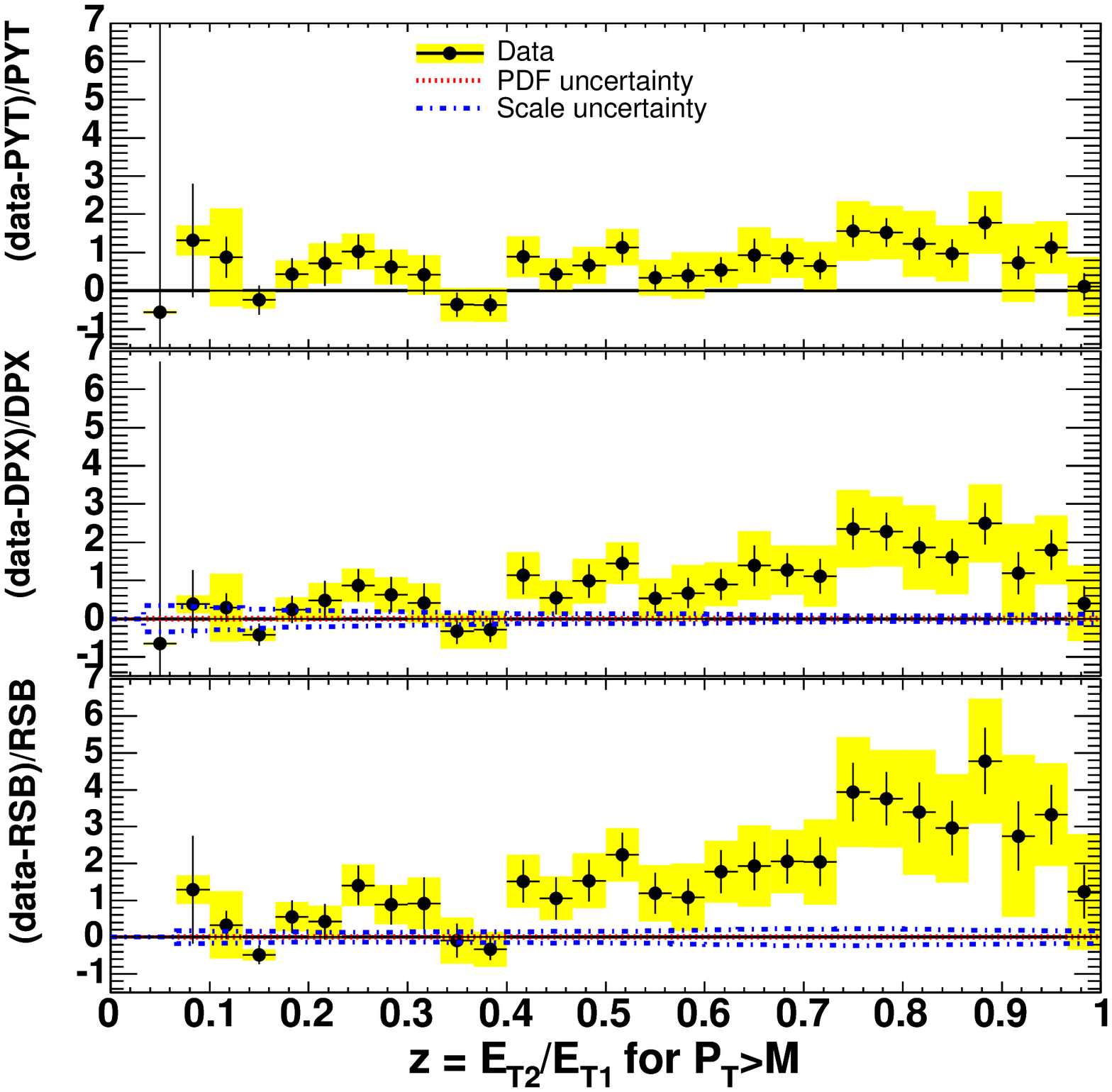}\hfill
\vspace*{0cm}
\includegraphics[width=9.0cm]{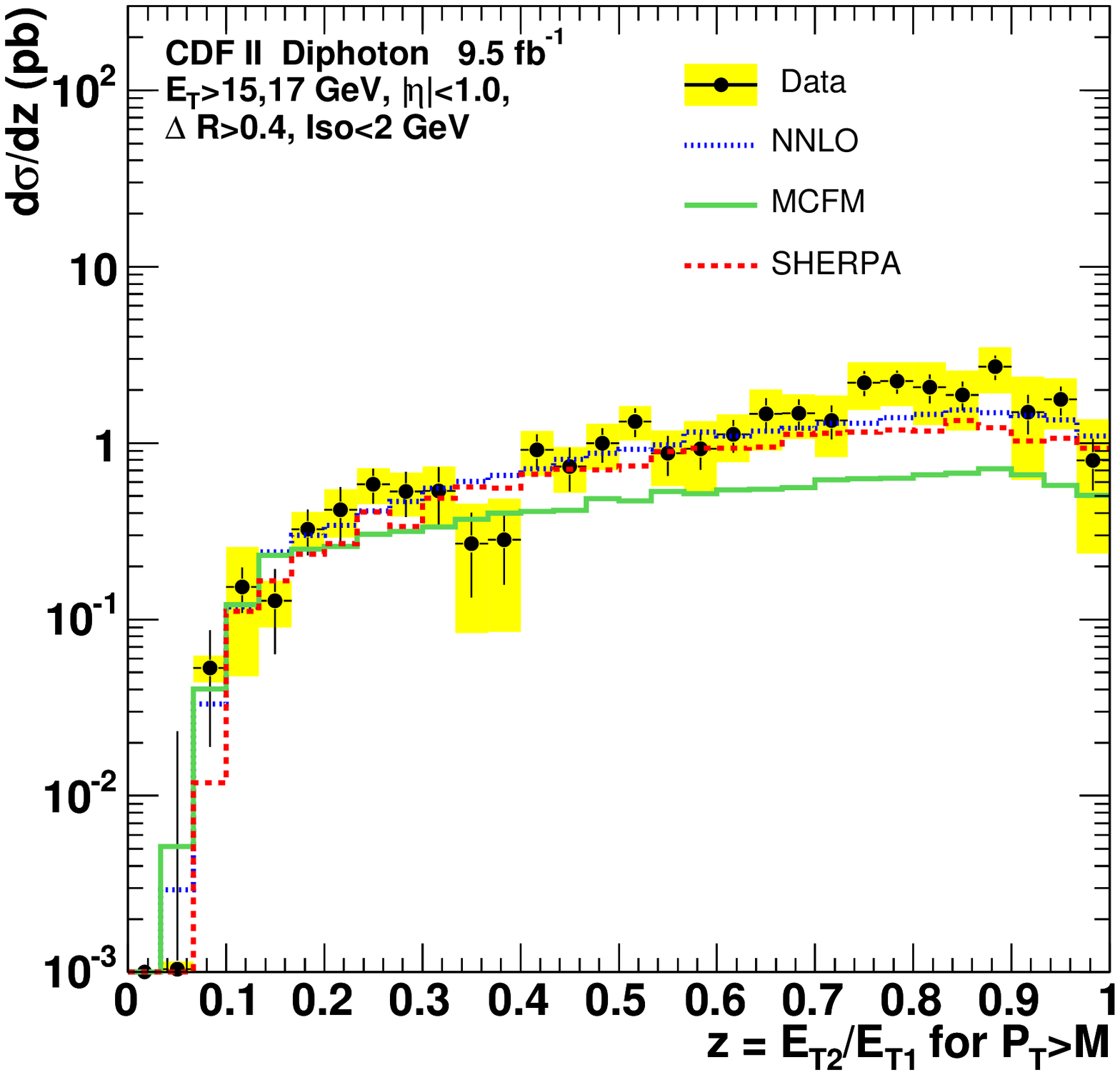}\hspace*{0cm}
\includegraphics[width=8.5cm]{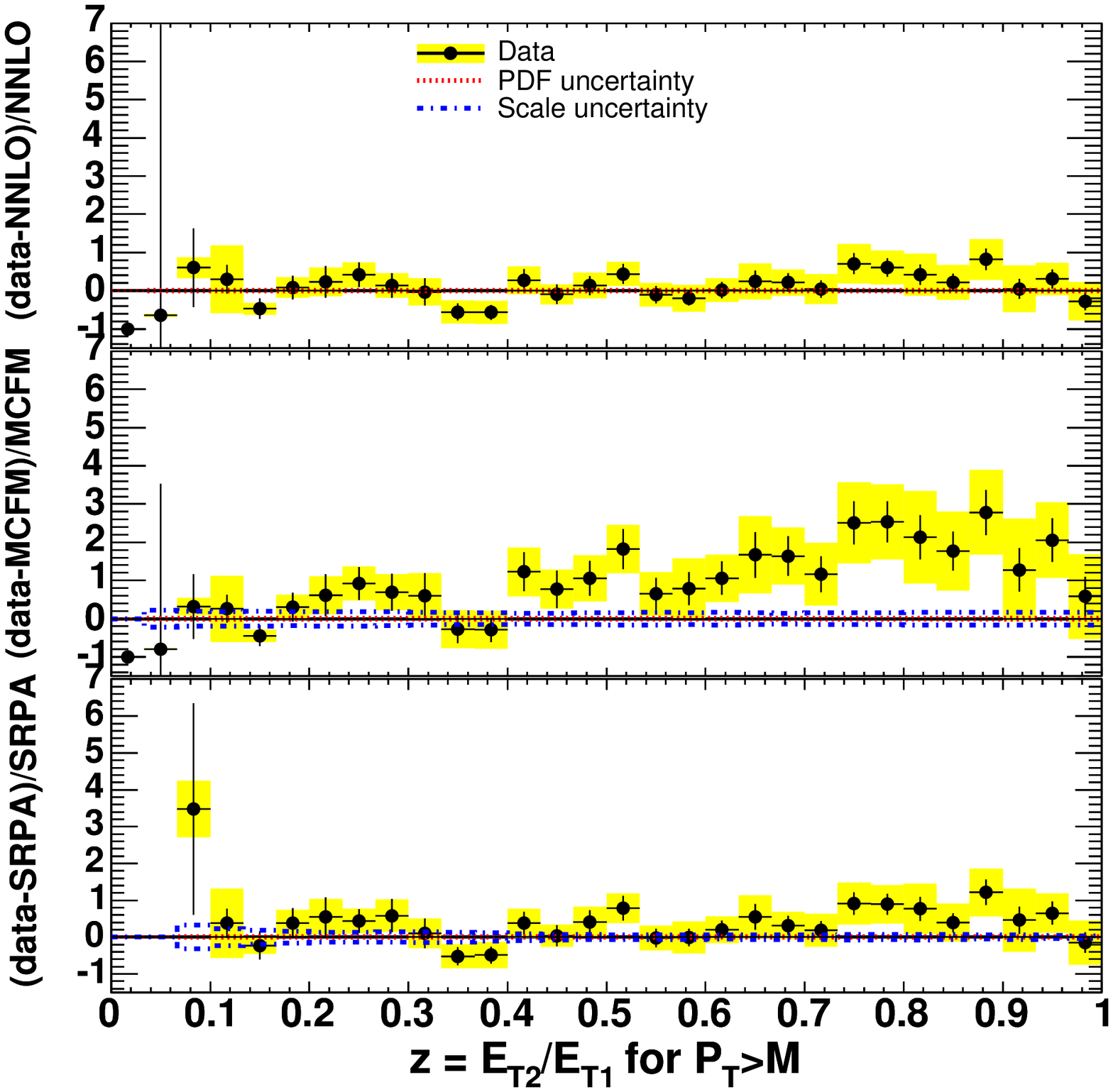}
\vspace*{-0.1cm}
\caption{The measured differential cross sections 
for $E_T^{\gamma 2}/E_T^{\gamma 1}$, 
when $P_T(\gamma\gamma) > M(\gamma\gamma)$,
compared with six 
theoretical predictions discussed in the text. The left windows show the
absolute comparisons and the right windows show the fractional deviations
of the data from the theoretical predictions. Note that the vertical axis
scales differ between fractional deviation plots.}
\end{figure*}

\pagebreak
\vfill

\begin{table}[!hp]
\begin{center}
\begin{tabular}{cccc}
\hline\hline
 Bin & Cross Section (pb) & Sherpa & NNLO   \\ \hline
  0.0333 -   0.0667 & $ 0.001 \pm  0.022 \pm  0.00011 $ &  0.000 &  0.003 \\
  0.0667 -   0.100 & $ 0.053 \pm  0.034 \pm  0.0091 $ &  0.012 &  0.033 \\
  0.100 -   0.133 & $ 0.15 \pm  0.044 \pm  0.11 $ &  0.11 &  0.12 \\
  0.133 -   0.167 & $ 0.128 \pm  0.064 \pm  0.038 $ &  0.166 &  0.242 \\
  0.167 -   0.200 & $ 0.325 \pm  0.094 \pm  0.082 $ &  0.234 &  0.300 \\
  0.200 -   0.233 & $ 0.42 \pm  0.14 \pm  0.13 $ &  0.27 &  0.34 \\
  0.233 -   0.267 & $ 0.58 \pm  0.13 \pm  0.14 $ &  0.41 &  0.41 \\
  0.267 -   0.300 & $ 0.53 \pm  0.15 \pm  0.15 $ &  0.34 &  0.47 \\
  0.300 -   0.333 & $ 0.54 \pm  0.20 \pm  0.20 $ &  0.49 &  0.55 \\
  0.333 -   0.367 & $ 0.27 \pm  0.14 \pm  0.18 $ &  0.56 &  0.61 \\
  0.367 -   0.400 & $ 0.28 \pm  0.13 \pm  0.20 $ &  0.55 &  0.65 \\
  0.400 -   0.433 & $ 0.91 \pm  0.21 \pm  0.26 $ &  0.66 &  0.72 \\
  0.433 -   0.467 & $ 0.74 \pm  0.20 \pm  0.21 $ &  0.71 &  0.81 \\
  0.467 -   0.500 & $ 1.00 \pm  0.22 \pm  0.29 $ &  0.70 &  0.88 \\
  0.500 -   0.533 & $ 1.33 \pm  0.25 \pm  0.30 $ &  0.74 &  0.92 \\
  0.533 -   0.567 & $ 0.87 \pm  0.22 \pm  0.30 $ &  0.90 &  0.98 \\
  0.567 -   0.600 & $ 0.92 \pm  0.22 \pm  0.41 $ &  0.93 &  1.15 \\
  0.600 -   0.633 & $ 1.12 \pm  0.24 \pm  0.34 $ &  0.94 &  1.10 \\
  0.633 -   0.667 & $ 1.46 \pm  0.33 \pm  0.55 $ &  0.95 &  1.17 \\
  0.667 -   0.700 & $ 1.47 \pm  0.29 \pm  0.42 $ &  1.12 &  1.21 \\
  0.700 -   0.733 & $ 1.34 \pm  0.29 \pm  0.51 $ &  1.14 &  1.29 \\
  0.733 -   0.767 & $ 2.20 \pm  0.35 \pm  0.67 $ &  1.15 &  1.29 \\
  0.767 -   0.800 & $ 2.24 \pm  0.34 \pm  0.62 $ &  1.19 &  1.40 \\
  0.800 -   0.833 & $ 2.07 \pm  0.39 \pm  0.80 $ &  1.17 &  1.46 \\
  0.833 -   0.867 & $ 1.87 \pm  0.35 \pm  0.70 $ &  1.35 &  1.54 \\
  0.867 -   0.900 & $ 2.70 \pm  0.42 \pm  0.79 $ &  1.22 &  1.49 \\
  0.900 -   0.933 & $ 1.50 \pm  0.38 \pm  0.88 $ &  1.03 &  1.43 \\
  0.933 -   0.967 & $ 1.76 \pm  0.33 \pm  0.57 $ &  1.07 &  1.35 \\
  0.967 -   1.00 & $ 0.80 \pm  0.26 \pm  0.56 $ &  0.93 &  1.10 \\
\hline\hline
\end{tabular}
\end{center}
\end{table}

\pagebreak
\vfill

\renewcommand{\plotname}{zQLSS01}
\begin{center}
{\LARGE $E_T^{\gamma 2}/E_T^{\gamma 1}$ for $P_T(\gamma\gamma) < M(\gamma\gamma)$}
\end{center}

\begin{figure*}[!hp]
\includegraphics[width=9.0cm]{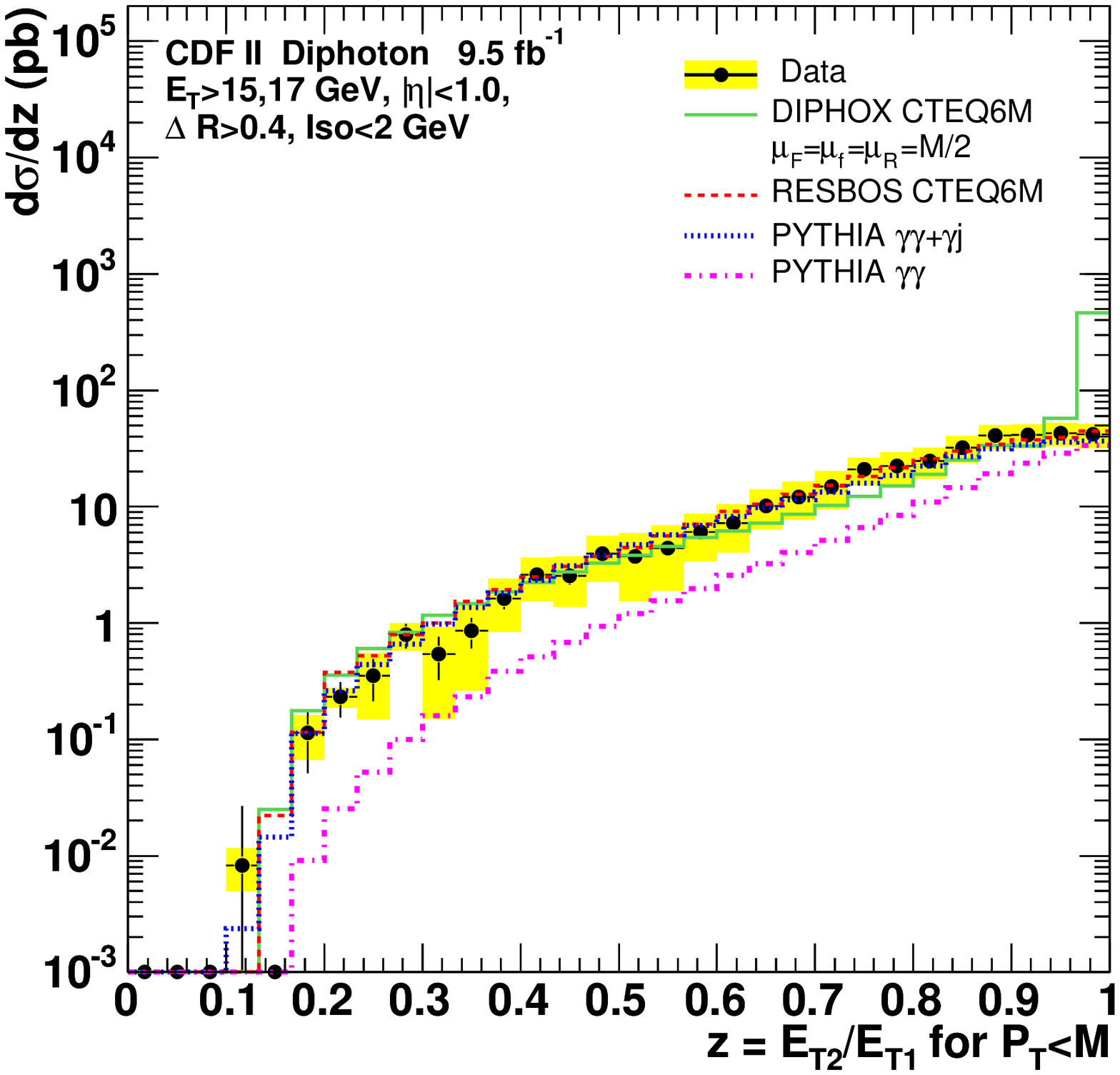}\hspace*{0cm}
\includegraphics[width=8.5cm]{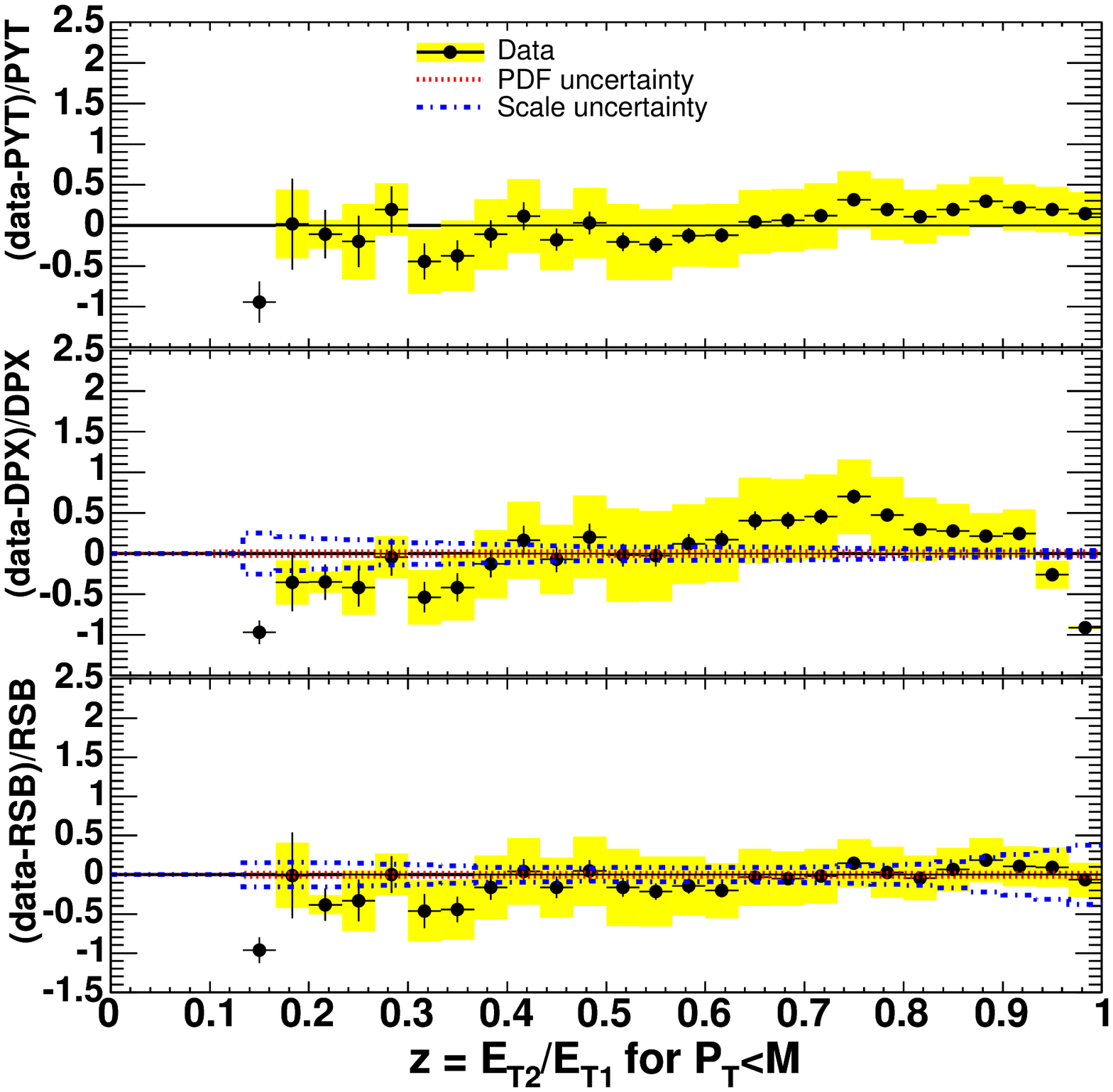}\hfill
\vspace*{0cm}
\includegraphics[width=9.0cm]{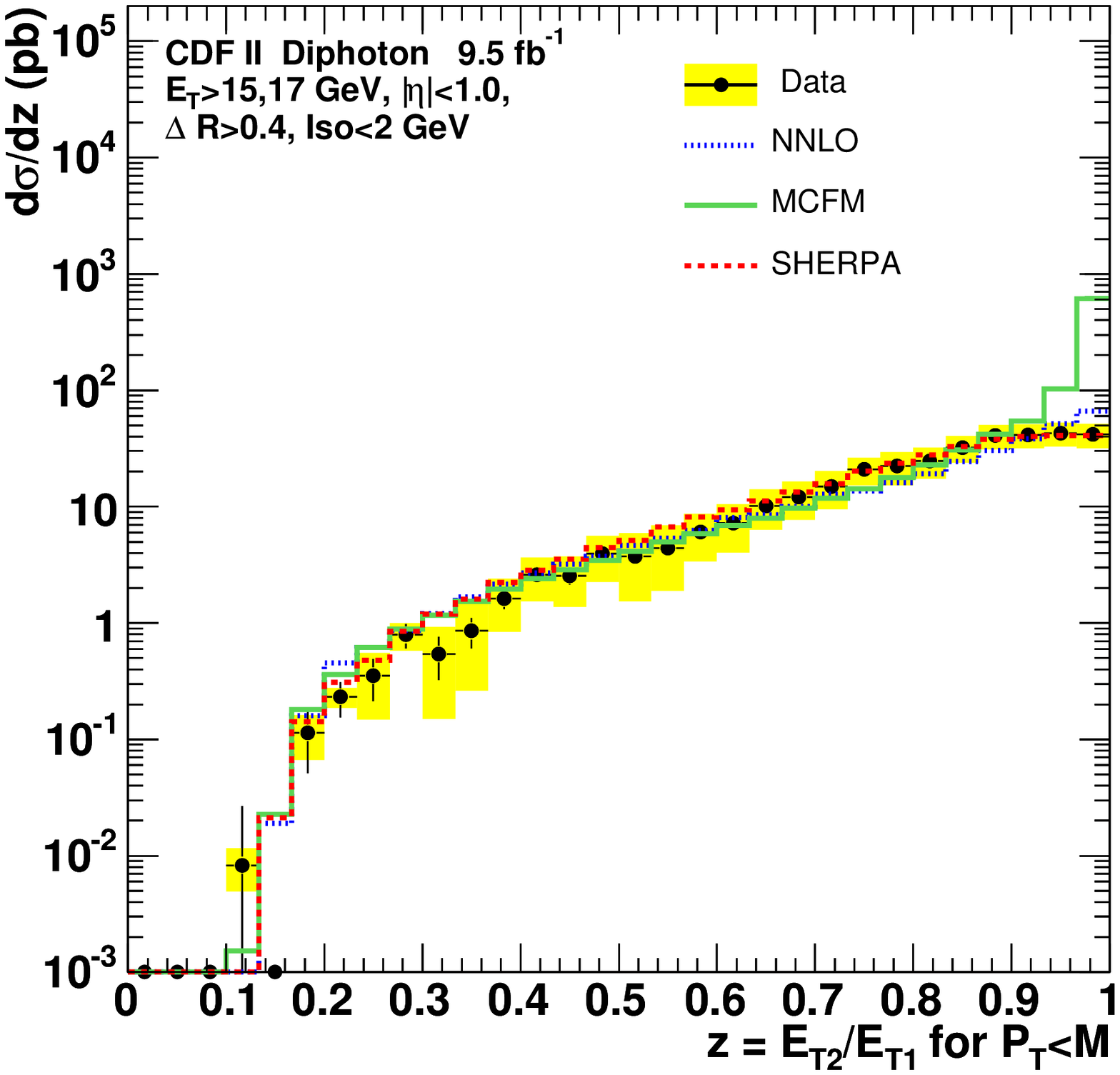}\hspace*{0cm}
\includegraphics[width=8.5cm]{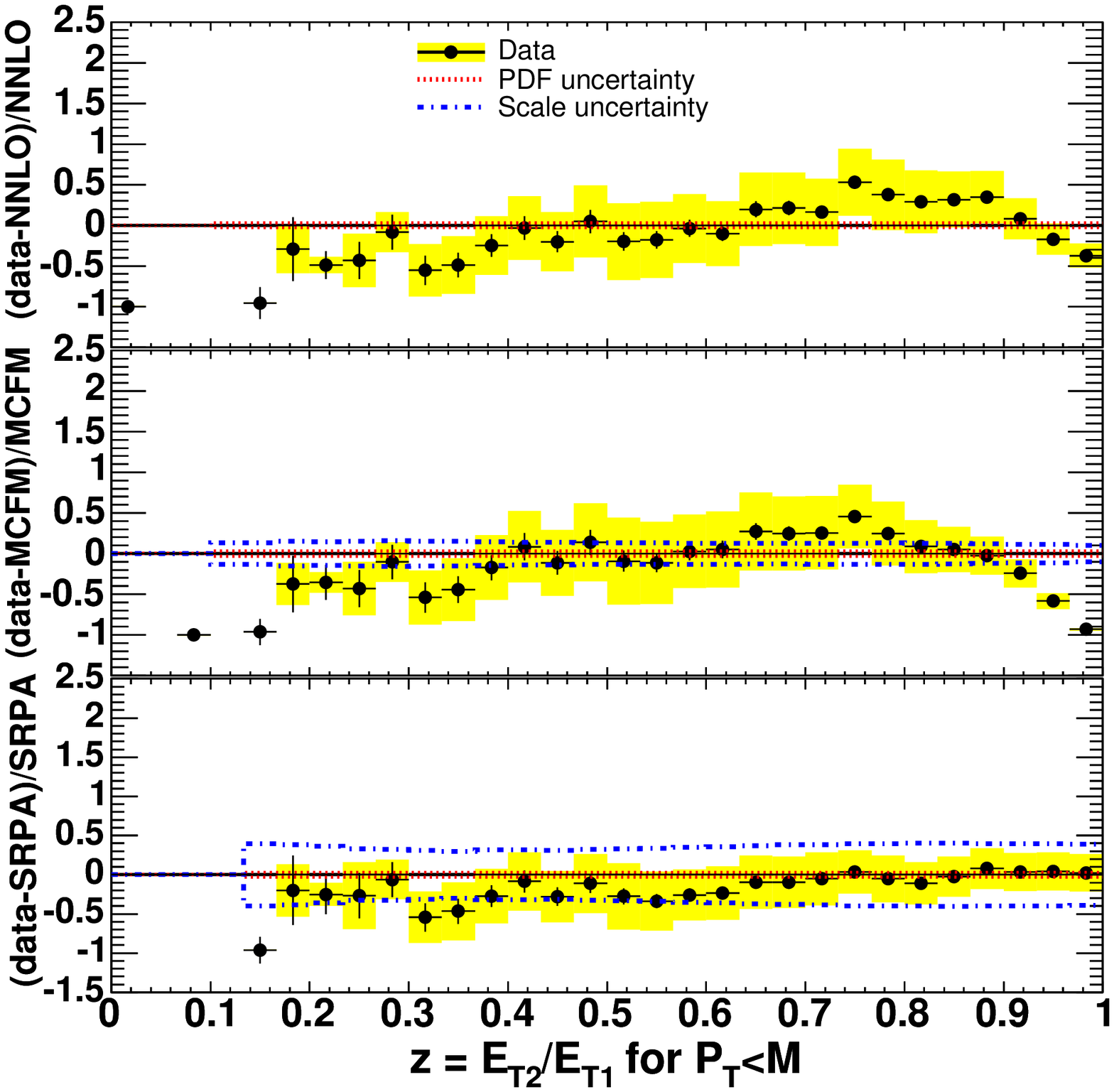}
\vspace*{-0.1cm}
\caption{The measured differential cross sections 
for $E_T^{\gamma 2}/E_T^{\gamma 1}$, 
when $P_T(\gamma\gamma) < M(\gamma\gamma)$,
compared with six 
theoretical predictions discussed in the text. The left windows show the
absolute comparisons and the right windows show the fractional deviations
of the data from the theoretical predictions. Note that the vertical axis
scales differ between fractional deviation plots.}
\end{figure*}

\pagebreak
\vfill

\begin{table}[!hp]
\begin{center}
\begin{tabular}{cccc}
\hline\hline
 Bin & Cross Section (pb) & Sherpa & NNLO   \\ \hline
  0.100 -   0.133 & $ 0.008 \pm  0.018 \pm  0.0033 $ &  0.000 &  0.001 \\
  0.133 -   0.167 & $ 0.0008 \pm  0.0037 \pm  0.000088 $ &  0.0213 &  0.0191 \\
  0.167 -   0.200 & $ 0.114 \pm  0.063 \pm  0.048 $ &  0.142 &  0.161 \\
  0.200 -   0.233 & $ 0.232 \pm  0.078 \pm  0.046 $ &  0.310 &  0.454 \\
  0.233 -   0.267 & $ 0.35 \pm  0.14 \pm  0.20 $ &  0.48 &  0.62 \\
  0.267 -   0.300 & $ 0.79 \pm  0.19 \pm  0.21 $ &  0.85 &  0.87 \\
  0.300 -   0.333 & $ 0.54 \pm  0.22 \pm  0.39 $ &  1.19 &  1.21 \\
  0.333 -   0.367 & $ 0.86 \pm  0.25 \pm  0.59 $ &  1.61 &  1.68 \\
  0.367 -   0.400 & $ 1.62 \pm  0.30 \pm  0.78 $ &  2.23 &  2.16 \\
  0.400 -   0.433 & $ 2.6 \pm  0.40 \pm  1.1 $ &  2.8 &  2.7 \\
  0.433 -   0.467 & $ 2.6 \pm  0.42 \pm  1.2 $ &  3.6 &  3.2 \\
  0.467 -   0.500 & $ 3.9 \pm  0.53 \pm  1.7 $ &  4.4 &  3.8 \\
  0.500 -   0.533 & $ 3.7 \pm  0.53 \pm  2.2 $ &  5.2 &  4.7 \\
  0.533 -   0.567 & $ 4.4 \pm  0.59 \pm  2.5 $ &  6.7 &  5.4 \\
  0.567 -   0.600 & $ 6.0 \pm  0.66 \pm  2.7 $ &  8.1 &  6.3 \\
  0.600 -   0.633 & $ 7.2 \pm  0.72 \pm  3.2 $ &  9.4 &  8.1 \\
  0.633 -   0.667 & $10.2 \pm  0.82 \pm  3.8 $ & 11.2 &  8.5 \\
  0.667 -   0.700 & $12.1 \pm  0.88 \pm  4.4 $ & 13.4 & 10.0 \\
  0.700 -   0.733 & $14.9 \pm  0.97 \pm  5.3 $ & 15.7 & 12.8 \\
  0.733 -   0.767 & $20.9 \pm  1.1 \pm  5.6 $ & 20.2 & 13.6 \\
  0.767 -   0.800 & $22.3 \pm  1.1 \pm  7.0 $ & 23.5 & 16.2 \\
  0.800 -   0.833 & $24.8 \pm  1.2 \pm  7.4 $ & 27.8 & 19.2 \\
  0.833 -   0.867 & $32.2 \pm  1.3 \pm  8.5 $ & 33.0 & 24.5 \\
  0.867 -   0.900 & $40.8 \pm  1.4 \pm  9.7 $ & 37.8 & 30.3 \\
  0.900 -   0.933 & $41.4 \pm  1.5 \pm  9.7 $ & 40.0 & 38.3 \\
  0.933 -   0.967 & $42.9 \pm  1.5 \pm 10.0 $ & 41.1 & 51.7 \\
  0.967 -   1.00 & $41.7 \pm  1.5 \pm  9.8 $ & 40.7 & 66.5 \\
\hline\hline
\end{tabular}
\end{center}
\end{table}

\pagebreak
\vfill
\renewcommand{\plotname}{CosSS01}

\begin{center}
{\LARGE $\cos\theta_{Collins-Soper}$}
\end{center}

\begin{figure*}[!hp]
\includegraphics[width=9.0cm]{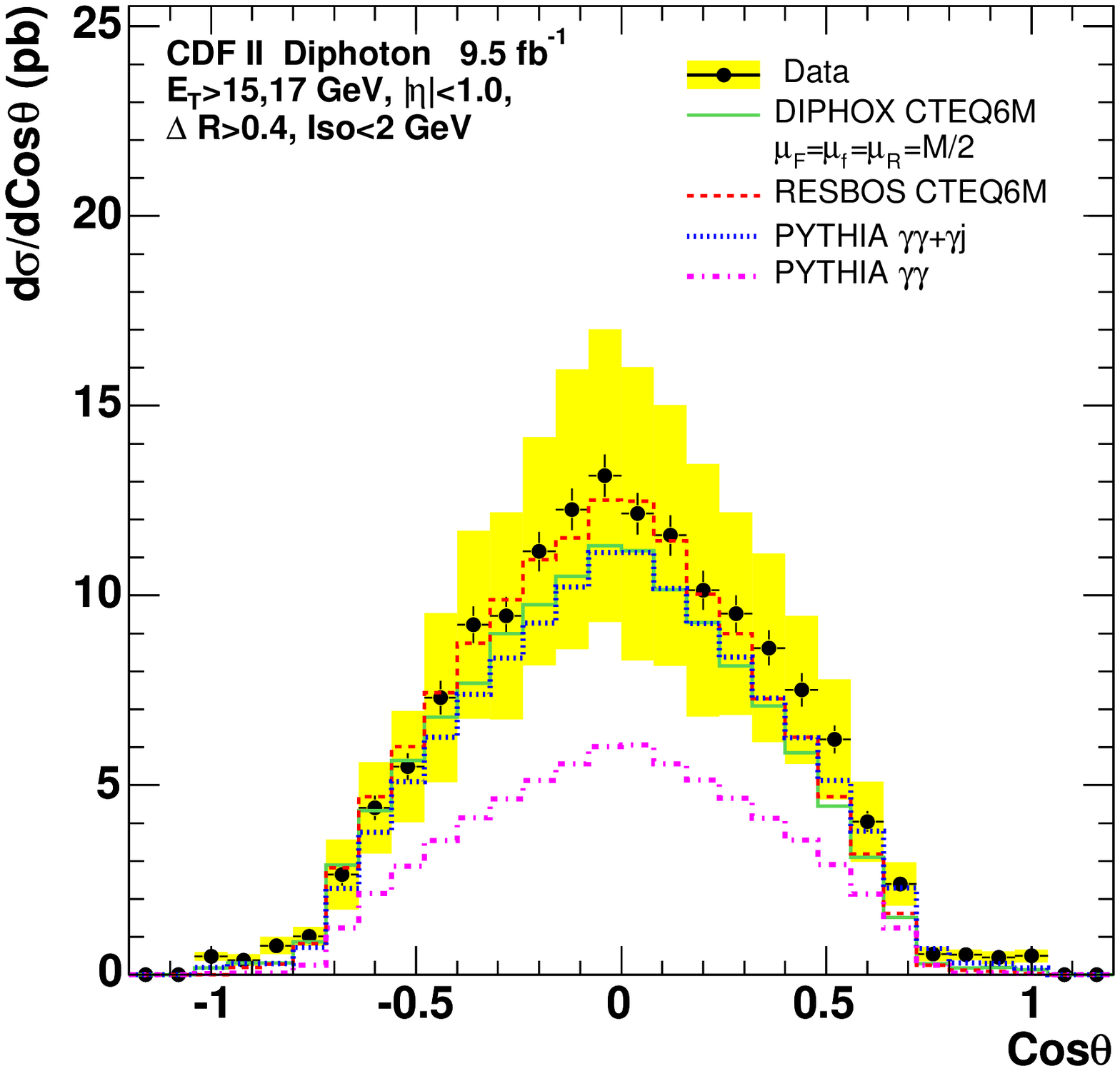}\hspace*{0cm}
\includegraphics[width=8.5cm]{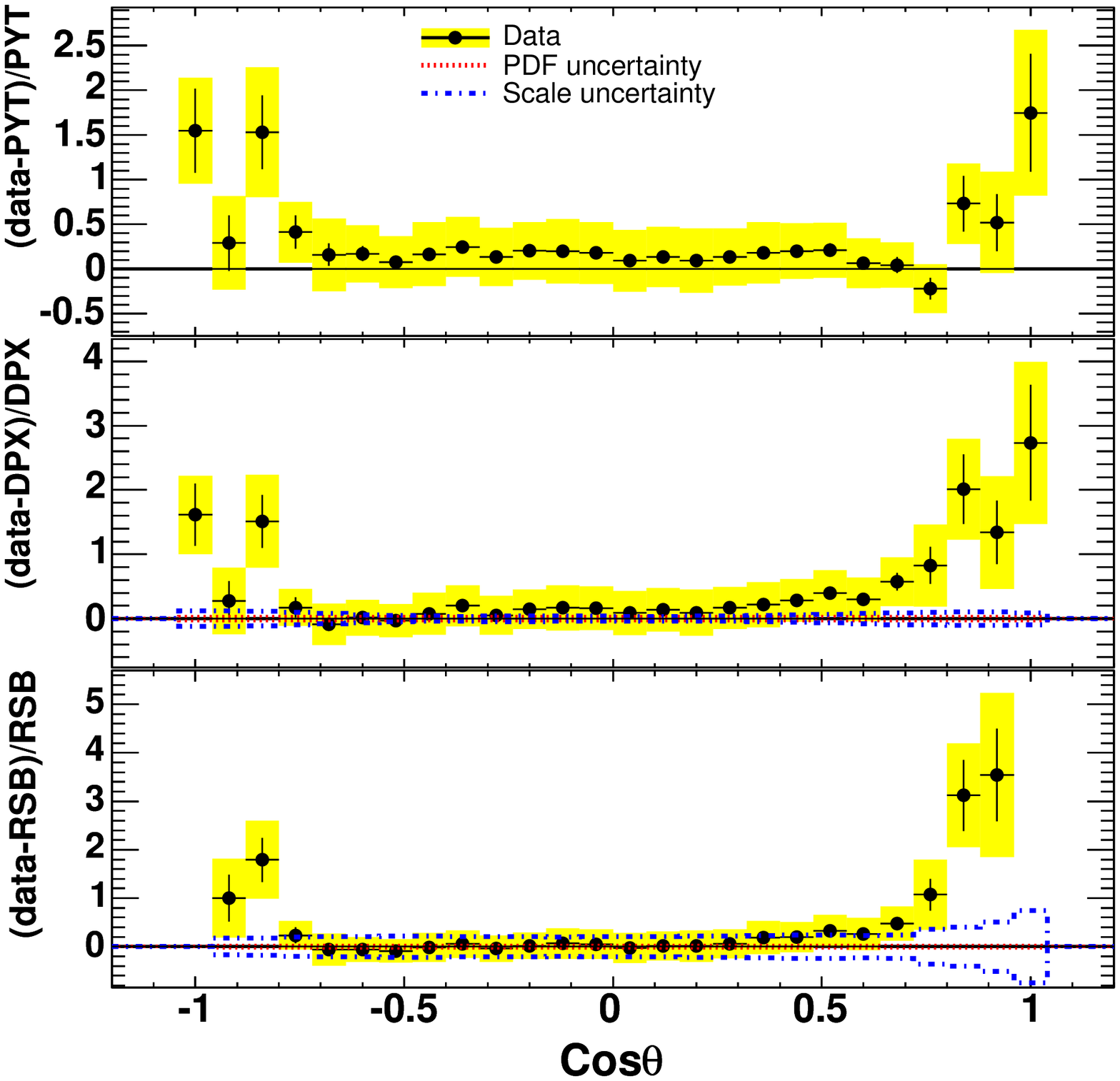}\hfill
\vspace*{0cm}
\includegraphics[width=9.0cm]{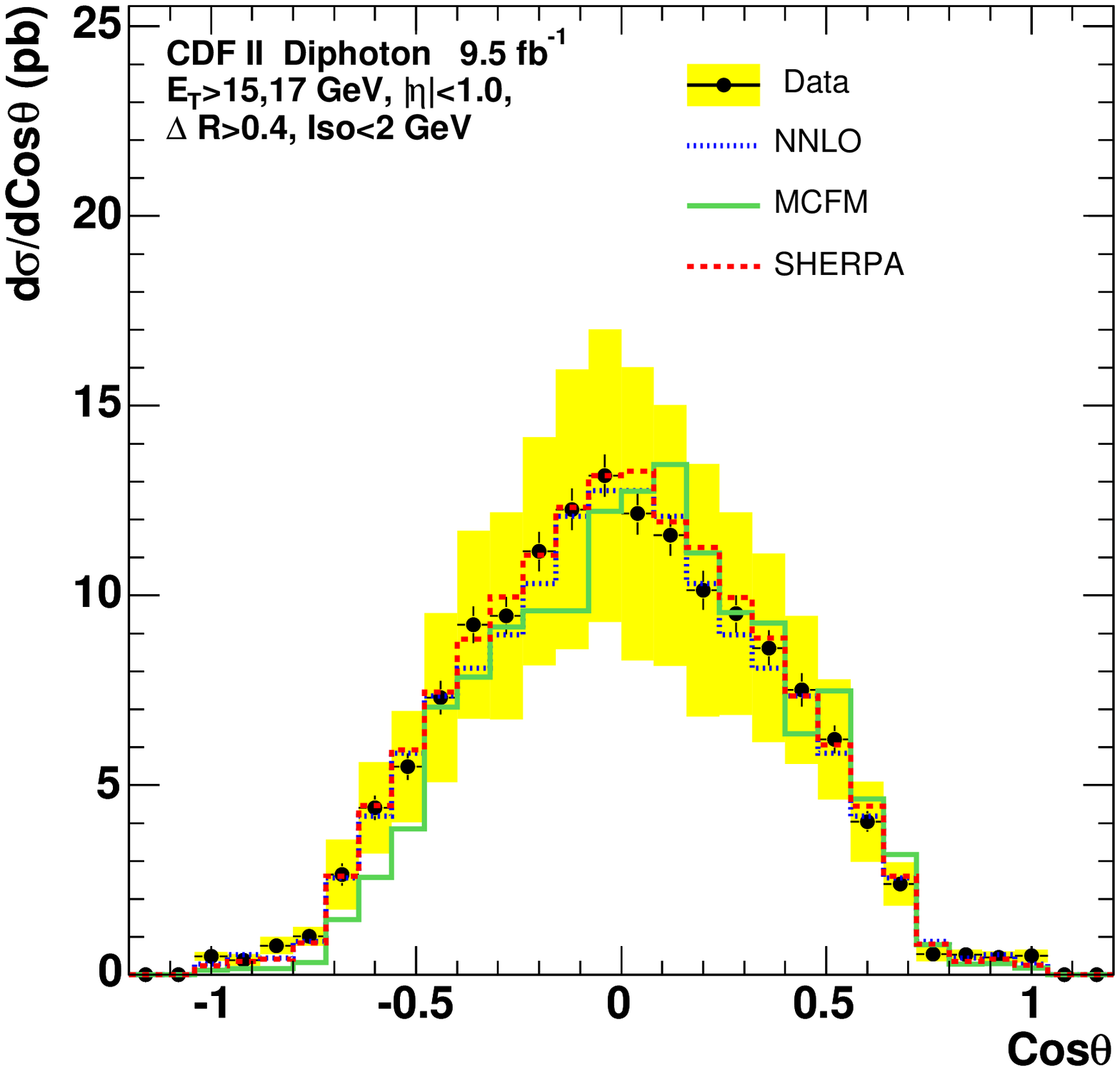}\hspace*{0cm}
\includegraphics[width=8.5cm]{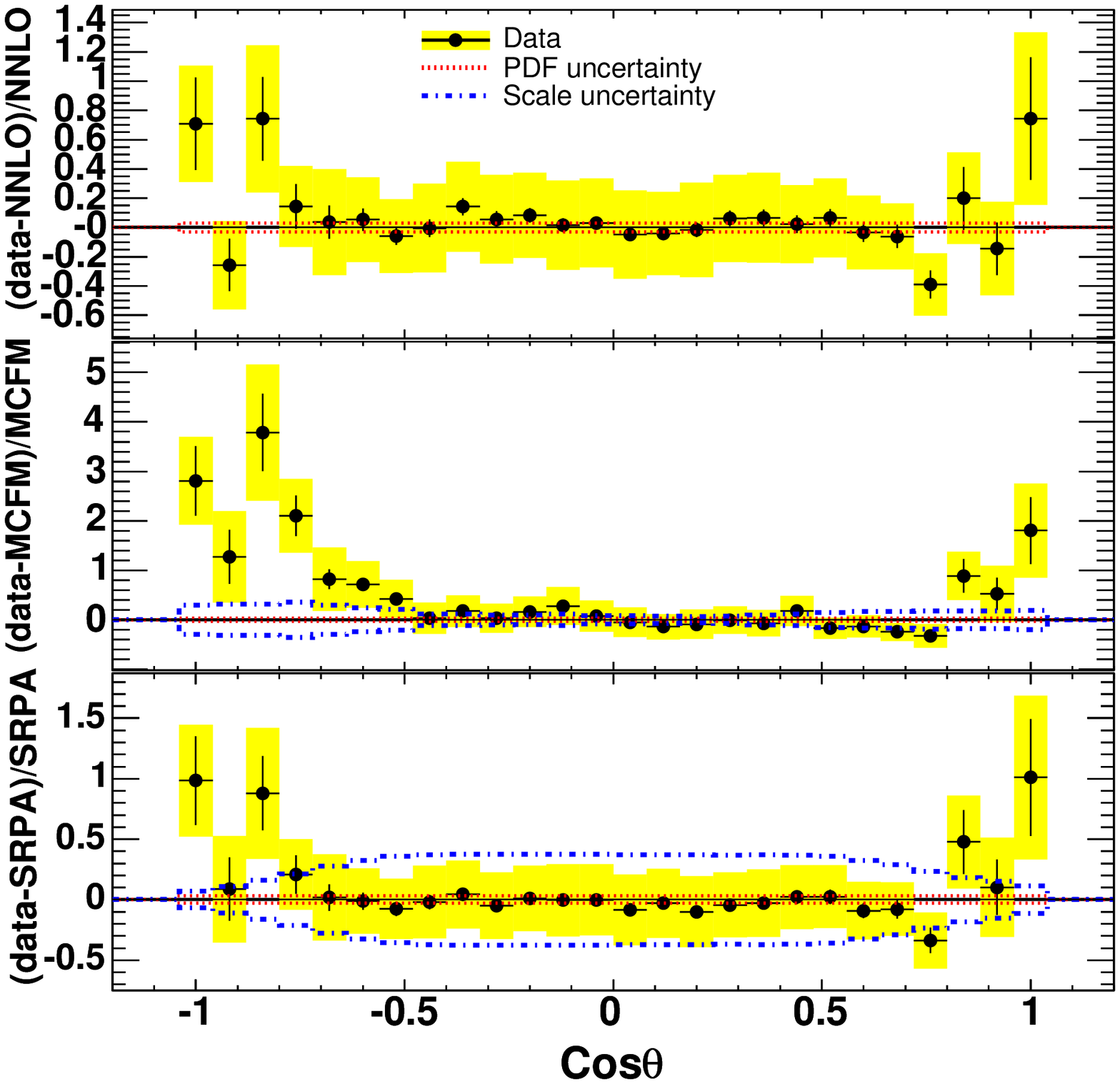}
\vspace*{-0.1cm}
\caption{The measured differential cross sections 
for the cosine of the polar angle of the leading 
photon in the Collins-Soper frame 
compared with six 
theoretical predictions discussed in the text. The left windows show the
absolute comparisons and the right windows show the fractional deviations
of the data from the theoretical predictions. Note that the vertical axis
scales differ between fractional deviation plots.}
\end{figure*}

\pagebreak
\vfill

\begin{table}[!hp]
\begin{center}
\begin{tabular}{cccc}
\hline\hline
 Bin & Cross Section (pb) & Sherpa & NNLO   \\ \hline
 -1.04 -  -0.960 & $ 0.49 \pm  0.090 \pm  0.11 $ &  0.25 &  0.29 \\
 -0.960 -  -0.880 & $ 0.39 \pm  0.095 \pm  0.16 $ &  0.36 &  0.53 \\
 -0.880 -  -0.800 & $ 0.77 \pm  0.13 \pm  0.22 $ &  0.41 &  0.44 \\
 -0.800 -  -0.720 & $ 1.01 \pm  0.13 \pm  0.24 $ &  0.84 &  0.89 \\
 -0.720 -  -0.640 & $ 2.64 \pm  0.29 \pm  0.93 $ &  2.60 &  2.55 \\
 -0.640 -  -0.560 & $ 4.4 \pm  0.32 \pm  1.2 $ &  4.5 &  4.2 \\
 -0.560 -  -0.480 & $ 5.5 \pm  0.35 \pm  1.5 $ &  5.9 &  5.8 \\
 -0.480 -  -0.400 & $ 7.3 \pm  0.44 \pm  2.2 $ &  7.5 &  7.3 \\
 -0.400 -  -0.320 & $ 9.2 \pm  0.48 \pm  2.5 $ &  8.9 &  8.1 \\
 -0.320 -  -0.240 & $ 9.5 \pm  0.49 \pm  2.7 $ & 10.0 &  9.0 \\
 -0.240 -  -0.160 & $11.2 \pm  0.52 \pm  3.0 $ & 11.1 & 10.3 \\
 -0.160 -  -0.0800 & $12.3 \pm  0.55 \pm  3.7 $ & 12.3 & 12.1 \\
 -0.0800 -   0.00 & $13.2 \pm  0.55 \pm  3.9 $ & 13.2 & 12.8 \\
  0.00 -   0.0800 & $12.2 \pm  0.55 \pm  3.9 $ & 13.3 & 12.8 \\
  0.0800 -   0.160 & $11.6 \pm  0.53 \pm  3.4 $ & 11.9 & 12.1 \\
  0.160 -   0.240 & $10.1 \pm  0.51 \pm  3.3 $ & 11.3 & 10.3 \\
  0.240 -   0.320 & $ 9.5 \pm  0.49 \pm  2.7 $ & 10.0 &  9.0 \\
  0.320 -   0.400 & $ 8.6 \pm  0.46 \pm  2.5 $ &  8.9 &  8.1 \\
  0.400 -   0.480 & $ 7.5 \pm  0.44 \pm  2.0 $ &  7.4 &  7.3 \\
  0.480 -   0.560 & $ 6.2 \pm  0.36 \pm  1.6 $ &  6.1 &  5.8 \\
  0.560 -   0.640 & $ 4.0 \pm  0.27 \pm  1.1 $ &  4.4 &  4.2 \\
  0.640 -   0.720 & $ 2.39 \pm  0.21 \pm  0.58 $ &  2.60 &  2.55 \\
  0.720 -   0.800 & $ 0.54 \pm  0.086 \pm  0.19 $ &  0.82 &  0.89 \\
  0.800 -   0.880 & $ 0.53 \pm  0.095 \pm  0.14 $ &  0.36 &  0.44 \\
  0.880 -   0.960 & $ 0.45 \pm  0.095 \pm  0.17 $ &  0.41 &  0.53 \\
  0.960 -   1.04 & $ 0.50 \pm  0.12 \pm  0.17 $ &  0.25 &  0.29 \\
\hline\hline
\end{tabular}
\end{center}
\end{table}

\pagebreak
\vfill

\renewcommand{\plotname}{CosQHSS01}
\begin{center}
{\LARGE $\cos\theta_{Collins-Soper}$ for $P_T(\gamma\gamma) > M(\gamma\gamma)$}
\end{center}

\begin{figure*}[!hp]
\includegraphics[width=9.0cm]{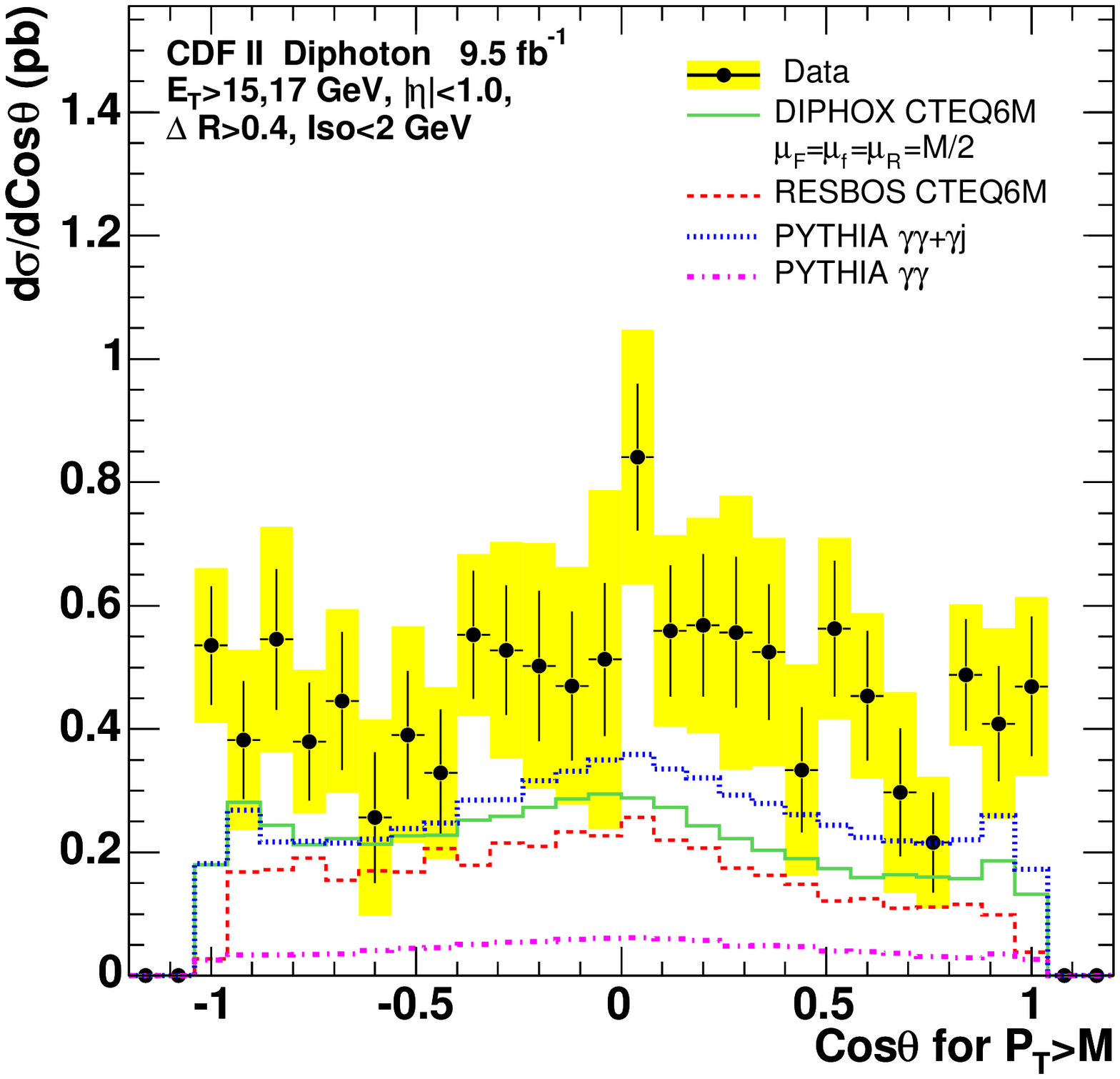}\hspace*{0cm}
\includegraphics[width=8.5cm]{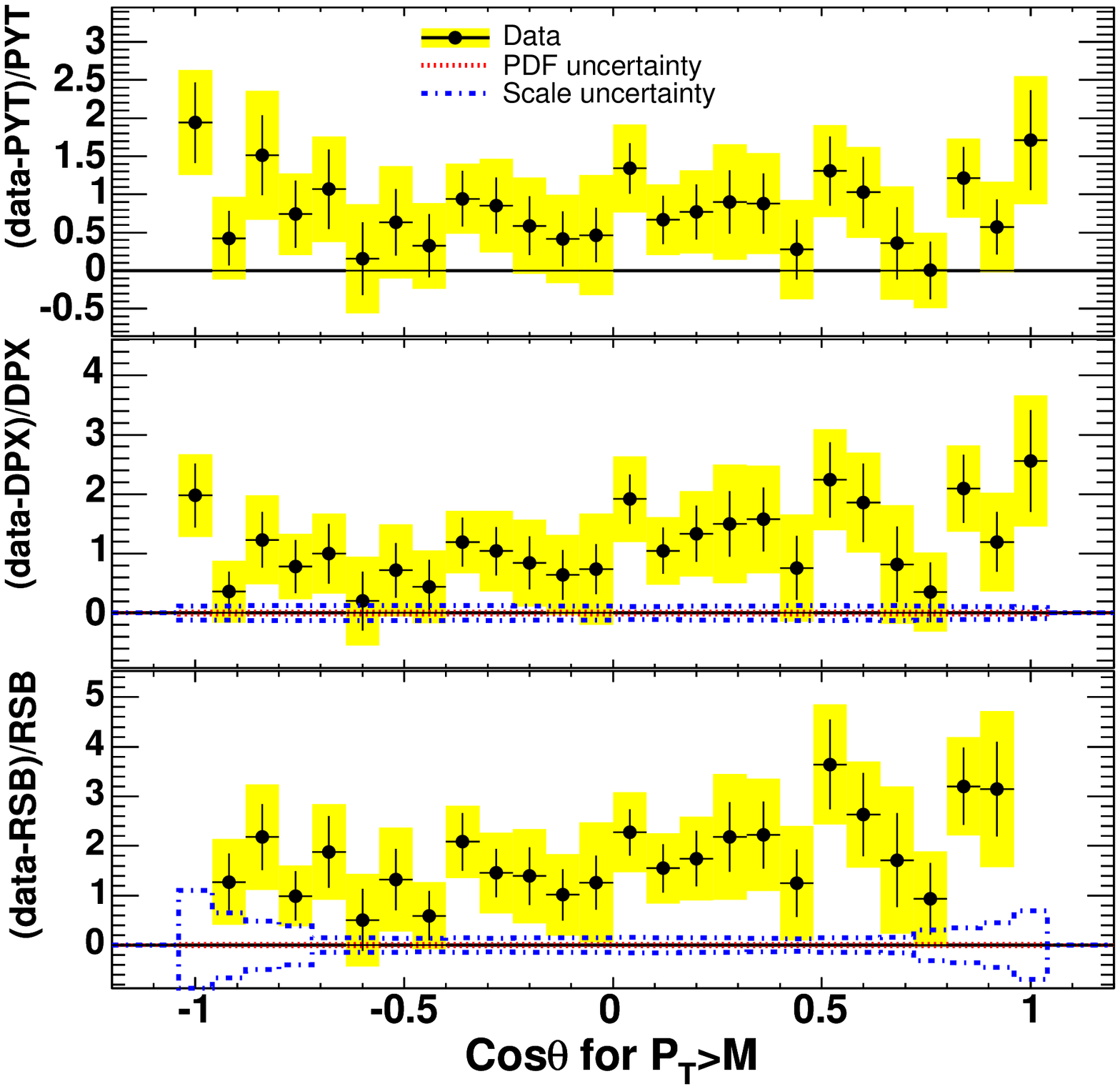}\hfill
\vspace*{0cm}
\includegraphics[width=9.0cm]{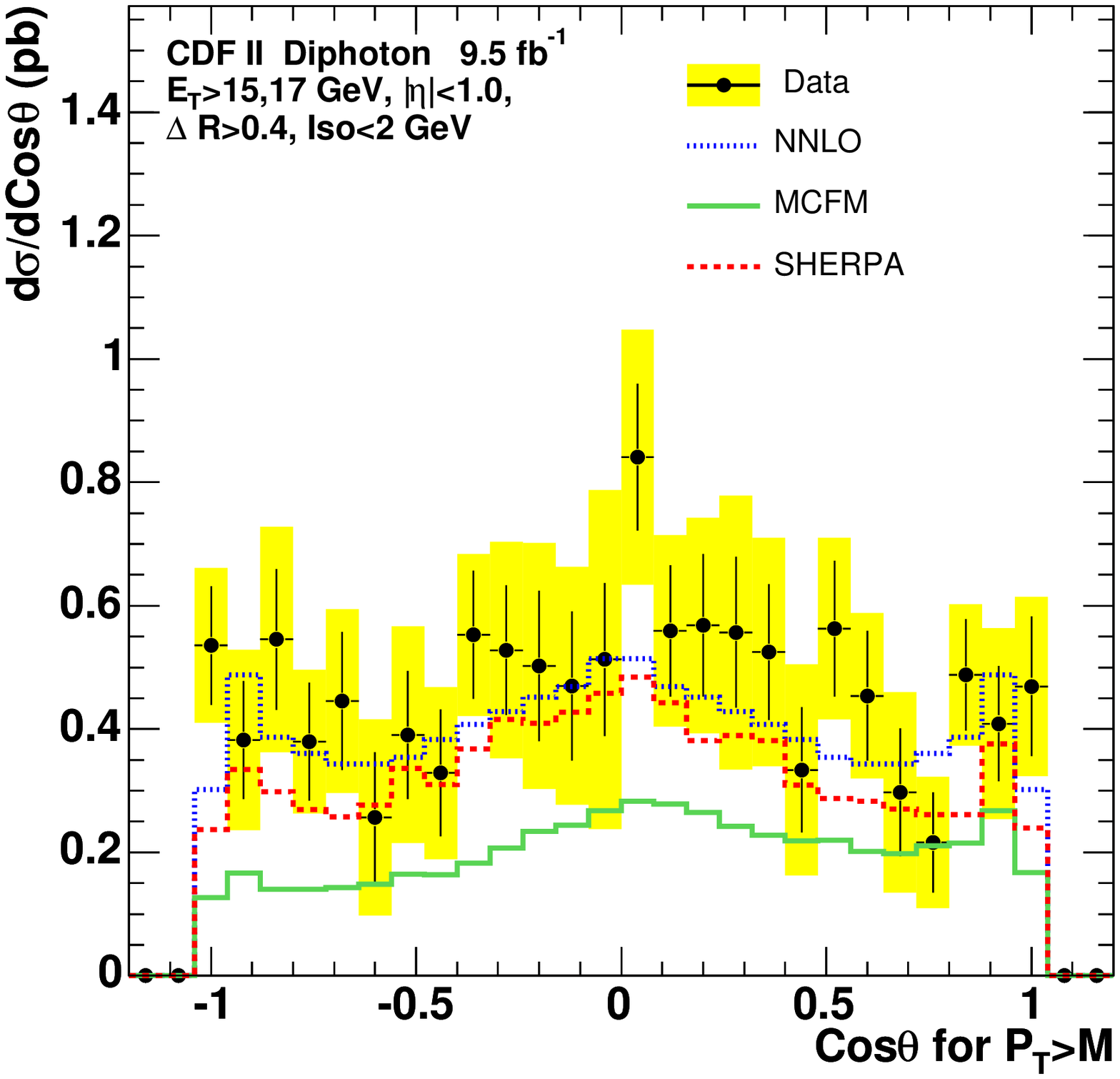}\hspace*{0cm}
\includegraphics[width=8.5cm]{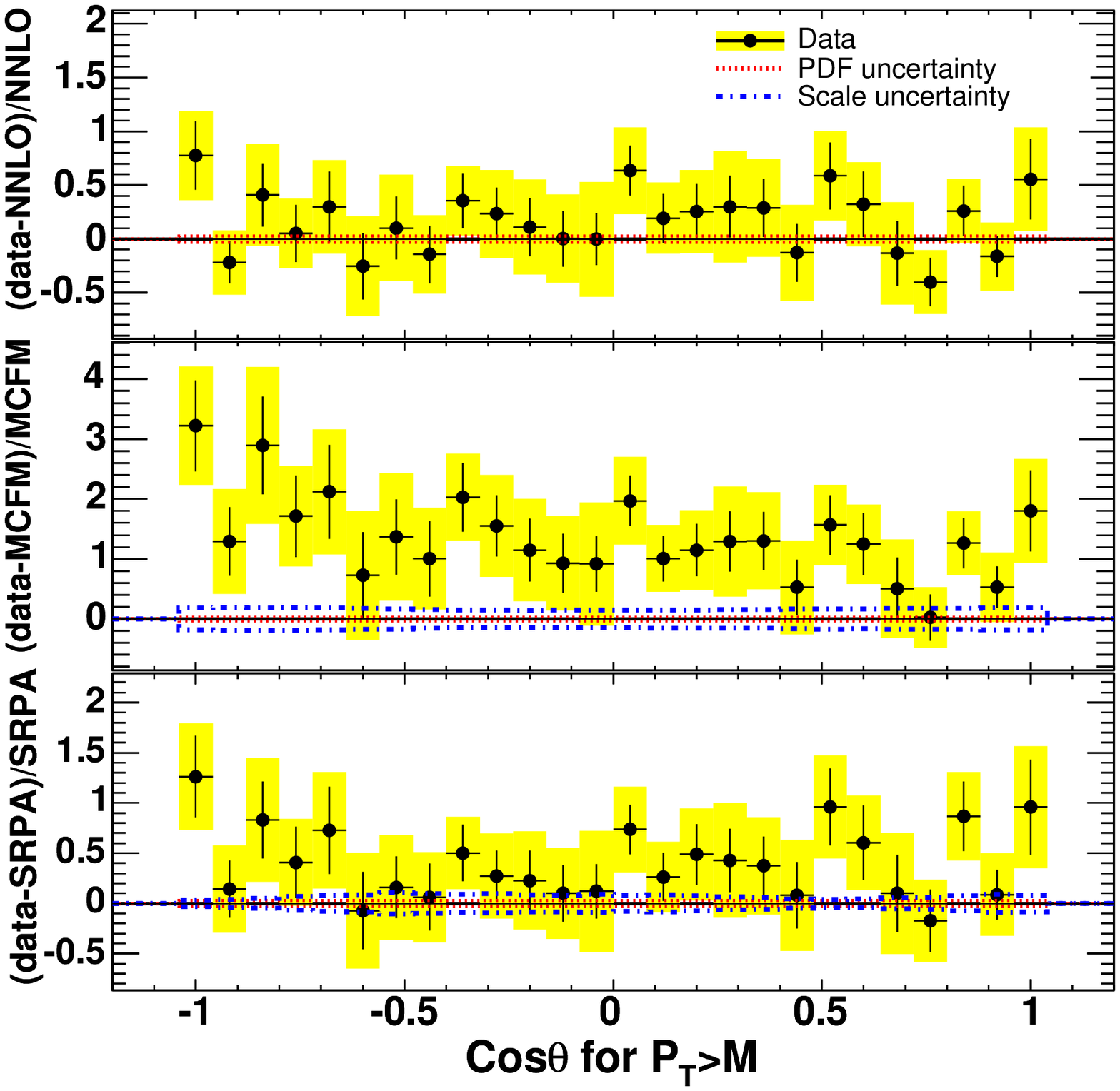}
\vspace*{-0.1cm}
\caption{The measured differential cross sections 
for the cosine of the polar angle of the leading 
photon in the Collins-Soper frame, 
when $P_T(\gamma\gamma) > M(\gamma\gamma)$, 
compared with six 
theoretical predictions discussed in the text. The left windows show the
absolute comparisons and the right windows show the fractional deviations
of the data from the theoretical predictions. Note that the vertical axis
scales differ between fractional deviation plots.}
\end{figure*}

\pagebreak
\vfill

\begin{table}[!hp]
\begin{center}
\begin{tabular}{cccc}
\hline\hline
 Bin & Cross Section (pb) & Sherpa & NNLO   \\ \hline
 -1.04 -  -0.960 & $ 0.54 \pm  0.096 \pm  0.13 $ &  0.24 &  0.30 \\
 -0.960 -  -0.880 & $ 0.38 \pm  0.096 \pm  0.15 $ &  0.33 &  0.49 \\
 -0.880 -  -0.800 & $ 0.55 \pm  0.11 \pm  0.18 $ &  0.30 &  0.39 \\
 -0.800 -  -0.720 & $ 0.38 \pm  0.096 \pm  0.12 $ &  0.27 &  0.36 \\
 -0.720 -  -0.640 & $ 0.45 \pm  0.11 \pm  0.15 $ &  0.26 &  0.34 \\
 -0.640 -  -0.560 & $ 0.26 \pm  0.11 \pm  0.16 $ &  0.28 &  0.34 \\
 -0.560 -  -0.480 & $ 0.39 \pm  0.10 \pm  0.18 $ &  0.34 &  0.35 \\
 -0.480 -  -0.400 & $ 0.33 \pm  0.10 \pm  0.14 $ &  0.31 &  0.38 \\
 -0.400 -  -0.320 & $ 0.55 \pm  0.10 \pm  0.13 $ &  0.37 &  0.41 \\
 -0.320 -  -0.240 & $ 0.53 \pm  0.11 \pm  0.18 $ &  0.42 &  0.43 \\
 -0.240 -  -0.160 & $ 0.50 \pm  0.12 \pm  0.20 $ &  0.41 &  0.45 \\
 -0.160 -  -0.0800 & $ 0.47 \pm  0.12 \pm  0.19 $ &  0.43 &  0.47 \\
 -0.0800 -   0.00 & $ 0.51 \pm  0.12 \pm  0.28 $ &  0.46 &  0.51 \\
  0.00 -   0.0800 & $ 0.84 \pm  0.12 \pm  0.21 $ &  0.48 &  0.51 \\
  0.0800 -   0.160 & $ 0.56 \pm  0.11 \pm  0.16 $ &  0.44 &  0.47 \\
  0.160 -   0.240 & $ 0.57 \pm  0.12 \pm  0.17 $ &  0.38 &  0.45 \\
  0.240 -   0.320 & $ 0.56 \pm  0.12 \pm  0.22 $ &  0.39 &  0.43 \\
  0.320 -   0.400 & $ 0.52 \pm  0.11 \pm  0.18 $ &  0.38 &  0.41 \\
  0.400 -   0.480 & $ 0.33 \pm  0.10 \pm  0.17 $ &  0.31 &  0.38 \\
  0.480 -   0.560 & $ 0.56 \pm  0.11 \pm  0.15 $ &  0.29 &  0.35 \\
  0.560 -   0.640 & $ 0.45 \pm  0.10 \pm  0.13 $ &  0.28 &  0.34 \\
  0.640 -   0.720 & $ 0.30 \pm  0.10 \pm  0.16 $ &  0.27 &  0.34 \\
  0.720 -   0.800 & $ 0.22 \pm  0.081 \pm  0.11 $ &  0.26 &  0.36 \\
  0.800 -   0.880 & $ 0.49 \pm  0.091 \pm  0.11 $ &  0.26 &  0.39 \\
  0.880 -   0.960 & $ 0.41 \pm  0.094 \pm  0.15 $ &  0.38 &  0.49 \\
  0.960 -   1.04 & $ 0.47 \pm  0.11 \pm  0.15 $ &  0.24 &  0.30 \\
\hline\hline
\end{tabular}
\end{center}
\end{table}

\pagebreak
\vfill

\renewcommand{\plotname}{CosQLSS01}
\begin{center}
{\LARGE $\cos\theta_{Collins-Soper}$ for $P_T(\gamma\gamma) < M(\gamma\gamma)$}
\end{center}

\begin{figure*}[!hp]
\includegraphics[width=9.0cm]{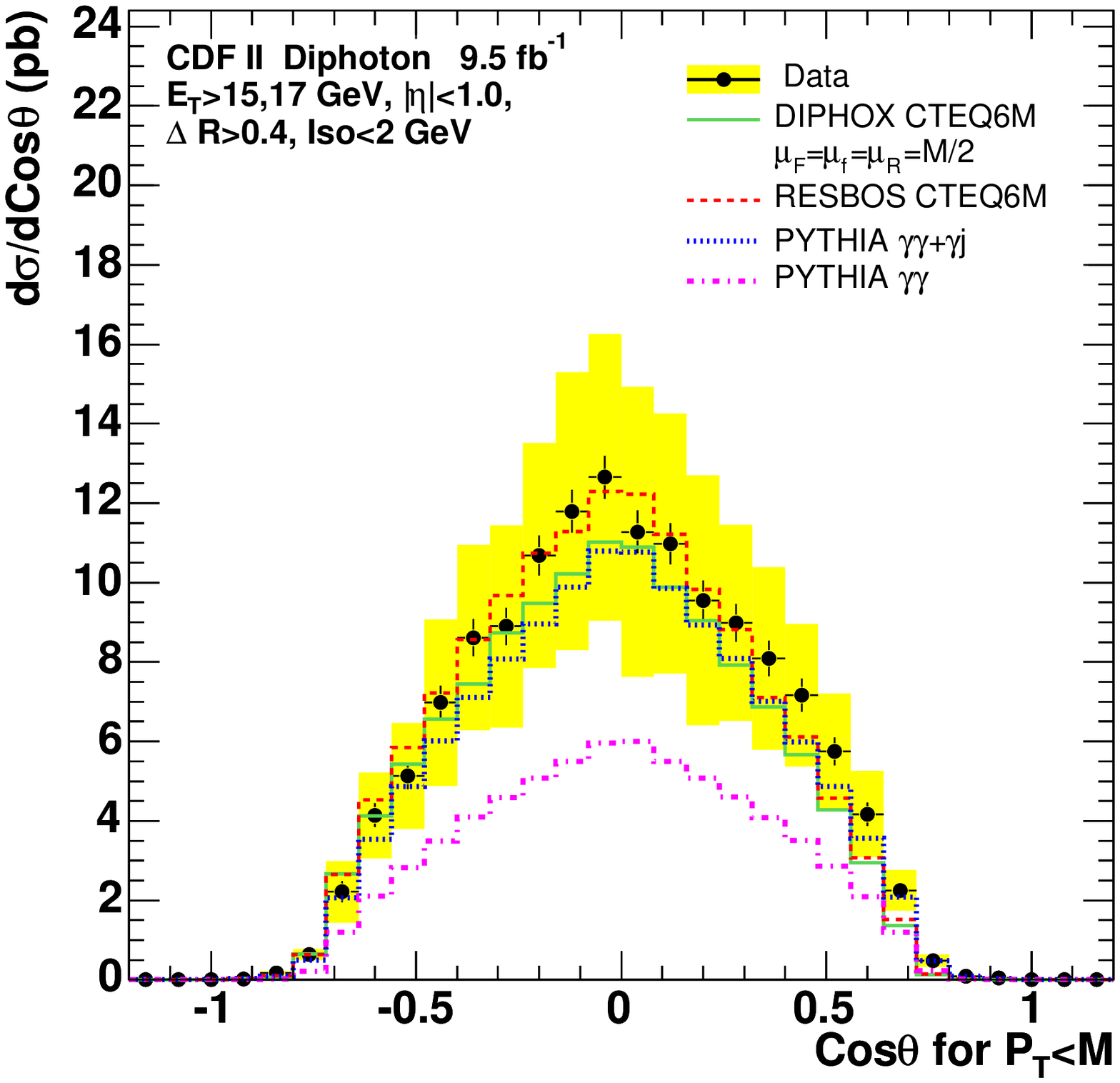}\hspace*{0cm}
\includegraphics[width=8.5cm]{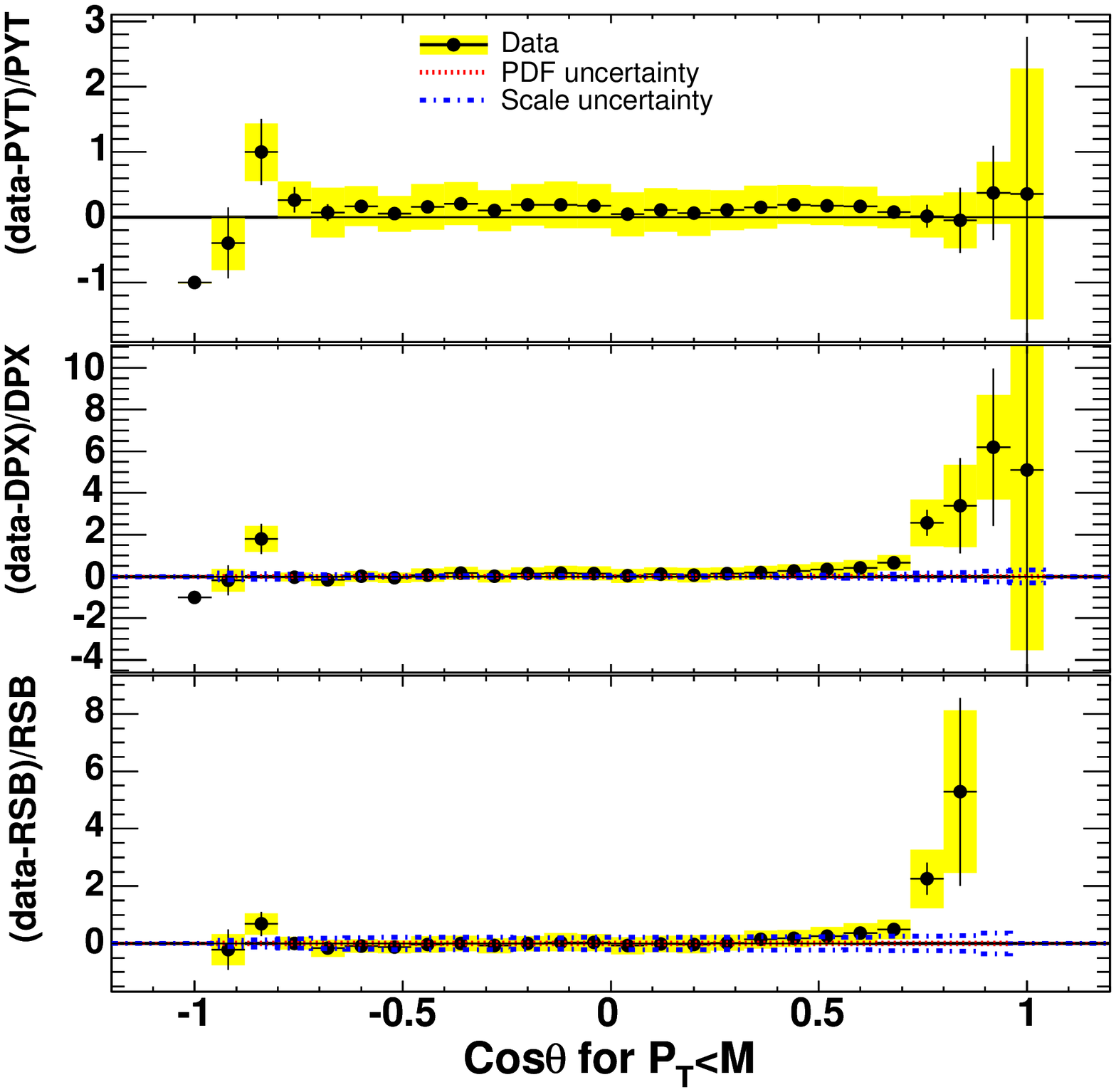}\hfill
\vspace*{0cm}
\includegraphics[width=9.0cm]{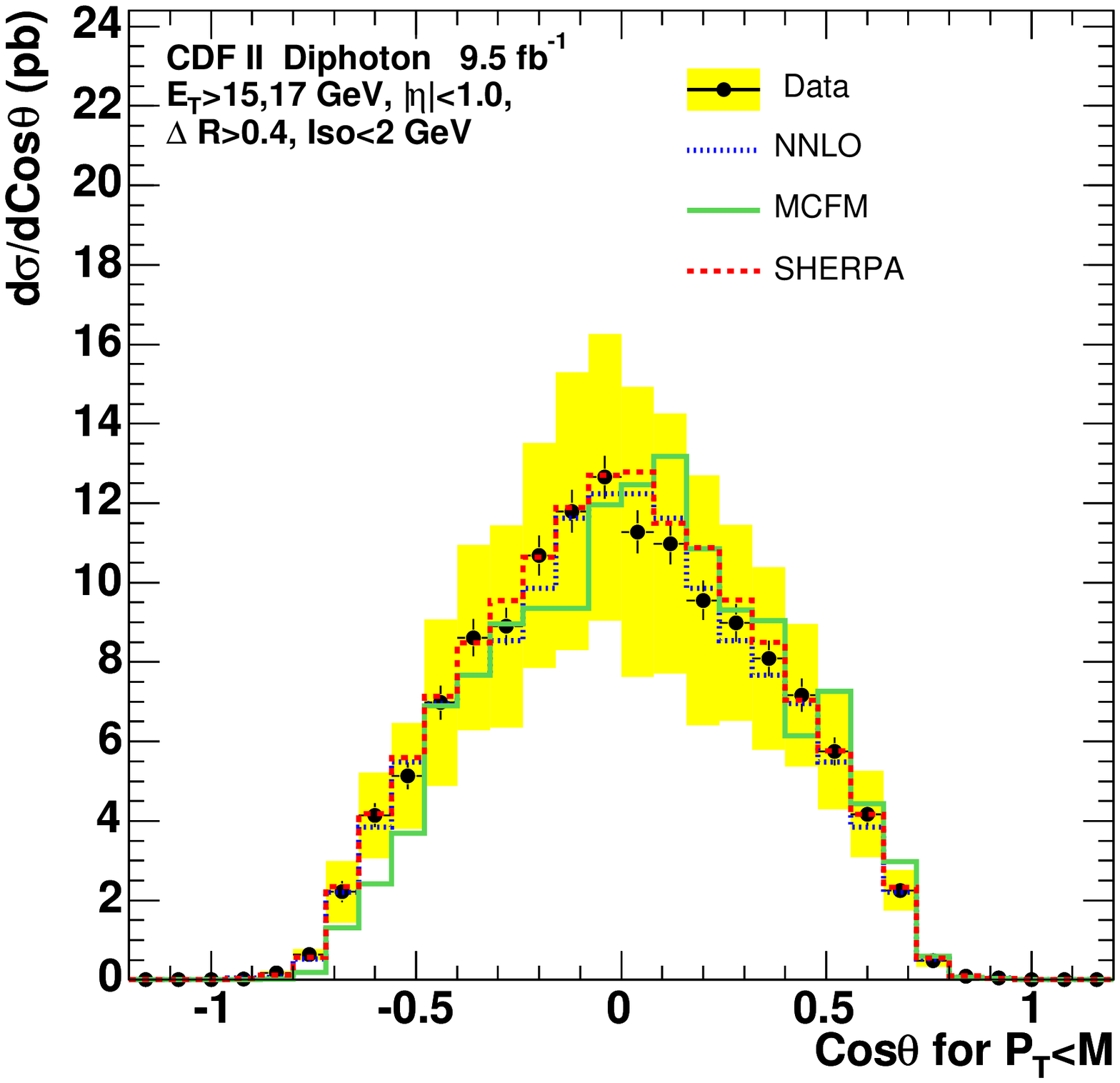}\hspace*{0cm}
\includegraphics[width=8.5cm]{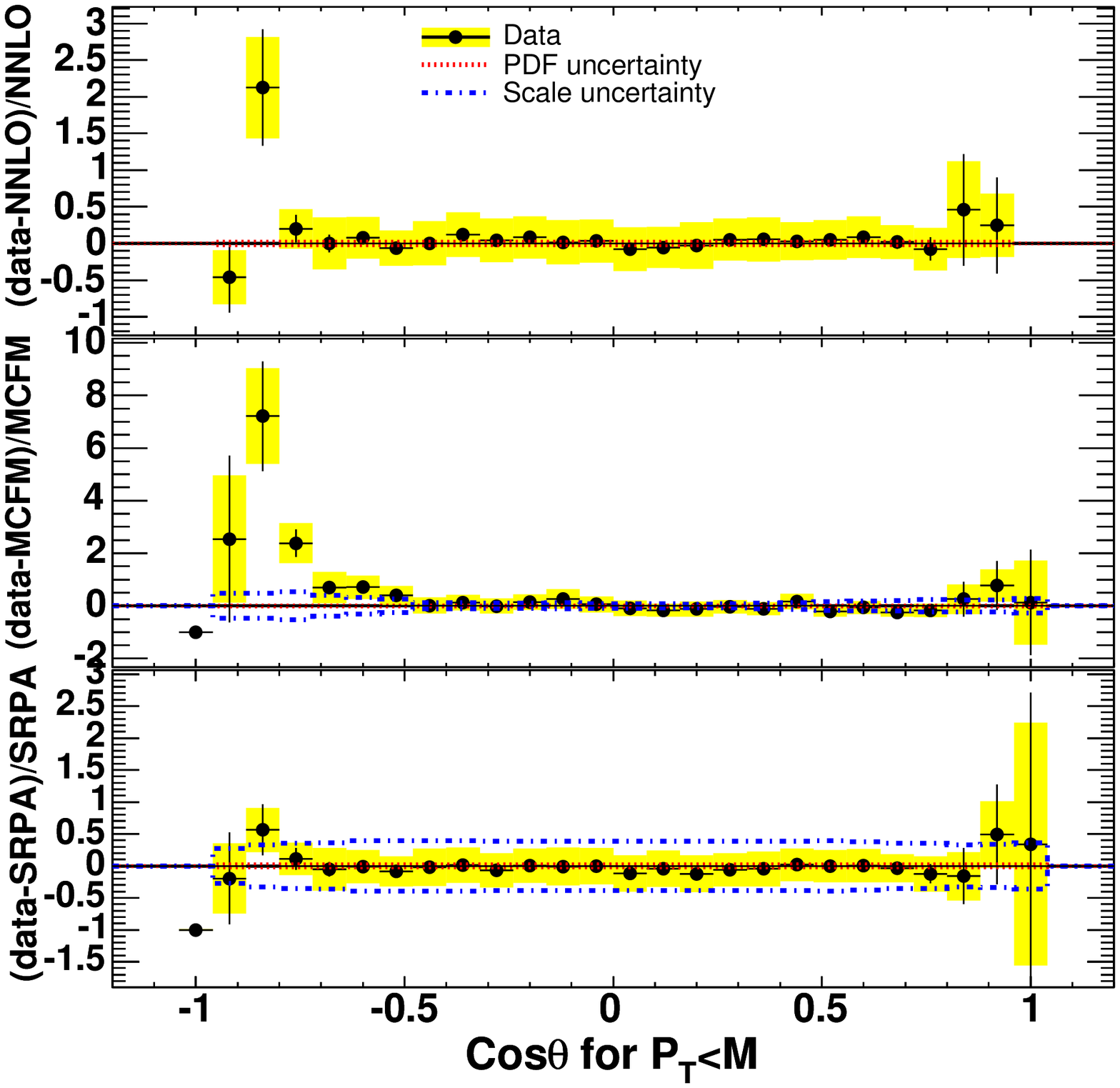}
\vspace*{-0.1cm}
\caption{The measured differential cross sections 
for the cosine of the polar angle of the leading 
photon in the Collins-Soper frame, 
when $P_T(\gamma\gamma) < M(\gamma\gamma)$, 
compared with six 
theoretical predictions discussed in the text. The left windows show the
absolute comparisons and the right windows show the fractional deviations
of the data from the theoretical predictions. Note that the vertical axis
scales differ between fractional deviation plots.}
\end{figure*}

\pagebreak
\vfill

\begin{table}[!hp]
\begin{center}
\begin{tabular}{cccc}
\hline\hline
 Bin & Cross Section (pb) & Sherpa & NNLO   \\ \hline
 -0.960 -  -0.880 & $ 0.022 \pm  0.020 \pm  0.015 $ &  0.028 &  0.041 \\
 -0.880 -  -0.800 & $ 0.177 \pm  0.045 \pm  0.039 $ &  0.113 &  0.057 \\
 -0.800 -  -0.720 & $ 0.63 \pm  0.098 \pm  0.14 $ &  0.57 &  0.53 \\
 -0.720 -  -0.640 & $ 2.21 \pm  0.27 \pm  0.78 $ &  2.34 &  2.21 \\
 -0.640 -  -0.560 & $ 4.1 \pm  0.30 \pm  1.1 $ &  4.2 &  3.8 \\
 -0.560 -  -0.480 & $ 5.1 \pm  0.34 \pm  1.3 $ &  5.6 &  5.5 \\
 -0.480 -  -0.400 & $ 7.0 \pm  0.43 \pm  2.1 $ &  7.1 &  7.0 \\
 -0.400 -  -0.320 & $ 8.6 \pm  0.47 \pm  2.3 $ &  8.5 &  7.7 \\
 -0.320 -  -0.240 & $ 8.9 \pm  0.47 \pm  2.5 $ &  9.5 &  8.5 \\
 -0.240 -  -0.160 & $10.7 \pm  0.51 \pm  2.8 $ & 10.6 &  9.9 \\
 -0.160 -  -0.0800 & $11.8 \pm  0.54 \pm  3.5 $ & 11.9 & 11.6 \\
 -0.0800 -   0.00 & $12.7 \pm  0.54 \pm  3.6 $ & 12.7 & 12.2 \\
  0.00 -   0.0800 & $11.3 \pm  0.53 \pm  3.6 $ & 12.8 & 12.2 \\
  0.0800 -   0.160 & $11.0 \pm  0.52 \pm  3.3 $ & 11.5 & 11.6 \\
  0.160 -   0.240 & $ 9.6 \pm  0.49 \pm  3.1 $ & 10.9 &  9.9 \\
  0.240 -   0.320 & $ 9.0 \pm  0.47 \pm  2.5 $ &  9.6 &  8.5 \\
  0.320 -   0.400 & $ 8.1 \pm  0.45 \pm  2.3 $ &  8.5 &  7.7 \\
  0.400 -   0.480 & $ 7.2 \pm  0.42 \pm  1.8 $ &  7.0 &  7.0 \\
  0.480 -   0.560 & $ 5.7 \pm  0.35 \pm  1.5 $ &  5.8 &  5.5 \\
  0.560 -   0.640 & $ 4.2 \pm  0.29 \pm  1.1 $ &  4.2 &  3.8 \\
  0.640 -   0.720 & $ 2.25 \pm  0.20 \pm  0.51 $ &  2.33 &  2.21 \\
  0.720 -   0.800 & $ 0.49 \pm  0.084 \pm  0.15 $ &  0.56 &  0.53 \\
  0.800 -   0.880 & $ 0.083 \pm  0.043 \pm  0.037 $ &  0.099 &  0.057 \\
  0.880 -   0.960 & $ 0.051 \pm  0.027 \pm  0.018 $ &  0.035 &  0.041 \\
  0.960 -   1.04 & $ 0.012 \pm  0.021 \pm  0.017 $ &  0.009 & -0.015 \\
\hline\hline
\end{tabular}
\end{center}
\end{table}

\pagebreak
\vfill
\renewcommand{\plotname}{DEtaSS01}

\begin{center}
{\LARGE $\eta_1-\eta_2$}
\end{center}

\begin{figure*}[!hp]
\includegraphics[width=9.0cm]{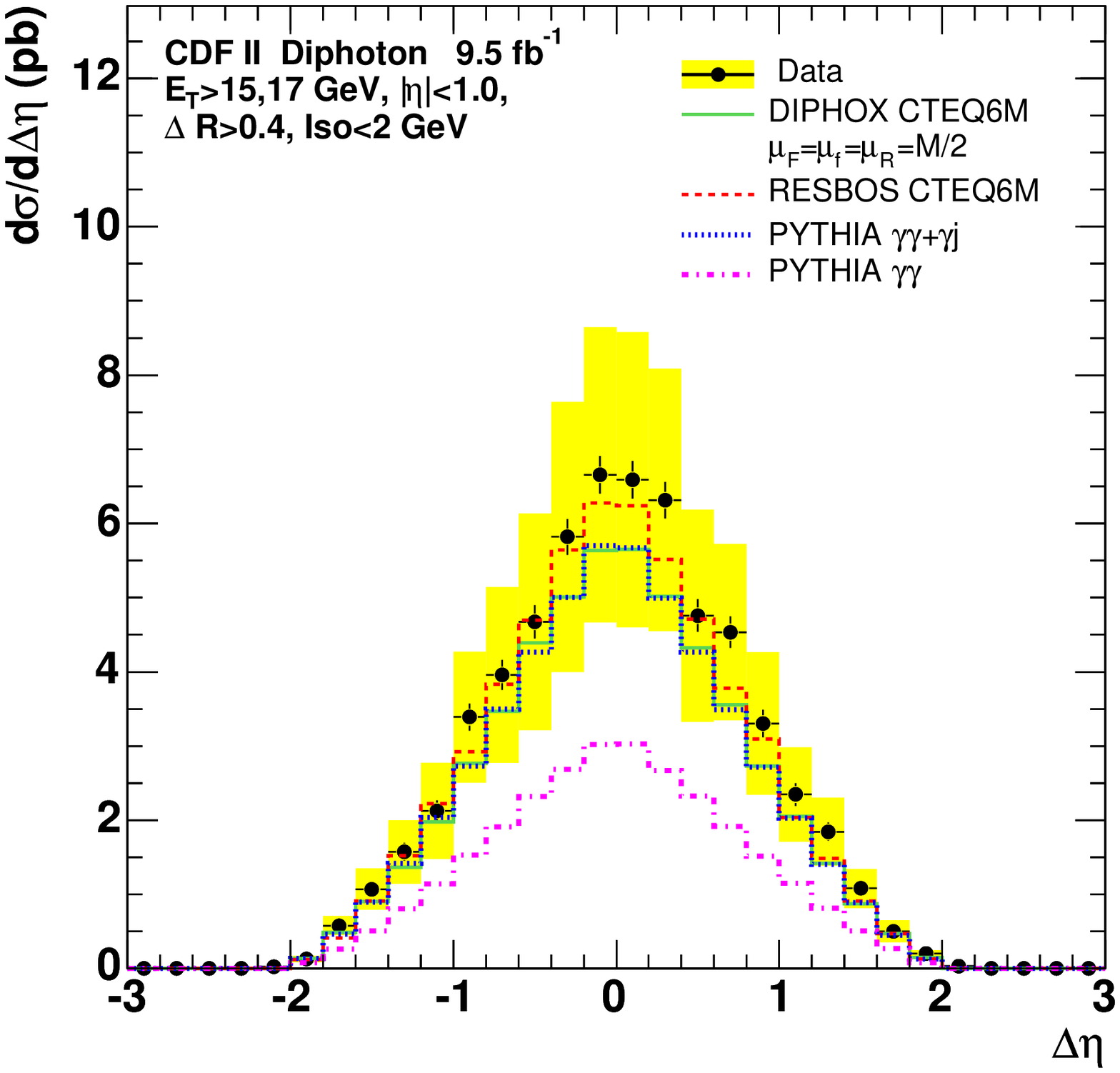}\hspace*{0cm}
\includegraphics[width=8.5cm]{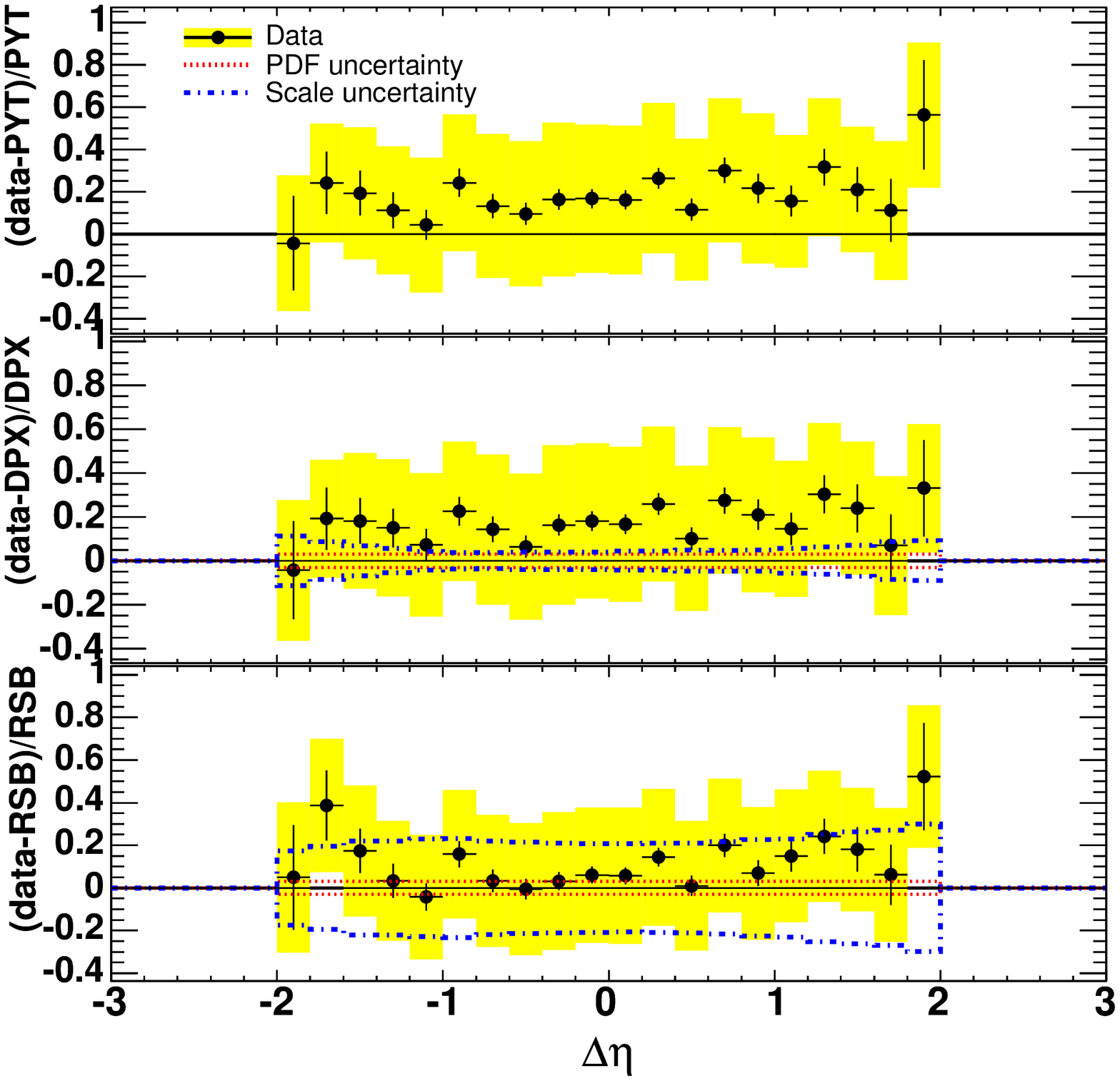}\hfill
\vspace*{0cm}
\includegraphics[width=9.0cm]{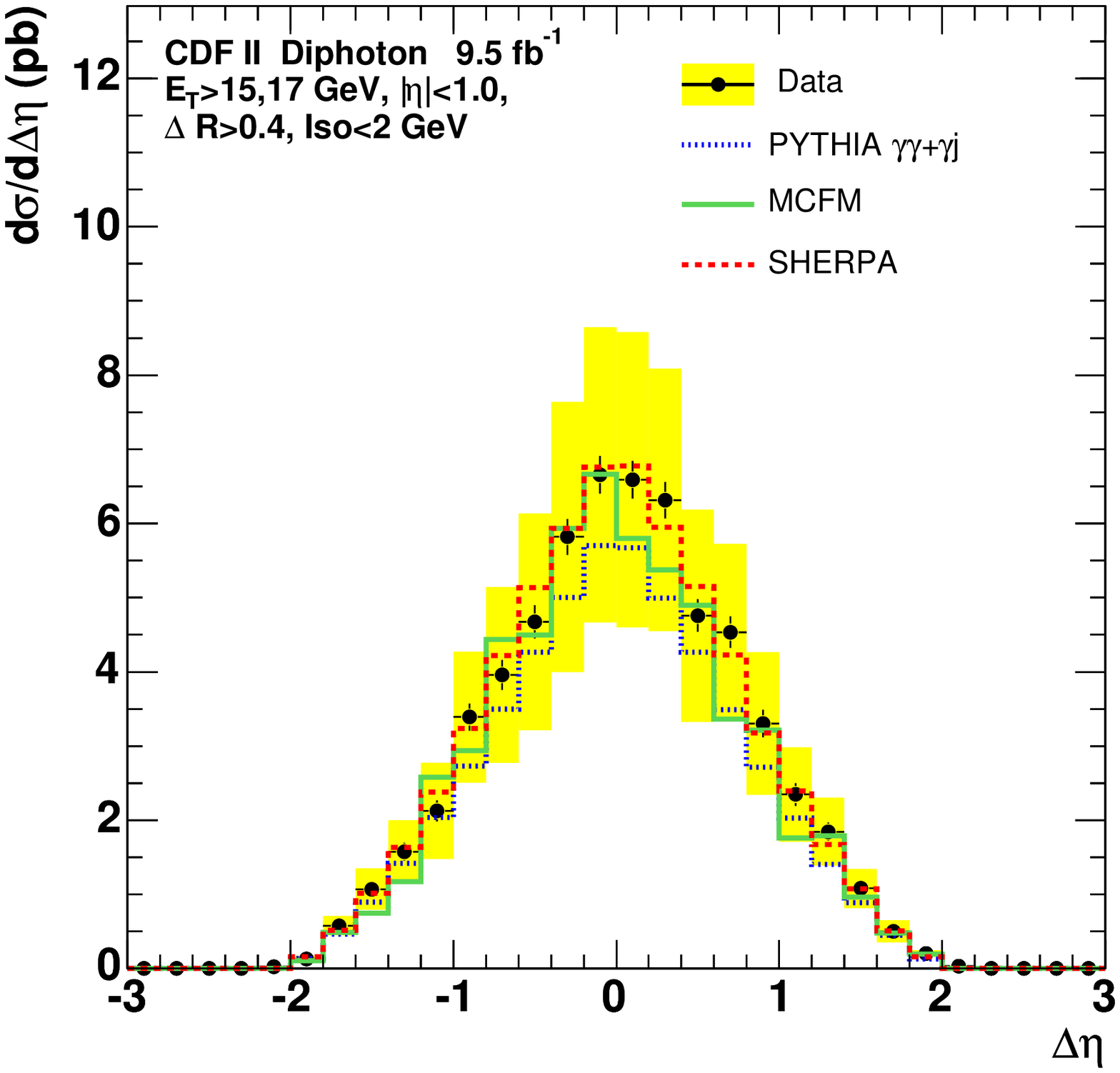}\hspace*{0cm}
\includegraphics[width=8.5cm]{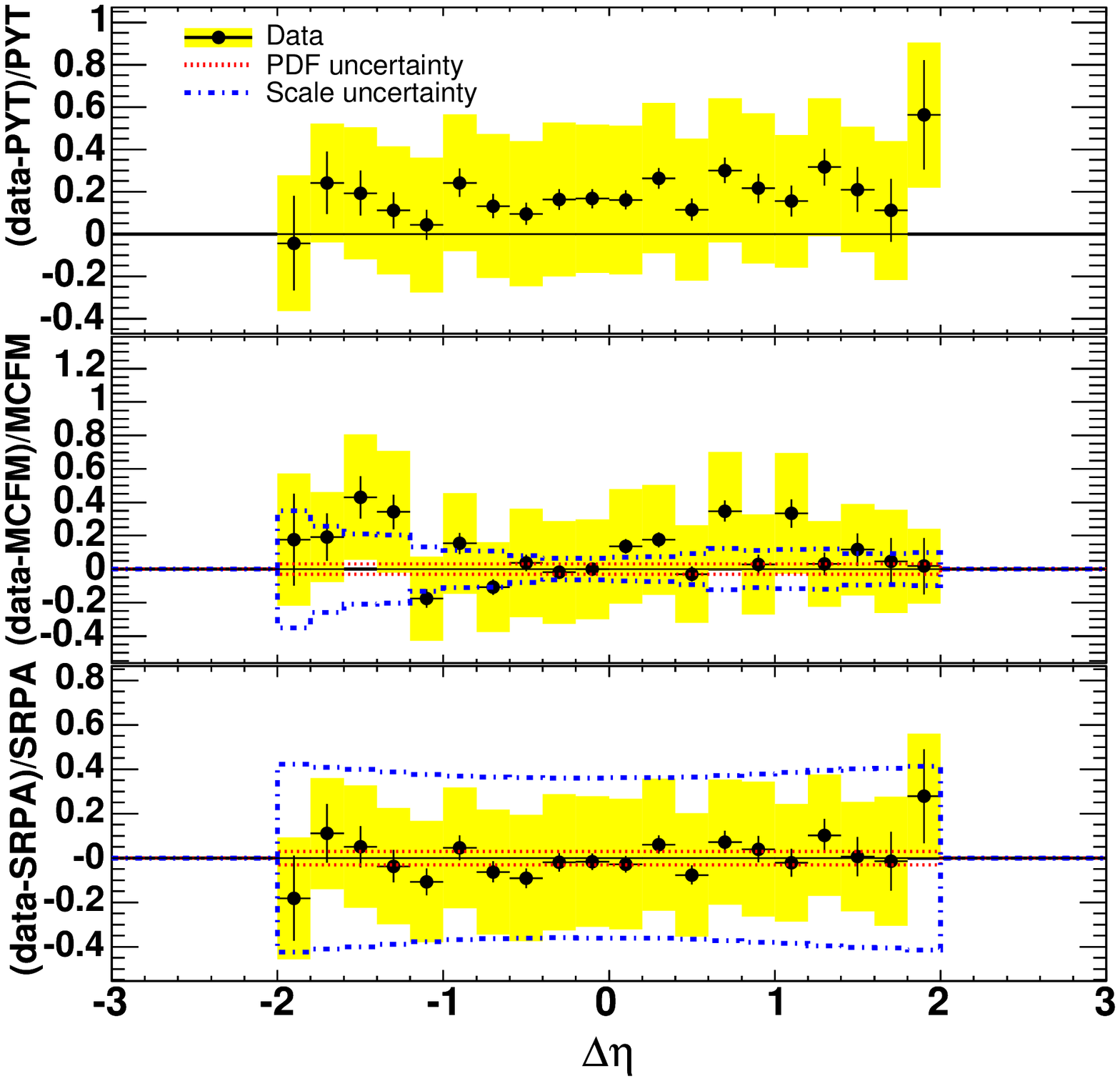}
\vspace*{-0.1cm}
\caption{The measured differential cross sections 
for $\eta_1-\eta_2$, the diffrence in photon pseudorapidities, 
compared with six 
theoretical predictions discussed in the text. The left windows show the
absolute comparisons and the right windows show the fractional deviations
of the data from the theoretical predictions. Note that the vertical axis
scales differ between fractional deviation plots.}
\end{figure*}

\pagebreak
\vfill

\begin{table}[!hp]
\begin{center}
\begin{tabular}{ccc}
\hline\hline
 Bin & Cross Section (pb) & Sherpa   \\ \hline
 -2.20 -  -2.00 & $ 0.026 \pm  0.018 \pm  0.0080 $ &  0.000 \\
 -2.00 -  -1.80 & $ 0.128 \pm  0.030 \pm  0.043 $ &  0.157 \\
 -1.80 -  -1.60 & $ 0.58 \pm  0.069 \pm  0.13 $ &  0.52 \\
 -1.60 -  -1.40 & $ 1.07 \pm  0.095 \pm  0.28 $ &  1.02 \\
 -1.40 -  -1.20 & $ 1.57 \pm  0.12 \pm  0.43 $ &  1.63 \\
 -1.20 -  -1.00 & $ 2.13 \pm  0.14 \pm  0.65 $ &  2.38 \\
 -1.00 -  -0.800 & $ 3.39 \pm  0.18 \pm  0.88 $ &  3.24 \\
 -0.800 -  -0.600 & $ 4.0 \pm  0.20 \pm  1.2 $ &  4.2 \\
 -0.600 -  -0.400 & $ 4.7 \pm  0.22 \pm  1.5 $ &  5.1 \\
 -0.400 -  -0.200 & $ 5.8 \pm  0.24 \pm  1.8 $ &  5.9 \\
 -0.200 -   0.00 & $ 6.7 \pm  0.25 \pm  2.0 $ &  6.8 \\
  0.00 -   0.200 & $ 6.6 \pm  0.25 \pm  2.0 $ &  6.8 \\
  0.200 -   0.400 & $ 6.3 \pm  0.24 \pm  1.8 $ &  6.0 \\
  0.400 -   0.600 & $ 4.8 \pm  0.22 \pm  1.4 $ &  5.1 \\
  0.600 -   0.800 & $ 4.5 \pm  0.21 \pm  1.2 $ &  4.2 \\
  0.800 -   1.00 & $ 3.30 \pm  0.19 \pm  0.96 $ &  3.18 \\
  1.00 -   1.20 & $ 2.35 \pm  0.15 \pm  0.64 $ &  2.40 \\
  1.20 -   1.40 & $ 1.84 \pm  0.12 \pm  0.46 $ &  1.67 \\
  1.40 -   1.60 & $ 1.08 \pm  0.095 \pm  0.26 $ &  1.07 \\
  1.60 -   1.80 & $ 0.50 \pm  0.067 \pm  0.15 $ &  0.51 \\
  1.80 -   2.00 & $ 0.201 \pm  0.033 \pm  0.044 $ &  0.157 \\
  2.00 -   2.20 & $ 0.031 \pm  0.017 \pm  0.0064 $ &  0.000 \\
\hline\hline
\end{tabular}
\end{center}
\end{table}

\pagebreak
\vfill
\renewcommand{\plotname}{QoMSS01}

\begin{center}
{\LARGE log$_{10}$ ($P_T(\gamma\gamma)/M(\gamma\gamma))$}
\end{center}

\begin{figure*}[!hp]
\includegraphics[width=9.0cm]{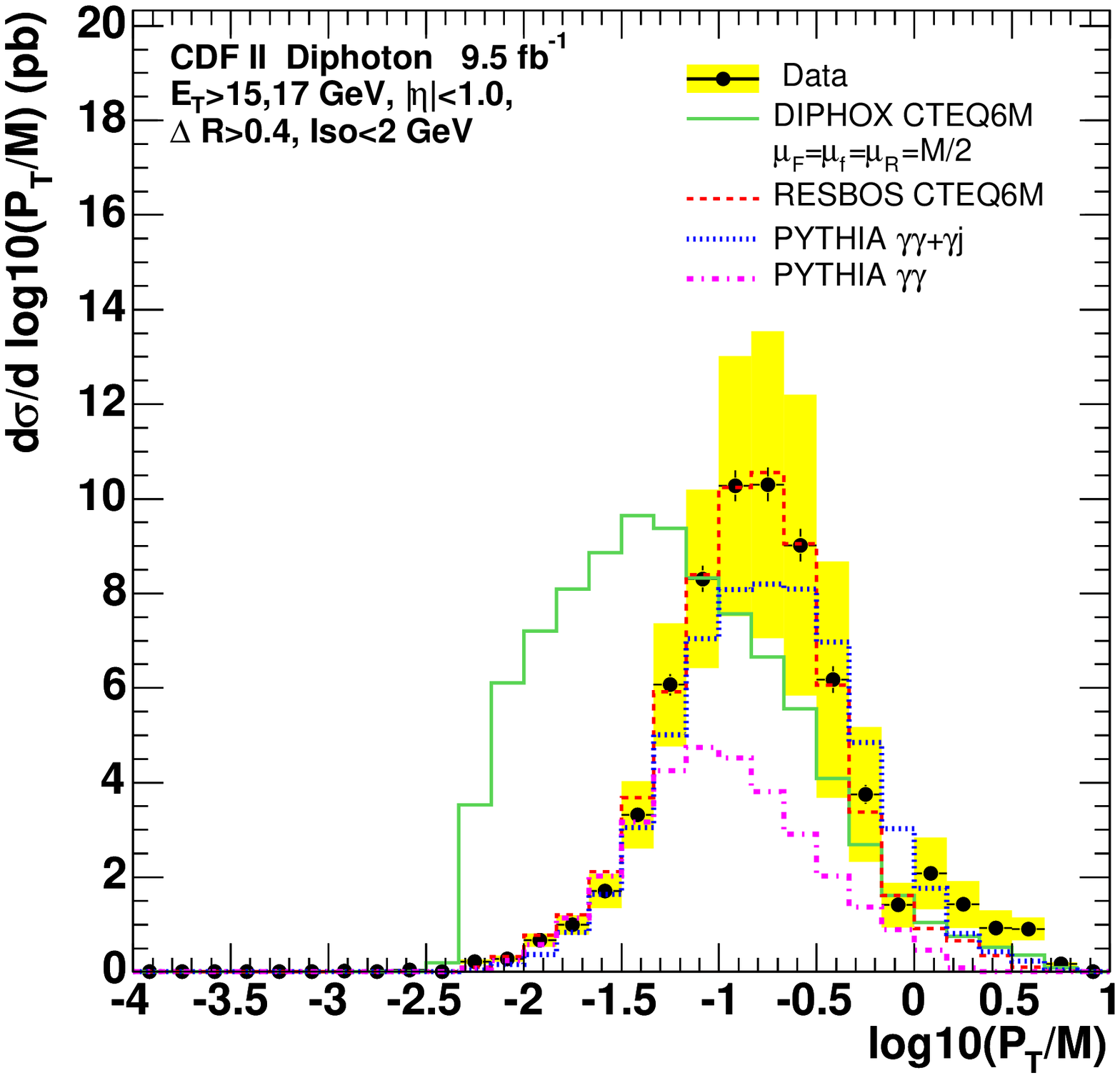}\hspace*{0cm}
\includegraphics[width=8.5cm]{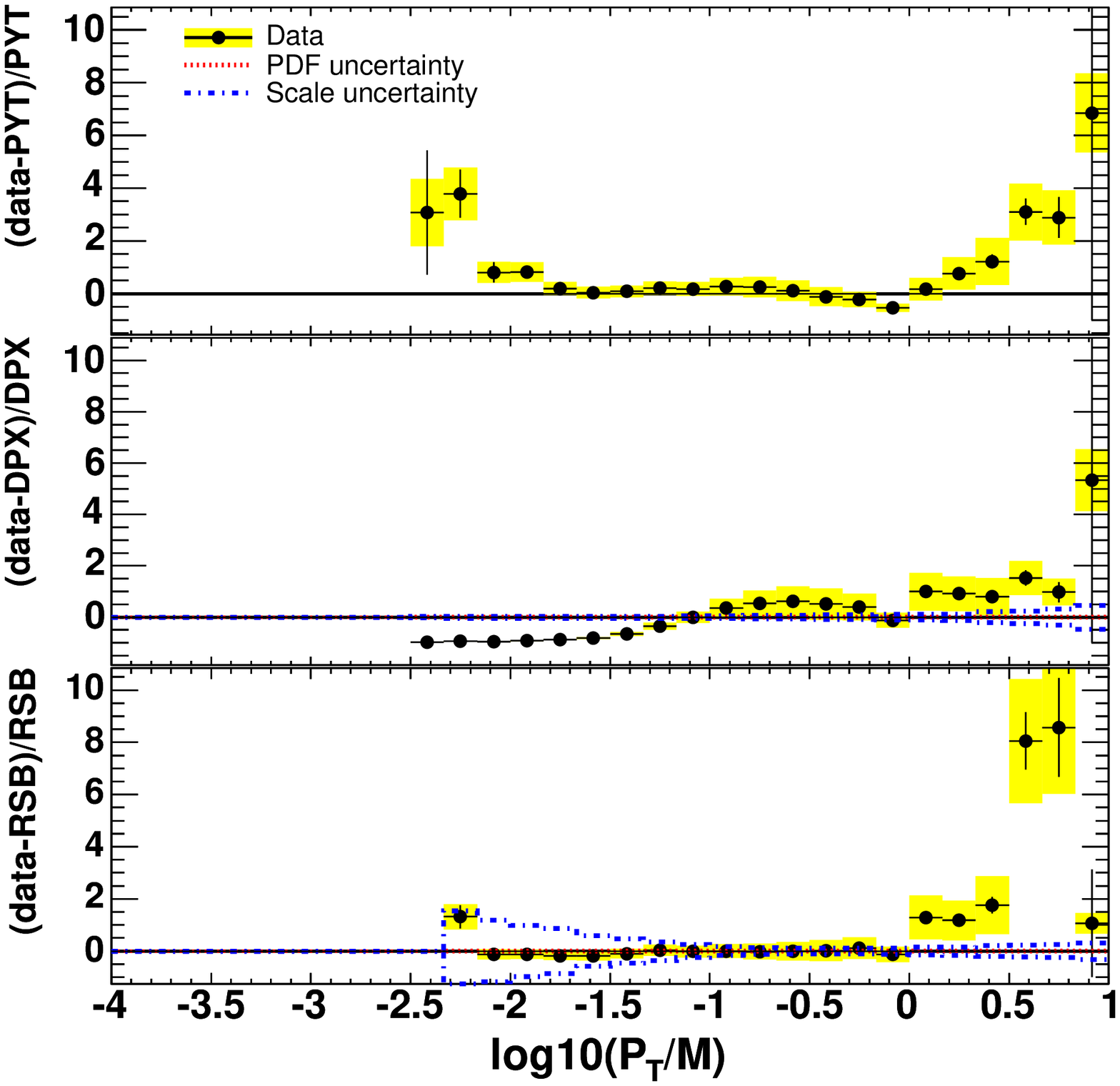}\hfill
\vspace*{0cm}
\includegraphics[width=9.0cm]{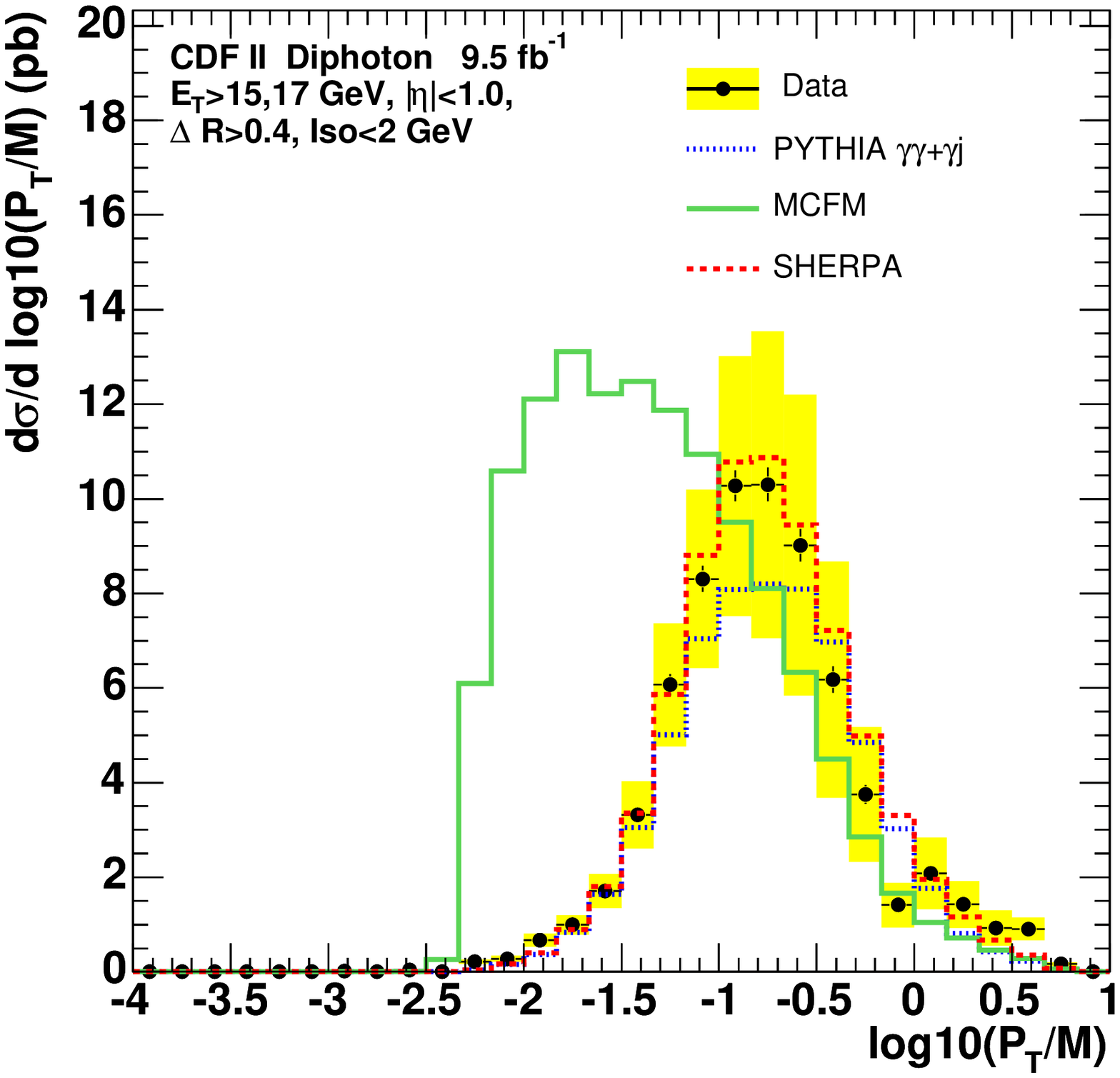}\hspace*{0cm}
\includegraphics[width=8.5cm]{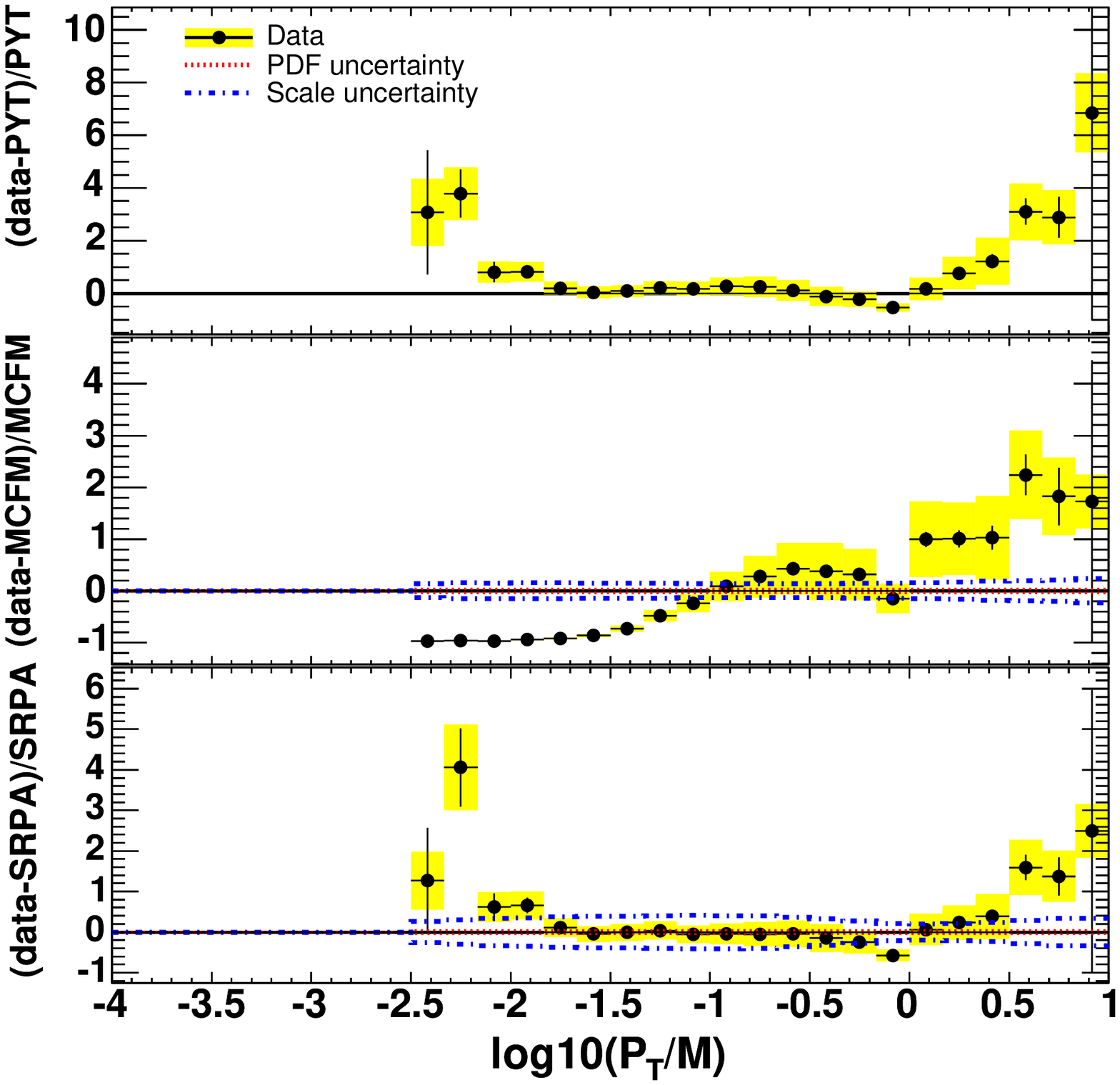}
\vspace*{-0.1cm}
\caption{The measured differential cross sections 
for log$_{10}$ ($P_T(\gamma\gamma)/M(\gamma\gamma))$  
compared with six 
theoretical predictions discussed in the text. The left windows show the
absolute comparisons and the right windows show the fractional deviations
of the data from the theoretical predictions. Note that the vertical axis
scales differ between fractional deviation plots.}
\end{figure*}

\pagebreak
\vfill

\begin{table}[!hp]
\begin{center}
\begin{tabular}{ccc}
\hline\hline
 Bin & Cross Section (pb) & Sherpa   \\ \hline
 -3.17 -  -3.00 & $ 0.0050 \pm  0.0050 \pm  0.0013 $ &  0.0000 \\
 -3.00 -  -2.83 & $ 0.0102 \pm  0.0074 \pm  0.0022 $ &  0.0000 \\
 -2.83 -  -2.67 & $ 0.002 \pm  0.011 \pm  0.00020 $ &  0.000 \\
 -2.67 -  -2.50 & $ 0.036 \pm  0.016 \pm  0.010 $ &  0.000 \\
 -2.50 -  -2.33 & $ 0.0064 \pm  0.0037 \pm  0.0020 $ &  0.0028 \\
 -2.33 -  -2.17 & $ 0.216 \pm  0.041 \pm  0.045 $ &  0.043 \\
 -2.17 -  -2.00 & $ 0.275 \pm  0.057 \pm  0.062 $ &  0.170 \\
 -2.00 -  -1.83 & $ 0.67 \pm  0.077 \pm  0.14 $ &  0.41 \\
 -1.83 -  -1.67 & $ 0.99 \pm  0.092 \pm  0.20 $ &  0.89 \\
 -1.67 -  -1.50 & $ 1.71 \pm  0.13 \pm  0.36 $ &  1.80 \\
 -1.50 -  -1.33 & $ 3.32 \pm  0.17 \pm  0.71 $ &  3.36 \\
 -1.33 -  -1.17 & $ 6.1 \pm  0.23 \pm  1.3 $ &  5.9 \\
 -1.17 -  -1.00 & $ 8.3 \pm  0.28 \pm  1.9 $ &  8.8 \\
 -1.00 -  -0.833 & $10.3 \pm  0.33 \pm  2.7 $ & 10.8 \\
 -0.833 -  -0.667 & $10.3 \pm  0.36 \pm  3.2 $ & 10.9 \\
 -0.667 -  -0.500 & $ 9.0 \pm  0.34 \pm  3.2 $ &  9.4 \\
 -0.500 -  -0.333 & $ 6.2 \pm  0.28 \pm  2.5 $ &  7.2 \\
 -0.333 -  -0.167 & $ 3.8 \pm  0.20 \pm  1.4 $ &  5.0 \\
 -0.167 -  -0.00 & $ 1.41 \pm  0.086 \pm  0.48 $ &  3.31 \\
 -0.00 -   0.167 & $ 2.08 \pm  0.15 \pm  0.76 $ &  1.96 \\
  0.167 -   0.333 & $ 1.43 \pm  0.12 \pm  0.50 $ &  1.16 \\
  0.333 -   0.500 & $ 0.93 \pm  0.11 \pm  0.37 $ &  0.67 \\
  0.500 -   0.667 & $ 0.91 \pm  0.11 \pm  0.24 $ &  0.35 \\
  0.667 -   0.833 & $ 0.164 \pm  0.033 \pm  0.044 $ &  0.069 \\
  0.833 -   1.00 & $ 0.0033 \pm  0.0033 \pm  0.00063 $ &  0.0009 \\
\hline\hline
\end{tabular}
\end{center}
\end{table}

\pagebreak
\vfill
\renewcommand{\plotname}{drSS01}

\begin{center}
{\LARGE $\Delta R(\gamma\gamma)$}
\end{center}

\begin{figure*}[!hp]
\includegraphics[width=9.0cm]{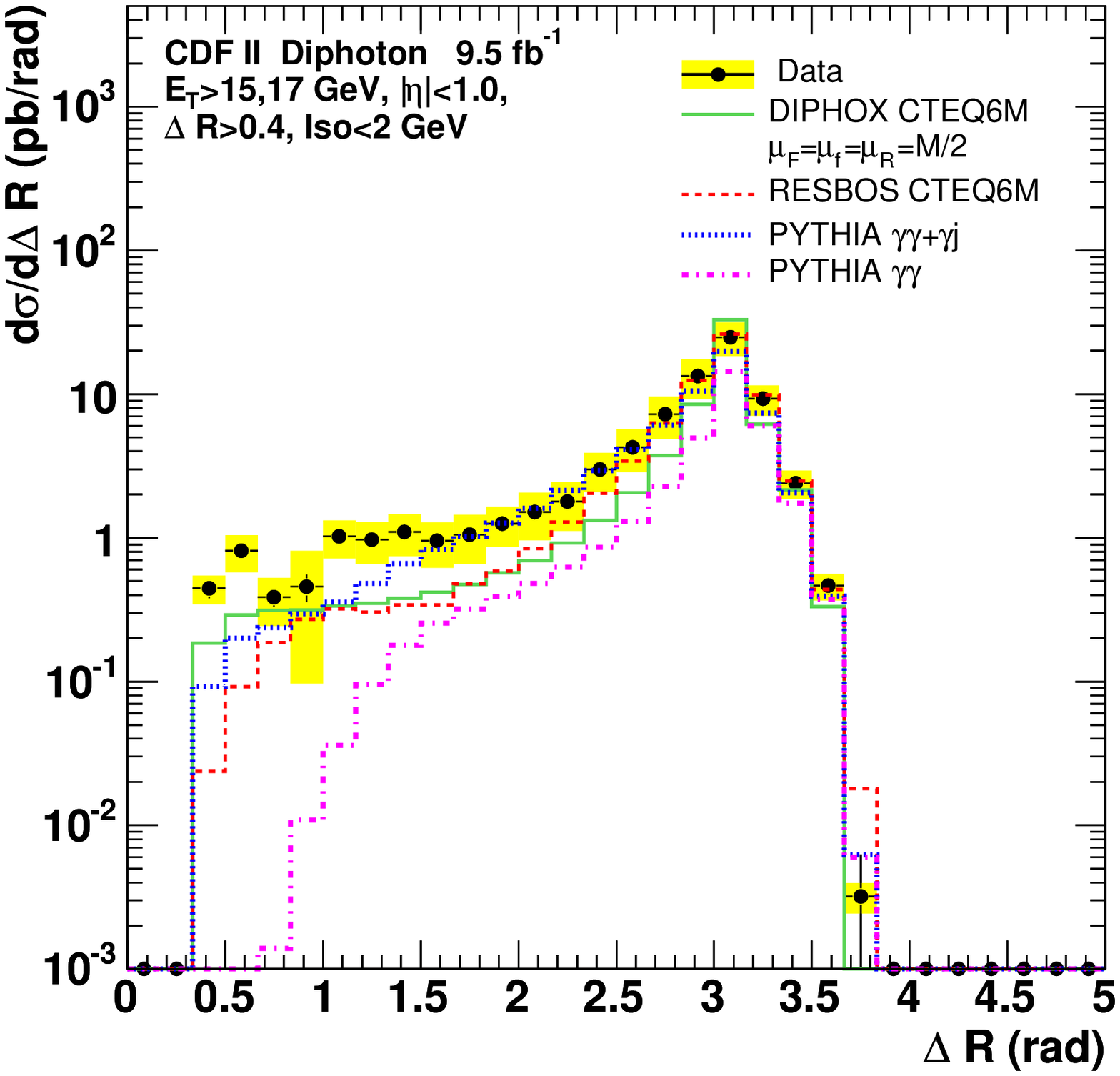}\hspace*{0cm}
\includegraphics[width=8.5cm]{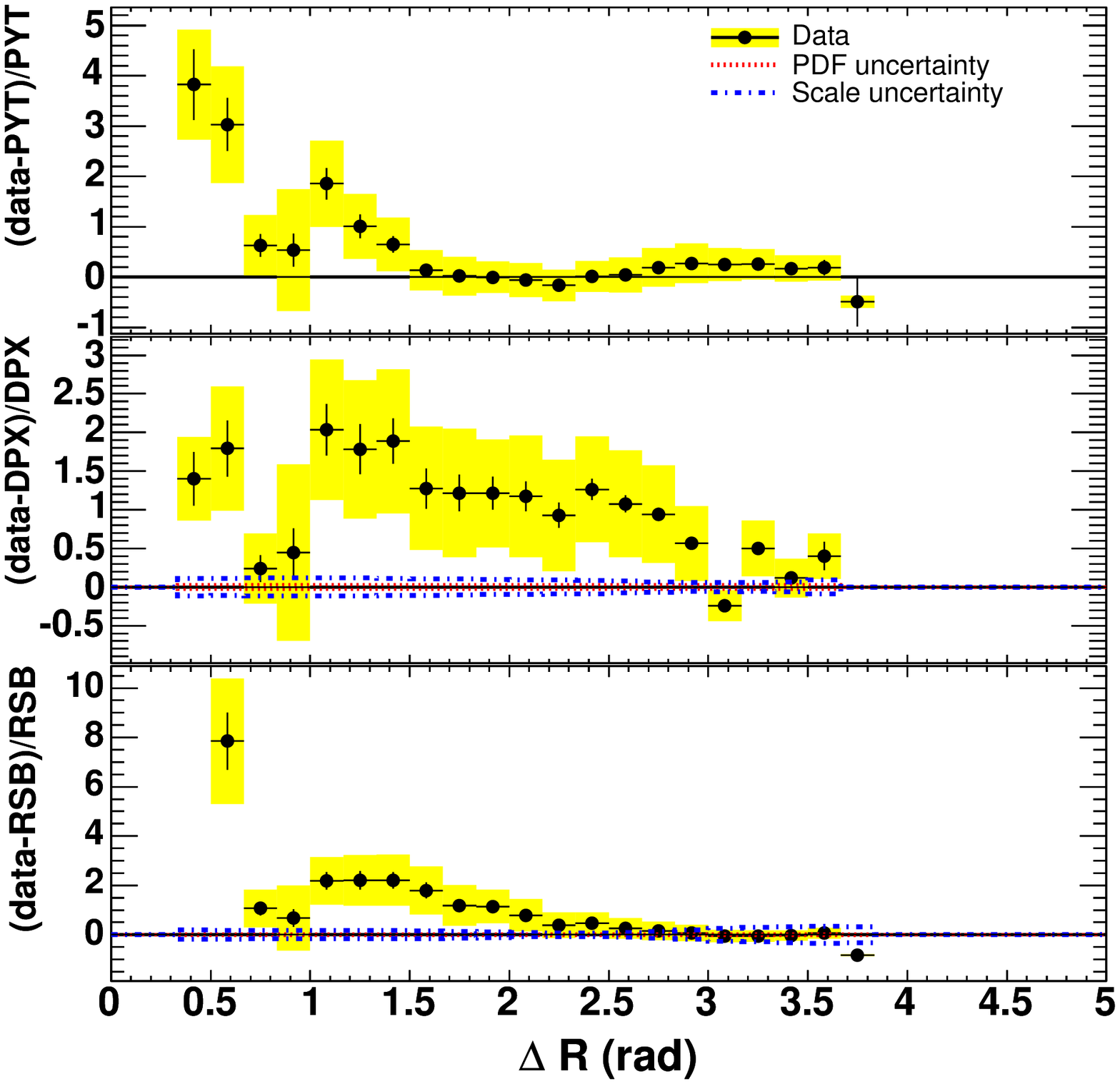}\hfill
\vspace*{0cm}
\includegraphics[width=9.0cm]{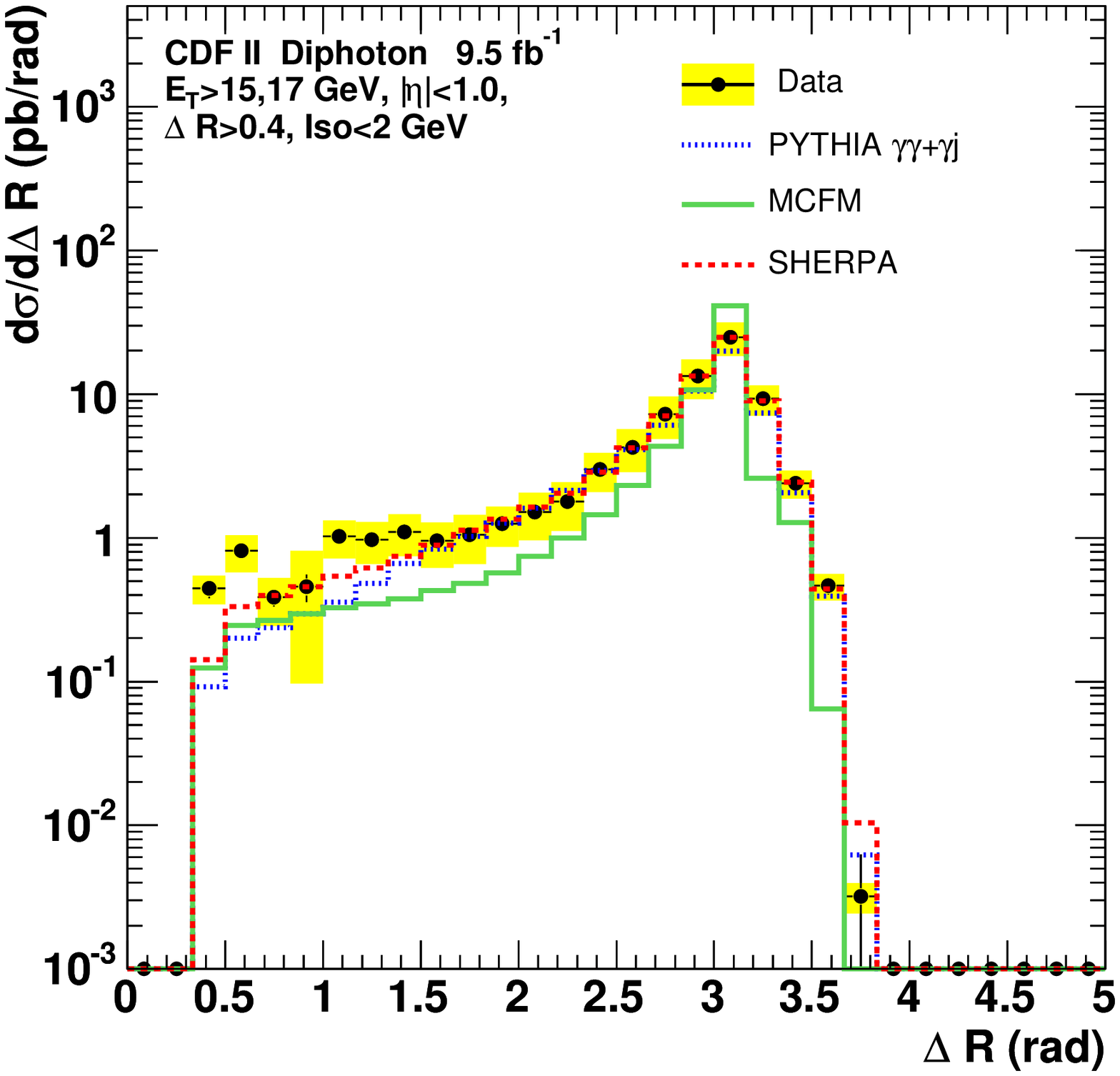}\hspace*{0cm}
\includegraphics[width=8.5cm]{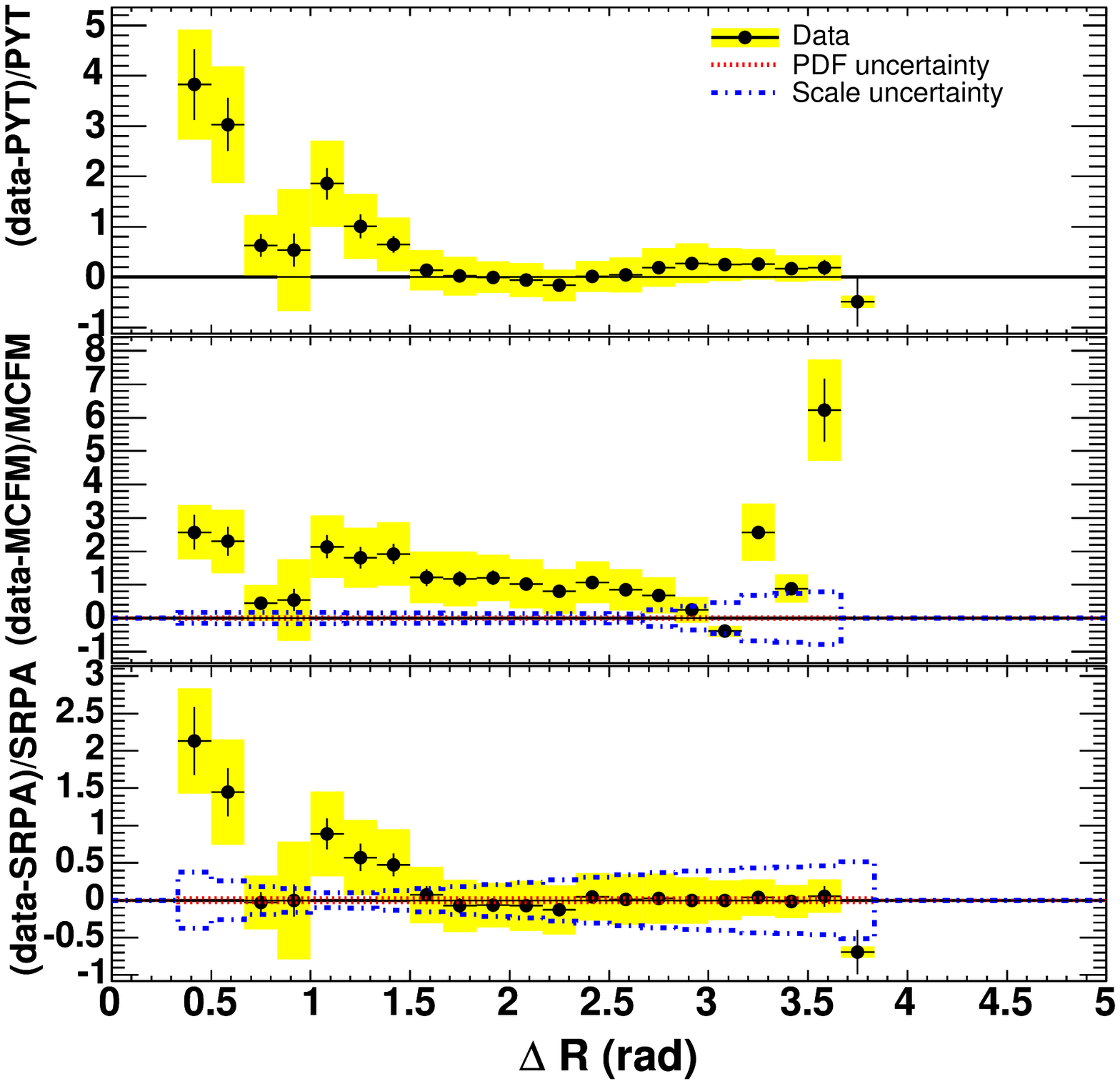}
\vspace*{-0.1cm}
\caption{The measured differential cross sections 
for $\Delta R(\gamma\gamma)$ where 
$\Delta R(\gamma\gamma)^2 = \Delta \Phi(\gamma\gamma)^2 + (\eta_1-\eta_2)^2$
compared with six 
theoretical predictions discussed in the text. The left windows show the
absolute comparisons and the right windows show the fractional deviations
of the data from the theoretical predictions. Note that the vertical axis
scales differ between fractional deviation plots.}
\end{figure*}

\pagebreak
\vfill

\begin{table}[!hp]
\begin{center}
\begin{tabular}{ccc}
\hline\hline
 Bin & Cross Section (pb) & Sherpa   \\ \hline
  0.333 -   0.500 & $ 0.45 \pm  0.065 \pm  0.10 $ &  0.14 \\
  0.500 -   0.667 & $ 0.81 \pm  0.11 \pm  0.23 $ &  0.33 \\
  0.667 -   0.833 & $ 0.39 \pm  0.054 \pm  0.14 $ &  0.40 \\
  0.833 -   1.00 & $ 0.46 \pm  0.098 \pm  0.36 $ &  0.46 \\
  1.00 -   1.17 & $ 1.02 \pm  0.11 \pm  0.31 $ &  0.54 \\
  1.17 -   1.33 & $ 0.97 \pm  0.11 \pm  0.31 $ &  0.62 \\
  1.33 -   1.50 & $ 1.10 \pm  0.11 \pm  0.35 $ &  0.74 \\
  1.50 -   1.67 & $ 0.95 \pm  0.11 \pm  0.33 $ &  0.89 \\
  1.67 -   1.83 & $ 1.05 \pm  0.11 \pm  0.39 $ &  1.13 \\
  1.83 -   2.00 & $ 1.26 \pm  0.12 \pm  0.40 $ &  1.34 \\
  2.00 -   2.17 & $ 1.51 \pm  0.13 \pm  0.55 $ &  1.63 \\
  2.17 -   2.33 & $ 1.78 \pm  0.15 \pm  0.67 $ &  2.05 \\
  2.33 -   2.50 & $ 2.98 \pm  0.18 \pm  0.90 $ &  2.86 \\
  2.50 -   2.67 & $ 4.3 \pm  0.23 \pm  1.4 $ &  4.2 \\
  2.67 -   2.83 & $ 7.3 \pm  0.31 \pm  2.4 $ &  7.1 \\
  2.83 -   3.00 & $13.3 \pm  0.42 \pm  4.1 $ & 13.3 \\
  3.00 -   3.17 & $24.9 \pm  0.55 \pm  6.6 $ & 24.9 \\
  3.17 -   3.33 & $ 9.3 \pm  0.33 \pm  2.2 $ &  9.0 \\
  3.33 -   3.50 & $ 2.40 \pm  0.15 \pm  0.54 $ &  2.45 \\
  3.50 -   3.67 & $ 0.465 \pm  0.061 \pm  0.098 $ &  0.440 \\
  3.67 -   3.83 & $ 0.0032 \pm  0.0031 \pm  0.00076 $ &  0.0104 \\
\hline\hline
\end{tabular}
\end{center}
\end{table}

\pagebreak
\vfill
\renewcommand{\plotname}{etSSV01}

\begin{center}
{\LARGE $E_T(\gamma)$, two entries per event}
\end{center}

\begin{figure*}[!hp]
\includegraphics[width=9.0cm]{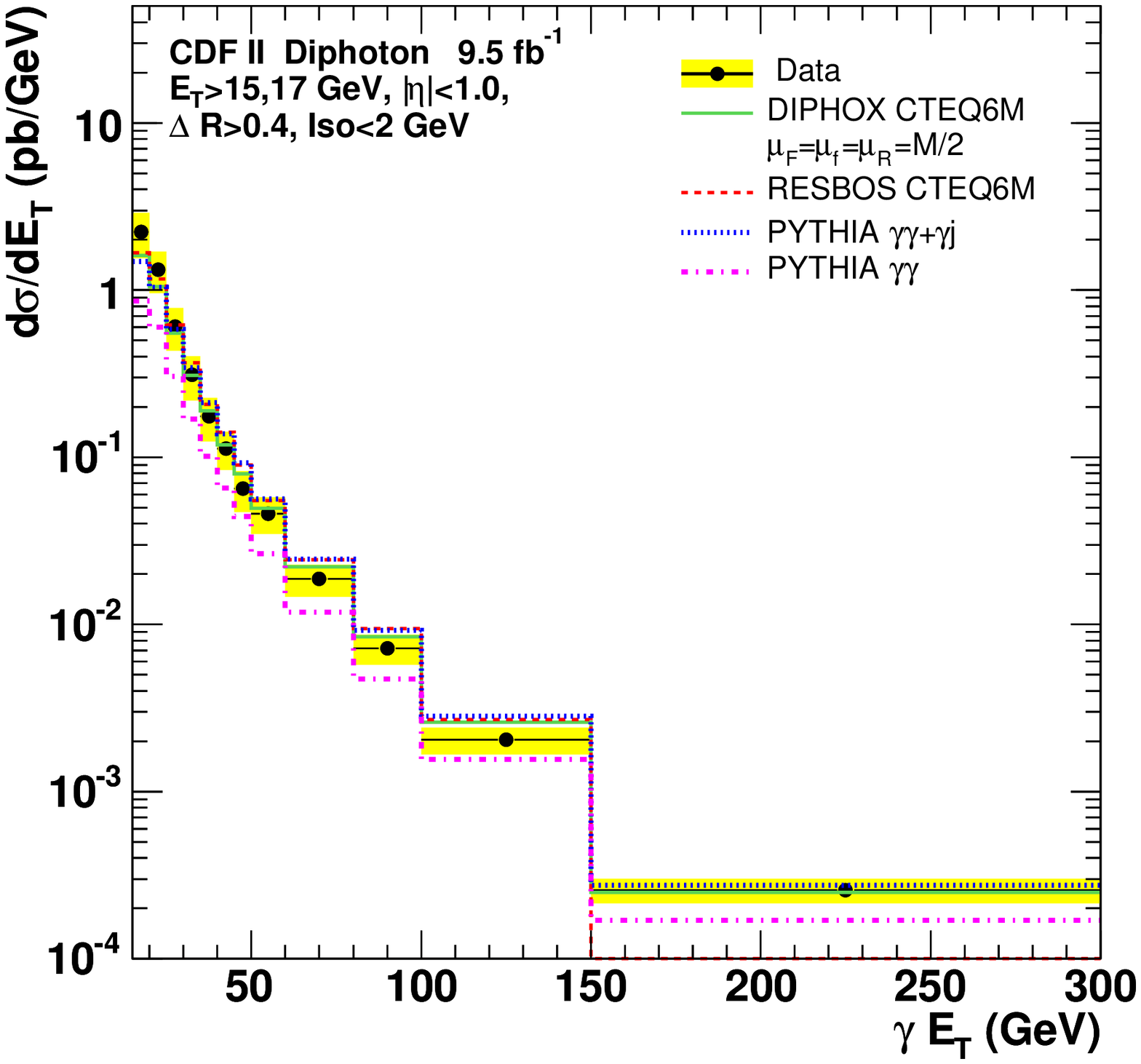}\hspace*{0cm}
\includegraphics[width=8.5cm]{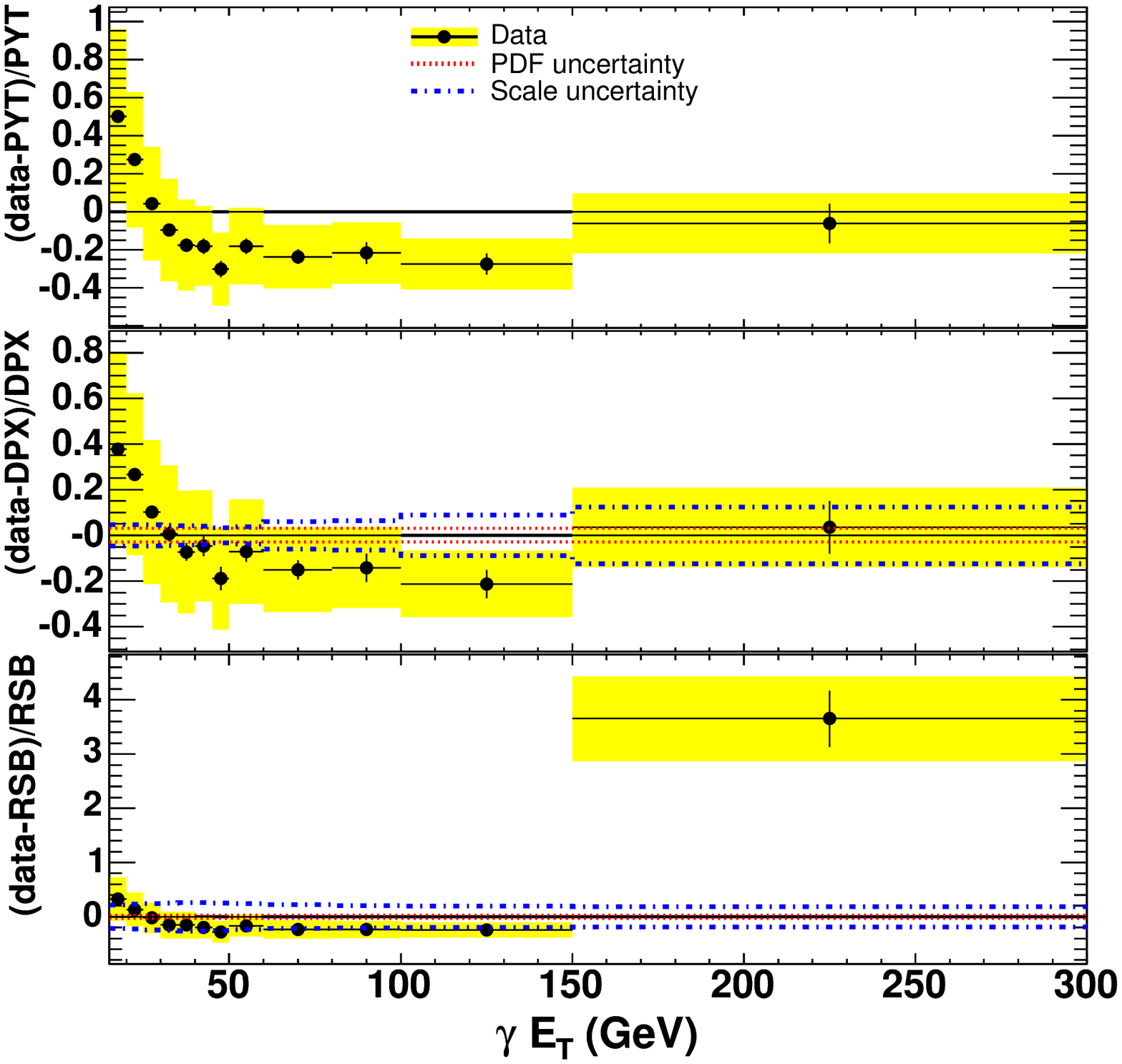}\hfill
\vspace*{0cm}
\includegraphics[width=9.0cm]{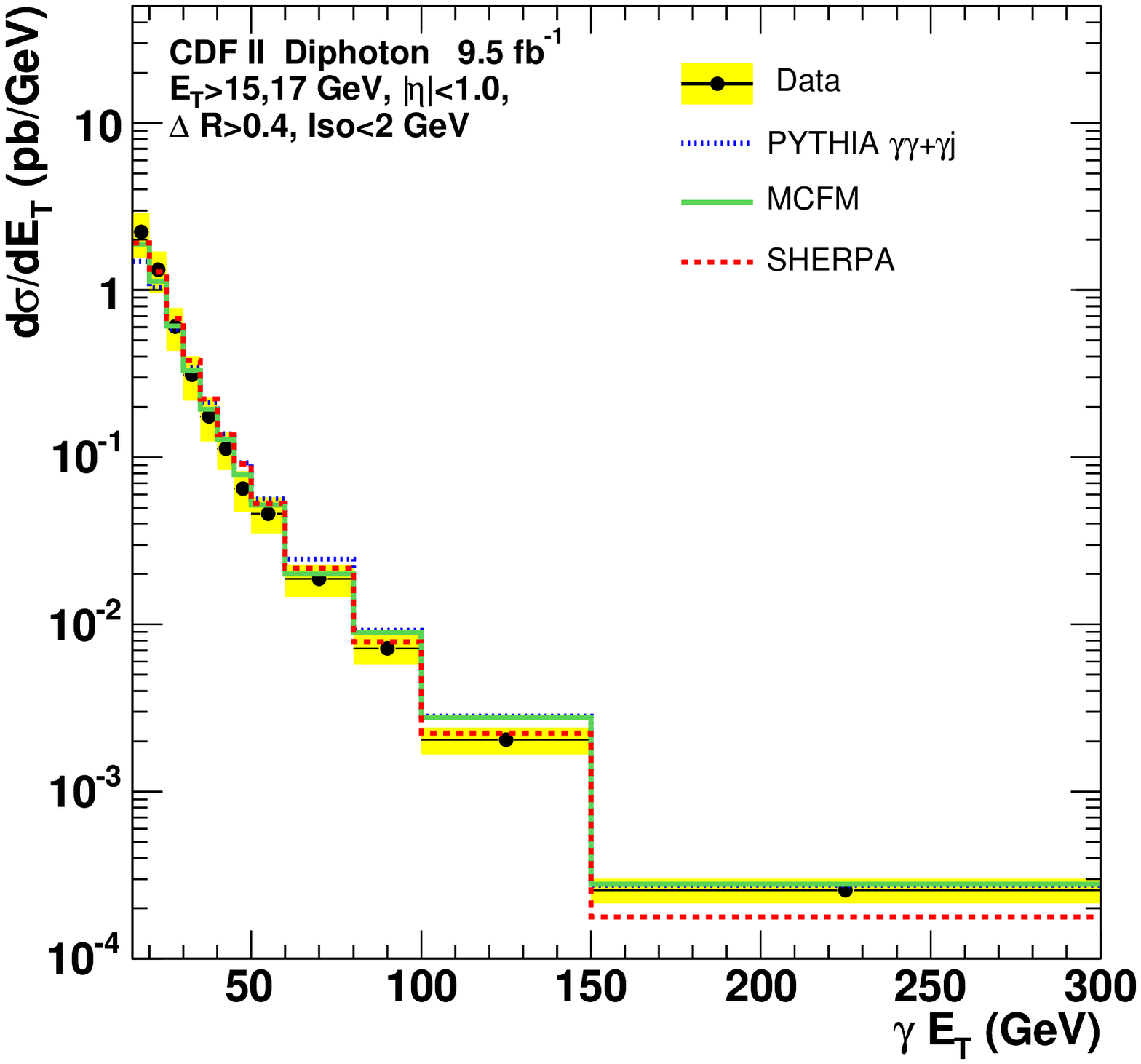}\hspace*{0cm}
\includegraphics[width=8.5cm]{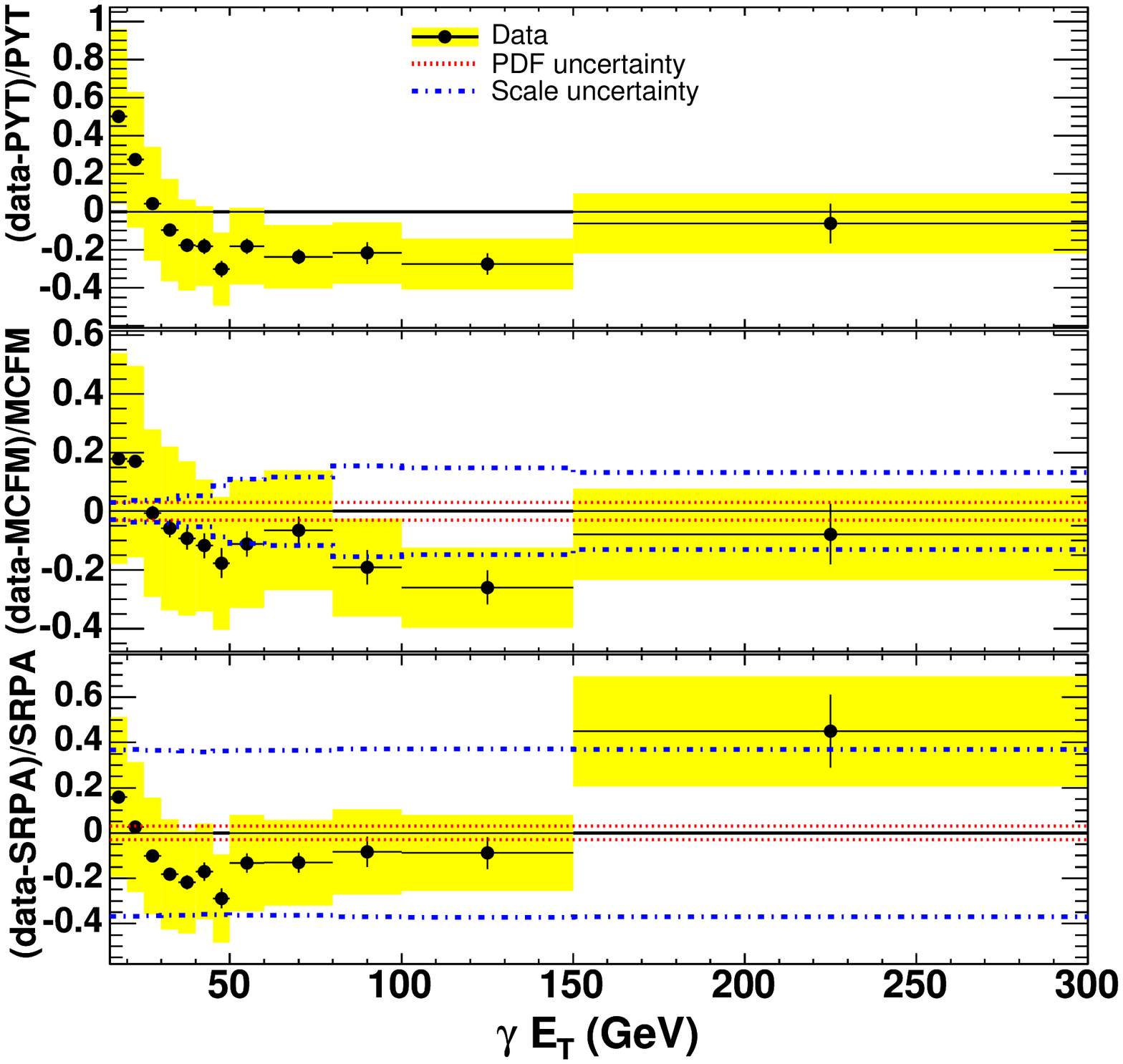}
\vspace*{-0.1cm}
\caption{The measured differential cross sections 
for the photon $E_T$ (two entries per event) 
compared with six 
theoretical predictions discussed in the text. The left windows show the
absolute comparisons and the right windows show the fractional deviations
of the data from the theoretical predictions. Note that the vertical axis
scales differ between fractional deviation plots.}
\end{figure*}

\pagebreak
\vfill

\begin{table}[!hp]
\begin{center}
\begin{tabular}{ccc}
\hline\hline
 Bin & Cross Section (pb) & Sherpa   \\ \hline
 15.0 -  20.0 & $ 2.22 \pm  0.033 \pm  0.68 $ &  1.92 \\
 20.0 -  25.0 & $ 1.32 \pm  0.023 \pm  0.37 $ &  1.29 \\
 25.0 -  30.0 & $ 0.61 \pm  0.015 \pm  0.17 $ &  0.68 \\
 30.0 -  35.0 & $ 0.311 \pm  0.010 \pm  0.092 $ &  0.380 \\
 35.0 -  40.0 & $ 0.175 \pm  0.0072 \pm  0.051 $ &  0.224 \\
 40.0 -  45.0 & $ 0.113 \pm  0.0054 \pm  0.029 $ &  0.136 \\
 45.0 -  50.0 & $ 0.065 \pm  0.0040 \pm  0.018 $ &  0.091 \\
 50.0 -  60.0 & $ 0.046 \pm  0.0023 \pm  0.011 $ &  0.053 \\
 60.0 -  80.0 & $ 0.0187 \pm  0.00094 \pm  0.0041 $ &  0.0216 \\
 80.0 - 100.0 & $ 0.0072 \pm  0.00053 \pm  0.0015 $ &  0.0079 \\
100.0 - 150.0 & $ 0.00204 \pm  0.00016 \pm  0.00038 $ &  0.00224 \\
150.0 - 300.0 & $ 0.000258 \pm  0.000029 \pm  0.000043 $ &  0.000178 \\
\hline\hline
\end{tabular}
\end{center}
\end{table}

\pagebreak
\vfill
\renewcommand{\plotname}{etaSS01}

\begin{center}
{\LARGE $\eta(\gamma)$, two entries per event}
\end{center}

\begin{figure*}[!hp]
\includegraphics[width=9.0cm]{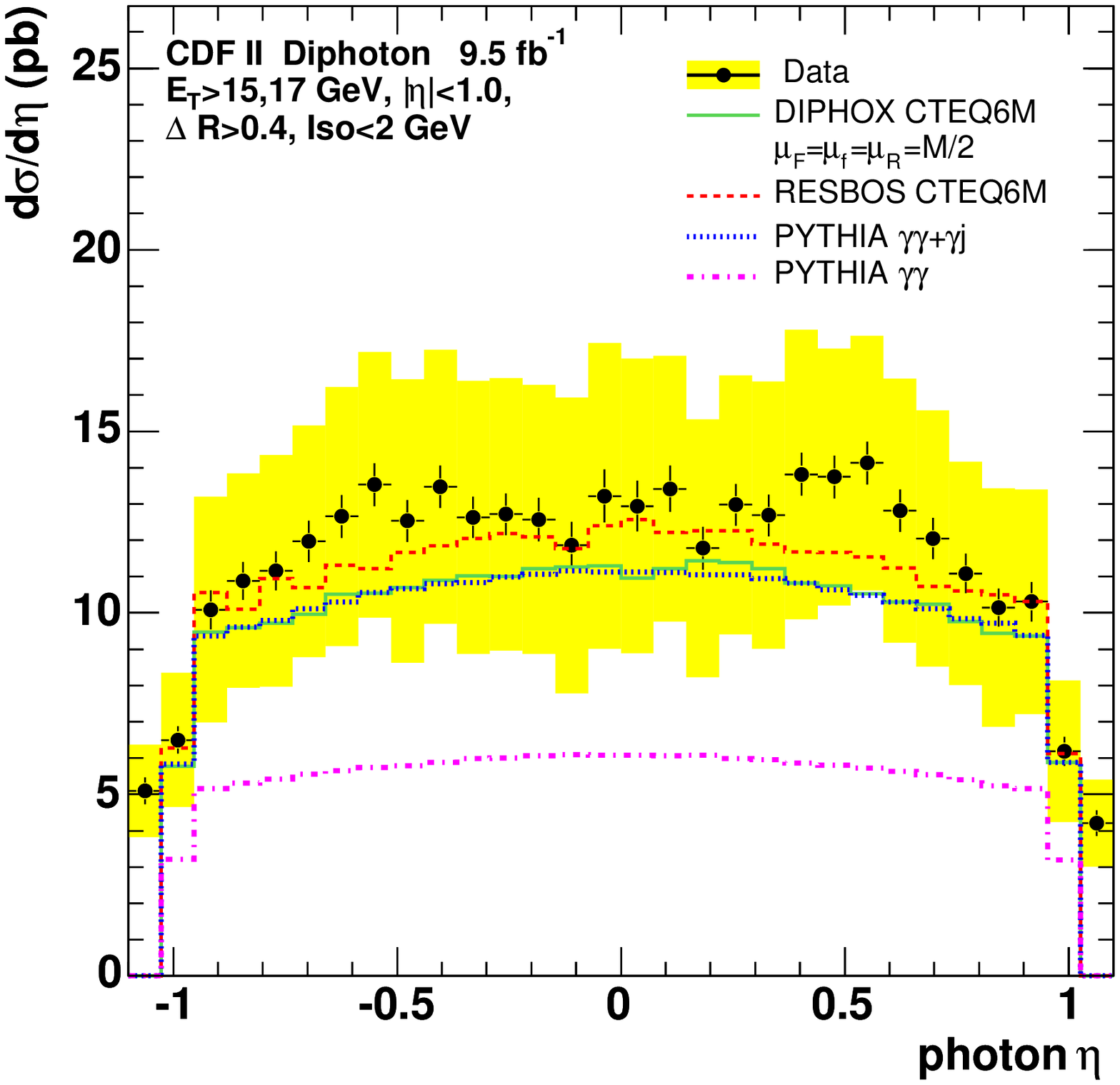}\hspace*{0cm}
\includegraphics[width=8.5cm]{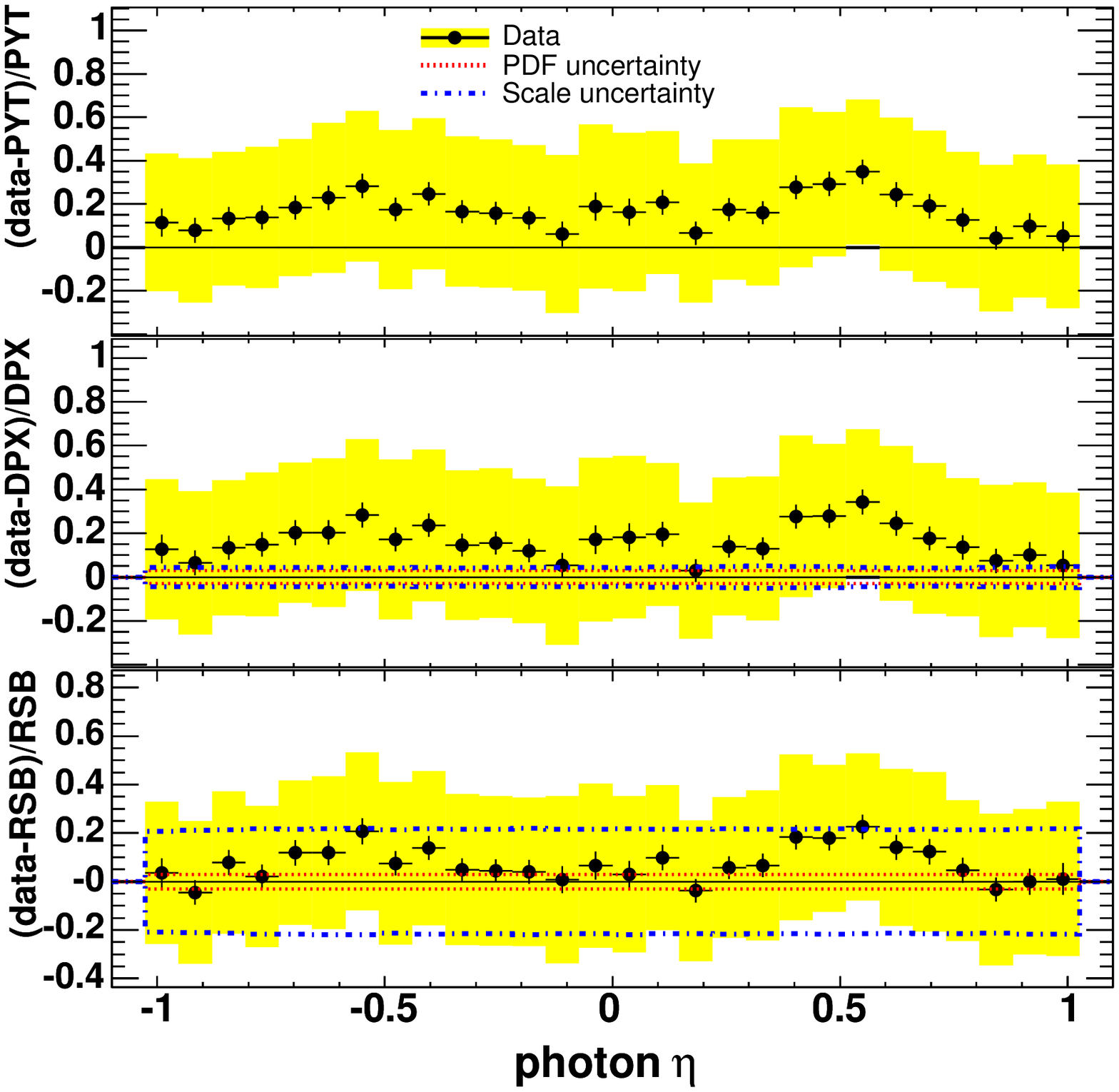}\hfill
\vspace*{0cm}
\includegraphics[width=9.0cm]{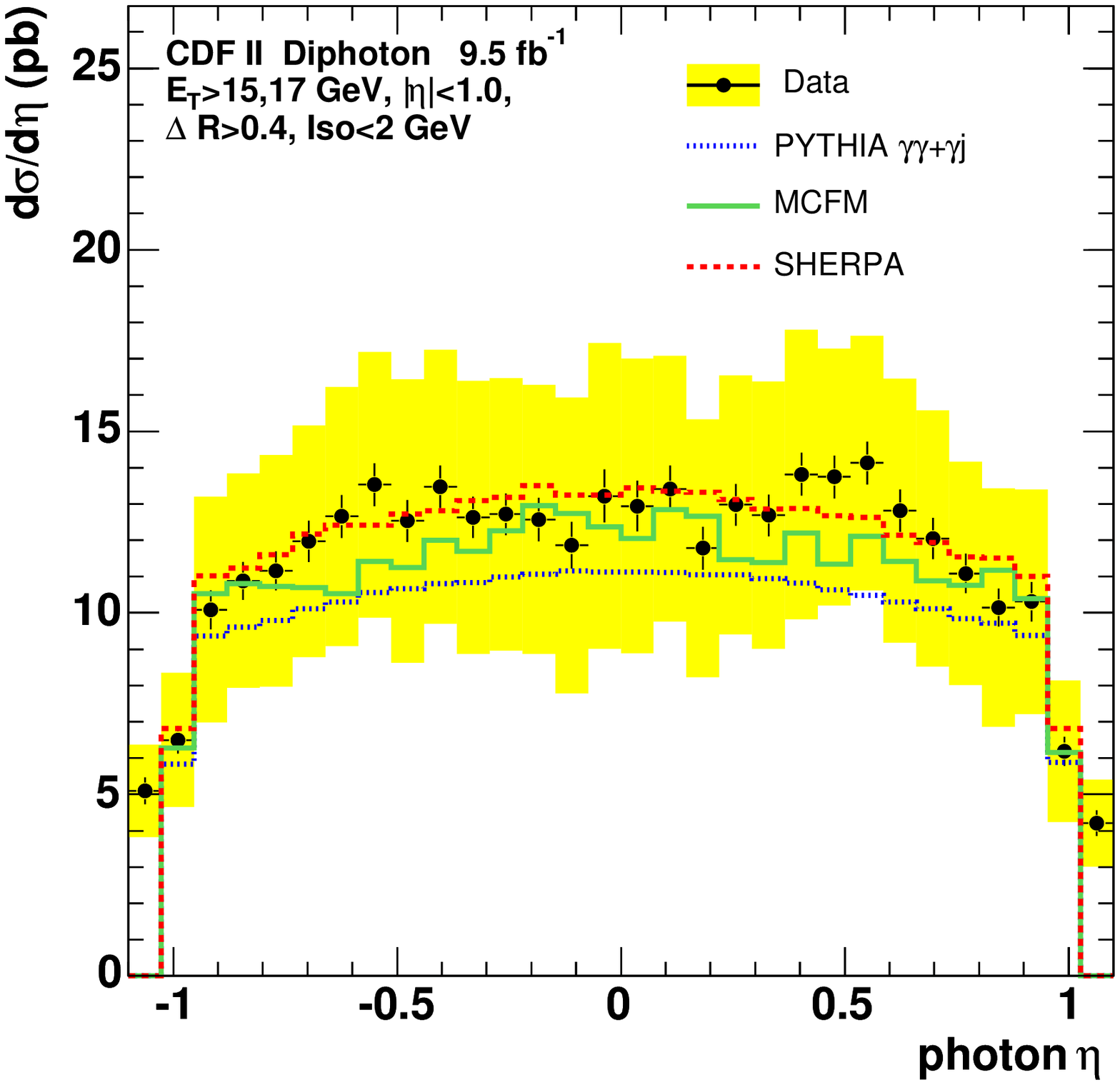}\hspace*{0cm}
\includegraphics[width=8.5cm]{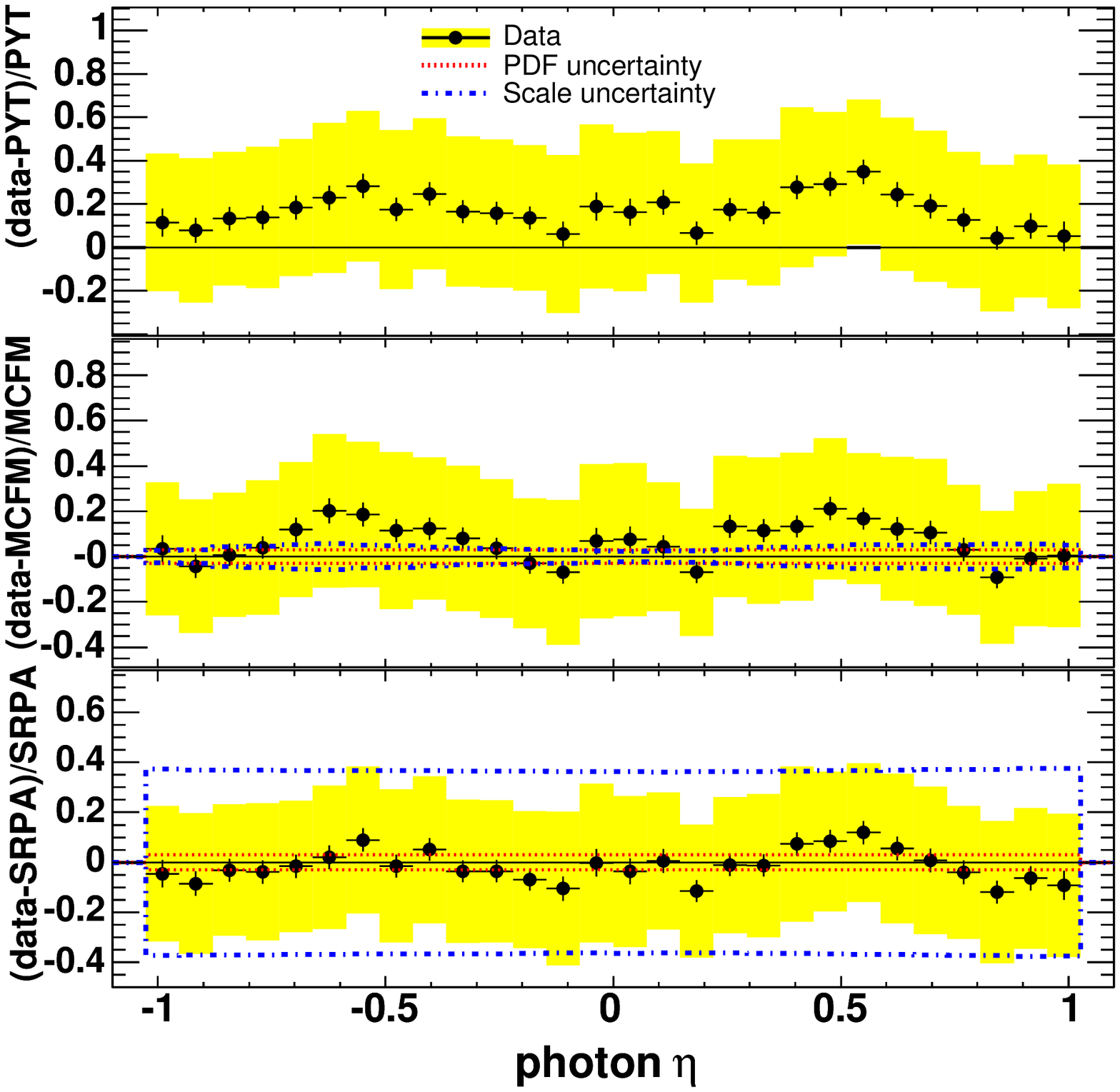}
\vspace*{-0.1cm}
\caption{The measured differential cross sections 
for the photon $\eta$ (two entries per event) 
compared with six 
theoretical predictions discussed in the text. The left windows show the
absolute comparisons and the right windows show the fractional deviations
of the data from the theoretical predictions. Note that the vertical axis
scales differ between fractional deviation plots.}
\end{figure*}

\pagebreak
\vfill

\begin{table}[!hp]
\begin{center}
\begin{tabular}{ccc}
\hline\hline
 Bin & Cross Section (pb) & Sherpa   \\ \hline
 -1.10 -  -1.03 & $ 5.1 \pm  0.37 \pm  1.3 $ &  0.0 \\
 -1.03 -  -0.953 & $ 6.5 \pm  0.37 \pm  1.8 $ &  6.8 \\
 -0.953 -  -0.880 & $10.1 \pm  0.54 \pm  3.1 $ & 11.0 \\
 -0.880 -  -0.807 & $10.9 \pm  0.52 \pm  3.0 $ & 11.2 \\
 -0.807 -  -0.733 & $11.2 \pm  0.54 \pm  3.2 $ & 11.6 \\
 -0.733 -  -0.660 & $12.0 \pm  0.57 \pm  3.2 $ & 12.2 \\
 -0.660 -  -0.587 & $12.7 \pm  0.59 \pm  3.6 $ & 12.4 \\
 -0.587 -  -0.513 & $13.5 \pm  0.59 \pm  3.7 $ & 12.4 \\
 -0.513 -  -0.440 & $12.5 \pm  0.58 \pm  3.9 $ & 12.7 \\
 -0.440 -  -0.367 & $13.5 \pm  0.58 \pm  3.8 $ & 12.8 \\
 -0.367 -  -0.293 & $12.6 \pm  0.57 \pm  3.8 $ & 13.1 \\
 -0.293 -  -0.220 & $12.7 \pm  0.58 \pm  3.8 $ & 13.2 \\
 -0.220 -  -0.147 & $12.6 \pm  0.60 \pm  3.7 $ & 13.5 \\
 -0.147 -  -0.0733 & $11.9 \pm  0.64 \pm  4.1 $ & 13.2 \\
 -0.0733 -  -0.00 & $13.2 \pm  0.73 \pm  4.2 $ & 13.2 \\
 -0.00 -   0.0733 & $12.9 \pm  0.70 \pm  4.1 $ & 13.4 \\
  0.0733 -   0.147 & $13.4 \pm  0.64 \pm  3.7 $ & 13.3 \\
  0.147 -   0.220 & $11.8 \pm  0.59 \pm  3.6 $ & 13.3 \\
  0.220 -   0.293 & $13.0 \pm  0.58 \pm  3.6 $ & 13.1 \\
  0.293 -   0.367 & $12.7 \pm  0.57 \pm  3.7 $ & 12.9 \\
  0.367 -   0.440 & $13.8 \pm  0.59 \pm  4.0 $ & 12.9 \\
  0.440 -   0.513 & $13.7 \pm  0.59 \pm  3.5 $ & 12.7 \\
  0.513 -   0.587 & $14.1 \pm  0.59 \pm  3.5 $ & 12.6 \\
  0.587 -   0.660 & $12.8 \pm  0.59 \pm  3.6 $ & 12.1 \\
  0.660 -   0.733 & $12.0 \pm  0.57 \pm  3.5 $ & 11.9 \\
  0.733 -   0.807 & $11.1 \pm  0.55 \pm  3.1 $ & 11.5 \\
  0.807 -   0.880 & $10.1 \pm  0.53 \pm  3.3 $ & 11.5 \\
  0.880 -   0.953 & $10.3 \pm  0.55 \pm  3.1 $ & 11.0 \\
  0.953 -   1.03 & $ 6.2 \pm  0.40 \pm  2.0 $ &  6.8 \\
  1.03 -   1.10 & $ 4.2 \pm  0.35 \pm  1.2 $ &  0.0 \\
\hline\hline
\end{tabular}
\end{center}
\end{table}